\documentclass[11pt,a4paper]{article}
\usepackage{jheppub,esint,shuffle,psfrag}
 \usepackage[utf8]{inputenc}

\usepackage{textgreek}
\usepackage{varioref}
\usepackage{amsmath,amsfonts,amsthm,mathrsfs}
\usepackage{enumerate}
\usepackage{fancyvrb}
\usepackage{verbatim}
\usepackage{wrapfig}
\usepackage{appendix}
\usepackage{amstext}
\usepackage{amssymb}
\usepackage{graphicx}
\usepackage{color}
\usepackage{varioref}
\usepackage{multirow,graphics}
\usepackage{epstopdf}
\usepackage{bbm}
\usepackage{slashed}
\usepackage{psfrag}
\usepackage{relsize}
\usepackage{bbm}
\usepackage{mathrsfs}
\usepackage{epsfig}

\usepackage{mathtools}

       \usepackage{dutchcal}
       \usepackage{dsfont}

\DeclareMathAlphabet{\mathpzc}{OT1}{pzc}{m}{it}

\usepackage{tikz}
\usepackage{subcaption}

\usepackage{mathtools}
\usepackage{bigints}

\usetikzlibrary{decorations}

\usepackage{amstext}
\usepackage{amssymb}
\usepackage{amsmath}
\usepackage{verbatim}
\usepackage[bbsymbols]{bboldx}

\usepackage{tikz}
\usetikzlibrary{math}
\usetikzlibrary{arrows,shapes,positioning}
\usetikzlibrary{decorations.markings}
\tikzstyle arrowstyle=[scale=1]
\tikzstyle directed=[postaction={decorate,decoration={markings,
    mark=at position .65 with {\arrow[arrowstyle]{stealth}}}}]
\tikzstyle reverse directed=[postaction={decorate,decoration={markings,
    mark=at position .65 with {\arrowreversed[arrowstyle]{stealth};}}}]

\newcommand{\bea}{\begin{eqnarray}}
\newcommand{\eea}{\end{eqnarray}}

\def\pa {\partial}

\def\la{\label}

\def\qq{{\mathcal q}}

\def\mI{\mathbbm{1}}
\def\qq{\mathfrak{q}}
\def\sgn{{\text{sgn}}}
\def\imp{\bbDelta}

\def\r {\rho}
\def\a {\alpha}
\def\b {\beta}
\def\pv {\text{p.v.}}

\numberwithin{equation}{section}

\def\<{\langle}
\def\>{\rangle}

\makeatletter
\@ifundefined{usebibtex}{} {}
\makeatother

\def\Tr{\text{Tr}~}

\newcommand\var{\varkappa}

\usepackage{esint}
\usepackage{breqn}
\def \Tr {\mathop{\rm Tr}\nolimits}

\def\q{\alg{q}}
\def\i{\textsl{\textsf{i}}}
\def\refl{\mathcal{r}}
\def\X{\mathsf{X}}
\def\T{\textsf{T}}
\def\q{\mathfrak{q}}

\title{Elliptic spin Ruijsenaars--Schneider model via the Poisson reduction of the affine Heisenberg double}

\author[a]{Gleb Arutyunov,}
\author[a]{~Lukas Hardi}
\author[a]{and\,~Davide Polvara}

\affiliation[a]{
II. Institut f\"ur Theoretische Physik, Universit\"at Hamburg,
Luruper Chaussee 149, 22761 Hamburg, Germany}

\emailAdd{gleb.arutyunov@desy.de}
\emailAdd{lukas.hardi@desy.de}
\emailAdd{davide.polvara@gmail.com}

\abstract{
We construct the Hamiltonian structure of the elliptic spin Ruijsenaars--Schneider model with dynamical inhomogeneities by Poisson reduction of the affine Heisenberg double extended by an affine oscillator manifold. This construction extends the Heisenberg double approach to spin Ruijsenaars--Schneider models from the hyperbolic to the elliptic case. We solve the moment map constraint and derive the Poisson algebras of the invariant Lax operator and spin variables. A realization with finitely many degrees of freedom is obtained by evaluating and localizing the invariant oscillators at distinct points of the affine parameter. Consistency requires these points to be dynamical so that their nontrivial Poisson brackets with the spin variables cancel derivative terms that would otherwise obstruct closure of the Poisson bracket on finitely many degrees of freedom. The resulting Poisson brackets and Hamiltonian coincide with those recently obtained from the elliptic Coulomb branch of the necklace quiver. Our construction thus provides an independent derivation of this Hamiltonian structure and establishes a direct link between Poisson reductions and Coulomb branches.

}

\begin{document}
\begin{flushright}\small{ZMP-HH/26-39}
\end{flushright}

\maketitle

\section{Introduction}
Constructing a Poisson algebra for the elliptic spin
Ruijsenaars--Schneider (RS) integrable model has been a
long-standing problem. The model describes the dynamics of
$N$ particles, each carrying $\ell$ spin components, interacting
through an elliptic potential with coupling constant $\gamma$.
Its rational and trigonometric/hyperbolic limits arise by
degenerating the underlying elliptic curve. The natural
variables are the particle positions $\q_i$ and the spin
variables $\mathbf{a}_{i\alpha}$ and $\mathbf{b}_{\alpha i}$, where
$i=1,\ldots,N$ labels the particles and
$\alpha=1,\ldots,\ell$ labels the spin components.
For $\ell=1$, the model reduces to the usual spinless RS
model, also known as the relativistic Calogero--Moser model.
The elliptic spin RS model was introduced in \cite{Krichever:1995zw} through
equations of motion governing the evolution of poles of
elliptic solutions of the non-abelian two-dimensional Toda
chain, and the equations were shown to admit a Lax representation.
Subsequent work sought to construct a symplectic structure
on the phase space parametrized by these variables and
a Hamiltonian whose flow reproduces the equations of motion. Recent developments in elliptic many-body systems also include spin generalizations of the Calogero--Inozemtsev model and models of interacting tops associated with the BC$_n$ root system. Lax representations for these systems were constructed in~\cite{Domanevsky:2026coz} using associative Yang--Baxter identities involving quantum $R$ matrices and boundary reflection matrices.

The construction of phase spaces of spinless RS models
by Poisson (or Hamiltonian) reduction has a long history.
Specifically, the rational RS model can be obtained as a reduction of the cotangent bundle $T^* GL_N$~\cite{Fock:1999ae,Arutyunov:1996cmb}. The trigonometric RS model arises by reducing either the cotangent bundle of the loop group $T^* GL_N(t)$ \cite{Gorsky:1993dq} or the Heisenberg double of $GL_N$ \cite{Feher:2008moa}. 
Similarly, 
the elliptic RS model can be obtained by reducing either the cotangent bundle of the two-dimensional loop group $T^* GL_N(z,\bar z)$ \cite{Arutyunov:1996vy} or the affine Heisenberg double of $GL_N$ \cite{Arutyunov:1996uw}. In each case, the reduction is performed with respect to the natural adjoint action of $GL_N$, equipped, where appropriate, with the Poisson--Lie structure given by the Sklyanin bracket \cite{Semenov-Tian-Shansky:1985mgd}.

The construction of spinless RS phase spaces by reduction extends naturally to spin RS models. The rational spin RS model 
was obtained in \cite{Arutyunov:1997ey} by
reducing the 
cartesian product  
$T^* GL_N \times (T^* \mathbb C^N)^{\times \ell}$, where $\mathbb{C}^N$ is the 
defining representation of $GL_N$.
This construction reduces the integration of the equations of motion to a matrix factorization problem. After reduction,
$T^* GL_N$
provides the particle coordinates and momenta,
while $(T^* \mathbb C^N)^{\times \ell}$
provides the spin variables.

The trigonometric spin RS phase space has been constructed by several methods. In \cite{Chalykh:2018wce}, 
it was obtained by quasi-Hamiltonian reduction as a multiplicative quiver variety \cite{crawley-boevey_shaw_multiplicative_2006,yamakawa_geometry_2008,jordan_quantized_2014} associated with the framed Jordan quiver of rank $N$ and $\ell$ flavors. These varieties are multiplicative analogues of Nakajima quiver varieties, which are special cases of Higgs branches. Another construction  \cite{Arutyunov:2019alw} 
combines the Heisenberg double used
in the spinless case with an oscillator manifold $\Sigma_{N,\ell}$ deforming $(T^* \mathbb C^N)^{\times \ell}$. Poisson reduction by the Poisson--Lie action of $GL_N$, equipped with the Sklyanin bracket, then yields the trigonometric spin RS phase space. The trigonometric real form was studied in \cite{Fairon:2020aoy}.

The remaining step is to construct the elliptic spin RS phase space by reduction. A Poisson structure in the natural variables of the elliptic spin RS model was recently obtained in \cite{Arutyunov:2026yjy} by  identifying its phase space with the elliptic Coulomb branch of the necklace quiver with $\ell$ nodes of rank $N$. From the perspective of quiver varieties,
this construction is mirror dual to the approach pursued here. 
The aim of the present paper is to derive the Poisson structure on the corresponding  reduced phase space that plays the role of an elliptic quiver variety associated with the framed Jordan quiver. We then compare the resulting structure  with the one obtained from the elliptic Coulomb branch. 

To this end, we follow the spinless construction and form the
cartesian product of the affine Heisenberg double of $GL_N$ with a
suitable affine oscillator manifold. The latter is constructed from
a spin-space $r$-matrix $\rho$ of rank $\ell$ satisfying the modified
classical Yang--Baxter equation. The resulting unreduced phase space
carries a natural Poisson--Lie action of the loop group $LGL_N$,
equipped with the Sklyanin bracket. Reduction by this action yields
particle positions and momenta, together with spectral-parameter-dependent spin vectors $\mathbf{a}_{i\alpha}(y)$ and $\mathbf{b}_{\alpha i}(y)$.

\paragraph{Main result.} Our main result is the explicit Poisson structure on the reduced phase
space, expressed in terms of the particle positions, momenta, and
spectral-parameter-dependent spin variables. To describe this structure
in the variables of the elliptic spin RS model, we choose a particular
spin space $r$-matrix $\rho$ and impose the relations
\begin{equation}
\label{eq:osc_on_impur}
    \mathbf{a}_{i\alpha} = \mathbf{a}_{i\alpha}(\imp^\alpha), \qquad \mathbf{b}_{\alpha i} \delta(y-\imp^\alpha) = \mathbf{b}_{\alpha i}(y),
\end{equation}
where $\imp^1,\dots,\imp^\ell$ are dynamical inhomogeneities, which serve as the generators of the maximal torus of the spin rotation symmetry, i.e.
\begin{equation}
    \frac{1}{\var} \,  \{ \imp^\alpha, \mathbf{a}_{i\beta} \} = -i(\delta_{\alpha\beta}-\delta_{\alpha1}) \mathbf{a}_{i\beta}, \qquad \frac{1}{\var} \,  \{ \imp^\alpha, \mathbf{b}_{\beta i} \} = i(\delta_{\alpha\beta}-\delta_{\alpha1}) \mathbf{b}_{\beta i}.
\end{equation}
These generators Poisson commute, $\{\imp^\alpha,\imp^\beta\}=0$, and 
also with coordinates: $\{\imp^\alpha,\q_i\}$=0.
Reduction then gives the following Poisson brackets among the spin
variables\footnote{The brackets are obtained in the gauge $\mathbf{a}_{i 1}=1$; the spin indices $\alpha$ and $\beta$ associated to the $\mathbf{a}$ oscillators are then assumed to be different from one in the brackets.}:
\begin{equation}
\label{Poisson_structure}
\begin{split}
\frac{1}{\var} \, \{\mathbf{a}_{a \alpha} \,,\mathbf{a}_{b \beta} \}=&+ i (1-\delta_{\alpha \beta})\varphi(\imp^{\beta \alpha}+\Delta,\mathfrak{q}_{ba})\mathbf{a}_{b \alpha} \mathbf{a}_{a \beta}+i (1+\delta_{\alpha \beta}) \varphi^{\text{reg}}(\mathfrak{q}_{ba})\mathbf{a}_{a \alpha} \mathbf{a}_{b \beta}  \\
&- i \varphi(\imp^{\beta1}+\Delta,\mathfrak{q}_{ba})\mathbf{a}_{a \alpha} \mathbf{a}_{a \beta}- i \varphi(\imp^{1\alpha}+\Delta,\mathfrak{q}_{ba})\mathbf{a}_{b \alpha} \mathbf{a}_{b \beta} \,,\\
\frac{1}{\var} \{\mathbf{b}_{\alpha a},\mathbf{b}_{\beta b}\}  =&+i (1+\delta_{\alpha \beta})  
  \varphi^{\text{reg}}( \mathfrak{q}_{ba})   \mathbf{b}_{\alpha a} \mathbf{b}_{\beta b}  +(1-\delta_{\alpha \beta}) i \mathbf{b}_{\alpha b} \mathbf{b}_{\beta a}   \varphi(\imp^{\alpha \beta}+\Delta, \mathfrak{q}_{ba})\\
&+\delta_{\alpha 1} \mathbb{L}^{\text{reg},\alpha}_{ba} \mathbf{b}_{\beta b} -\delta_{\beta 1} \mathbb{L}^{\text{reg},\beta}_{ab} \mathbf{b}_{\alpha a}   \,,\\
\frac{1}{\var} \, \{\mathbf{a}_{a \alpha}, \mathbf{b}_{\beta b}\}=&+i \varphi(\imp^{1\alpha}+\Delta,\mathfrak{q}_{ba})   \mathbf{a}_{b \alpha} \mathbf{b}_{\beta b}  -i \varphi^{\text{reg}}(\qq_{ba}) \mathbf{a}_{a \alpha} \mathbf{b}_{\beta b} \\
&-\delta_{\alpha \beta} \mathbb{L}^{\text{reg},\alpha}_{ab} +\delta_{\beta 1} \mathbb{L}^{\text{reg},1}_{ab} \mathbf{a}_{a \alpha} \,,
\end{split}
\end{equation}
where we defined
\begin{equation}
\mathbb{L}^{\text{reg},\alpha}_{ba}:= \frac{1}{i}  \sum_{\gamma \ne \alpha}  \varphi(\imp^{\alpha\gamma}+\Delta, \qq_{ba} - \eta) \mathbf{a}_{b \gamma} \mathbf{b}_{\gamma a}+\frac{1}{i}  \varphi^{\text{reg}}(\qq_{ba} - \eta) \mathbf{a}_{b \alpha} \mathbf{b}_{\alpha a} \,,
\end{equation}
and we have used the Kronecker elliptic function $\varphi$ and its regular part $\varphi^{\text{reg}}$ defined in~\eqref{eq:app_varphi} and \eqref{eq:wreg}, respectively. 
In particular, $\varphi^{\mathrm{reg}}(s)
=\zeta(s)-\zeta(\pi)s/\pi$ for nonzero arguments away from the period
lattice, with the finite-part convention
$\varphi^{\mathrm{reg}}(0)=0$. The shifted argument
$\mathfrak{q}_{ba}-\eta$ is not set to zero when $a=b$.

Remarkably, these brackets agree with those obtained from the elliptic Coulomb
branch in \cite{Arutyunov:2026yjy}. Thus the
reduction provides an independent derivation of the Poisson structure
in the natural variables of the elliptic spin RS model. In particular,
the reduced bracket reproduces its equations of motion with dynamical
inhomogeneities:
\begin{equation}\label{eq:eoms}
    \begin{aligned}
        \dot{\mathfrak{q}}_i
        ={}& -\sum_{\rho=1}^\ell f_{ii}^\rho, \\[-4pt]
        \dot{\mathbf{a}}_{i\alpha}
        ={}& \sum_{j=1}^N \sum_{\rho=1}^\ell (V(\imp^{1\rho},\mathfrak{q}_{ij}) f_{ij}^\rho \mathbf{a}_{i\alpha} - V(\imp^{\alpha\rho},\mathfrak{q}_{ij}) f_{ij}^\rho \mathbf{a}_{j\alpha}), \\[-2pt]
        \dot{\mathbf{b}}_{\alpha i}
        ={}& \sum_{j=1}^N \sum_{\rho=1}^\ell (V(\imp^{\alpha\rho},\mathfrak{q}_{ji}) f_{ji}^\rho \mathbf{b}_{\alpha j} - V(\imp^{1\rho},\mathfrak{q}_{ij}) f_{ij}^\rho \mathbf{b}_{\alpha i}).
    \end{aligned}
\end{equation}
Here $f_{ij}^\alpha \coloneq \mathbf{a}_{i\alpha} \mathbf{b}_{\alpha j}$, $\imp^{\alpha\beta} \coloneq \imp^\alpha - \imp^\beta$, $\mathfrak{q}_{ij} \coloneq \mathfrak{q}_i-\mathfrak{q}_j$, and $\mathbf{a}_{i1} = 1$, while the potential is given by 
\begin{equation}
    V(z,w) \coloneq
    \begin{cases}
        \varphi(z+\Delta,w) - \varphi(z+\Delta,w-\eta), & z \neq 0, \\
        \varphi^{\text{reg}}(w) - \varphi^{\text{reg}}(w-\eta), & z = 0.
    \end{cases}
\end{equation}
with $\eta$ the coupling constant. Importantly, (\ref{eq:eoms}) reduce to the equations of motion of Krichever and Zabrodin \cite{Krichever:1995zw} in the homogeneous limit $\imp^\alpha \to 0$, up to conventions\footnote{The explicit map involves the changes $t\to -t$ and $\eta\to -\eta$,
together with the constant shift $-\zeta(\pi)\eta/\pi$ in the
potential $V$. One must also account for the scalar gauge terms
$-\lambda_i a_i$ and $+\lambda_i b_i$ in the Krichever--Zabrodin
equations. These terms implement time-dependent
rescalings of the spin variables; in our parametrization, the
corresponding freedom is fixed by $\mathbf a_{i1}=1$.}.
Equations (\ref{eq:eoms}) are Hamiltonian with respect to the Poisson brackets (\ref{Poisson_structure}).
The Hamiltonian and our convention for its flow are
\begin{equation}
H_0=\sum_{i=1}^{N}\sum_{\rho=1}^{\ell}f_{ii}^{\rho},
\qquad
\dot F=\frac{1}{i\varkappa}\{F,H_0\}
\label{eq:spin-RS-Hamiltonian}
\end{equation}
for any observable \(F\). In particular, this convention gives
\(\dot \q_i=-\sum_{\rho=1}^{\ell}f_{ii}^{\rho}\), as in the first
equation of (\ref{eq:eoms}); applying it to the spin variables yields the
remaining equations in (\ref{eq:eoms}).

\paragraph{Gauge-theoretic interpretation.}
The agreement between the reduced Poisson structure constructed
here and that of the elliptic Coulomb branch in \cite{Arutyunov:2026yjy} also admits
a gauge-theoretic interpretation. It relates two moduli spaces associated with two theories: five-dimensional
$\mathscr N=1$ necklace quiver gauge theory on
$\mathbb R^3\times T^2$, with $\ell$ nodes of rank $N$, and
four-dimensional $GL_N$-Chern--Simons theory on
$\mathbb R \times S^1\times E$, with $\ell$ line defects extended along $\mathbb R$.
Here $E$ is the elliptic curve with periods $2\pi$ and $2\Delta$.

To explain this relation, we reinterpret the affine Heisenberg
double with generating matrices $A(x)$ and $B(x)$ as well as the oscillator manifold as building blocks for
the moduli space of certain multiplicative $GL_N$-Higgs bundles, following \cite{Elliott:2018yqm}.
Let $\mathscr L_w$ denote the line bundle on $E$ with
quasi-periodicity multiplier $e^{iw}$.
A degree-zero $\eta$-twisted multiplicative $GL_N$-Higgs bundle on $E$ with minimal punctures $\{ \imp^\alpha \}_{\alpha=1}^\ell$ consists of a rank-$N$
holomorphic vector bundle $\mathscr V\to E$, described by
a clutching function $A(x)$ defined on the annulus $[-\pi,\pi] \times (-\Delta,\Delta)$, together with a meromorphic invertible map
\[
B(x):\mathscr V\longrightarrow\mathscr V\otimes \mathscr L_\eta.
\]
The map $B(x)$ is the $\eta$-twisted multiplicative Higgs field
and is allowed to have simple poles at $\imp^\alpha$ with rank-one
residue. Note that $\eta$-twisted multiplicative $GL_N$-Higgs bundles are special cases of (untwisted) $PGL_N$-Higgs bundles. On the generic locus considered here, diagonalizing
the clutching data gives
\[
\mathscr V\cong\bigoplus_{i=1}^{N} \mathscr L_{-\mathfrak q_i}.
\]
In this decomposition, the multiplicative Higgs field is represented by
the Lax matrix $\mathbb L(x)$. The moment map equation in our Poisson reduction shows that the entries $\mathbb L_{ij}(x)$ are
meromorphic sections of $\mathscr L_{\eta - \mathfrak q_{ij}}$ and that the oscillator variables parametrize
the rank-one residues. Taking the quotient by changes
of trivialization yields the Poisson reduction studied
in this paper.

The moduli space of multiplicative Higgs bundles was identified in \cite{Elliott:2018yqm} with the
phase space of four-dimensional Chern--Simons theory
with line defects. Our comparison with the elliptic
Coulomb branch therefore realizes, at the level of
Poisson structures, a particular instance of a duality
between four-dimensional Chern--Simons theory with line defects and five-dimensional necklace quiver gauge theory. A string-theoretic explanation of this duality, based on a brane construction,
was given in \cite{Maruyoshi:2020cwy}.

\paragraph{Organization of the paper.} The paper is organized as follows. Section~\ref{sec:AffineHD_osc} introduces the affine Heisenberg double and its associated phase space of affine oscillators. Starting from the Cartesian product of the two manifolds, we fix the moment map corresponding to the action of the Poisson--Lie group on the double. We show that the Lax operator of the elliptic spin RS models corresponds to a particular moment-map solution in which the invariant oscillators are specialized to certain inhomogeneities, as in~\eqref{eq:osc_on_impur}. 
In section~\ref{sec:Poisson_inv} we derive the Poisson brackets for the quantities that are invariant under the Poisson-Lie group action. In section~\ref{sec:finite_dim} we provide a realization of the brackets of these
invariant objects in which we remove the dependence on the spectral parameter from the spin oscillators. We show that such a realization requires the inhomogeneities $\{\imp^\alpha \}^\ell_{\alpha=1}$ to be dynamical and matches the results of~\cite{Arutyunov:2026yjy} obtained from the Coulomb branch of the necklace quiver gauge theory. We conclude in Section~\ref{sec:conclusions}, highlighting the main results and open problems. 
Technical details of the calculations are provided in the appendices.

\section{Affine Heisenberg double and oscillator manifolds}
\label{sec:AffineHD_osc}
We first introduce the Poisson manifolds that serve as initial phase spaces for the reduction procedure 
leading to the integrable models of Ruijsenaars--Schneider (RS) type. These are the affine Heisenberg double and the affine oscillator manifold, the latter accounting for 
additional spin degrees of freedom.

\subsection{Affine Heisenberg double}

We describe the affine Heisenberg double $\widehat{D}$ as a Poisson manifold generated by Fourier modes
of two $GL_N$-valued fields $A(x)$ and $B(x)$ on the circle, where $x\in [-\pi, \pi]$. 
Its Poisson structure is specified on the generators by the following brackets
\bea
\la{AFD}
\begin{aligned}
\frac{1}{\varkappa}\{ A_1(x), A_2(y) \}
= &-\, r_{\mp}(x-y)\,A_1(x)A_2(y)
- A_1(x)A_2(y)\, r_{\pm}(x-y)\\
&+ A_1(x)r_-(x-y+2\Delta)A_2(y)
+ A_2(y)r_+(x-y-2\Delta)A_1(x)\, ,  \\
\frac{1}{\varkappa}\{ B_1(x), B_2(y) \}
=& -\, r_{\mp}(x-y)\,B_1(x)B_2(y)
- B_1(x)B_2(y)\, r_{\pm}(x-y) \\
&+ B_1(x)r_-(x-y)B_2(y)
+ B_2(y)r_+(x-y)B_1(x)\, ,  \\
\frac{1}{\varkappa}\{ A_1(x), B_2(y) \}
=& -\, r_-(x-y)\,A_1(x)B_2(y)
- A_1(x)B_2(y)r_-(x-y+2\Delta) \\
&+ A_1(x)r_-(x-y+2\Delta)B_2(y)
+ B_2(y)\,r_+(x-y)A_1(x)\, ,   \\
\frac{1}{\varkappa}\{ B_1(x), A_2(y) \}
= &-\, r_+(x-y)\,B_1(x)A_2(y)
-B_1(x)A_2(y)r_+(x-y-2\Delta) \\
&+ B_1(x)r_-(x-y)A_2(y)+ A_2(y)r_+(x-y-2\Delta)B_1(x)
 \, .
\end{aligned}
\eea
In (\ref{AFD}) the variable $\var$ is a deformation parameter that connects $\widehat{D}$ to the cotangent bundle of the central extension of the loop group $LG\!: S^{1}\to G$, where 
$G=GL_N$. The classical $r$-matrices $r(x)$ and $r_{\pm}(x)$ are  defined in Appendix~\ref{app:matrix_bblocks}.  Their construction involves either a Cauchy principal value prescription or regularization parameter $\varepsilon$ 
ensuring that the Poisson brackets are well-defined as distributions on the full interval $x, y \in [-\pi, +\pi]$. 
The parameter $\Delta$, with $\Im (\Delta)>0$, plays the role of the central charge.  Setting the left-hand side of (\ref{AFD}) to zero Fourier modes 
and setting to zero all non-zero Fourier modes on the right-hand side yields the Poisson structure of the finite-dimensional Heisenberg double. From a physical point of view,
this procedure corresponds to the Kaluza--Klein reduction. We assume that $A(x)$ and $B(x)$ extend analytically to the complex plane 
and are analytic in the strip $|\Im (x)|< \Im (\Delta)$.
For non-real arguments, the kernels $r_\pm(z)$ appearing in these
brackets are understood as the meromorphic continuation
\begin{equation}
\widetilde r(z)=r-\frac{i}{2}C_{12}\cot\frac{z}{2},
\end{equation}
rather than as their Fourier series evaluated outside their domains
of convergence. The subscripts $\pm$ specify the boundary values on
the real axis,
\[
r_\pm(x)=\lim_{\varepsilon\to0^+}
\widetilde r(x\mp i\varepsilon),
\]
in the distributional sense. The same convention applies to shifted
r-matrices in the subsequent formulas.

The Poisson manifold (\ref{AFD}) is endowed with the action of the Poisson-Lie loop group $LG$ such that the map $LG\times \widehat{D}\to \widehat{D}$ is Poisson. The Poisson-Lie structure on $LG$
is given by the Sklyanin bracket
\bea
\la{PLG}
\{h_1(x),h_2(y)\}= - \var [r_{\pm}(x-y),h_1(x)h_2(y)]\, , ~~~h(x)\in LG\, .
\eea
The occurrence  of $r_{\pm}(x-y)$ in (\ref{AFD}) and (\ref{PLG}) means that any consistent choice of upper or lower signs is possible and defines the same Poisson structures.
On the generators of the double, the action of $LG$ is given by 
\begin{equation}
\label{eq:group_action}
\begin{split}
&A(x) \to h(x - \Delta) A(x) h^{-1}(x + \Delta)\,,\\
&B(x) \to h(x - \Delta) B(x) h^{-1}(x - \Delta)\,.
\end{split}
\end{equation}
This action is generated by the non-abelian moment map $M\in LG\simeq LG^*$:
\bea
\label{eq:mom_map_HD}
M(x)=B(x+\Delta)A^{-1}(x-\Delta)B^{-1}(x-\Delta)A(x-\Delta)\, .
\eea
Using the Poisson brackets~\eqref {AFD}, one finds that
\begin{equation}
\label{eq:Pois_Mx_My_to_be_checked}
\begin{split}
\frac{1}{\var} \{ M_1(x), A_2(y)\} &= - r_+(x-y+ \Delta) M_1(x) A_2(y) + A_2(y)  r_+(x-y-\Delta) M_1(x) \\
&+M_1(x) r_-(x-y +\Delta)  A_2(y) - M_1(x) A_2(y)r_-(x-y -\Delta) \,,\\
\frac{1}{\var} \{ M_1(x), B_2(y)\} &= -[ r_+(x-y+ \Delta)M_1(x), B_2(y) ] + [ M_1(x) r_-(x-y +\Delta) , B_2(y) ]\,.
\end{split}
\end{equation}
It follows from these relations that the moment map satisfies the Poisson relations of the Semenov-Tian-Shansky bracket 
\bea
\la{br_ST}
\begin{aligned}
\frac{1}{\var}\{ M_1(x), M_2(y) \}=&- r_+(x-y) M_1(x) M_2 (y)  - M_1(x)  M_2 (y) r_-(x-y) \\
&+ M_1(x) r_-(x-y) M_2 (y)+ M_2 (y) r_+(x-y) M_1(x)  
\end{aligned}
\eea
The functions $\textsl{m}_k(x)=\Tr M^k(x)$ are central for this bracket.

\subsection{Affine oscillator manifold}
\label{sec:affine_osc_manifold}

The affine oscillator manifold $\widehat{\Sigma}$ is generated by Fourier modes of two matrix-valued fields $a(x)$ and $b(x)$,
where $a(x)$ and $b(x)$ are rectangular $N\times \ell$ and $\ell\times N$ matrices with components $a_{i\a}$ and $b_{\a i}$, where $i=1,\ldots, N$ and $\a=1,\ldots, \ell$.
We endow $\widehat{\Sigma}$ with the structure of the Poisson manifold by defining on the generators the following Poisson bracket
\bea
\label{eq:starting_osc_br}
\begin{aligned}
\frac{1}{\var}\{a_1(x),a_2(y)\}&=+r(x-y)a_1(x)a_2(y)-a_1(x)a_2(y)\rho(x-y)\, \\
\frac{1}{\var}\{b_1(x),b_2(y)\}&=+b_1(x)b_2(y)r(x-y)-\rho(x-y)b_1(x)b_2(y)\, \\
\frac{1}{\var}\{a_1(x),b_2(y)\}&=-b_2(y)r_+(x-y)a_1(x)+a_1(x)\rho_-(x-y)b_2(y)-\frac{2\pi}{\var} C_{12}^{\rm rec}\delta(x-y)\, ,\\
\frac{1}{\var}\{b_1(x),a_2(y)\}&=-b_1(x)r_-(x-y)a_2(y)+a_2(y)\rho_+(x-y)b_1(x)+\frac{2\pi}{\var} C_{12}^{\rm rec}\delta(x-y)\, .
\end{aligned}
\eea
Here the matrix $\rho(x)$ acts on $\mathbb{C}^\ell \otimes \mathbb{C}^\ell$.
We consider two admissible choices for $\rho(x)$: $\hat{\rho}(x)$ and $\check{\rho}(x)$, given respectively by
\begin{equation}
\begin{split}
\hat{\rho}(x) &= \frac{1}{2} \sum^\ell_{\a, \b=1} \sgn_{\beta \alpha} E_{\a \b} \otimes E_{\b \a}  -\frac{i}{2} \, \pv \, \cot \frac{x}{2} \sum^\ell_{\alpha, \beta=1} E_{\alpha \beta} \otimes E_{\beta \alpha}   \,,\\
\check{\rho}(x)&=  \pi \delta(x) \sum^\ell_{\a, \b=1} \sgn_{\beta \alpha} E_{\a \b} \otimes E_{\b \a} - \frac{i}{2} \, \pv \, \cot \frac{x}{2} \sum^\ell_{\alpha=1} E_{\alpha \alpha} \otimes E_{\alpha \alpha}  \,.
\end{split}
\end{equation}
Similar definitions apply to $\rho_{\pm}(x)$, which can be both $\hat{\rho}_{\pm}(x)$ or $\check{\rho}_{\pm}(x)$.
These matrices, together with $r(x)$ and $r_\pm(x)$, are defined in appendix~\ref{app:definitions}.

For both choices of $\hat{\rho}$ and $\check{\rho}$, the brackets above form a Poisson algebra. We provide a detailed check of the Jacobi identity of these brackets in appendix~\ref{app:Jacobi_oscillators}.
We also introduce the constant matrix $C_{12}^{\rm rec}$, which we refer to as the rectangular split Casimir, analogous to the split Casimir $C_{12}$;
see Appendix~\ref{app:matrix_bblocks}. 

The Poisson Lie group $LG$ acts on $\widehat{\Sigma}$,  as 
\bea
a(x)\to h(x)a(x)\, , ~~~~b(x)\to b(x)h^{-1}(x) \,,
\eea
where $h(x)$ must satisfy the same Sklyanin bracket~\eqref{PLG} used to define the group action on the double.
This action $LG\times \widehat{\Sigma}\to \widehat{\Sigma}$ is a mapping of Poisson manifolds and it is generated by the non-abelian moment map
\bea
\label{eq:oscil_mom_map}
\omega(x)=\mI + \varkappa a(x)b(x)\, .
\eea
In particular, the Poisson relations of $\omega(x)$ with affine oscillators are 
\bea
\begin{aligned}
\frac{1}{\var}\{\omega_1(x),a_2(y)\}&=+\big(r_+(x-y)\omega_1(x)-\omega_1(x)r_-(x-y)\big)a_2(y)\, ,\\
\frac{1}{\var}\{\omega_1(x),b_2(y)\}&=- b_2(y)\big(r_+(x-y)\omega_1(x)-\omega_1(x)r_-(x-y)\big)\, .
\end{aligned}
\eea
From here we find the Poisson relations among the entries of $\omega$
\bea
\label{eq:Pois_br_mom_map}
\begin{aligned}
\frac{1}{\varkappa}\{\omega_1(x),\omega_2(y)\}=&+r_+(x-y)\omega_1(x)\omega_2(y)+\omega_1(x)\omega_2(y)r_-(x-y)\\
&-\omega_1(x)r_-(x-y)\omega_2(y)-\omega_2(y)r_+(x-y)\omega_1(x)\, .
\end{aligned}
\eea
We notice that up to an overall sign, $\omega$ satisfies the same Poisson algebra~\eqref{br_ST} of the moment map of the Heisenberg double. This is the same feature already observed in~\cite{Arutyunov:2019alw}.

\subsection{Solutions of the moment map}

Given $u(x)$ in the dual Poisson-Lie group $G^*$, we use the factorization conventions of~\cite{Arutyunov:2019alw} on a domain where $u(x)=u_+ (x) u_-^{-1} (x)$ exists. The product induced by the dual Poisson–Lie group is
\begin{equation}
u(x) \star v(x)=u_+(x) v(x) u_-^{-1}(x).
\end{equation}
The moment map $\omega$ factorizes inside the double to $\omega(x) = \omega_+ (x) \omega^{-1}_-(x)$ and its inverse with respect to the $\star$ product is $\mathcal{N}(x) =\omega^{-1}_+ (x) \omega_-(x)$.  
Using the affine analogue of the product construction for Poisson--Lie moment maps~\cite{FlaschkaRatiu1996}, we combine the two moment maps through multiplication in the dual group. In the factorization convention introduced above, this gives
\begin{equation}
\mathcal{M}(x)= M(x) \star \mathcal{N}(x)
=M_+(x) \mathcal{N}(x) M_-^{-1}(x) \,,
\end{equation}
and fixing the central level 
\begin{equation}
\mathcal{M}(x) = \gamma \mathbf{1}
\end{equation}
is equivalent to requiring
\begin{equation}
M(x)= \gamma \omega(x) \,.
\end{equation}
This yields
\begin{equation}
\label{eq:mom_map1eq}
B(y+\Delta)A^{-1}(y-\Delta)B^{-1}(y-\Delta)A(y-\Delta) = \gamma \omega(y) \,.
\end{equation}

We work locally on a generic domain of phase space on which we can parameterise $A(x)$ and $B(x)$ as follows
\begin{equation}
\label{eq:D_mathcalL_parameterization}
A(y) = T(y-\Delta) Q T^{-1}(y+\Delta) \,, \qquad B(y) = T(y-\Delta) L(y) T^{-1}(y-\Delta) \,,
\end{equation}
where $Q$ is a diagonal matrix independent of the loop coordinate, while its entries remain dynamical variables. 
The continuous ambiguity of~\eqref{eq:D_mathcalL_parameterization} is a constant diagonal transformation of $T(x)$. After choosing a local ordering of the eigenvalues of $Q$, the normalization
\begin{equation}
\label{eq:app_mirabolic_condition_T}
T(0) e =e \,, \qquad e=(1, \,1, \, \dots, \,1,\, 1) \in \mathbb{R}^N
\end{equation}
fixes the residual diagonal freedom wherever this normalization is nonsingular.
We restrict to group transformations~\eqref{eq:group_action} leaving $Q$ unchanged and transforming $T$ while preserving the condition~\eqref{eq:app_mirabolic_condition_T}.
Then, under gauge transformations, we have
\begin{equation}
T(x) \to T'(x)=h(x) T(x) d^{(T)}_h \,.
\end{equation}
where $d^{(T)}_h$ is a constant diagonal matrix fixed by imposing
\begin{equation}
  e=T'(0) e=   h(0) T(0) d^{(T)}_h e \implies d^{(T)}_h e = T^{-1}(0) h^{-1}(0) e \,.
\end{equation}
This constraint is equivalent to
\begin{equation}
(d^{(T)}_h)_{ii} = \sum^N_{k=1} (T^{-1}(0) h^{-1}(0))_{ik}\,.
\end{equation}
Combining the transformations for $T$ and $B$ we have that $L(x)$ transforms as follows
\begin{equation}
\label{eq:app_gauge_tr_L}
L(x) \to L'(x)= (d_h^{(T)})^{-1} L(x) d_h^{(T)} \,.
\end{equation}

Using the parameterisation~\eqref{eq:D_mathcalL_parameterization} in the moment map equation~\eqref{eq:mom_map1eq} we get
\begin{equation}
L(y+\Delta) - \gamma Q^{-1} L(y-\Delta) Q =  \gamma \var T^{-1}(y) a(y) b(y) T(y) Q^{-1} L(y-\Delta) Q \,.
\end{equation}
Now we define a diagonal matrix $t(x)$ with the following entries
\begin{equation}
\label{eq:ti_x}
t_i(x) = (T^{-1}(x) a(x))_{i 1} \,.
\end{equation}
We restrict to the patch on which the entries $t_i(x)$ are nonzero, so that the invariant quantities involving $t_i^{-1}(x)$ are well defined.
Under the group action, we have
\begin{equation}
t_i(x) \to  t'_i(x) = (d_h^{(T)})_{ii}^{-1} t_i(x) \,.
\end{equation}
From this fact, we see that the L operator
\begin{equation}
\label{eq:full_inv_L}
\mathbb{L}(x,y):= t^{-1}(x) L(y) t(x) \,,
\end{equation}
is a full invariant object, no matter the choice of $x$.

Using the expression in~\eqref{eq:full_inv_L}, we obtain
\begin{equation}
\label{eq:D_mathcalL_parameterization2}
\mathbb{L}(x,y+\Delta) - \gamma Q^{-1} \mathbb{L}(x,y-\Delta) Q =  \gamma \var t^{-1}(x) T^{-1}(y) a(y) b(y) T(y) Q^{-1} L(y-\Delta) Q t(x) \,.
\end{equation}
From this constraint, we read the definitions for the invariant spin oscillators to be
\begin{equation}
\label{eq:a_inv_spin}
\mathbf{a}_{i \alpha}(x,y):= (t^{-1}(x) T^{-1}(y) a(y))_{i \alpha}
\end{equation}
and 
\begin{equation}
\label{eq:b_inv_spin}
\begin{split}
\mathbf{b}_{\alpha i}(x,y): &= (b(y) T(y) Q^{-1} L(y-\Delta) Q t(x))_{\alpha i} \,.
\end{split}
\end{equation}
These are similar to the definitions in equation (5.11) of~\cite{Arutyunov:2019alw}, but here shifts appear due to the presence of the spectral parameter. 
Moreover we are working with the gauge choice
\begin{equation}
\mathbf{a}_{i 1}(x,x)=1 \,.
\end{equation}
Plugging these expressions into~\eqref{eq:D_mathcalL_parameterization2}, we obtain
\begin{equation}
\mathbb{L}(x,y+\Delta) - \gamma Q^{-1} \mathbb{L}(x, y-\Delta) Q = \gamma \var  \mathbf{a}(x,y) \mathbf{b}(x,y)
\end{equation}
To solve this equation, we define
\begin{equation}
\label{eq:coupling_redef}
\tilde{\varkappa} = \gamma \, \varkappa \,,
\end{equation}
and
\begin{equation}
Q_{ij} := \delta_{ij} Q_i \,, \qquad Q_i := e^{i \qq_i} \,, \qquad \qq_{ij} := \qq_i - \qq_j \,, \qquad \gamma := e^{i \eta} \,.
\end{equation}
Then the moment map constraint in components becomes
\begin{equation}
\label{eq:mom_map_constr_L_comp}
\mathbb{L}_{ij}(x,y+\Delta) - e^{-i (\qq_{ij} - \eta)}  \mathbb{L}_{ij}(x, y-\Delta) =  \tilde{\varkappa}  f_{ij}(x,y) \,,
\end{equation}
where
\begin{equation}
f_{ij}(x,y):=  \sum^\ell_{\alpha=1} \mathbf{a}_{i \a}(x,y) \mathbf{b}_{\a j}(x,y) \,.
\end{equation}

To solve~\eqref{eq:mom_map_constr_L_comp} we should first find a solution for the Green's function
\begin{equation}
\label{eq:eq_for_Green_fun}
\mathbb{G}_{ij}(y+\Delta-i0) -  \mathbb{G}_{ij}(y-\Delta+i0) e^{-i (\qq_{ij} -\eta)} = \delta(y) \,.
\end{equation}
Using property~\eqref{eq:varphi_period_property}, we notice that a solution for this function is provided by
\begin{equation}
\mathbb{G}_{ij}(y) = \frac{1}{2 \pi i} \, \varphi(y, \qq_{ij}- \eta) \,,
\end{equation}
with $\varphi$ defined in~\eqref{eq:app_varphi}.
The deformation parameter $\eta$ ensures that the Kronecker Delta $\delta_{ij}$ is always absent from~\eqref{eq:varphi_period_property}.
Once such a Green's function is found, we can write a general solution to the moment map as
\begin{equation}
\label{eq:Loperator_general_sol}
\mathbb{L}_{ij}(x,y) = \tilde{\varkappa} \int^{+\pi}_{-\pi} \frac{dt}{2 \pi i} \,  \varphi(y-t, \qq_{ij}- \eta) f_{ij}(x, t) \,.
\end{equation}
We now evaluate the first loop argument of the invariant fields at a common point $x=\imp^1$, defining
\begin{equation}
\label{eq:gauge_a}
\mathbf{a}_{i \alpha}(y):= \mathbf{a}_{i \alpha}(\imp^1, y) \,, \qquad \mathbf{b}_{\alpha i}(y) := \mathbf{b}_{\alpha i}(\imp^1, y) \,, \qquad \mathbb{L}_{ij}(y) := \mathbb{L}_{ij}(\imp^1,y) \,.
\end{equation}
The original dependence on the first argument is retained when computing the Poisson brackets, since derivatives with respect to this argument can contribute in the coincidence limit. Section \ref{sec:removing_first_loop} explains why the evaluation point must be treated as a dynamical variable and specifies its brackets with the oscillator fields.

On the space~\eqref{eq:gauge_a} we have
\begin{equation}
\label{eq:Loperator_general_sol2}
\mathbb{L}_{ij}(y) = \tilde{\varkappa} \int^{+\pi}_{-\pi} \frac{dt}{2 \pi i} \,  \varphi(y-t, \qq_{ij}- \eta) f_{ij}(t) \,.
\end{equation}

\subsection{Finite dimensional representations for the oscillator algebra}
\label{sec:finite_dim_osc_alg}

Despite having provided a solution to the moment map constraint, equation~\eqref{eq:Loperator_general_sol2} still describes a too large space of functions. Most of these functions do not correspond to the Lax operator of RS models. Indeed, the invariant spin oscillators $\mathbf{a}_{i \alpha}(x)$ and $\mathbf{b}_{ \alpha i}(x)$ still depend on a loop parameter and are defined in terms of infinitely many Fourier modes. The spin oscillators of the RS models introduced in~\cite{Krichever:1995zw} have instead a finite number of degrees of freedom.
We should then ask ourselves the following question: ``How do we single out $2 \ell  N$ degrees of freedom $\mathbf{a}_{i \alpha}$ and $\mathbf{b}_{ \alpha i}$ (this time independent of $x$) from these infinitely many modes?'
In this subsection, we explain two ways to isolate this finite number of degrees of freedom, the former leading to the spin hyperbolic RS models, while the latter to the elliptic models.

\paragraph{Hyperbolic spin RS models.}

Hyperbolic spin RS models can be obtained through a zero-mode truncation, from which we recover the reduction performed in~\cite{Arutyunov:2019alw}.
In particular, we perform the following mode expansion on the interval $[-\pi, +\pi]$
\begin{equation}
A(x) = \sum_{n=-\infty}^{+\infty} A^{(n)} \frac{1}{\sqrt{2 \pi}} \, e^{i n x} \iff A^{(n)} = \int^{+ \pi}_{-\pi} dx A_{ij}(x) \frac{e^{-i n x}}{\sqrt{2 \pi}} \,.
\end{equation}
We recall that all functions are periodic on this interval.
Performing such an expansion on the first line in~\eqref{AFD}, we get
\begin{equation}
    \begin{split}
\frac{1}{\var} \{A^{(0)}_1, A^{(0)}_2 \} = &- r_{\mp} A^{(0)}_1 A^{(0)}_2 - A^{(0)}_1 A^{(0)}_2 r_\pm + A^{(0)}_2 r_{+} A^{(0)}_1 + A^{(0)}_1 r_-  A^{(0)}_2\\
&+ \sum^{+\infty}_{n=1} e^{2 i n \Delta} \left( A^{(-n)}_2 C_{12} A^{(+n)}_1  + A^{(-n)}_1  C_{12} A^{(+n)}_2 \right) 
    \end{split}
\end{equation}
coupling infinitely many modes. The definitions for the different finite-dimensional $r$-matrices $r$ and $r_\pm$ are provided in Appendix~\ref {app:matrix_bblocks}.
By setting all nonzero modes $A^{(n\ne 0)}=0$, we obtain the Poisson algebra of a finite dimensional double
\begin{equation}
    \begin{split}
\frac{1}{\var} \{A^{(0)}_1, A^{(0)}_2 \} = &- r_{\mp} A^{(0)}_1 A^{(0)}_2 - A^{(0)}_1 A^{(0)}_2 r_\pm + A^{(0)}_2 r_{+} A^{(0)}_1 + A^{(0)}_1 r_-  A^{(0)}_2 \,.
    \end{split}
\end{equation}
The other brackets for the double are obtained similarly; it turns out they correspond to replacing $r_\pm(x) \to r_\pm$ in~\eqref{AFD}. A similar truncation on the oscillator manifold leads to the oscillator brackets proposed in~\cite{Arutyunov:2019alw}.

If we restrict to the zero modes of the oscillators, then equation~\eqref{eq:Loperator_general_sol2} becomes
\begin{equation}
\mathbb{L}_{ij}(y) = \frac{\tilde{\var}}{\sqrt{2 \pi}} \int^{+\pi}_{-\pi} \frac{dt}{2 \pi i} \,  \varphi(y-t, \qq_{ij}- \eta) f^{(0)}_{ij}\,, \qquad f^{(0)}_{ij} = \frac{1}{\sqrt{2 \pi}} \sum_{\alpha} \mathbf{a}^{(0)}_{i \alpha} \mathbf{b}^{(0)}_{\alpha j}  \,.
\end{equation}
Using the periodicity properties of $\varphi$ the integral can be performed exactly. Indeed, shifting the integration line from $(-\pi, +\pi)$ to $(-\pi+\Delta+ i \varepsilon, +\pi+\Delta+ i \varepsilon)$ (where $\varepsilon$ is a positive infinitesimal parameter) and subtracting the associated residue of the pole coming from the shift we obtain
\begin{equation}
\begin{split}
\int^{+\pi}_{-\pi} \frac{dt}{2 \pi i} \,  \varphi(y-t, \qq_{ij}- \eta) &= \int^{+\pi}_{-\pi} \frac{dt}{2 \pi i} \,  \varphi(y-t+\Delta+i\varepsilon, \qq_{ij}- \eta) +1 \,.
\end{split}
\end{equation}
From the periodicity of $\varphi$ we know that
\begin{equation}
\varphi(y-t+\Delta+i\varepsilon, \qq_{ij}- \eta)=e^{-i (\qq_{ij}- \eta)} \varphi(y-t-\Delta+i\varepsilon, \qq_{ij}- \eta)
\end{equation}
and the integral above can be written as
\begin{equation}
\begin{split}
\int^{+\pi}_{-\pi} \frac{dt}{2 \pi i} \,  \varphi(y-t, \qq_{ij}- \eta)&= \int^{+\pi}_{-\pi} \frac{dt}{2 \pi i} \,  e^{-i (\qq_{ij}- \eta)}\varphi(y-t-\Delta+i\varepsilon, \qq_{ij}- \eta) +1\,.
\end{split}
\end{equation}
We can now restore the integration of the first term on the RHS of the expression above to the real line without crossing any poles:
\begin{equation}
\int^{+\pi}_{-\pi} \frac{dt}{2 \pi i} \,  \varphi(y-t, \qq_{ij}- \eta)= e^{i (\eta-\qq_{ij})} \int^{+\pi}_{-\pi} \frac{dt}{2 \pi i} \,  \varphi(y-t, \qq_{ij}- \eta) +1 \,.
\end{equation}
From this equation, we obtain
\begin{equation}
\begin{split}
\int^{+\pi}_{-\pi} \frac{dt}{2 \pi i} \,  \varphi(y-t, \qq_{ij}- \eta) = \frac{1}{1-e^{i(\eta-\qq_{ij})}} \,.
\end{split}
\end{equation}
We obtain
\begin{equation}
\mathbb{L}^{(0)}_{ij} =  \frac{\tilde{\var}}{1-e^{i(\eta-\qq_{ij})}} f^{(0)}_{ij} \,.
\end{equation}
This corresponds to the Lax operator of the hyperbolic spin RS models\footnote{Strictly speaking, the Lax operator is trigonometric for real values of $\qq_i$ and hyperbolic for imaginary values.}. One can check by a direct computation (as done in~\cite{Arutyunov:2019alw}) that $H_k= \text{Tr} \left( (\mathbb{L}^{(0)})^k \right)$ are in involution, and $H_1$ is the Hamiltonian of the spin hyperbolic RS models.

\paragraph{Elliptic spin RS models with inhomogeneities.}

To obtain the elliptic spin RS models, we restrict to the following representations of the invariant spin oscillators:
\begin{equation}
\label{eq:repres_a_b}
\mathbf{b}_{\a j}(y) = \mathbf{b}_{\a j} \delta(y - \imp^{\alpha}) \,,
\end{equation}
where $\mathbf{b}_{\a j}$ is independent of the loop coordinate.
In the expressions above, the parameters $\{\imp^\alpha \}^\ell_{\alpha=1}$ correspond to inhomogeneities. These inhomogeneities are real numbers, distinct modulo $2 \pi$. By replacing the representation~\eqref{eq:repres_a_b} in~\eqref{eq:Loperator_general_sol2}, we obtain
\begin{equation}
\label{eq:Loperator_specific_sol}
\mathbb{L}_{ij}(y)=\tilde{\varkappa}\sum_\alpha \frac{1}{2 \pi i} \,  \varphi(y-\imp^\alpha, \qq_{ij}- \eta) \mathbf{a}_{i \a}\mathbf{b}_{\a j} \,, 
\end{equation}
where
\begin{equation}
\mathbf{a}_{i \alpha}:= \mathbf{a}_{i \alpha}(\imp^\alpha) \,.
\end{equation}
We could not find a realization of the oscillator algebra allowing us to localize all the inhomogeneities to the same point. Moreover, to avoid derivatives of the oscillators arising by localizing the oscillators at the inhomogeneities' positions,
\begin{equation}
 \{ \mathbf{a}_{i \alpha}(y),\mathbf{a}_{\beta j}(t) \} \to \{ \mathbf{a}_{i \alpha}(\imp^\alpha),\mathbf{a}_{\beta j}(\imp^\beta) \} \,,
\end{equation}
it turns out that these inhomogeneities must be dynamical variables, satisfying the Poisson brackets obtained in~\cite{Arutyunov:2026yjy}. As already mentioned in the introduction, the brackets on the reduced phase space reproduce the equations of motion of the elliptic spin RS models of~\cite{Krichever:1995zw}, supplemented by dynamical variables. While at the level of the equations of motion it is possible to take the limit in which all the inhomogeneities coincide, it turns out that in this homogeneous limit the algebra is no longer Poisson. In particular, the brackets for the invariant oscillators diverge because of terms $\varphi(\imp^{\beta\alpha}+\Delta, \qq_{ba})$ that grow like $1/\imp^{\beta \alpha}$. Moreover, the constraint $\imp^\alpha=\imp^\beta$, with $\alpha \ne \beta$, does not form a Poisson ideal. Simply taking the regular part of these divergent brackets does not form a Poisson structure since it does not satisfy the Jacobi identity, as observed in~\cite{Arutyunov:2026yjy}.

In the following sections and associated appendices, we will perform the explicit derivation of the Poisson brackets on the reduced phase space. After these brackets are derived for arbitrary values of the loop parameters, we will replace the representations~\eqref{eq:repres_a_b} and show how singularities and derivatives with respect to the loop parameter cancel from the brackets, allowing us to restrict the algebra to a finite number of degrees of freedom. For these cancellations to be possible, it will be essential that the inhomogeneities are dynamical.

\section{Poisson brackets of invariant objects}
\label{sec:Poisson_inv}

In this section, we derive the Poisson brackets of the quantities invariant under the group actions, namely the $\mathbb{L}$ operator defined in~\eqref{eq:full_inv_L} and the invariant spin oscillators~\eqref{eq:a_inv_spin} and~\eqref{eq:b_inv_spin}. We compute these brackets by expressing the invariant quantities in terms of the elements $A(x)$ and $B(x)$ of the original affine Heisenberg double and the oscillator variables on the symplectic manifold introduced in section~\ref{sec:affine_osc_manifold}. The derivation does not require imposing the moment map constraint.

\subsection{TT bracket}
\label{sec:variations_and_Green_elliptic}

We start by writing the matrices $A(x)$ and $B(x)$ using the parameterisation~\eqref{eq:D_mathcalL_parameterization}. We are interested in finding the Poisson bracket
\begin{equation}
\label{eq:TT_br_expr1}
\{T_{ij}(x) , T_{kl}(y) \} = \int^{+\pi}_{-\pi} ds  \, \int^{+\pi}_{-\pi} ds' \, \frac{\delta T_{ij}(x)}{\delta A_{ab}(s)} \frac{\delta T_{kl}(y)}{\delta A_{cd}(s')} \, \{A_{ab}(s), A_{cd}(s')\} 
\end{equation}
To obtain such a bracket first we need to evaluate the variation $\delta T/\delta A$.
Let us take the variation of the first equation in~\eqref{eq:D_mathcalL_parameterization}; defining 
\begin{equation}
X(x)= T^{-1}(x) \delta T(x) \,, \qquad Y(x) = T^{-1}(x-\Delta) \delta A(x)  T(x+ \Delta) \,,
\end{equation}
we obtain
\begin{equation}
\label{eq:variation_dTD_dA_new}
Y_{ij}(x) = X_{ij}(x-\Delta) Q_j - Q_i X_{ij}(x+\Delta) +\delta_{ij} \, \delta Q_{i} \,.
\end{equation}
We assumed $\delta Q_{ij}$ to be diagonal, since it is the variation of a diagonal matrix.
Integrating both sides of this equation on the circle, we obtain
\begin{equation}
\delta Q_i = \frac{1}{2 \pi} \, \int^{+\pi}_{-\pi} dx\,  Y_{ii}(x) \,.
\end{equation}
Using property~\eqref{eq:varphi_period_property} one can show that a solution to~\eqref{eq:variation_dTD_dA_new} is provided by
\begin{equation}
X_{ij}(x) = \a_{i} \delta_{ij} - \int^{+\pi}_{-\pi}\frac{\text{d}z}{2 \pi i} \, \frac{1}{Q_i} \varphi(x-z, \qq_{ij}) \, Y_{ij}(z) \,.
\end{equation}
The constant coefficients $\{\alpha_i\}^N_{i=1}$ are fixed by demanding~\eqref{eq:app_mirabolic_condition_T}, from which we obtain
\begin{equation}
\sum^N_{k=1} \, X_{ik}(0) =0 \implies \alpha_i = \sum^N_{k=1} \int^{+\pi}_{-\pi} \frac{dz}{2 \pi i} \, \frac{1}{Q_i} \varphi(-z, \qq_{ik}) Y_{ik}(z) \,.
\end{equation}
Then the required solution is given by
\begin{equation}
X_{ij}(t) = \frac{1}{2 \pi i} \int^{+\pi}_{-\pi} \frac{dz}{Q_i} \left(\delta_{ij} \sum^N_{k=1} \, \varphi(-z, \mathfrak{q}_{ik}) Y_{ik}(z) - \varphi(t-z, \mathfrak{q}_{ij}) Y_{ij}(z) \right) \,.
\end{equation}
From this solution, we can read off the variation of $T$ with respect to $A$:
\begin{equation}
\label{eq:TovA_variation}
\begin{split}
\frac{\delta T_{ij}(x)}{\delta A_{mn}(y)} = - \frac{1}{2 \pi i} \sum^N_{a=1} &\biggl(e^{-i \mathfrak{q}_a} \varphi(x-y, \mathfrak{q}_{aj}) T_{ia}(x) T^{-1}_{am}(y-\Delta) T_{nj}(y+\Delta)\\
&-e^{-i \mathfrak{q}_j} \varphi(-y, \mathfrak{q}_{ja}) T_{ij}(x) T^{-1}_{jm}(y-\Delta) T_{na}(y+\Delta) \biggr)
\end{split}
\end{equation}
We can now plug this variation in~\eqref{eq:TT_br_expr1} and compute the integral. Interestingly, using the analyticity properties of the $\varphi$ function, together with the assumption that $T(x)$ has neither poles nor cuts on the analyticity strip $|\Im(x)|<\Im(\Delta)$, the integral can be computed exactly in a closed form by the Cauchy residue theorem.  After a quite long computation, which we detail in Appendix~\ref{app:derivation_TT_bracket}, we arrive at the following Poisson bracket between $T$'s
\begin{equation}
\label{eq:brackets_TT_final}
    \begin{split}
     \frac{1}{\var} \{T_1(x), T_2(y) \}&= T_1(x) T_2(y) r_{\mathfrak{q}}(x,y) - T_1(x) T_2(y) \refl_{\pi}(x,y)\\
     &+i T_1(x) T_2(y) \sum_{ijk} \left(\X_{jk}-\X_{ik} \right) E_{ii} \otimes E_{jj} \,.
    \end{split}
\end{equation}
In the expression above, we defined
\begin{equation}
\label{eq:defXY}
\X= T^{-1}(0) \partial T(0) \,.
\end{equation}
These derivative terms arise from double poles contained in the contour adopted to perform the integration with Cauchy theorem.
The different r-matrices are listed in Appendix~\ref{app:list_of_r_mat}.
We checked that this bracket is both skew-symmetric and satisfies the Jacobi identity, providing in this way a consistent Poisson structure. The verification of the Jacobi identity relies on the shifted classical Yang-Baxter equation satisfied by $r_{\mathfrak{q}}(x,y)$. Namely, if we define
\begin{equation}
\mathcal{R}_{ab}^\qq=r_{\qq}(x_a,x_b)_{ab} + i \frac{\partial}{\partial x_a}-i \frac{\partial}{\partial x_b}
\end{equation}
then we obtain
\begin{equation}
[\mathcal{R}_{13}^\qq, \mathcal{R}_{12}^\qq] +[\mathcal{R}_{23}^\qq, \mathcal{R}_{13}^\qq]+[\mathcal{R}_{23}^\qq, \mathcal{R}_{12}^\qq]
= 0 \,.
\end{equation}
As far as we know, this is a new Poisson bracket. Evaluating the unreduced bracket~\eqref{eq:brackets_TT_final} on the gauge slice $T(x)=\mathbb{I}$ reproduces the result of~\cite{Arutyunov:1996uw}.

\subsection{LL bracket}

Once the TT and TB brackets are known, the derivation of the remaining brackets is a matter of an algebraic computation.
Performing a long but straightforward computation, we find the following bracket between invariant $\mathbb{L}$ operators:
\begin{equation}
\label{eq:LL_bracket_with_index_arbitraryxz}
\begin{split}
&\frac{1}{\var} \{ \mathbb{L}_{ab}(x,y), \mathbb{L}_{cd}(z,t) \} =\\
&\scriptstyle{+\mathbb{L}_{ab}(x,y) \mathbb{L}_{cd}(z,t) \left(i t^{-1}_b(x) t^{-1}_d(z) \varphi(z-x+\Delta, \mathfrak{q}_{db}) t_d(x) t_b(z) + i t^{-1}_a(x) t^{-1}_c(z) \varphi(z-x+\Delta, \mathfrak{q}_{ca}) t_c(x) t_a(z) \right)} \\
&\scriptstyle{+\mathbb{L}_{ab}(x,y) \mathbb{L}_{cd}(z,t) \left(-i t^{-1}_b(x) t^{-1}_c(z) \varphi(z-x+\Delta, \mathfrak{q}_{cb}) t_c(x) t_b(z) + i t^{-1}_a(x) t^{-1}_d(z) \varphi(x-z+\Delta, \mathfrak{q}_{ad}) t_d(x) t_a(z) \right)}\\
&\scriptstyle{+ i \mathbb{L}_{ad}(x,y)\mathbb{L}_{cb}(z,t)  t^{-1}_{d}(x) t^{-1}_b(z)  \ \varphi(y-t+\Delta, \mathfrak{q}_{db}) t_{b}(x) t_{d}(z)+i \mathbb{L}_{cb}(x,y)\mathbb{L}_{ad}(z,t)
         t^{-1}_{a}(x) t^{-1}_c(z) \varphi(t-y+\Delta,\mathfrak{q}_{ca}) t_{c}(x) t_{a}(z)} \\
&\scriptstyle{+\mathbb{L}_{ab}(x,y) \mathbb{L}_{bd}(z,t) t_b^{-1}(x) t^{-1}_c(z) i \varphi(t-x, \mathfrak{q}_{cb})  t_c(x) t_b(z) -\mathbb{L}_{ab}(x,y) \mathbb{L}_{ad}(z,t)   t_a^{-1}(x) t^{-1}_c(z)  i \varphi(t-x, \mathfrak{q}_{ca})  t_c(x) t_a(z)   } \\
&\scriptstyle{+ \mathbb{L}_{cb}(x,y) \mathbb{L}_{cd}(z,t)  t_a^{-1}(x) t_c^{-1}(z) i\varphi(y-z, \mathfrak{q}_{ac}) t_a(z) t_c(x)   -\mathbb{L}_{db}(x,y) \mathbb{L}_{cd}(z,t) t_a^{-1}(x) t_d^{-1}(z) i\varphi(y-z, \mathfrak{q}_{ad}) t_a(z) t_d(x)  }\,.
  \end{split}
\end{equation}
We notice that for arbitrary values of the loop parameters $x$, $y$, $z$ and $t$, the bracket above does not close on the $\mathbb{L}$ operators alone. Instead, on the RHS also appear the elements $t_i$ defined in~\eqref{eq:ti_x}. To remove these elements, we need to take the limit in which $x$ and $z$ are evaluated at coincident points $x=z=\imp^1$.
Even if multiple terms on the RHS of the expression above are singular for $x=z$, it turns out that all singularities cancel each other and the final expression is finite. Remarkably, also derivative terms arising from taking the limit $x \to \imp^1$ and $z \to \imp^1$ cancel each other, leading to the following formula written entirely in terms of $\mathbb{L}$ operators
\begin{equation}
\label{eq:LL_bracket_main}
\begin{split}
\frac{1}{\var} \{ \mathbb{L}_{ab}(y), \mathbb{L}_{cd}(t) \} &=\mathbb{L}_{ab}(y) \mathbb{L}_{cd}(t) \left(i  \varphi^{\text{reg}}(\mathfrak{q}_{db})  + i  \varphi^{\text{reg}}(\mathfrak{q}_{ca}) -i  \varphi^{\text{reg}}(\mathfrak{q}_{cb})  + i  \varphi^{\text{reg}}(\mathfrak{q}_{ad})  \right) \\
&+ i \mathbb{L}_{ad}(y)\mathbb{L}_{cb}(t)   \ \varphi(y-t+\Delta, \mathfrak{q}_{db}) +i \mathbb{L}_{cb}(y)\mathbb{L}_{ad}(t) \varphi(t-y+\Delta,\mathfrak{q}_{ca})  \\
&+\mathbb{L}_{ab}(y) \mathbb{L}_{bd}(t) i \varphi(t-\imp^1, \mathfrak{q}_{cb})   -\mathbb{L}_{ab}(y) \mathbb{L}_{ad}(t)     i \varphi(t-\imp^1, \mathfrak{q}_{ca})    \\
&+ \mathbb{L}_{cb}(y) \mathbb{L}_{cd}(t)   i \varphi(y-\imp^1, \mathfrak{q}_{ac})    -\mathbb{L}_{db}(y) \mathbb{L}_{cd}(t)  i \varphi(y-\imp^1, \mathfrak{q}_{ad})   \,.
  \end{split}
\end{equation}
The bracket above can also be written using the matrix notation
\begin{equation}
\label{eq:Poiss_bracket_inv_L_magic_limit}
\begin{split}
\frac{1}{\var} \{ \mathbb{L}_{1}(y), \mathbb{L}_{2}(t) \}&= r_{12}(y;t)  \mathbb{L}_{1}(y)\mathbb{L}_{2}(t)-\mathbb{L}_{1}(y) \mathbb{L}_{2}(t) \underline{r}_{12}(y;t) \\
&+\mathbb{L}_1(y) \overline{r}_{21}(t;y)  \mathbb{L}_2(t)-\mathbb{L}_2(t) \overline{r}_{12}(y;t) \mathbb{L}_1(y) \,,
  \end{split}
\end{equation}
with the different $r$ matrices written in Appendix~\ref{app:list_of_r_mat}.

The quadratic Poisson algebra~\eqref{eq:Poiss_bracket_inv_L_magic_limit} implies that the spectral invariants of the Lax operator are in involution. 
Indeed the explicit expressions in Appendix~\ref{app:list_of_r_mat} satisfy
\begin{equation}
r_{12}(y;t)-\underline{r}_{12}(y;t)
+\bar{r}_{21}(t;y)-\bar{r}_{12}(y;t)=0.
\end{equation}
Defining
\begin{equation}
I_m(y):=\operatorname{Tr}\bigl(\mathbb{L}(y)^m\bigr),
\qquad m\geq1,
\end{equation}
and applying the Leibniz rule and cyclic property of the trace, we obtain
\begin{equation}
\begin{aligned}
&\frac{1}{\var} \{I_m(y),I_n(t) \}
=\frac{mn}{\var}\operatorname{Tr}_{12}
\left[
\mathbb{L}_1(y)^{m-1}\mathbb{L}_2(t)^{n-1}
\{ \mathbb L_1(y),\mathbb L_2(t) \}
\right] \\
&=mn \, \operatorname{Tr}_{12}
\left[
\bigl(
r_{12}(y;t)-\underline{r}_{12}(y;t)
+\bar{r}_{21}(t;y)-\bar{r}_{12}(y;t)
\bigr)
\mathbb L_1(y)^m\mathbb L_2(t)^n
\right] =0.
\end{aligned}
\end{equation}
Here $\text{Tr}_{12}$ denotes the trace over both auxiliary spaces. Thus the functions $I_m(y)$ form a Poisson-commutative family for all positive integers $m$ and arbitrary values of the spectral parameter $y$.

The spectral invariants $I_m(y)$ can therefore be regarded as generating functions of Poisson-commuting charges. In particular, taking residues preserves their involution: locally in phase space, each residue can be expressed as an integral over a fixed contour enclosing the corresponding pole. More explicitly, taking the moment map solution~\eqref{eq:Loperator_specific_sol}, for $\alpha=1, 2, \dots, \ell$, we can define
\begin{equation}
H_\alpha = \oint_{\Gamma^\alpha} I_1(y) dy \,, 
\end{equation}
where $\Gamma^\alpha$ is a small circle surrounding the pole of the Lax at $y=\Delta+\imp^\alpha$. The Poisson bracket can be taken under the integral, and we observe
\begin{equation}
 \{ H_\alpha , I_m(t) \} =\oint_{\Gamma^\alpha} \, dy \,  \{I_1(y), I_m(t) \} =0 \,.
\end{equation}
This remains valid when the pole positions are dynamical, provided no pole crosses the contour. In particular, we obtain
\begin{equation}
 \{H_\alpha, H_\beta \} =0 \,.
\end{equation}
As we will show in the next section, up to an overall multiplicative constant, the Ruijsenaars–Schneider Hamiltonian can be identified with
\begin{equation}
H_0 = \sum^\ell_{\alpha=1} H_\alpha
\end{equation}
and commutes with the spectral invariants. These invariants thus provide conserved quantities for the Hamiltonian flow generated by $H_0$.

\subsection{Brackets of invariant oscillators}
\label{sec:Brackets_inv_osc}

As for the $\mathbb{L}$ operators, the brackets of the invariant oscillators also do not close on the invariant objects $\mathbb{L}(x,y)$, $\mathbf{a}(x,y)$ and $\mathbf{b}(x,y)$. The different brackets for the double-loop invariant spin oscillators are listed in~\eqref{eq:aa_inv_br_doubleloop}, \eqref{eq:ab_inv_br_doubleloop} and~\eqref{eq:bb_inv_br_doubleloop}, respectively. To allow these brackets to close on the invariant $\mathbb{L}$ and spin operators, we take the limit $x=z=\imp^1$, exactly as we did for the $\mathbb{L}$ operators. Namely, we take the limit
\begin{equation}
\lim_{x \to \imp^1} \lim_{z \to \imp^1} \, \frac{1}{\var} \, \{ \mathbf{a}_{i \alpha}(x,y), \mathbf{a}_{j \beta}(z,t) \}
\end{equation}
It is possible to show that all the double-loop oscillator brackets are finite in this limit, even though terms considered separately might be singular. Let us show in some more detail how the limit is computed. We explain here the logic adopted in performing the limit, while we refer the reader to Appendices~\ref{app:aa_collin_lim}, \ref{app:ab_collin_lim} and~\ref{app:bb_collin_lim} for the full derivation.

\subsection{Removing the first loop variable in the oscillators}
\label{sec:removing_first_loop}

\paragraph{Coincidence limit and derivatives.}

Let us consider the $\mathbf{a}$-$\mathbf{a}$ bracket~\eqref{eq:aa_inv_br_doubleloop}. We consider here only the terms which are singular at the point $x=z=\imp^1$:
\begin{equation}
\label{eq:aa_inv_br_doubleloopsing}
\begin{split}
\frac{1}{\var} \, \{\mathbf{a}_{a \alpha}(x,y) \,,\mathbf{a}_{b \beta}(z,t) \}&=  t^{-1}_a(x) t^{-1}_b(z)    t_b(x) t_a(z)    
  i \varphi(z-x+\Delta,\mathfrak{q}_{ba})\mathbf{a}_{a \alpha}(x,y) \mathbf{a}_{b \beta}(z,t)  \\
&-   \mathbf{a}_{a \gamma} (x,x)\mathbf{a}_{b \delta}(z,z) \rho_{\gamma 1, \delta 1}(x-z) \mathbf{a}_{a \alpha}(x,y) \mathbf{a}_{b \beta}(z,t)+\dots\,.
\end{split}
\end{equation}
The ellipses contain all the remaining terms which are regular in the limit. Setting $z=\imp^1$, $x=\imp^1+\epsilon$ and taking the limit $\epsilon \to 0$, it turns out that the singularity of $\varphi(z-x+\Delta,\mathfrak{q}_{ba})$ cancels against the singularity of $\rho_{\gamma 1, \delta 1}(x-z)$ generating a finite result. 
The result is the same if we replace $\rho=\hat{\rho}$ or $\rho=\check{\rho}$, as defined in section~\ref{sec:affine_osc_manifold}.
Even if the result is finite, expanding the RHS of~\eqref{eq:aa_inv_br_doubleloopsing} around the singular values of $\varphi$ and $\rho$ produces derivatives of the oscillators. Keeping these derivatives into account, we end up with the following $\mathbf{a}$-$\mathbf{a}$ bracket at coincident points
\begin{equation}
\label{eq:aa_bracket_final_components_main}
\begin{split}
&\frac{1}{\var} \, \{\mathbf{a}_{a \alpha}(x,y) \,,\mathbf{a}_{b \beta}(z,t) \}\bigl|_{x=z=\imp^1}=\\
&  + i \varphi(t-y+\Delta,\mathfrak{q}_{ba})\mathbf{a}_{b \alpha}(y) \mathbf{a}_{a \beta}(t) - i \varphi(t-\imp^1+\Delta,\mathfrak{q}_{ba})\mathbf{a}_{a \alpha}(y) \mathbf{a}_{a \beta}(t)\\
&- i \varphi(\imp^1-y+\Delta,\mathfrak{q}_{ba})\mathbf{a}_{b \alpha}(y) \mathbf{a}_{b \beta}(t)+i \varphi^{\text{reg}}(\mathfrak{q}_{ba})\mathbf{a}_{a \alpha}(y) \mathbf{a}_{b \beta}(t)\\
&- \mathbf{a}_{a \gamma}(y)\mathbf{a}_{b \delta}(t) \rho_{\gamma \alpha, \delta \beta}(y-t) +\mathbf{a}_{a \beta} (\imp^1,\imp^1)\mathbf{a}_{b 1}(t)   \rho_{\beta 1, 1 \beta}(\imp^1-t) \mathbf{a}_{a \alpha}( y)\\
&+ \mathbf{a}_{a 1} (y)\mathbf{a}_{b \alpha}(\imp^1,\imp^1)   \rho_{1 \alpha, \alpha 1}(y-\imp^1) \mathbf{a}_{b \beta}(t)\\
&-i \frac{d}{dx} ((\mathbf{a}_{a \alpha}(x,y) )\Bigr|_{x=\imp^1}   \mathbf{a}_{b \beta}(t)+i \frac{d}{dx} (\mathbf{a}_{b \beta}(x,t) )\Bigr|_{x=\imp^1}  \mathbf{a}_{a \alpha}(y)\,.
\end{split}
\end{equation}
In the expression above, we are using the previously introduced definition $\mathbf{a}_{i \alpha}(y) \equiv \mathbf{a}_{i \alpha}(\imp^1, y)$.
Similarly, we obtain the following Poisson brackets for the other oscillators at coincident points:
\begin{equation}
\begin{split}
&\frac{1}{\var} \, \{\mathbf{a}_{a \alpha}(x,y), \mathbf{b}_{\beta b}(z,t)\}\bigl|_{x=z=\imp^1}=\\
&+i \varphi(\imp^1-y+\Delta,\mathfrak{q}_{ba})   \mathbf{a}_{b \alpha}(y) \mathbf{b}_{\beta b}(t) -i \varphi^{\text{reg}}(\qq_{ba}) \mathbf{a}_{a \alpha}(y) \mathbf{b}_{\beta b}(t)\\
&+\mathbf{a}_{a \gamma}(y) \rho(y-t)_{-\gamma \alpha, \beta \gamma} \mathbf{b}_{\gamma b}(t)-  \mathbf{a}_{a 1} (y)\mathbf{a}_{b \alpha}(\imp^1,\imp^1) \rho_{1 \alpha, \alpha 1}(y-\imp^1) \mathbf{b}_{\beta b}(t)\\
&-\frac{2 \pi }{\var} \delta_{\alpha \beta} Q_b Q^{-1}_a  \mathbb{L}_{ab}(t-\Delta) \delta(y-t)+\frac{2 \pi }{\var} \delta_{1 \beta} Q_b Q^{-1}_a  \mathbb{L}_{ab}(t-\Delta) \delta(\imp^1-t)\mathbf{a}_{a \alpha}(y)\\
&-\mathbf{a}_{a \gamma}(\imp^1,\imp^1) \rho(\imp^1-t)_{-\gamma 1, \beta \delta} \mathbf{b}_{\delta b}(t) \mathbf{a}_{a \alpha}(y)\\
&+i \mathbf{a}_{a\alpha}(y) \frac{d}{dx} \mathbf{b}_{\beta b}(x,t)\Bigl|_{x=\imp^1} +i \mathbf{b}_{\beta b}(t) \frac{d}{dx} \mathbf{a}_{a \alpha}(x,y)\Bigl|_{x=\imp^1}  \, ,
\end{split}
\end{equation}
\begin{equation}
\begin{split}
&\frac{1}{\var} \{\mathbf{b}_{\alpha a}(x,y),\mathbf{b}_{\beta b}(z,t)\}\bigl|_{x=z=\imp^1} =\\
&+i \mathbf{b}_{\alpha b}(y) \mathbf{b}_{\beta a}(t)   \varphi(y-t+\Delta, \mathfrak{q}_{ba}) +i \mathbf{b}_{\alpha a}(y) \varphi^{\text{reg}}(\mathfrak{q}_{ba})    \mathbf{b}_{\beta b}(t) \\
&-\rho_{\alpha \gamma, \beta \delta}(y-t)
\mathbf{b}_{\gamma a}(y) \mathbf{b}_{\delta b}(t)+\mathbf{b}_{\beta b}(t)  \mathbf{a}_{b \gamma}(\imp^1,\imp^1) \rho(y-\imp^1)_{+ \alpha \gamma, \gamma 1} \mathbf{b}_{\gamma a}(y)  \\
&+\mathbf{b}_{\alpha a}(y)  \mathbf{a}_{a \gamma}(\imp^1,\imp^1)  \rho(\imp^1-t)_{-\gamma 1, \beta \gamma} \mathbf{b}_{\gamma b}(t)\\
&+\frac{2 \pi}{\var} \delta_{\alpha 1}  \delta(y-\imp^1)  \, \mathbf{b}_{\beta b}(t) Q^{-1}_b    \mathbb{L}_{ba}(\imp^1, y-\Delta) Q_a\\
&-\frac{2 \pi }{\var} \delta_{1 \beta} \mathbf{b}_{\alpha a}(y) Q^{-1}_a \mathbb{L}_{ab}(\imp^1,t-\Delta) Q_b \delta(\imp^1-t)\\
&+i \frac{d}{dz}\mathbf{b}_{\alpha a}(z,y) \Bigl|_{z=\imp^1} \mathbf{b}_{\beta b}(t)-i\mathbf{b}_{\alpha a}(\imp^1,y) \frac{d}{dz}\mathbf{b}_{\beta b}(z,t) \Bigl|_{z=\imp^1}\,.
\end{split}
\end{equation}
After taking the collinear limit $x=z=\imp^1$, all the brackets close on the invariants $\mathbb{L}$, $\mathbf{a}$ and $\mathbf{b}$. The price to pay is that derivative terms arise in the limit.

\paragraph{The importance of being dynamical.}

To obtain a finite number of degrees of freedom for the oscillator brackets, it is fundamental to remove the derivative terms. If $\{\mathbf{a}, \mathbf{a}\}$ contains derivatives of the oscillators, say $\mathbf{a}'$, then the bracket $\{\mathbf{a}, \mathbf{a}'\}$ should also be considered and, in general, might produce second-order derivatives $\mathbf{a}''$. Higher-order derivatives then propagate in the brackets; as a matter of fact, these derivatives are a manifestation that the loop variables ($x$ and $z$ in the expressions above) cannot be removed. The oscillator brackets cannot then be closed on a finite number of degrees of freedom. A way to prevent derivative terms is to promote $\imp^1$ to be a dynamical variable, which therefore satisfies non-trivial Poisson brackets with the oscillators. In the following, we present the logic of promoting $\imp^1$ to be a dynamical variable. We invite the reader to look at Appendix~\ref{app:dyn_imp} for a more complete derivation.
Here the dynamical evaluation point is introduced as part of the
finite-dimensional realization; we do not identify it with a specified
function on the original unreduced phase space. Its brackets with the
oscillators are constrained by requiring the evaluated brackets to close
without derivatives with respect to the loop parameters.

Suppose $\imp^1$ to be dynamical; then the expansion of the bracket~\eqref{eq:aa_inv_br_doubleloop} around $x=z=\imp^1$ is given by
\begin{equation}
\begin{split}
\{\mathbf{a}_{a \alpha}(x,y), \mathbf{a}_{b \beta}(z,t) \} &=\{\mathbf{a}_{a \alpha}(\imp^1,y), \mathbf{a}_{b \beta}(\imp^1,t) \} + \frac{d}{dx} \mathbf{a}_{a \alpha}(x,y)\bigl|_{x=\imp^1} \{(x-\imp^1), \mathbf{a}_{b \beta}(\imp^1,t) \}\\
&+  \{\mathbf{a}_{a \alpha}(\imp^1,y), (z-\imp^1) \} \frac{d}{dz} \mathbf{a}_{b \beta}(z,t)\bigl|_{z=\imp^1} + \dots \,.
\end{split}
\end{equation}
Because $\imp^1$ does not trivially commute with the oscillators, from the expression above we read
\begin{equation}
\begin{split}
\{\mathbf{a}_{a \alpha}&(\imp^1,y), \mathbf{a}_{b \beta}(\imp^1,t) \} =\{\mathbf{a}_{a \alpha}(x,y), \mathbf{a}_{b \beta}(z,t) \}\bigr|_{x=z=\imp^1} \\
&+ \frac{d}{dx} \mathbf{a}_{a \alpha}(x,y)\bigl|_{x=\imp^1} \{\imp^1, \mathbf{a}_{b \beta}(\imp^1,t) \}- \{ \imp^1,\mathbf{a}_{a \alpha}(\imp^1,y) \} \frac{d}{dz} \mathbf{a}_{b \beta}(z,t)\bigl|_{z=\imp^1}
\end{split}
\end{equation}
The first term on the RHS of the expression above is exactly~\eqref{eq:aa_bracket_final_components_main}.
Requiring that for $\alpha \ne 1$ it holds
\begin{equation}
\label{eq:D1_a_comm_main}
\frac{1}{\var} \{\imp^1, \mathbf{a}_{a \alpha}(y) \} = + i \mathbf{a}_{a \alpha}(y) \,,
\end{equation}
then we obtain (for $\alpha, \beta \ne 1$)
\begin{equation}
\label{eq:aa_bracket_final_components_main_dyn}
\begin{split}
&\frac{1}{\var} \, \{\mathbf{a}_{a \alpha}(\imp^1,y) \,,\mathbf{a}_{b \beta}(\imp^1,t) \}=\\
&  + i \varphi(t-y+\Delta,\mathfrak{q}_{ba})\mathbf{a}_{b \alpha}(y) \mathbf{a}_{a \beta}(t) - i \varphi(t-\imp^1+\Delta,\mathfrak{q}_{ba})\mathbf{a}_{a \alpha}(y) \mathbf{a}_{a \beta}(t)\\
&- i \varphi(\imp^1-y+\Delta,\mathfrak{q}_{ba})\mathbf{a}_{b \alpha}(y) \mathbf{a}_{b \beta}(t)+i \varphi^{\text{reg}}(\mathfrak{q}_{ba})\mathbf{a}_{a \alpha}(y) \mathbf{a}_{b \beta}(t)\\
&- \mathbf{a}_{a \gamma}(y)\mathbf{a}_{b \delta}(t) \rho_{\gamma \alpha, \delta \beta}(y-t) +\mathbf{a}_{a \beta} (\imp^1,\imp^1)\mathbf{a}_{b 1}(t)   \rho_{\beta 1, 1 \beta}(\imp^1-t) \mathbf{a}_{a \alpha}( y)\\
&+ \mathbf{a}_{a 1} (y)\mathbf{a}_{b \alpha}(\imp^1,\imp^1)   \rho_{1 \alpha, \alpha 1}(y-\imp^1) \mathbf{a}_{b \beta}(t)\,.
\end{split}
\end{equation}
Notice the difference between~\eqref{eq:aa_bracket_final_components_main} and~\eqref{eq:aa_bracket_final_components_main_dyn}. In the former bracket, $x$ and $z$ are assumed to be non-dynamical and equal to $\imp^1$, while in the latter the limit is taken assuming $\imp^1$ to be a dynamical variable.

The normalization $\mathbf{a}_{i1}(\imp^1)=1$ requires
\begin{equation}
\{\imp^1,\mathbf{a}_{i1}(\imp^1) \}=0.
\end{equation}
This condition constrains the bracket at the evaluation point; its extension to arbitrary values of the remaining loop coordinate must also be specified. We choose the multiplicative action
\begin{equation}
\label{eq:D1_ab_comm_main}
\frac{1}{\var} \, \{\imp^1,\mathbf{a}_{i\alpha}(y) \}
=i(1-\delta_{\alpha 1})\mathbf{a}_{i\alpha}(y),
\qquad
\frac{1}{\var} \{\imp^1,\mathbf{b}_{\alpha i}(y) \}
=-i(1-\delta_{\alpha1})\mathbf{b}_{\alpha i}(y).
\end{equation}
This choice respects the normalization and leaves each product
$\mathbf{a}_{i\alpha}(y)\mathbf{b}_{\alpha j}(y)$ invariant under the action of $\imp^1$. 
Due to this fact, the $\mathbb{L}$-$\mathbb{L}$ bracket, which is already closed on the $\mathbb{L}$ operators alone (see eq.~\eqref{eq:LL_bracket_main}), remains unchanged and additional unwanted derivatives are not produced in that bracket.

The choice~\eqref{eq:D1_ab_comm_main} supplies the chain-rule contributions needed to cancel the derivative terms when both spin indices differ from $1$. In particular, we obtain
\begin{equation}
\label{eq:aa_main_xD1_dyn}
\begin{split}
&\frac{1}{\var} \, \{\mathbf{a}_{a \alpha}(y) \,,\mathbf{a}_{b \beta}(t) \}=- \mathbf{a}_{a \beta}(y)\mathbf{a}_{b \alpha}(t) \rho_{\beta \alpha, \alpha \beta}(y-t)\\
&  + i \varphi(t-y+\Delta,\mathfrak{q}_{ba})\mathbf{a}_{b \alpha}(y) \mathbf{a}_{a \beta}(t) - i \varphi(t-\imp^1+\Delta,\mathfrak{q}_{ba})\mathbf{a}_{a \alpha}(y) \mathbf{a}_{a \beta}(t)\\
&- i \varphi(\imp^1-y+\Delta,\mathfrak{q}_{ba})\mathbf{a}_{b \alpha}(y) \mathbf{a}_{b \beta}(t)+i \varphi^{\text{reg}}(\mathfrak{q}_{ba})\mathbf{a}_{a \alpha}(y) \mathbf{a}_{b \beta}(t)\\
&+\mathbf{a}_{a \beta} (\imp^1)\mathbf{a}_{b 1}(t)   \rho_{\beta 1, 1 \beta}(\imp^1-t) \mathbf{a}_{a \alpha}(y)- \mathbf{a}_{a 1} (y)\mathbf{a}_{b \alpha}(\imp^1)   \rho_{\alpha 1,1 \alpha}(\imp^1-y) \mathbf{a}_{b \beta}(t)\\
&-i \delta_{\beta 1} \frac{d}{dx} ((\mathbf{a}_{a \alpha}(x,y) )\Bigr|_{x=\imp^1}   \mathbf{a}_{b \beta}(\imp^1,t)+i \delta_{\alpha 1} \mathbf{a}_{a \alpha}(\imp^1,y) \frac{d}{dx} (\mathbf{a}_{b \beta}(x,t) )\Bigr|_{x=\imp^1}  \,,
\end{split}
\end{equation}
\begin{equation}
\label{eq:ab_main_xD1_dyn}
\begin{split}
&\frac{1}{\var} \, \{\mathbf{a}_{a \alpha}(y), \mathbf{b}_{\beta b}(t)\}=+\mathbf{a}_{a \gamma}(y) \rho(y-t)_{-\gamma \alpha, \beta \gamma} \mathbf{b}_{\gamma b}(t)\\
&+i \varphi(\imp^1-y+\Delta,\mathfrak{q}_{ba})   \mathbf{a}_{b \alpha}(y) \mathbf{b}_{\beta b}(t) -i \varphi^{\text{reg}}(\qq_{ba}) \mathbf{a}_{a \alpha}(y) \mathbf{b}_{\beta b}(t)\\
&-  \mathbf{a}_{a 1} (y)\mathbf{a}_{b \alpha}(\imp^1) \rho_{1 \alpha, \alpha 1}(y-\imp^1) \mathbf{b}_{\beta b}(t) -\mathbf{a}_{a \gamma}(\imp^1) \rho(\imp^1-t)_{-\gamma 1, \beta \delta} \mathbf{b}_{\delta b}(t) \mathbf{a}_{a \alpha}(y)\\
&-\frac{2 \pi }{\var} \delta_{\alpha \beta} Q_b Q^{-1}_a  \mathbb{L}_{ab}(t-\Delta) \delta(y-t)+\frac{2 \pi }{\var} \delta_{1 \beta} Q_b Q^{-1}_a  \mathbb{L}_{ab}(t-\Delta) \delta(\imp^1-t)\mathbf{a}_{a \alpha}(y)\\
&+i \delta_{\beta 1}  \frac{d}{dx} \mathbf{a}_{a \alpha}(x,y)\Bigl|_{x=\imp^1} \mathbf{b}_{\beta b}(\imp^1 ,t) +i \delta_{\alpha 1} \mathbf{a}_{a\alpha}(\imp^1 ,y) \frac{d}{dx} \mathbf{b}_{\beta b}(x,t)\Bigl|_{x=\imp^1}  \,,
\end{split}
\end{equation}
\begin{equation}
\label{eq:bb_main_xD1_dyn}
\begin{split}
&\frac{1}{\var} \{\mathbf{b}_{\alpha a}(y),\mathbf{b}_{\beta b}(t)\} =-\rho_{\alpha \beta, \beta \alpha}(y-t)
\mathbf{b}_{\beta a}(y) \mathbf{b}_{\alpha b}(t) \\
&+i \mathbf{b}_{\alpha b}(y) \mathbf{b}_{\beta a}(t)   \varphi(y-t+\Delta, \mathfrak{q}_{ba})   +i \mathbf{b}_{\alpha a}(y) \varphi^{\text{reg}}(\mathfrak{q}_{ba})    \mathbf{b}_{\beta b}(t) \\
&+\mathbf{b}_{\beta b}(t)  \mathbf{a}_{b \gamma}(\imp^1) \rho(y-\imp^1)_{ + \alpha \gamma, \gamma 1} \mathbf{b}_{\gamma a}(y)  -\mathbf{b}_{\alpha a}(y)  \mathbf{a}_{a \gamma}(\imp^1)  \rho(t-\imp^1)_{+ \beta \gamma, \gamma 1} \mathbf{b}_{\gamma b}(t)\\
&+\frac{2 \pi}{\var} \delta_{\alpha 1}  \delta(y-\imp^1)  \, \mathbf{b}_{\beta b}(t) Q^{-1}_b    \mathbb{L}_{ba}(y-\Delta) Q_a -\frac{2 \pi }{\var} \delta_{1 \beta} \mathbf{b}_{\alpha a}(y) Q^{-1}_a \mathbb{L}_{ab}(t-\Delta) Q_b \delta(\imp^1-t)\\
&+i \delta_{\beta 1}\frac{d}{dx}\mathbf{b}_{\alpha a}(x,y) \Bigl|_{x=\imp^1} \mathbf{b}_{\beta b}(\imp^1,t)-i \delta_{\alpha 1} \mathbf{b}_{\alpha a}(\imp^1,y) \frac{d}{dx}\mathbf{b}_{\beta b}(x,t) \Bigl|_{x=\imp^1} \,.
\end{split}
\end{equation}
One might be concerned that for $\alpha$ or $\beta$ equal to $1$ the derivatives are still there. However, it turns out that these residual derivatives play an important role in the localization of the second loop variable. We will come back to this point in the next section.
A complete computation of the collinear limit $x=z=\imp^1$ and derivation of formulas~\eqref{eq:aa_main_xD1_dyn}, \eqref{eq:ab_main_xD1_dyn} and~\eqref{eq:bb_main_xD1_dyn} is provided in Appendices~\ref{app:br_inv_spins} and~\ref{app:dyn_imp}.

\section{Finite dimensional realisation of the invariant oscillators}
\label{sec:finite_dim}

Following Appendix~\ref{app:fin_dim_with_imp}, in this section we provide a finite-dimensional realization of oscillator brackets~\eqref{eq:aa_main_xD1_dyn}, \eqref{eq:ab_main_xD1_dyn} and~\eqref{eq:bb_main_xD1_dyn}.
The finite-dimensional realization involves additional choices beyond
the solution of the moment-map constraint: a localization ansatz for
the invariant oscillators and prescriptions for their singular
products. We adopt the regularization described below and in
Appendix~\ref{app:fin_dim_with_imp}. Within this realization, cancellation
of derivative terms imposes nontrivial Poisson brackets between the
localization points and the spin variables.
As already mentioned, such a finite-dimensional realization is provided by the following representation of the $\mathbf{b}$ oscillators
\begin{equation}
\label{eq:b_representation}
\mathbf{b}_{\alpha a}(x) \equiv \mathbf{b}_{\alpha a} \delta_\epsilon(x-\imp^\alpha) \,,
\end{equation}
and by localising the $\mathbf{a}$ oscillators as follows
\begin{equation}
\mathbf{a}_{a \alpha} \equiv \mathbf{a}_{a \alpha}(\imp^\alpha) \,.
\end{equation}
The parameters $\{\imp^\alpha\}^\ell_{\alpha=1}$ are the inhomogeneities previously introduced in Section~\ref{sec:finite_dim_osc_alg} to solve the moment map. The function $\delta_\epsilon$ in~\eqref{eq:b_representation} is given by:
\begin{equation}
\label{eq:reg_Dirac_Delta_main}
\delta_\epsilon(x)= \frac{1}{2 \pi} \frac{\sinh \epsilon}{\cosh \epsilon - \cos{x}} \,,
\end{equation}
and in the limit $\epsilon \to 0$ becomes a Dirac delta function of period $2 \pi$. The importance of keeping the regulator $\epsilon$ in~\eqref{eq:reg_Dirac_Delta_main} comes from the fact that certain terms on the RHS of~\eqref{eq:ab_main_xD1_dyn} and~\eqref{eq:bb_main_xD1_dyn} are singular in the limit $\epsilon \to 0$. 
We should then expand the bracket in an $\epsilon$ expansion, show that at the order $1/\epsilon$ all terms that are separately singular sum to zero, and only at the end send $\epsilon \to 0$. A detailed computation of this mechanism is explained in Appendix~\ref{app:fin_dim_with_imp}. 
Here, we only provide the main steps of the derivation. In particular, we will show that such a localization procedure is possible only if we assume that $\rho(x)= \check{\rho}(x)$, while it fails for $\rho(x)= \hat{\rho}(x)$.
The explicit expressions for these matrices are provided in Appendix~\ref{app:rho_matrices}.

\subsection{Complete removal of the loop parameters}

Now we show how it is possible to localize the brackets of the spin oscillators at the inhomogeneity points. Performing such a procedure for all the brackets is quite long, and here we only consider a particular instance of the $\mathbf{b}$-$\mathbf{b}$ bracket. In particular, we focus on the case $\alpha, \beta \ne 1$.
In such a situation, the last two lines in~\eqref{eq:bb_main_xD1_dyn} vanish.
Moreover, both $\hat{\rho}(y-\imp^1)_{ + \alpha \gamma, \gamma 1}$ and $\check{\rho}(y-\imp^1)_{ + \alpha \gamma, \gamma 1}$ are proportional to $\delta_{\alpha 1}$ (and similarly $\rho(t-\imp^1)_{+ \beta \gamma, \gamma 1}$, with either $\rho=\hat{\rho}$ or $\rho=\check{\rho}$, is proportional to $\delta_{\beta 1}$). Due to this fact, the third line in~\eqref{eq:bb_main_xD1_dyn} also cancels, and for $\alpha, \beta \ne 1$, we can write 
\begin{equation}
\label{eq:beforeexpansion_bb_Diraceps}
\begin{split}
&\frac{1}{\var} \{\mathbf{b}_{\alpha a}(y),\mathbf{b}_{\beta b}(t)\} =-\rho_{\alpha \beta, \beta \alpha}(y-t)
\mathbf{b}_{\beta a}(y) \mathbf{b}_{\alpha b}(t) \\
&+i \mathbf{b}_{\alpha b}(y) \mathbf{b}_{\beta a}(t)   \varphi(y-t+\Delta, \mathfrak{q}_{ba})   +i \mathbf{b}_{\alpha a}(y) \varphi^{\text{reg}}(\mathfrak{q}_{ba})    \mathbf{b}_{\beta b}(t) \,.
\end{split}
\end{equation}
Since we assume $\imp^\alpha$ and $\imp^\beta$ to be dynamical, when we replace the representation~\eqref{eq:b_representation} into the bracket above, we need to take into account that
\begin{equation}
\label{eq:expansion_bb_Diraceps}
\begin{split}
&\frac{1}{\var} \{\mathbf{b}_{\alpha a}(y),\mathbf{b}_{\beta b}(t)\}=\\
&+\frac{1}{\var} \{\mathbf{b}_{\alpha a},\mathbf{b}_{\beta b}\} \delta_\epsilon(y-\imp^\alpha) \delta_\epsilon(t-\imp^\beta)+\frac{1}{\var} \{\imp^\alpha,\imp^\beta\} \mathbf{b}_{\alpha a} \mathbf{b}_{\beta b} \delta'_\epsilon(y-\imp^\alpha) \delta'_\epsilon(t-\imp^\beta)\\
&-\frac{1}{\var} \{\imp^\alpha ,\mathbf{b}_{\beta b}\} \mathbf{b}_{\alpha a} \delta'_\epsilon(y-\imp^\alpha) \delta_\epsilon(t-\imp^\beta)+\frac{1}{\var} \{\imp^\beta ,\mathbf{b}_{\alpha a}\} \mathbf{b}_{\beta b} \delta_\epsilon(y-\imp^\alpha) \delta'_\epsilon(t-\imp^\beta)
\end{split}
\end{equation}
Combining~\eqref{eq:beforeexpansion_bb_Diraceps} and~\eqref{eq:expansion_bb_Diraceps} we see that
\begin{equation}
\label{eq:afterexpansion_bb_Diracepsn}
\begin{split}
&\frac{1}{\var} \{\mathbf{b}_{\alpha a},\mathbf{b}_{\beta b}\} \delta_\epsilon(y-\imp^\alpha) \delta_\epsilon(t-\imp^\beta)=-\rho_{\alpha \beta, \beta \alpha}(y-t)
\mathbf{b}_{\beta a} \mathbf{b}_{\alpha b}\delta_\epsilon(y-\imp^\beta) \delta_\epsilon(t-\imp^\alpha) \\
&+\Bigl(i \mathbf{b}_{\alpha b} \mathbf{b}_{\beta a}   \varphi(y-t+\Delta, \mathfrak{q}_{ba})    +i \mathbf{b}_{\alpha a} \varphi^{\text{reg}}(\mathfrak{q}_{ba})    \mathbf{b}_{\beta b} \Bigr) \delta_\epsilon(y-\imp^\alpha) \delta_\epsilon(t-\imp^\beta) \\
&-\frac{1}{\var} \{\imp^\alpha,\imp^\beta\} \mathbf{b}_{\alpha a} \mathbf{b}_{\beta b} \delta'_\epsilon(y-\imp^\alpha) \delta'_\epsilon(t-\imp^\beta)\\
&+\frac{1}{\var} \{\imp^\alpha ,\mathbf{b}_{\beta b}\} \mathbf{b}_{\alpha a} \delta'_\epsilon(y-\imp^\alpha) \delta_\epsilon(t-\imp^\beta)-\frac{1}{\var} \{\imp^\beta ,\mathbf{b}_{\alpha a}\} \mathbf{b}_{\beta b} \delta_\epsilon(y-\imp^\alpha) \delta'_\epsilon(t-\imp^\beta)
\end{split}
\end{equation}
The Derivative of the regularized Dirac delta function might look suspicious, and it might seem tempting to assume $\imp^\alpha$ and $\imp^\beta$ commute trivially with the oscillators.
However, this is not the case, and in a moment we will explain why.

Let us first evaluate the expression above for $\alpha \ne \beta$ (and both different from $1$, as already specified).
In this case, all functions on the RHS of~\eqref{eq:afterexpansion_bb_Diracepsn} are well defined and we can take the limit $\epsilon \to 0$ smoothly, getting
\begin{equation}
\label{eq:afterexpansion_bb_Diraceps}
\begin{split}
&\frac{1}{\var} \{\mathbf{b}_{\alpha a},\mathbf{b}_{\beta b}\}_{\alpha \ne \beta} \delta(y-\imp^\alpha) \delta(t-\imp^\beta)=-\rho_{\alpha \beta, \beta \alpha}(\imp^{\beta \alpha})
\mathbf{b}_{\beta a} \mathbf{b}_{\alpha b}\delta(y-\imp^\beta) \delta(t-\imp^\alpha) \\
&+\Bigl(i \mathbf{b}_{\alpha b} \mathbf{b}_{\beta a}   \varphi(\imp^{\alpha \beta}+\Delta, \mathfrak{q}_{ba})    +i \mathbf{b}_{\alpha a} \varphi^{\text{reg}}(\mathfrak{q}_{ba})    \mathbf{b}_{\beta b} \Bigr) \delta(y-\imp^\alpha) \delta(t-\imp^\beta) \\
&-\frac{1}{\var} \{\imp^\alpha,\imp^\beta\}_{\alpha \ne \beta} \mathbf{b}_{\alpha a} \mathbf{b}_{\beta b} \delta'(y-\imp^\alpha) \delta'(t-\imp^\beta)\\
&+\frac{1}{\var} \{\imp^\alpha ,\mathbf{b}_{\beta b}\}_{\alpha \ne \beta} \mathbf{b}_{\alpha a} \delta'(y-\imp^\alpha) \delta(t-\imp^\beta)-\frac{1}{\var} \{\imp^\beta ,\mathbf{b}_{\alpha a}\}_{\alpha \ne \beta} \mathbf{b}_{\beta b} \delta(y-\imp^\alpha) \delta'(t-\imp^\beta) \,,
\end{split}
\end{equation}
where we use the definition $\imp^{\alpha \beta} \equiv \imp^\alpha-\imp^\beta$.
We start by noting that the matrix element $\rho_{\alpha \beta, \beta \alpha}$ exchanges the inhomogeneities inside the Dirac delta function. On the LHS of~\eqref{eq:afterexpansion_bb_Diraceps} we have indeed
$\delta(y-\imp^\alpha) \delta(t-\imp^\beta)$, while the $\rho$ matrix comes multiplied by $\delta(y-\imp^\beta) \delta(t-\imp^\alpha)$. The only possibility for the equality above to be satisfied is then $\rho_{\alpha \beta, \beta \alpha}(\imp^{\beta \alpha})=0$, for $\alpha \ne \beta$. From the expressions in~\eqref{eq:app_hatrho_comp} and~\eqref{eq:app_check_rho_comp} we immediately see that such a condition is verified by $\check{\rho}$ and not by $\hat{\rho}$. The reduction to a finite number of degrees of freedom in the presence of inhomogeneities then requires the matrix $\check{\rho}$. From~\eqref{eq:afterexpansion_bb_Diraceps}, we are then forced to set
\begin{equation}\{\imp^\alpha,\imp^\beta\}_{\alpha \ne \beta}=\{\imp^\alpha ,\mathbf{b}_{\beta b}\}_{\alpha \ne \beta}=0\,,
\end{equation}
and we get
\begin{equation}
\label{eq:bb_aneb_noDir}
\frac{1}{\var} \{\mathbf{b}_{\alpha a},\mathbf{b}_{\beta b}\}_{1 \ne \alpha \ne \beta \ne 1} =i \mathbf{b}_{\alpha b} \mathbf{b}_{\beta a}   \varphi(\imp^{\alpha \beta}+\Delta, \mathfrak{q}_{ba})    +i \mathbf{b}_{\alpha a} \mathbf{b}_{\beta b} \varphi^{\text{reg}}(\mathfrak{q}_{ba})      \,.
\end{equation}

Now we consider the case $\alpha=\beta$, for which the bracket~\eqref{eq:afterexpansion_bb_Diracepsn} becomes
\begin{equation}
\label{eq:bb_with_Dir_alphaeqbeta}
\begin{split}
&\frac{1}{\var} \{\mathbf{b}_{\alpha a},\mathbf{b}_{\alpha b}\} \delta_\epsilon(y-\imp^\alpha) \delta_\epsilon(t-\imp^\alpha)= \\
&+\mathbf{b}_{\alpha a} \mathbf{b}_{\alpha b}\Bigl(-\check{\rho}_{\alpha \alpha, \alpha \alpha}(y-t)+i    \varphi(y-t+\Delta, \mathfrak{q}_{ba})    +i  \varphi^{\text{reg}}(\mathfrak{q}_{ba})     \Bigr) \delta_\epsilon(y-\imp^\alpha) \delta_\epsilon(t-\imp^\alpha) \\
&+\frac{1}{\var} \{\imp^\alpha ,\mathbf{b}_{\alpha b}\} \mathbf{b}_{\alpha a} \delta'_\epsilon(y-\imp^\alpha) \delta_\epsilon(t-\imp^\alpha)-\frac{1}{\var} \{\imp^\alpha ,\mathbf{b}_{\alpha a}\} \mathbf{b}_{\alpha b} \delta_\epsilon(y-\imp^\alpha) \delta'_\epsilon(t-\imp^\alpha)
\end{split}
\end{equation}
In such a situation, the matrix $\check{\rho}$ plays the important role of generating derivatives of Dirac delta functions
in the bracket. Taking both $\varphi(y-t+\Delta, \mathfrak{q}_{ba})$ and  $\check{\rho}$ in the principal value prescription, we obtain the following result in the $\epsilon \to 0$ limit:
\begin{equation}
\begin{split}
&\lim_{\epsilon \to 0^+} \, \Bigl(-\check{\rho}_{\alpha \alpha, \alpha \alpha}(y-t)+i    \varphi(y-t+\Delta, \mathfrak{q}_{ba})     \Bigr) \delta_\epsilon(y-\imp^\alpha) \delta_\epsilon(t-\imp^\alpha) = \\
&+i  
  \varphi^{\text{reg}}( \mathfrak{q}_{ba})    \delta(y-\imp^\alpha)\delta(t-\imp^\alpha)-i \left(\delta'(y-\imp^\alpha) \delta(t-\imp^\alpha) - \delta(y-\imp^\alpha) \delta'(t-\imp^\alpha) \right) \,.
\end{split}
\end{equation}
Without the $\rho$ matrix, the second term on the RHS of the expression above would have had a $1/2$ coefficient.
We invite the reader to look at Appendix~\ref{app:bb_bracket_finite} for a more detailed derivation of the limit.
From such a limit, we notice that all the derivatives of the Dirac delta functions disappear from~\eqref{eq:bb_with_Dir_alphaeqbeta} if we impose
\begin{equation}
\frac{1}{\var} \, \{\imp^\alpha, \mathbf{b}_{\alpha b} \}_{\alpha \ne 1} = +i \mathbf{b}_{\alpha b} \,.
\end{equation}
Then by matching the coefficients of the Dirac delta functions on the LHS and RHS of~\eqref{eq:bb_with_Dir_alphaeqbeta} we obtain
\begin{equation}
\label{eq:bb_with_Dir_alphaeqbeta2}
\frac{1}{\var} \{\mathbf{b}_{\alpha a},\mathbf{b}_{\alpha b}\}_{\alpha \ne 1} = 
2 \mathbf{b}_{\alpha a} \mathbf{b}_{\alpha b}  i  \varphi^{\text{reg}}(\mathfrak{q}_{ba})      
\end{equation}

Similar results can be obtained for the remaining oscillator brackets and for the cases $\alpha, \beta=1$. A complete treatment is reported in Appendix~\ref{app:fin_dim_with_imp}. In the following subsection we summarise the obtained results for the different brackets and index choices.

\subsection{Summary of the results}

After using the representation~\eqref{eq:b_representation}, evaluating the $\mathbf{a}$ oscillators at the inhomogeneities and taking the $\epsilon \to 0^+$ limit, 
the localized brackets for the invariant oscillators turn out to be 
\begin{equation}
\label{eq:oscillator_final_inhom}
\begin{split}
\frac{1}{\var} \, \{\mathbf{a}_{a \alpha} \,,\mathbf{a}_{b \beta} \}=&+ i (1-\delta_{\alpha \beta})\varphi(\imp^{\beta \alpha}+\Delta,\mathfrak{q}_{ba})\mathbf{a}_{b \alpha} \mathbf{a}_{a \beta}+i (1+\delta_{\alpha \beta}) \varphi^{\text{reg}}(\mathfrak{q}_{ba})\mathbf{a}_{a \alpha} \mathbf{a}_{b \beta}  \\
&- i \varphi(\imp^{\beta1}+\Delta,\mathfrak{q}_{ba})\mathbf{a}_{a \alpha} \mathbf{a}_{a \beta}- i \varphi(\imp^{1\alpha}+\Delta,\mathfrak{q}_{ba})\mathbf{a}_{b \alpha} \mathbf{a}_{b \beta} \,,\\
\frac{1}{\var} \{\mathbf{b}_{\alpha a},\mathbf{b}_{\beta b}\}  =&+i (1+\delta_{\alpha \beta})  
  \varphi^{\text{reg}}( \mathfrak{q}_{ba})   \mathbf{b}_{\alpha a} \mathbf{b}_{\beta b}  +(1-\delta_{\alpha \beta}) i \mathbf{b}_{\alpha b} \mathbf{b}_{\beta a}   \varphi(\imp^{\alpha \beta}+\Delta, \mathfrak{q}_{ba})\\
&+\delta_{\alpha 1} \mathbb{L}^{\text{reg},\alpha}_{ba} \mathbf{b}_{\beta b} -\delta_{\beta 1} \mathbb{L}^{\text{reg},\beta}_{ab} \mathbf{b}_{\alpha a}   \,,\\
\frac{1}{\var} \, \{\mathbf{a}_{a \alpha}, \mathbf{b}_{\beta b}\}=&+i \varphi(\imp^{1\alpha}+\Delta,\mathfrak{q}_{ba})   \mathbf{a}_{b \alpha} \mathbf{b}_{\beta b}  -i \varphi^{\text{reg}}(\qq_{ba}) \mathbf{a}_{a \alpha} \mathbf{b}_{\beta b} \\
&-\delta_{\alpha \beta} \mathbb{L}^{\text{reg},\alpha}_{ab} +\delta_{\beta 1} \mathbb{L}^{\text{reg},1}_{ab} \mathbf{a}_{a \alpha} \,,
\end{split}
\end{equation}
where we defined
\begin{equation}
\label{eq:Lreg_ba_alpha}
\mathbb{L}^{\text{reg},\alpha}_{ba} =\frac{1}{i}  \sum_{\gamma \ne \alpha}  \varphi(\imp^{\alpha\gamma}+\Delta, \qq_{ba} - \eta) \mathbf{a}_{b \gamma} \mathbf{b}_{\gamma a}+\frac{1}{i}  \varphi^{\text{reg}}(\qq_{ba} - \eta) \mathbf{a}_{b \alpha} \mathbf{b}_{\alpha a} \,.
\end{equation}
The normalization in~\eqref{eq:Lreg_ba_alpha} absorbs the common factor
$\widetilde{\var}/2\pi$ appearing in the regular part of $\mathbb{L}_{ba}(u)$ when expanding $u$ around $\Delta+\imp^\alpha$: 
\begin{equation}
\mathbb{L}_{ba}(\Delta+\imp^\alpha+u)=\frac{\widetilde{\var}}{2\pi i}
\frac{\mathbf{a}_{b\alpha}\mathbf{b}_{\alpha a}}{u}
+
\frac{\widetilde{\var}}{2\pi}
\mathbb{L}^{\mathrm{reg},\alpha}_{ba}+O(u).
\end{equation}
As already remarked, the spin indices $\alpha, \beta$ are assumed to be different from one in the $\mathbf{a}$ oscillators, since $\mathbf{a}_{i1}=1$. The brackets above are then always well defined.

The inhomogeneities are dynamical, pairwise distinct modulo $2\pi$,
and solve
\begin{equation}
\label{eq:Delta_a_b_br}
   \frac{1}{\var} \, \{ \imp^\alpha, \mathbf{a}_{i\beta} \} = -i(\delta_{\alpha\beta}-\delta_{\alpha1}) \mathbf{a}_{i\beta}, \qquad \frac{1}{\var} \, \{ \imp^\alpha, \mathbf{b}_{\beta i} \} = i(\delta_{\alpha\beta}-\delta_{\alpha1}) \mathbf{b}_{\beta i} \,, \qquad \frac{1}{\var} \, \{ \imp^\alpha, \imp^\beta \}=0 \,.
\end{equation}
It is possible to check that replacing the localized oscillators into the $\mathbb{L}$ operator~\eqref{eq:Loperator_specific_sol}, and using the relations~\eqref{eq:oscillator_final_inhom}, we obtain the Poisson bracket~\eqref{eq:LL_bracket_main}. This is a self-consistent check of the correctness of our derivation.
Finally, following a standard route, we also obtain
\begin{equation}
\label{eq:bq_br}
\frac{1}{\var} \{\qq_i, \qq_j \} = \frac{1}{\var} \{\mathbf{a}_{k\alpha}, \qq_j \} =\frac{1}{\var} \{\imp^\alpha, \qq_j \} =0 \,, \qquad \frac{1}{\var} \{\mathbf{b}_{\alpha k}, \qq_j \} =i \delta_{kj} \mathbf{b}_{\alpha k} \,,
\end{equation}
obtaining in this manner the complete Poisson brackets for the elliptic spin RS models in the presence of inhomogeneities. The results obtained in this paper match those derived in~\cite{Arutyunov:2026yjy}, as detailed in Appendix~\ref{app:AH_comparison}.

The Hamiltonian corresponding to that of~\cite{Arutyunov:2026yjy} is
\begin{equation}
H_0=\sum_{i=1}^{N}\sum_{\alpha=1}^{\ell}
\mathbf a_{i\alpha}\mathbf b_{\alpha i}.
\end{equation}
Using~\eqref{eq:Loperator_specific_sol}, it can be expressed as a sum of residues of the Lax matrix:
\begin{equation}
H_0=
\frac{2\pi i}{\widetilde\var}
\sum_{\alpha=1}^{\ell}
\operatorname{Res}_{y=\imp^\alpha+\Delta}
\operatorname{Tr}\mathbb L(y).
\end{equation}
With the convention 
\begin{equation}
\dot{F}=\frac{1}{i \var} \, \{F,H_0 \} \,,
\end{equation}
equations~\eqref{eq:Delta_a_b_br} and~\eqref{eq:bq_br} give
\begin{equation}
\dot{\qq}_i=- \sum_{\alpha=1}^{\ell}
\mathbf{a}_{i\alpha}\mathbf{b}_{\alpha i},
\qquad
\dot\imp^\alpha=0.
\end{equation}
The inhomogeneities therefore remain constant along this Hamiltonian flow, although they have nontrivial Poisson brackets with the spin variables. Under the identification with~\cite{Arutyunov:2026yjy}, the resulting dynamics reproduces the elliptic spin RS equations with inhomogeneities, up to the constant rescaling of time associated with the normalization of the Poisson bracket.

The comparison in Appendix~\ref{app:AH_comparison} identifies the
resulting finite-dimensional brackets with those of~\cite{Arutyunov:2026yjy}. This provides a check of the Jacobi identity of the 
resulting Poisson algebra, distinct from a direct verification of the intermediate
distributional manipulations.

\subsection{On the dimensionality of the phase space}

The generic symplectic leaves are expected to have dimension $2N \ell$, in agreement with~\cite{Arutyunov:2026yjy}. We identify below $\ell$ Casimir functions. Assuming their functional independence on a generic locus, they imply the upper bound $\operatorname{rank}P\leq 2N\ell$ for the Poisson bivector $P$ that we will introduce below.

Let
\begin{equation}
D(u) = \text{det}\,  \mathbb{L}(u+\Delta) 
\end{equation}
be the determinant of the Lax operator.
From the solution of the moment map we know that $\mathbb{L}(u+\Delta)$ has $\ell$ singularities on the interval $(-\pi, +\pi)$ and the residue at each singularity has rank one, namely for $u \simeq \imp^\alpha$
\begin{equation}
\mathbb{L}(u+\Delta) = \frac{R_\alpha}{u-\imp^\alpha} \,, \qquad \text{rank}\,  R_\alpha=1 \,.
\end{equation}
Because of this fact, the determinant $D(u)$ has $\ell$ simple poles at $u=\imp^1, \imp^2, \dots, \imp^\ell$. The quasi-periodicity of $\mathbb{L}$ also implies it has $\ell$ zeros that we label by $u=\nu_1, \nu_2, \dots, \nu_\ell$. We can then write the determinant as
\begin{equation}
D(u)=A \frac{\prod\limits^\ell_{\gamma=1} \theta(u-\nu_\gamma)}{\prod\limits^\ell_{\gamma=1} \theta(u-\imp^\gamma)} \,.
\end{equation}
Here
\begin{equation}
\theta(u):= e^{-\frac{\zeta(\pi)}{2 \pi} u^2} \, \sigma(u) \,,
\end{equation}
is the normalized\footnote{The normalization is introduced to satisfy the convention $\varphi(u+\Delta, s)= \frac{\theta(u+s)}{\theta(u) \theta(s)}$.} odd Jacobi theta function, and $A$ is a dynamical quantity independent of $u$ which can be written in terms of the invariant oscillators. Defining
\begin{equation}
\mathcal{F}(u):= D(u) \prod^\ell_{\gamma=1} \theta(u-\imp^\gamma) \,,
\end{equation}
one can check that any element $X$ of the Poisson algebra satisfies
\begin{equation}
\label{eq:FX_comm}
\{\mathcal{F}(u), X \}= g_X \mathcal{F}(u) \,,
\end{equation}
where $g_X$ depends on the chosen element but it does not depends on $u$. Let us pick up a generic regular point $u_*$. Because of~\eqref{eq:FX_comm}, then it holds
\begin{equation}
\Big\{\frac{\mathcal{F}(u)}{\mathcal{F}(u_*)}, X \Big\} =0 \,.
\end{equation}
Expanding the quantity above around a zero of $\mathcal{F}$, say $\nu_\alpha \in \{\nu_1, \dots, \nu_\ell \}$, we get
\begin{equation}
0=\Big\{\frac{\mathcal{F}(\nu_\alpha)}{\mathcal{F}(u_*)}, X \Big\} - \frac{\mathcal{F}'(\nu_\alpha)}{\mathcal{F}(u_*)} \{\nu_\alpha, X \} \,,
\end{equation}
from which it follows that
\begin{equation}
\{\nu_\alpha, X \} =0 \,, \qquad \alpha=1, 2, \dots, \ell \,.
\end{equation}
We have identified in this manner $\ell$ Casimirs, each one associated to the position of a zero of $D(u)$. The symplectic phase space is obtained by fixing these Casimirs. The initial number of independent coordinates (having fixed $\mathbf{a}_{i 1}=1$) is 
\begin{equation}
\underbrace{N}_{\qq_i} \, + \, \underbrace{N (\ell-1)}_{\mathbf{a}_{i \alpha}} \, + \,\underbrace{N \ell}_{\mathbf{b}_{\alpha i}} \, + \, \underbrace{\ell}_{\imp^\alpha} = 2N \ell + \ell \,.
\end{equation}
At regular points where these $\ell$ Casimirs are functionally independent, fixing their values defines a submanifold of dimension $2N \ell$, which is even as expected for a symplectic space. 

However, we should remark that the existence of $\ell$ Casimirs does not automatically ensure that the submanifold obtained by fixing their values is a symplectic leaf. If we label by $z^A$ the coordinates in the Poisson space, then we can introduce the matrix 
\begin{equation}
P^{A B} := \{z^A, z^B \} \,,
\end{equation}
and define the Poisson bivector as
\begin{equation}
P=\frac{1}{2} P^{AB} \frac{\partial}{\partial z_A} \wedge \frac{\partial}{\partial z_B} \,. 
\end{equation}
Clearly, a Casimir belongs to the kernel of $P^{AB}$:
\begin{equation}
\{z^A, C\} = P^{AB} \partial_B C =0 \,.
\end{equation}
Starting from a Poisson space having dimensionality $d=2N \ell + \ell$, the existence of $\ell$ independent Casimirs sets a lower bound on the dimensionality of the kernel of $P^{AB}$, and an upper bound on its rank:
\begin{equation}
\text{dim} (\text{ker} \, P^{AB}) \ge \ell \implies \text{rank} P^{AB} \le 2 N \ell \,.
\end{equation}
The submanifold obtained by fixing the values of the $\ell$ Casimirs is locally a symplectic leaf if the Poisson bivector $P^{AB}$ has rank $2N \ell$; otherwise, it is further foliated by lower-dimensional symplectic leaves.
Establishing that these level sets are generically symplectic therefore requires showing that $\operatorname{rank}P=2N\ell$. The numerical calculations of~\cite{Arutyunov:2026yjy} establish this rank for the cases examined with $N,\ell\leq3$, supporting the expected generic dimension. The Casimir argument alone does not establish saturation of the rank bound for arbitrary $N$ and $\ell$.

It is also worth remarking that the quasi-periodicity of $\mathbb{L}$ requires the following constraint on its poles and zeros:
\begin{equation}
\sum^\ell_{\gamma=1} \nu_\gamma = \sum^\ell_{\gamma=1} \imp^\gamma +N \eta \qquad \text{mod} \,2 \pi \,.
\end{equation}
Because of this, we can identify the central element 
\begin{equation}
\mathcal{C}:=\sum^\ell_{\gamma=1} \imp^\gamma \,.
\end{equation}
This element is a function of $\nu_\gamma$, and not an extra Casimir.

\section{Conclusions}
\label{sec:conclusions}
We have constructed the Poisson structure of the elliptic spin
Ruijsenaars--Schneider model with dynamical inhomogeneities by reducing
the affine Heisenberg double coupled to an affine oscillator manifold.
After solving the moment-map constraint, we derived brackets for the
invariant Lax operator and spin variables. Localizing the invariant
oscillators at distinct dynamical points then produced a
finite-dimensional Poisson algebra and its Hamiltonian. The resulting
brackets agree with those obtained in \cite{Arutyunov:2026yjy} from the elliptic
Coulomb branch of the necklace quiver. Our reduction thus gives an
independent derivation of that Hamiltonian structure.

The dynamical character of the inhomogeneities is essential: their
brackets with the spin variables cancel the derivative terms that
arise when the affine oscillators are localized. The equations of
motion nevertheless admit a homogeneous limit
\(\imp^\alpha\to 0\), recovering the Krichever--Zabrodin equations~\cite{Krichever:1995zw}. The same substitution does not define a Poisson bracket
for the homogeneous model. Indeed, the constraints identifying the
inhomogeneities do not form a Poisson ideal; for example,
\begin{equation}
\lim_{\imp^2 \to \imp^1} \,\frac{1}{\var} \, \{\imp^1-\imp^2, \mathbf{a}_{j 2}\}= 2i  \mathbf{a}_{j 2} \ne 0 \,.
\end{equation}
Thus the homogeneous limit of the dynamics must be distinguished from
a reduction of its Hamiltonian structure.

This distinction points to the principal open problem: to construct
Poisson brackets for the natural variables of the {\it homogeneous} elliptic
spin RS model. A further reduction of the inhomogeneous phase space
or another finite-dimensional realization of the invariant oscillator
algebra may provide such brackets. The choice 
\(\hat\rho\), discussed
in Appendix G, removes certain singularities but does not cancel the
derivative terms generated by localization. More generally, the
zero-mode realization connects the affine construction to the
hyperbolic spin RS model, while localization at distinct dynamical
points gives the inhomogeneous elliptic model. It would be valuable
to classify other finite-dimensional realizations and determine
which of them admit an integrable Hamiltonian interpretation.

The agreement with the elliptic Coulomb-branch brackets also invites
a more direct geometric comparison of the two phase spaces, including
their symplectic leaves and nongeneric loci. 
It would also be interesting to quantize the affine oscillator algebra
and the associated reduction, with the aim of constructing a quantum
counterpart of the elliptic spin RS model. Independently, at the
classical level, one may ask whether the Poisson structures derived
here can be applied to the Ruijsenaars chain of \cite{zabrodin_field_2022, trunina_lax_2022, matushko_matrix_2026}.

\section*{Acknowledgements}

We are grateful to S.~Frolov for valuable discussions.
We acknowledge support by the DFG -- SFB 1624 -- ``Higher structures, moduli spaces and integrability'' -- 506632645. G.A. acknowledges support by the DFG under Germany's Excellence Strategy -- EXC 2121 ``Quantum Universe'' -- 390833306.

\appendix

\section{Definitions}
\label{app:definitions}

\subsection{Matrices on the finite-dimensional space}
\label{app:matrix_bblocks}

We define $E_{ij}$ to be a $N \times N$ matrix being $1$ on the $i^{\text{th}}$ row and $j^{\text{th}}$ column and zero everywhere else:
\begin{equation}
(E_{ij})_{k l} = \delta_{i k} \delta_{jl} \,.
\end{equation}
We also define the following matrices
\begin{equation}
\label{eq:F_matrix_def}
F_{ij}= E_{ii}-E_{ij} \,,
\end{equation}
which satisfies
\begin{equation}
F_{ij} e =0 \,, \qquad \text{for} \qquad e= (1, 1, \dots, 1) \,. 
\end{equation} 
The identity and permutation matrices can be written in terms of these elements as
\begin{equation}
I = \sum^N_{i, j=1} E_{ii} \otimes E_{jj}\, \qquad C_{12}= \sum_{i,j=1}^N E_{i j} \otimes E_{j i}
\end{equation}
and acts on $\mathbb{C}^N \otimes \mathbb{C}^N$ as 
\begin{equation}
I (u \otimes v)= u \otimes v\,, \qquad C_{12} (u \otimes v)= v \otimes u\,.
\end{equation}
We also refer to the permutation operator $C_{12}$ as the split Casimir.
Following~\cite{Arutyunov:1996uw} we also define the following matrices
\begin{equation}
\begin{split}
r_+&=+\frac{1}{2} \sum_{i} E_{ii} \otimes E_{ii} + \sum_{i<j} E_{ij} \otimes E_{ji}\,,\\
r_-&=-\frac{1}{2} \sum_{i} E_{ii} \otimes E_{ii} - \sum_{i>j} E_{ij} \otimes E_{ji} = -C_{12} r_+ C_{12} \,,\\
r&=\frac{1}{2} \left( r_+ + r_- \right)\,.
\end{split}
\end{equation}
The matrices above satisfy
\begin{equation}
r_+ - r_-= C_{12}\,, \qquad C_{12} r_+ C_{12}=-r_-\,.
\end{equation}

Let us now define the matrices acting on the space 
\begin{equation}
\Sigma_{N, \ell}= \text{Mat}_{N, \ell}(\mathbb{C}) \times \text{Mat}_{\ell, N}(\mathbb{C})
\end{equation}
Analogously to before, we define the rectangular matrices $E_{i \alpha}$ and $E_{\alpha i}$ ($\alpha=1, \dots, \ell$, $i=1, \dots, N$) having as components
\begin{equation}
(E_{i \alpha})_{j \beta}= \delta_{i j} \delta_{\alpha \beta} \qquad \text{and} \qquad (E_{\alpha i})_{\beta j}= \delta_{\alpha \beta} \delta_{i j}\,.
\end{equation}
The identity and permutation operators on this space are defined by
\begin{equation}
I^s = \sum^\ell_{\a , \b=1} E_{\a \a} \otimes E_{\b \b}\, \qquad C^s_{12}= \sum_{\a,\b=1}^\ell E_{\a \b} \otimes E_{\b \a}\,,
\end{equation}
while the ``rectangular split Casimirs'' are
\begin{equation}
C^{\text{rec}}_{12}= \sum^N_{i=1} \sum^\ell_{\alpha=1} E_{i \alpha} \otimes E_{\alpha i}\,, \qquad
C^{\text{rec}}_{21}= \sum^N_{i=1} \sum^\ell_{\alpha=1} E_{\alpha i} \otimes E_{i \alpha}\,.
\end{equation}

The analogue of the $r$ matrices in this space are 
\begin{equation}
\label{eq:app_rho_functions}
\begin{split}
\rho_+&=+\frac{1}{2} \sum_{\a} E_{\a \a} \otimes E_{\a \a} + \sum_{\a<\b} E_{\a \b} \otimes E_{\b \a}\,,\\
\rho_-&=-\frac{1}{2} \sum_{\a} E_{\a \a} \otimes E_{\a \a} - \sum_{\a> \b} E_{\a \b} \otimes E_{\b \a} \,,\\
\rho&=\frac{1}{2} \left( \rho_+ + \rho_- \right)\,,
\end{split}
\end{equation}
and satisfy
\begin{equation}
\rho_+ - \rho_-= C^s_{12}\,, \qquad C^s_{12} \rho_+ C^s_{12}=-\rho_-\,.
\end{equation}

The $r$ and $\rho$ matrices solve the following modified classical Yang-Baxter equations
\begin{equation}
\label{eq:app_r_rho_YBE}
\begin{split}
&[r_{23}, r_{12}] + [r_{23}, r_{13}]+ [r_{13}, r_{12}] = \frac{1}{4} [C_{23}, C_{12}]\,,\\
&[\rho_{23}, \rho_{12}] + [\rho_{23}, \rho_{13}]+ [\rho_{13}, \rho_{12}] = \frac{1}{4} [C^s_{23}, C^s_{12}] \,.
\end{split}
\end{equation}

\subsection{r-matrices on the affine space and YBE}

We introduce the following function of $x \in \mathbb{R}$
\begin{equation}
\label{eq:app_series_without_regulator}
\begin{split}
r_\pm(x)&=r_\pm \pm C_{12} \sum^{+\infty}_{n=1} e^{\mp i n x }\,,
\end{split}
\end{equation}
satisfying the condition
\begin{equation}
\label{eq:app_rpm_rhopm_properties}
r_+(x)-r_-(x)=2 \pi C_{12} \delta(x)\,, \qquad C_{12} r_+(x) C_{12} = - r_-(-x)\,.
\end{equation}
This definition follows from~\cite{Arutyunov:1996uw}.
We also introduce
\begin{equation}
\label{eq:r_rho_half_sums}
r(x)=\frac{1}{2} \left( r_+(x)+r_-(x) \right)\,.
\end{equation}
From the property above, it follows that
\begin{equation}
\label{eq:r_rho_plus_interms_of_r_rho}
r_{\pm}(x) = r(x) \pm \pi C_{12} \delta(x)\,,
\end{equation}
where the Dirac delta function is 
\begin{equation}
\delta(x) = \frac{1}{2\pi} \, \sum_{n \in \mathbb{Z}} e^{i n x} \,.
\end{equation}

\paragraph{Summing the series.}
To check the properties of the Poisson algebra, one wants to be able to resum the series in~\eqref{eq:app_series_without_regulator}. This is not possible for $x \in \mathbb{R}$ unless we put a regulator $\varepsilon$. This is not the same regulator used to define the representations of the oscillator algebra in~\eqref{eq:repres_a_b}. In the present situation $\varepsilon$ is an infinitesimal parameter that we immediately send to zero after resumming the series. In this respect we assume $\varepsilon \ll \epsilon$. We will often refer to this infinitesimal parameter $i \varepsilon$ by $i 0$.
Then we have
\begin{equation}
\label{eq:app_series_with_regulator_new}
r_\pm(x)=r_\pm \pm C_{12} \sum^{+\infty}_{n=1} e^{n (\mp i x -\varepsilon)}=r_\pm \pm C_{12} \frac{e^{\mp i x -\varepsilon}}{1- e^{\mp i x -\varepsilon}}\,.
\end{equation}
In the limit $\varepsilon \to 0$ we obtain
\begin{equation}
\label{eq:app_series_with_regulator_new2}
r_\pm(x)=r_\pm \pm C_{12} \pv \frac{e^{\mp i x }}{1- e^{\mp i x}} \pm \pi C_{12} \delta(x) \,.
\end{equation}
It turns out that
\begin{equation}
r(x)=r_\pm \pm C_{12} \pv \frac{e^{\mp i x }}{1- e^{\mp i x}} \,.
\end{equation}
No matter the choice of upper or lower sign, we get the same value.
Taking the half sum of the two functions above, we get
\begin{equation}
r(x)=r - \frac{i}{2} C_{12} \pv \cot \frac{x}{2} \,,
\end{equation}
and verify that~\eqref{eq:r_rho_plus_interms_of_r_rho} is satisfied. 

One can also define a function
\begin{equation}
\widetilde{r}(x)=r - \frac{i}{2} C_{12} \cot \frac{x}{2} \,.
\end{equation}
The function $\widetilde r(z)$ is meromorphic, with poles at
$z\in2\pi\mathbb{Z}$. On the real axis, $r_\pm(x)$ are its
distributional boundary values, while $r(x)$ is their symmetric
average, equivalently the principal-value distribution:
\begin{equation}
r_\pm(x) = \widetilde{r}(x \mp i \varepsilon) \,, \qquad r(x) = \pv \, \widetilde{r}(x) = \frac{1}{2}(\widetilde{r}(x - i \varepsilon)+\widetilde{r}(x + i \varepsilon)) \,.
\end{equation}

In components, we have
\begin{equation}
 \widetilde{r}_{a b, c d}(x)= \frac{1}{2} \delta_{a d} \delta_{b c} \left( \sgn_{b a} - i \cot \frac{x}{2} \right) \,.
\end{equation}

\paragraph{Yang-Baxter for $r(x)$.}

The matrix $r(x)$ is skew-symmetric:
\begin{equation}
C_{12}r(x) C_{12}=- r(-x) \,.
\end{equation}
Moreover, it is possible to show that it violates the classical Yang-Baxter equation (cYBE) by local contact terms. Let us show this fact explicitly. For separate points $x_1 \ne x_2 \ne x_3$, the matrix $r(x)$ satisfies the classical Yang-Baxter equation:
\begin{equation}
[r_{23}(x_{23}), r_{12}(x_{12})] + [r_{23}(x_{23}), r_{13}(x_{13})]+ [r_{13}(x_{13}), r_{12}(x_{12})] = 0\,.
\end{equation}

To keep track of coincident points, one should use the following distributional identity
\begin{multline}
\label{eq:pvCot_identity}
\pv \, \cot \frac{x_{23}}{2} \pv \,  \cot \frac{x_{12}}{2} - \pv \, \cot \frac{x_{23}}{2} \pv \,  \cot \frac{x_{13}}{2} - \pv \, \cot \frac{x_{13}}{2} \pv \,  \cot \frac{x_{12}}{2}\\
=1 - (2 \pi)^2 \delta(x_{12}) \delta(x_{23}) \,.
\end{multline}
Let us isolate the product of $\cot$ in the classical YB equation. Defining
\begin{equation}
T \equiv [C_{12},C_{23}] = [C_{23},C_{13}] =[C_{13},C_{12}] \,,
\end{equation}
we obtain
\begin{equation}
\begin{split}
&[r_{23}(x_{23}), r_{12}(x_{12})] + [r_{23}(x_{23}), r_{13}(x_{13})]+ [r_{13}(x_{13}), r_{12}(x_{12})]= \dots \\
&+ T \frac{1}{4}\left(\pv \, \cot \frac{x_{23}}{2} \pv \,  \cot \frac{x_{12}}{2} - \pv \, \cot \frac{x_{23}}{2} \pv \,  \cot \frac{x_{13}}{2} - \pv \, \cot \frac{x_{13}}{2} \pv \,  \cot \frac{x_{12}}{2}\right)\,,
\end{split}
\end{equation}
where the ellipses contain all the other terms (which are not products of two cotangents).
Using the result in~\eqref{eq:pvCot_identity}, and the fact that for distinct points $r(x)$ obeys the classical YBE, we conclude that
\begin{equation}
\label{eq:app_r_cYBE}
[r_{23}(x_{23}), r_{12}(x_{12})] + [r_{23}(x_{23}), r_{13}(x_{13})]+ [r_{13}(x_{13}), r_{12}(x_{12})]= [C_{23},C_{12}] \pi^2 \delta(x_{12}) \delta(x_{23}) \,.
\end{equation}

\subsection{\texorpdfstring{$\rho$}{rho}-matrices on the affine space and their YBE}
\label{app:rho_matrices}

On the affine space, we introduce two different $\rho$ matrices, acting on spin indices. We describe these matrices below.

\paragraph{The matrix $\hat{\rho}(x)$.}

We introduce a matrix
\begin{equation}
\hat{\rho}(x) = \rho -\frac{i}{2} C^s_{12} \pv \, \cot \frac{x}{2} \,,
\end{equation}
and
\begin{equation}
\label{eq:hatrhopm}
\hat{\rho}_\pm(x)= \hat{\rho}(x) \pm \pi  C^s_{12} \delta(x) \,.
\end{equation}
In components, we have
\begin{equation}
\label{eq:app_hatrho_comp}
\hat{\rho}_{\gamma \rho, \delta \lambda}(x)= \frac{1}{2} \delta_{\gamma \lambda} \delta_{\rho \delta} \left( \sgn_{\rho \gamma} - i \, \pv \, \cot \frac{x}{2} \right) 
\end{equation}
This is the analogue of the matrix $r(x)$ previously introduced.
By analogy with the result in the previous section, we obtain
\begin{equation}
\label{eq:cYBE_hatrhox}
\begin{split}
&[\hat{\rho}_{23}(x_{23}), \hat{\rho}_{12}(x_{12})] + [\hat{\rho}_{23}(x_{23}), \hat{\rho}_{13}(x_{13})]+ [\hat{\rho}_{13}(x_{13}), \hat{\rho}_{12}(x_{12})]=[C^s_{23},C^s_{12}] \pi^2 \delta(x_{12}) \delta(x_{23})\,.
\end{split}
\end{equation}

\paragraph{The matrix $\check{\rho}(x)$.}

We also introduce a matrix
\begin{equation}
\check{\rho}(x)= 2 \pi \delta(x) \rho - \frac{i}{2} \, \pv \, \cot \frac{x}{2} D \,, \qquad D \equiv \sum^\ell_{\alpha=1} E_{\alpha \alpha} \otimes E_{\alpha \alpha}
\end{equation}
together with
\begin{equation}
\label{eq:checkrhopm}
\check{\rho}_\pm(x)=\check{\rho}(x) \pm \pi C^s_{12} \delta(x) \,.
\end{equation}
In components, we have
\begin{equation}
\label{eq:app_check_rho_comp}
\check{\rho}_{\gamma \rho, \delta \lambda}(x)=  \pi \delta(x) \delta_{\gamma \lambda} \delta_{\rho \delta} \sgn_{\rho \gamma} - \frac{i}{2} \, \pv \, \cot \frac{x}{2} \delta_{\gamma \rho} \delta_{\gamma \delta} \delta_{\gamma \lambda} \,. 
\end{equation}

Let us show what type of YBE this $\rho$ matrix obeys. We split the cYBE into three pieces:
\begin{equation}
\begin{split}
&[\check{\rho}_{23}(x_{23}), \check{\rho}_{12}(x_{12})] + [\check{\rho}_{23}(x_{23}), \check{\rho}_{13}(x_{13})]+ [\check{\rho}_{13}(x_{13}), \check{\rho}_{12}(x_{12})]=\\
&\text{cYBE}^{(1)}_{\check{\rho}}+\text{cYBE}^{(2)}_{\check{\rho}}+\text{cYBE}^{(3)}_{\check{\rho}}\,.
\end{split}
\end{equation}
The first piece contains products of cotangents, the second contains a single cotangent, and the third piece contains no cotangents:
\begin{equation}
\begin{split}
\text{cYBE}^{(1)}_{\check{\rho}}= &-\frac{1}{4}  \pv \, \cot \frac{x_{23}}{2} \, \pv \, \cot \frac{x_{12}}{2} [D_{23}, D_{12}] \\
&-\frac{1}{4} \pv \, \cot \frac{x_{23}}{2} \, \pv \, \cot \frac{x_{13}}{2} [D_{23}, D_{13}] -\frac{1}{4} \pv \, \cot \frac{x_{13}}{2} \, \pv \, \cot \frac{x_{12}}{2} [D_{13}, D_{12}] \,,\\
\text{cYBE}^{(2)}_{\check{\rho}}= & -i  \pi \pv \, \cot \frac{x_{12}}{2} \delta(x_{23}) [\rho_{23}, D_{12}] -i  \pi \pv \, \cot \frac{x_{23}}{2}  \delta(x_{12}) [D_{23}, \rho_{12}]\\
&-i  \pi \pv \, \cot \frac{x_{13}}{2}\delta(x_{23}) [\rho_{23}, D_{13}] -i  \pi \pv \, \cot \frac{x_{23}}{2}\delta(x_{13}) [D_{23}, \rho_{13}]\\
&-i  \pi \pv \, \cot \frac{x_{12}}{2} \delta(x_{13}) [\rho_{13}, D_{12}] -i  \pi \pv \, \cot \frac{x_{13}}{2}\delta(x_{12}) [D_{13}, \rho_{12}]\,,\\
\text{cYBE}^{(3)}_{\check{\rho}}= & (2 \pi)^2 \delta(x_{23}) \delta(x_{12}) \left(\rho_{23} \rho_{12} + \rho_{23} \rho_{13}+ \rho_{13} \rho_{12}
- \rho_{12} \rho_{13} - \rho_{12} \rho_{23}- \rho_{13} \rho_{23} \right)
\end{split}
\end{equation}
Using that
\begin{equation}
[D_{23}, D_{12}]=[D_{23}, D_{13}]=[D_{13}, D_{12}]=0 \,,
\end{equation}
one immediately gets 
\begin{equation}
\text{cYBE}^{(1)}_{\check{\rho}}=0 \,.
\end{equation}
Mixed terms also vanish. Indeed, they can be written as
\begin{equation}
\begin{split}
\text{cYBE}^{(2)}_{\check{\rho}}= & -i  \pi \pv \, \cot \frac{x_{12}}{2} \delta(x_{23}) [\rho_{23}, D_{12}+D_{13}] \\
&-i  \pi \pv \, \cot \frac{x_{13}}{2}  \delta(x_{12}) [D_{13}+D_{23}, \rho_{12}]\\
& -i  \pi \pv \, \cot \frac{x_{12}}{2}\delta(x_{13}) [ \rho_{13},D_{12}+D_{23}]
\end{split}
\end{equation}
and it holds
\begin{equation}
[\rho_{23}, D_{12}+D_{13}]=[D_{13}+D_{23},\rho_{12}]=[\rho_{13}, D_{12}+D_{23}]=0 \implies \text{cYBE}^{(2)}_{\check{\rho}}=0 \,.
\end{equation}
Finally using~\eqref{eq:app_r_rho_YBE} we obtain
\begin{equation}
\text{cYBE}^{(3)}_{\check{\rho}}= [C^s_{23}, C^s_{12}]  \pi^2 \delta(x_{23}) \delta(x_{12}) \,.
\end{equation}
We conclude that
\begin{equation}
\label{eq:cYBE_checkrhox}
[\check{\rho}_{23}(x_{23}), \check{\rho}_{12}(x_{12})] + [\check{\rho}_{23}(x_{23}), \check{\rho}_{13}(x_{13})]+ [\check{\rho}_{13}(x_{13}), \check{\rho}_{12}(x_{12})]=[C^s_{23}, C^s_{12}] \pi^2 \delta(x_{23}) \delta(x_{12}) \,.
\end{equation}
The matrices $\hat{\rho}(x)$ and $\check{\rho}(x)$ are both skew-symmetric and obey the same type of modified Yang-Baxter equation.

\subsection{Elliptic Green's function}
\label{app:elliptic_Green}

Throughout the paper, we make use of the following elliptic function
\begin{equation}
\label{eq:app_varphi}
\varphi (x, s) \equiv \frac{\sigma(s + x -\Delta)}{\sigma(x -\Delta)\sigma(s)} \, e^{-
\frac{\zeta(\pi)}{\pi} (x- \Delta) s} \,, \qquad s \ne 0 \,.
\end{equation}
In the expression above, $\sigma$ and $\zeta$ are the Weierstrass $\sigma$ and $\zeta$ functions with periods equal to $2 \pi$ and $2 \Delta$\footnote{Note that the two periods cannot be both real and $\Delta$ must have a nonzero imaginary part.}. Labelling by $2 w_1$ and $2 w_2$ the periods of $\sigma$ (in our case $w_1=\pi$ and $w_2= \Delta$) it holds that
\begin{equation}
\sigma(x+2 w_i)=-e^{2 \zeta(w_i)\, (x+w_i)}\sigma(x)\,.
\end{equation}
Using this fact, one can check that
\begin{equation}
\varphi(x + 2\pi, s)= \varphi(x,s) \,, \qquad \varphi(x, s+ 2\pi)= \varphi(x,s) \,.
\end{equation}
Moreover, taking into account also the relation
\begin{equation}
2 \Delta \frac{\zeta(\pi)}{\pi} - 2 \zeta(\Delta)=i
\end{equation}
we get
\begin{equation}
\varphi(x + 2 \Delta, s)= e^{-is} \, \varphi(x,s) \,, \qquad \varphi(x, s+ 2\Delta)= e^{i(\Delta-x)} \varphi(x,s) \,.
\end{equation}

Since $\sigma(0)=0$, then the elliptic function~\eqref{eq:app_varphi} is ill defined for $s=0$.
We define $\varphi(x, 0)$ to be the regular part of~\eqref{eq:app_varphi} when expanding around $s=0$. We have
\begin{equation}
\varphi(x, s) = \frac{1}{s} +\varphi(x,0) + O(s) \,,
\end{equation}
where
\begin{equation}
\varphi(x,0) \equiv \zeta(x-\Delta) - \frac{\zeta(\pi)}{\pi} (x-\Delta)\,.
\end{equation}
This elliptic function solves the following property
\begin{equation}
\label{eq:varphi_period_property}
\lim_{\varepsilon \to 0} \left( \varphi(x+\Delta-i \varepsilon, \qq_{ij}) - e^{-i \qq_{ij}} \varphi(x-\Delta+i \varepsilon, \qq_{ij}) \right) = 2 \pi i \delta(x) - i \delta_{ij}\,,
\end{equation}
where $\qq_{ij} = \qq_i-\qq_j$.

\paragraph{Singular points.}

The Weierstrass $\sigma(x)$ function has no poles, but zeros at the lattice points
\begin{equation}
\label{eq:lattice_points}
x= 2 \pi m + 2 \Delta n\, \qquad m, n \in \mathbb{Z}\,.
\end{equation}
For $s\ne0$, expanding around $\Delta$ we obtain
\begin{equation}
\label{eq:varphis_around_Delta_subleading}
\begin{split}
\varphi(x+\Delta,s)\bigl|_{x=0}= &+\frac{1}{x} + \left( \zeta(s) - s \frac{\zeta(\pi)}{\pi} \right)\\
&+ x \left( \frac{\zeta(s)^2}{2} - \frac{\wp(s)}{2} - \frac{s}{\pi} \zeta(\pi) \zeta(s) + \frac{s^2 \zeta(\pi)^2}{2 \pi^2} \right)  + O(x^2)\,.
\end{split}
\end{equation}
For $s=0$, we have instead
\begin{equation}
\label{eq:varphi0_around_Delta_subleading}
\varphi(x+\Delta,0)= \zeta(x) - 
\frac{\zeta(\pi)}{\pi} x \,,
\end{equation}
and the expansion leads to
\begin{equation}
\varphi(x+\Delta,0)\bigl|_{x=0}= \frac{1}{x} - 
\frac{\zeta(\pi)}{\pi} x + O(x^3)\,.
\end{equation}
We define the regular part as a function of its argument:
\begin{equation}
\label{eq:wreg}
\varphi^{\mathrm{reg}}(s)
:= \zeta(s)-\frac{\zeta(\pi)}{\pi}s,
\qquad s\notin\Lambda,
\qquad
\Lambda=2\pi\mathbb Z+2\Delta\mathbb Z.
\end{equation}
At $s=0$, we adopt the finite-part prescription
\begin{equation}
\varphi^{\mathrm{reg}}(0)
:=\lim_{s\to0}
\left(
\zeta(s)-\frac{\zeta(\pi)}{\pi}s-\frac1s
\right)=0.
\end{equation}
Thus $\varphi^{\mathrm{reg}}(\mathfrak q_{lb})$ vanishes for $l=b$,
whereas a shifted argument must be evaluated before applying any
coincident-argument prescription. In particular,
\begin{equation}
\varphi^{\mathrm{reg}}(\mathfrak q_{ba}-\eta)
=
\zeta(\mathfrak q_{ba}-\eta)
-\frac{\zeta(\pi)}{\pi}(\mathfrak q_{ba}-\eta),
\end{equation}
provided $\mathfrak q_{ba}-\eta\notin\Lambda$.

We also notice that
\begin{equation}
\label{eq:waround_Delta_general_q_expansion}
\varphi(x+\Delta,\mathfrak{q}_{lb})\bigl|_{x \simeq 0}= \frac{1}{x} + \varphi^{\text{reg}}(\mathfrak{q}_{lb}) + O(x)\,.
\end{equation}
The expansion around $-\Delta$ is obtained by using the periodicity~\eqref{eq:varphi_period_property}, where $\varepsilon$ can be set to zero and the Dirac delta function removed:
\begin{equation}
\label{eq:w_xtomDelta_expansion}
\varphi(x-\Delta, \qq_{lb})\bigl|_{x \simeq 0} = e^{i \qq_{lb}} \left(\frac{1}{x} + \varphi^{\text{reg}}(\qq_{lb}) + i \delta_{lb} \right) + O(x)
\end{equation}
The Green's function $\varphi$ also satisfies the following properties
\begin{equation}
\label{eq:w_skew_sym}
\begin{split}
&\varphi(-x,-\qq_{ij}) = -e^{-i \qq_{ij}} \, \varphi(x, \qq_{ij})+ i \delta_{ij} \,,\\
&\varphi(-x+\Delta,-\qq_{ij}) = -  \varphi(x+\Delta, \qq_{ij}) \,.
\end{split}
\end{equation}

\subsection{List of r-matrices}
\label{app:list_of_r_mat}
In this appendix, we list the main r-matrices used in the paper. We already introduced $r(x)$ and $\rho(x)$. 

\paragraph{r-matrices in TT bracket.}
In the TT bracket, the following elliptic r-matrices appear:
\begin{equation}
\label{eq:rqxy_def}
\begin{split}
    r_{\mathfrak{q}}(x,y)&= \sum_{i j} i \varphi(y-x+\Delta,\mathfrak{q}_{ji}) E_{ij} \otimes E_{ji}- \sum_{i j} i \varphi^{\text{reg}}(\mathfrak{q}_{ij}) E_{ii} \otimes E_{jj}\\
    &- \sum_{ij} i \varphi(-x+\Delta, \mathfrak{q}_{ji}) E_{ij} \otimes E_{jj} + \sum_{ij} i \varphi(-y+\Delta, \mathfrak{q}_{ji}) E_{jj} \otimes E_{ij}  \,,
\end{split}
\end{equation}
\begin{equation}
\label{eq:rq_def}
\begin{split}
    r_{\mathfrak{q}}&=-i\sum_{i  j} \varphi^{\text{reg}}(\mathfrak{q}_{ij}) F_{ij} \otimes F_{ji}\,.
\end{split}
\end{equation}
The bracket also contains the following trigonometric r-matrices
\begin{equation}
\refl(x,y)=T^{-1}_1(x) T^{-1}_2(y) r(x-y) T_1(x) T_2(y) \,,
\end{equation}
\begin{equation}
\refl= T^{-1}_1(0) T^{-1}_2(0) r T_1(0) T_2(0) \,,
\end{equation}
\begin{equation}
\begin{split}
&\refl_{\pi}(x,y)=\refl(x,y) \\
    &- \sum_{nilk} \refl(0,y)_{in, kl} E_{ii} \otimes E_{kl} -\sum_{nijk} \refl(x,0)_{ij, kn} E_{ij} \otimes E_{kk} + \sum_{nmik} \refl_{in, km} E_{ii} \otimes E_{kk} \,,
    \end{split}
\end{equation}
\begin{equation}
\begin{split}
    &\refl_\pi=\refl - \sum_{nijk} \refl_{in, kl} E_{ii} \otimes E_{kl} -\sum_{nijk} \refl_{ij, kn} E_{ij} \otimes E_{kk} + \sum_{nmik} \refl_{in, km} E_{ii} \otimes E_{kk} \,.
    \end{split}
\end{equation}

\paragraph{r-matrices in LL bracket.}
Below we list the $r$ matrices entering the $\mathbb{L}$-$\mathbb{L}$ bracket~\eqref{eq:Poiss_bracket_inv_L_magic_limit}:
\begin{equation}
\begin{split}
r_{12}(y;t) &= i \sum_{i j}   \varphi(t-y+\Delta, \mathfrak{q}_{ji})  E_{ij} \otimes E_{ji} + i \sum_{i j} \varphi(y-\imp^1, \mathfrak{q}_{ij})  E_{ij} \otimes E_{jj}\\
&- i \sum_{i j}  \varphi(t-\imp^1, \mathfrak{q}_{ji})  E_{ii} \otimes E_{ji}+ i \sum_{i j}  \varphi^{\text{reg}}(\mathfrak{q}_{ji})  E_{ii} \otimes E_{jj} \,,
\end{split}
\end{equation}
\begin{equation}
\begin{split}
\underline{r}_{12}(y;t) =&- i \sum_{i j}   \varphi^{\text{reg}}(\mathfrak{q}_{ji})  E_{ii} \otimes E_{jj}- i \sum_{i j}  \varphi(y-t+\Delta, \mathfrak{q}_{ji})  E_{ji} \otimes E_{ij}  \,,
\end{split}
\end{equation}
\begin{equation}
\begin{split}
\overline{r}_{12}(y;t) =& i \sum_{i j}  \varphi(y-\imp^1, \mathfrak{q}_{ij})  E_{ij} \otimes E_{jj} -i \sum_{i j}  \varphi^{\text{reg}}(\mathfrak{q}_{ij})  E_{ii} \otimes E_{jj}\,,
\end{split}
\end{equation}
\begin{equation}
\begin{split}
\overline{r}_{21}(y;t) &=P \, \overline{r}_{12}(y;t) \, P=-i \sum_{i j}  \varphi^{\text{reg}}(\mathfrak{q}_{ij})  E_{jj} \otimes E_{ii}+ i \sum_{i j}  \varphi(y-\imp^1, \mathfrak{q}_{ij})  E_{jj} \otimes E_{ij} \,.
\end{split}
\end{equation}

\section{Jacobi identity for the oscillators}
\label{app:Jacobi_oscillators}
In this section, we show how the starting oscillator brackets solve the Jacobi identity.

\subsection{Checking Jacobi for a-a-a}
By applying the oscillator algebra~\eqref{eq:starting_osc_br} together with the Leibniz rule one can check that
\begin{equation}
\begin{split}
&\frac{1}{\var^2}\{a_1(x_1), \{a_2(x_2), a_3(x_3) \} \} + \frac{1}{\var^2} \{a_3(x_3), \{a_1(x_1), a_2(x_2) \} \} + \frac{1}{\var^2} \{a_2(x_2), \{a_3(x_3), a_1(x_1) \} \}\\
=&+\Bigl( [r_{23}(x_{23}), r_{12}(x_{12})] + [r_{23}(x_{23}), r_{13}(x_{13})]+ [r_{13}(x_{13}), r_{12}(x_{12})]\Bigr) a_1(x_1) a_2(x_2) a_3(x_3)\\
&-a_1(x_1) a_2(x_2) a_3(x_3) \Bigl( [\rho_{23}(x_{23}), \rho_{12}(x_{12})] + [\rho_{23}(x_{23}), \rho_{13}(x_{13})]+ [\rho_{13}(x_{13}), \rho_{12}(x_{12})]\Bigr) \,.
\end{split}
\end{equation}
In the expression above the $\rho$ matrix can either be $\hat{\rho}(x)$ or $\check{\rho}(x)$, as defined in Appendix~\ref{app:rho_matrices}.
All the terms mixing $r$ and $\rho$ cancel by skew-symmetry.
The $r$ and $\rho$ matrices are assumed in the Pv prescription. Using the modified classical Yang-Baxter equations for the $r(x)$ and $\rho(x)$ matrices (see~\eqref{eq:app_r_cYBE}, \eqref{eq:cYBE_hatrhox} and~\eqref{eq:cYBE_checkrhox}) we obtain
\begin{equation}
\begin{split}
&\frac{1}{\var^2}\{a_1(x_1), \{a_2(x_2), a_3(x_3) \} \} +\frac{1}{\var^2} \{a_3(x_3), \{a_1(x_1), a_2(x_2) \} \} + \frac{1}{\var^2}\{a_2(x_2), \{a_3(x_3), a_1(x_1) \} \}\\
&=\pi^2 \delta(x_{12}) \delta(x_{23}) \left( [C_{23}, C_{12}] a_1(x_1) a_2(x_1) a_3(x_1) - a_1(x_1) a_2(x_1) a_3(x_1) [C^s_{23}, C^s_{12}] \right) \,.
\end{split}
\end{equation}
We can now use the same properties observed in~\cite{Arutyunov:2019alw}:
\begin{equation}
C_{12} a_1(x) a_2(x)= a_1(x) a_2(x)  C^s_{12} \,,
\end{equation}
to verify that
\begin{multline}
\frac{1}{\var^2}\{a_1(x_1), \{a_2(x_2), a_3(x_3) \} \} + \frac{1}{\var^2}\{a_3(x_3), \{a_1(x_1), a_2(x_2) \} \} \\
+ \frac{1}{\var^2}\{a_2(x_2), \{a_3(x_3), a_1(x_1) \} \}=0 \,.
\end{multline}
This happens thanks to a conspiracy between the $r(x)$ and $\rho(x)$ matrices. 

\subsection{Checking Jacobi for a-a-b}

Let us check now the constraints coming from imposing Jacobi with different oscillators. For simplicity we set $\var=1$. 
We have
\begin{equation}
\begin{split}
&\{ \{a_1, a_2 \}, b_3  \} - \{\{ a_1, b_3 \} , a_2  \} + \{ \{a_2, b_3 \}, a_1  \}=\\
&+r_{12} a_1 \{ a_2 , b_3 \} + r_{12}  \{ a_1 , b_3 \} a_2 - a_1 \{ a_2, b_3 \} \rho_{12}- \{ a_1 , b_3 \} a_2 \rho_{12}\\
&  + b_3 r_{+ 13} \{a_1, a_2 \}+ \{ b_3 , a_2 \} r_{+ 13}a_1- a_1 \rho_{- 13} \{b_3, a_2 \}-  \{a_1 , a_2 \} \rho_{- 13} b_3 \\
&  - b_3 r_{+ 23} \{a_2, a_1 \}- \{ b_3 , a_1 \} r_{+ 23}a_2+ a_2 \rho_{- 23} \{b_3, a_1 \}+  \{a_2 , a_1 \} \rho_{- 23} b_3 \,,
\end{split}
\end{equation}
where the coordinates $x_1$, $x_2$ and $x_3$ are assumed:
\begin{equation}
r_{12} \equiv r_{12}(x_1-x_2)\,, \qquad r_{13} \equiv r_{13}(x_1-x_3)\,, \qquad  \dots \,.
\end{equation}
The matrices $\rho_\pm(x)$ can either be $\hat{\rho}_\pm(x)$ or $\check{\rho}_\pm(x)$. The following argument equally applies to both of them.
Performing the computation of the remaining Poisson brackets, we obtain
\begin{equation}
\begin{split}
&\{ \{a_1, a_2 \}, b_3  \} - \{\{ a_1, b_3 \} , a_2  \} + \{ \{a_2, b_3 \}, a_1  \}=\\
&+b_3 \Bigl(r_{+ 13} r_{12} +r_{+ 23} r_{+13}+r_{+ 23} r_{12} -r_{12}r_{+ 23} - r_{12}r_{+ 13} - r_{+13}r_{+ 23} \Bigr)  a_1 a_2\\
&-a_1 a_2 \Bigl(\rho_{- 13} \rho_{12} +\rho_{- 23} \rho_{-13}+\rho_{- 23} \rho_{12} -\rho_{12} \rho_{- 23} - \rho_{12} \rho_{- 13} - \rho_{-13} \rho_{- 23} \Bigr)  b_3\\
&+2 \pi \delta(x_{23}) \Bigl( a_1 C^{\text{rec}}_{23} \rho_{12}+C^{\text{rec}}_{23} r_{+ 13} a_1 -r_{12}  C^{\text{rec}}_{23} a_1 - a_1 \rho_{- 13} C^{\text{rec}}_{23}  \Bigr)\\
&+2 \pi \delta(x_{13}) \Bigl( a_2 \rho_{- 23} C^{\text{rec}}_{13}+a_2 C^{\text{rec}}_{13}  \rho_{12} -r_{12} C^{\text{rec}}_{13}  a_2  - C^{\text{rec}}_{13} r_{+ 23} a_2 \Bigr)
\end{split}
\end{equation}
Substituting $r_+(x)$ in terms of $r(x)$ and $\rho_-(x)$ in terms of $\rho(x)$ thanks to~\eqref{eq:r_rho_plus_interms_of_r_rho}, \eqref{eq:hatrhopm} and~\eqref{eq:checkrhopm} one finds
\begin{equation}
\begin{split}
&\{ \{a_1, a_2 \}, b_3  \} - \{\{ a_1, b_3 \} , a_2  \} + \{ \{a_2, b_3 \}, a_1  \}=\\
&+b_3 \Bigl(r_{13} r_{12} +r_{23} r_{13}+r_{23} r_{12} -r_{12}r_{ 23} - r_{12}r_{13} - r_{13}r_{23} \Bigr)  a_1 a_2\\
&-a_1 a_2 \Bigl(\rho_{13} \rho_{12} +\rho_{23} \rho_{13}+\rho_{23} \rho_{12} -\rho_{12} \rho_{23} - \rho_{12} \rho_{13} - \rho_{13} \rho_{23} \Bigr)  b_3+\mathcal{C}\,,
\end{split}
\end{equation}
where $\mathcal{C}$ is the contact term contribution
\begin{equation}
\label{eq:Jacobi_aab_C_term}
\begin{split}
&\mathcal{C}= {\color{purple} 2 \pi \delta(x_{23}) \Bigl( a_1 C^{\text{rec}}_{23} \rho_{12}+C^{\text{rec}}_{23} r_{13} a_1 -r_{12}  C^{\text{rec}}_{23} a_1 - a_1 \rho_{13} C^{\text{rec}}_{23}  \Bigr) }\\
&{\color{purple} +2 \pi \delta(x_{13}) \Bigl( a_2 \rho_{23} C^{\text{rec}}_{13}+a_2 C^{\text{rec}}_{13}  \rho_{12} -r_{12} C^{\text{rec}}_{13}  a_2  - C^{\text{rec}}_{13} r_{23} a_2 \Bigr) }\\
&{\color{blue} +2 \pi^2 \delta(x_{13}) \delta(x_{23}) \Bigl( C^{\text{rec}}_{23} C_{13} a_1  + a_1 C^s_{13} C^{\text{rec}}_{23} -a_2 C^s_{23} C^{\text{rec}}_{13}  - C^{\text{rec}}_{13} C_{23} a_2\Bigr) }\\
&{\color{blue}+ \pi b_3 \Bigl[\delta(x_{13}) \delta(x_{23}) \pi \bigl( C_{23} C_{13} - C_{13} C_{23} \bigr) + \delta(x_{13}) \bigl( C_{13} r_{12} + r_{23} C_{13} - r_{12} C_{13} - C_{13} r_{23} \bigr) }\\
&{\color{blue}+ \delta(x_{23}) \bigl( C_{23} r_{13} + C_{23} r_{12} - r_{12} C_{23} - r_{13} C_{23} \bigr)\Bigr] a_1 a_2}\\
&{\color{blue}+ \pi a_1 a_2 \Bigl[ \delta (x_{13}) \delta(x_{23}) \pi \bigl(C^s_{13} C^s_{23} - C^s_{23} C^s_{13} \bigr) + \delta(x_{13}) \bigl( C^s_{13} \rho_{12} + \rho_{23} C^s_{13} - C^s_{13} \rho_{23} - \rho_{12} C^s_{13} \bigr)}\\
&{\color{blue}+\delta(x_{23}) \bigl( C^s_{23} \rho_{13} + C^s_{23} \rho_{12} - \rho_{12} C^s_{23} - \rho_{13} C^s_{23} \bigr)\Bigr] b_3 }\,.
\end{split}
\end{equation}
In the expression above, we split the contact term into a purple and a blue contribution. The former comes entirely from the explicit Dirac delta functions in the third and fourth rows of the Poisson brackets~\eqref{eq:starting_osc_br}. The latter is due both to the explicit Dirac delta functions in the Poisson brackets~\eqref{eq:starting_osc_br} and to the Dirac delta functions contained in $r_+(x)$ and $\rho_-(x)$. 

Let us analyze this contact term in some detail.
It is simple to check that the first and second rows in~\eqref{eq:Jacobi_aab_C_term} vanish separately. In particular the different contributions cancel in pairs.
Consider, for example, the first and last term in the first row. We have
\begin{equation}
\begin{split}
&a_1(x_1) C^{\text{rec}}_{23} \rho_{12}(x_{12})- a_1(x_1) \rho_{13} (x_{13}) C^{\text{rec}}_{23} =\\
&a_1(x_1) C^{\text{rec}}_{23} \rho_{12}(x_{12})- a_1(x_1) C^{\text{rec}}_{23} \rho_{12} (x_{13})  C^{\text{rec}}_{32} C^{\text{rec}}_{23}=\\
&a_1(x_1) C^{\text{rec}}_{23} \rho_{12}(x_{12})- a_1(x_1) C^{\text{rec}}_{23} \rho_{12} (x_{13})=0\,.
\end{split}
\end{equation}
In the last equality, we are taking $x_2=x_3$ since there is an overall Dirac delta function $\delta(x_{23})$.
We conclude that the purple term in~\eqref{eq:Jacobi_aab_C_term} vanishes.
It is simple to check that the third row in~\eqref{eq:Jacobi_aab_C_term} cancels because
\begin{equation}
\begin{split}
&\delta(x_{13}) \delta(x_{23}) \left( C^{\text{rec}}_{23} C_{13} a_1(x) -a_2(y) C^s_{23} C^{\text{rec}}_{13} \right)=\\
&\delta(x_{13}) \delta(x_{23}) \left( C^{\text{rec}}_{23} C_{13} a_1(z) -a_2(z) C^s_{23} C^{\text{rec}}_{13} \right)=0\,,
\end{split}
\end{equation}
and similarly for the other two terms. 
All terms with a single Dirac delta function in the fourth, fifth, sixth, and seventh rows also cancel in pairs. For example, the seventh row can be written as
\begin{equation}
\begin{split}
&C^s_{23} \rho_{13} + C^s_{23} \rho_{12} - \rho_{12} C^s_{23} - \rho_{13} C^s_{23}=\\
&C^s_{23} \rho_{13} C^s_{32} C^s_{23} + C^s_{23} \rho_{12} C^s_{32} C^s_{23} - \rho_{12} C^s_{23} - \rho_{13} C^s_{23}=\\
& \rho_{12}  C^s_{23} +  \rho_{13}  C^s_{23} - \rho_{12} C^s_{23} - \rho_{13} C^s_{23}=0\,.
\end{split}
\end{equation}

The terms containing two Dirac delta functions in the fourth and sixth rows do not cancel, and we have
\begin{equation}
\label{eq:double_Delta_contribution_matrix_formalism}
\begin{split}
\mathcal{C}&=\pi^2 \delta (x_{13}) \delta(x_{23}) \Bigl( b_3 [C_{23}, C_{13}] a_1 a_2 -  a_1 a_2 [ C^s_{23},C^s_{13}] b_3 \Bigr)\\
&=\pi^2 \delta (x_{13}) \delta(x_{23}) \Bigl( b_3 [C_{12}, C_{23}] a_1 a_2 -  a_1 a_2 [ C^s_{12},C^s_{23}] b_3 \Bigr)\,.
\end{split}
\end{equation}
The Jacobiator for different oscillators is then given by
\begin{equation}
\begin{split}
&\{ \{a_1, a_2 \}, b_3  \} - \{\{ a_1, b_3 \} , a_2  \} + \{ \{a_2, b_3 \}, a_1  \}=\\
&b_3 \Bigl([r_{13}, r_{12}] +[r_{23}, r_{13}]+[r_{23}, r_{12}] \Bigr)  a_1 a_2-a_1 a_2 \Bigl([\rho_{13}, \rho_{12}] +[\rho_{23}, \rho_{13}]+[\rho_{23}, \rho_{12}] \Bigr)  b_3\\
&+\pi^2 \delta (x_{13}) \delta(x_{23}) \Bigl( b_3 [C_{12}, C_{23}] a_1 a_2 -  a_1 a_2 [ C^s_{12},C^s_{23}] b_3 \Bigr)\,.
\end{split}
\end{equation}
Using the previously obtained results~\eqref{eq:app_r_cYBE},\eqref{eq:cYBE_hatrhox} and~\eqref{eq:cYBE_checkrhox}, we see that
\begin{equation}
\{ \{a_1, a_2 \}, b_3  \} - \{\{ a_1, b_3 \} , a_2  \} + \{ \{a_2, b_3 \}, a_1  \}=0 \,.
\end{equation}

The Jacobi identity for the remaining brackets can be verified similarly.

\section{Poisson brackets for mirabolic invariants}
\label{app:mirabolic_invariants}

In this appendix, we derive the TT bracket~\eqref{eq:brackets_TT_final} starting from~\eqref{eq:TT_br_expr1}. Based on the obtained result, we construct first the bracket between $B$ and $T$, and then the bracket between the mirabolic $L$ operators
\begin{equation}
L(x)=T^{-1}(x-\Delta) B(x) T(x-\Delta) \,.
\end{equation}
These are important results, needed for the derivation of the Poisson brackets of the full invariant objects.

\subsection{Evaluating the brackets for T's}
\label{app:derivation_TT_bracket}

After replacing the variation~\eqref{eq:TovA_variation} into~\eqref{eq:TT_br_expr1} the TT bracket can be written in the following form
\begin{equation}
\label{eq:simplified_T_Pois_bracket}
\begin{split}
\frac{1}{\var} \, \{ T_{ij}(x), T_{kl}(y) \} = &+T_{ia}(x) T_{kb}(y) f_{abjl} (x,y)+T_{ia}(x) T_{kl}(y) g_{ajl}(x)\\
&-T_{ij}(x) T_{ka}(y) g_{alj}(y)+ T_{ij}(x) T_{kl}(y) m_{ijkl} \,.
\end{split}
\end{equation}
where
\begin{equation}
\label{eq:fabjl_eq1bis}
    \begin{split}
        &f_{abjl} (x,y)= -\frac{1}{4 \pi^2} \, \int^{+ \pi}_{- \pi} ds \, \int^{+ \pi}_{- \pi} ds' \varphi(y-s', \mathfrak{q}_{bl})  \\
        &\biggl[ \varphi(x-s,\mathfrak{q}_{aj}) e^{-i \mathfrak{q}_{bl}} \left(T^{-1}_{ac}(s+\Delta) T^{-1}_{bp}(s'-\Delta) r_{cn, pd}(s-s'+2 \Delta) T_{nj}(s+\Delta) T_{dl}(s'-\Delta) \right) \\
        &- e^{-i \mathfrak{q}_{bl}} e^{-i \mathfrak{q}_{aj}} \varphi(x-s,\mathfrak{q}_{aj}) \left(T^{-1}_{ac}(s-\Delta) T^{-1}_{bp}(s'-\Delta) r_{cn, pd}(s-s'+i\epsilon) T_{nj}(s-\Delta) T_{dl}(s'-\Delta) \right)\\
        &+e^{-i \mathfrak{q}_{aj}} \varphi(x-s, \mathfrak{q}_{aj}) \left(T^{-1}_{ac}(s-\Delta) T^{-1}_{bp}(s'+\Delta) r_{cn, pd}(s-s'-2 \Delta) T_{nj}(s-\Delta) T_{dl}(s'+\Delta) \right)\\
        &-\varphi(x-s, \mathfrak{q}_{aj}) \left(T^{-1}_{ac}(s+\Delta) T^{-1}_{bp}(s'+\Delta) r_{cn, pd}(s-s'- i \epsilon) T_{nj}(s+\Delta) T_{dl}(s'+\Delta) \right)\biggr]\,,
    \end{split}
\end{equation}
\begin{equation}
\label{eq:g_ajbis}
    \begin{split}
        &g_{ajl} (x)=  -\frac{1}{4 \pi^2} \, \int^{+ \pi}_{- \pi} ds \, \int^{+ \pi}_{- \pi} ds' \varphi(-s', \mathfrak{q}_{lb})  \\
        &\biggl[ \varphi(x-s,\mathfrak{q}_{aj})  e^{-i \mathfrak{q}_{aj}} e^{-i \mathfrak{q}_{lb}} \left(T^{-1}_{am}(s-\Delta) T^{-1}_{lp}(s'-\Delta) r_{mc, pd}(s-s'+ i \epsilon) T_{cj}(s-\Delta) T_{db}(s'-\Delta) \right) \\
        &+\varphi(x-s,\mathfrak{q}_{aj})  \left(T^{-1}_{ac}(s+\Delta) T^{-1}_{ld}(s'+\Delta) r_{cn, dq}(s-s'- i \epsilon) T_{nj}(s+\Delta) T_{qb}(s'+\Delta) \right)\\
        &-\varphi(x-s,\mathfrak{q}_{aj}) e^{-i \mathfrak{q}_{aj}}  \left(T^{-1}_{am}(s-\Delta) T^{-1}_{ld}(s'+\Delta) r_{mc, dq}(s-s'-2 \Delta) T_{cj}(s-\Delta) T_{qb}(s'+\Delta) \right)\\
        &-\varphi(x-s,\mathfrak{q}_{aj}) e^{-i \mathfrak{q}_{lb}}  \left(T^{-1}_{ac}(s+\Delta) T^{-1}_{lp}(s'-\Delta) r_{cn, pd}(s-s'+2 \Delta) T_{nj}(s+\Delta) T_{db}(s'-\Delta) \right) \biggr]\,,
    \end{split}
\end{equation}
\begin{equation}
\label{eq:starting_m_term}
\begin{split}
    m_{ijkl}&= -\frac{1}{4 \pi^2} \, \int^{+ \pi}_{- \pi} ds \, \int^{+ \pi}_{- \pi} ds' \varphi(-s', \mathfrak{q}_{lb})\\
    &\biggl[-\varphi(-s,\mathfrak{q}_{ja}) e^{-i \mathfrak{q}_{ja}} e^{-i \mathfrak{q}_{lb}} T_{jm}^{-1}(s-\Delta) T^{-1}_{lp}(s'-\Delta) r_{mc,pd}(s-s'+i\epsilon) T_{ca}(s-\Delta) T_{db}(s'-\Delta)\\
    &-\varphi(-s,\mathfrak{q}_{ja}) T_{jm}^{-1}(s+\Delta) T^{-1}_{lp}(s'+\Delta) r_{mc,pd}(s-s'-i\epsilon) T_{ca}(s+\Delta) T_{db}(s'+\Delta)\\
    &+\varphi(-s,\mathfrak{q}_{ja}) e^{-i \mathfrak{q}_{ja}} T_{jm}^{-1}(s-\Delta) T^{-1}_{lp}(s'+\Delta) r_{mc,pd}(s-s'-2\Delta) T_{ca}(s-\Delta) T_{db}(s'+\Delta) \\
    &+\varphi(-s,\mathfrak{q}_{ja}) e^{-i \mathfrak{q}_{lb}} T_{jm}^{-1}(s+\Delta) T^{-1}_{lp}(s'-\Delta) r_{mc,pd}(s-s'+2\Delta) T_{ca}(s+\Delta) T_{db}(s'-\Delta)\biggr]\,.
\end{split}
\end{equation}
Here $\epsilon$ is an infinitesimal parameter that we send to zero after integrating.
In the following, we show the computation of $f_{abjl}(x,y)$. In the computation, we assume $T(x)$ and $T^{-1}(x)$ to be periodic on the interval $(-\pi, + \pi)$ and holomorphic on the strip $|\Im(x)|< \Im(\Delta)$. The quantities in~\eqref{eq:g_ajbis} and~\eqref{eq:starting_m_term} can be computed similarly.

\paragraph{The f term.}

After proper shifts of $s$ by $\pm \Delta$ the expression in~\eqref{eq:fabjl_eq1bis} can be written as
\begin{equation}
\label{eq:fabjl_eq2}
    \begin{split}
        &f_{abjl} (x,y)= -\frac{1}{4 \pi^2}  \, \int^{+ \pi}_{- \pi} ds'  \varphi(y-s', \mathfrak{q}_{bl})  \\
        &\biggl[+ e^{-i \mathfrak{q}_{bl}} \int^{+ \pi+ \Delta+i \epsilon}_{- \pi+ \Delta+i \epsilon} ds \, \varphi(x-s+\Delta,\mathfrak{q}_{aj})\\
        &\left(T^{-1}_{ac}(s) T^{-1}_{bp}(s'-\Delta) r_{cn, pd}(s-s'+\Delta) T_{nj}(s) T_{dl}(s'-\Delta) \right) \\
        &- e^{-i \mathfrak{q}_{bl}} \int^{+ \pi- \Delta+i \epsilon}_{- \pi- \Delta+i \epsilon} ds \, e^{-i \mathfrak{q}_{aj}} \varphi(x-s-\Delta,\mathfrak{q}_{aj}) \\
        &\left(T^{-1}_{ac}(s) T^{-1}_{bp}(s'-\Delta) r_{cn, pd}(s-s'+\Delta) T_{nj}(s) T_{dl}(s'-\Delta) \right)\\
        &+\int_{-\pi-\Delta-i\epsilon}^{+\pi-\Delta-i\epsilon} ds \,  e^{-i \mathfrak{q}_{aj}} \varphi(x-s-\Delta, \mathfrak{q}_{aj})\\
        &\left(T^{-1}_{ac}(s) T^{-1}_{bp}(s'+\Delta) r_{cn, pd}(s-s'- \Delta) T_{nj}(s) T_{dl}(s'+\Delta) \right)\\
        &-\int_{-\pi+\Delta-i\epsilon}^{+\pi+\Delta-i\epsilon} ds \, \varphi(x-s+\Delta, \mathfrak{q}_{aj})\\
        &\left(T^{-1}_{ac}(s) T^{-1}_{bp}(s'+\Delta) r_{cn, pd}(s-s' - \Delta) T_{nj}(s) T_{dl}(s'+\Delta) \right)\biggr] \,.
    \end{split}
\end{equation}
By a shift of $\Delta$ we do not mean that we are shifting the contours, but we are just writing the integrals as
\begin{equation}
\int^{+\pi}_{-\pi} ds F(s\pm \Delta)=\int^{+\pi \pm \Delta}_{-\pi \pm \Delta} ds F(s) \,.
\end{equation}
The additional $i \epsilon$ term in the shift is harmless for all the functions that do not have poles on the real line; we only need to be careful to shifts by $i \epsilon$ in the function $r(x)$. 
The function $\varphi$ has poles at $\pm \Delta$, and they are also away from the real line. At this point, we want to restore the original integration on the real line, by which we mean an actual shift of the integration contour. Let us show how this can be done for the first term above. We can write
\begin{equation}
\begin{split}
        &\int^{+ \pi+ \Delta+i \epsilon}_{- \pi+ \Delta+i \epsilon} ds \, \varphi(x-s+\Delta, \mathfrak{q}_{aj}) \left(T^{-1}_{ac}(s) T^{-1}_{bp}(s'-\Delta) r_{cn, pd}(s-s'+\Delta) T_{nj}(s) T_{dl}(s'-\Delta) \right)\\
        &=\int_\Gamma ds \, \varphi(x-s+\Delta, \mathfrak{q}_{aj}) \left(T^{-1}_{ac}(s) T^{-1}_{bp}(s'-\Delta) r_{cn, pd}(s-s'+\Delta) T_{nj}(s) T_{dl}(s'-\Delta) \right) - \text{Res}_1 \,,
        \end{split}
\end{equation}
where $\Gamma$ is the green integration path depicted in figure~\ref{fig:TPoisContour} and we should subtract all the residues contained in the counterclockwise oriented parallelogram $P^{(+)}_{\text{up}}$ having vertices at
\begin{equation}
P^{(+)}_{\text{up}}: \qquad -\pi+i\epsilon  \, , \qquad +\pi+i\epsilon  \,, \qquad +\pi+\Delta +i \epsilon \, , \qquad -\pi + \Delta +i \epsilon \,.
\end{equation}
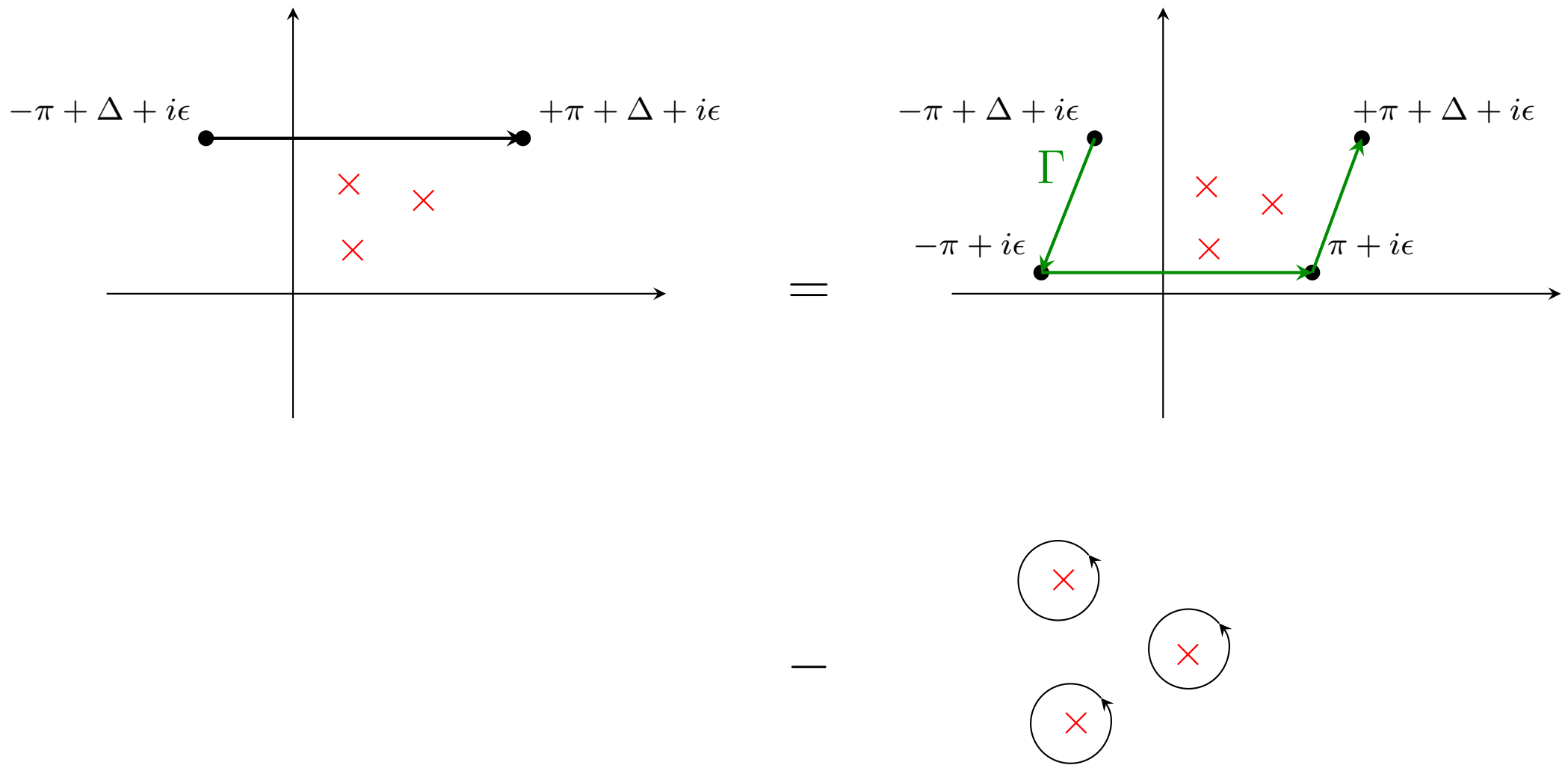
\begin{figure}
\begin{center}

\begin{tikzpicture}[scale=1.1, >=stealth]


\draw[->] (-4.8,0) -- (-0.3,0);
\draw[->] (-3.3,-1.0) -- (-3.3,2.3);

\draw[->, thick]
  (-4.0,1.25) -- (-1.45,1.25);

\fill (-4.0,1.25) circle (1.8pt);
\fill (-1.45,1.25) circle (1.8pt);

\node[above left] at (-4.0,1.25)
  {\scriptsize $-\pi+\Delta+i\epsilon$};

\node[above right] at (-1.45,1.25)
  {\scriptsize $+\pi+\Delta+i\epsilon$};

\node[red] at (-2.85,0.88) { $\times$};
\node[red] at (-2.82,0.35) {$\times$};
\node[red] at (-2.25,0.75) {$\times$};


\node at (0.85,0) {\Large $=$};


\draw[->] (2.0,0) -- (6.9,0);
\draw[->] (3.7,-1.0) -- (3.7,2.3);

\fill (3.15,1.25) circle (1.8pt);
\fill (5.30,1.25) circle (1.8pt);

\node[above left] at (3.15,1.25)
  {\scriptsize $-\pi+\Delta+i\epsilon$};

\node[above right] at (5.10,1.25)
  {\scriptsize $+\pi+\Delta+i\epsilon$};

\fill (2.72,0.17) circle (1.8pt);
\fill (4.90,0.17) circle (1.8pt);

\node[above left] at (2.72,0.17)
  {\scriptsize $-\pi+i\epsilon$};

\node[above right] at (4.90,0.17)
  {\scriptsize $\pi+i\epsilon$};

\draw[green!55!black, thick, ->]
  (2.72,0.17) -- (4.90,0.17);
\draw[green!55!black, thick, ->]
  (3.15,1.25) -- (2.72,0.17);
\draw[green!55!black, thick, ->]
  (4.90,0.17) -- (5.30,1.25);

\node[green!55!black] at (2.80,1.02) {\large $\Gamma$};

\node[red] at (4.05,0.86) { $\times$};
\node[red] at (4.07,0.36) { $\times$};
\node[red] at (4.58,0.72) { $\times$};


\node at (0.85,-3) {\Large $-$};

\draw[->] (3.1,-2.1)
  arc[start angle=40,end angle=400,radius=0.32];

\node[red] at (2.9,-2.3) { $\times$};

\draw[->] (4.15,-2.65)
  arc[start angle=40,end angle=400,radius=0.32];

\node[red] at (3.9,-2.9) { $\times$};

\draw[->] (3.2,-3.25)
  arc[start angle=40,end angle=400,radius=0.32];

\node[red] at (3,-3.45) { $\times$};

\end{tikzpicture}

\end{center}
\caption{\label{fig:TPoisContour}%
Contour deformation to restore the integral on the interval $(-\pi, +\pi)$. The integral on the interval $(- \pi + \Delta+ i \epsilon, + \pi + \Delta+ i \epsilon)$ (on the LHS) is equal to the integral over $(- \pi + i \epsilon, + \pi + i \epsilon)$ minus the residues of poles intersecting the contour during the deformation.}
\end{figure}
Let us call this parallelogram `up', since it is located in the positive imaginary part of the complex plane. The sign $(+)$ corresponds to the shift of $+i\epsilon$.
Similarly we define a counterclockwise oriented parallelogram $P^{(-)}_{\text{up}}$ having vertices at
\begin{equation}
P^{(-)}_{\text{up}}: \qquad -\pi-i\epsilon  \, , \qquad +\pi-i\epsilon  \,, \qquad +\pi+\Delta -i \epsilon \, , \qquad -\pi + \Delta -i \epsilon \,.
\end{equation}
We also define the following parallelograms on the lower half of the complex plane
\begin{equation}
P^{(+)}_{\text{down}}: \qquad -\pi-\Delta + i \epsilon \, , \qquad +\pi-\Delta + i \epsilon \,, \qquad +\pi+ i \epsilon  \, , \qquad -\pi+ i \epsilon   \,,
\end{equation}
\begin{equation}
P^{(-)}_{\text{down}}: \qquad -\pi-\Delta - i \epsilon \, , \qquad +\pi-\Delta - i \epsilon \,, \qquad +\pi - i \epsilon  \, , \qquad -\pi- i \epsilon   \,.
\end{equation}
The functions $r(s)$, $T(s)$ and $\varphi(s,q)$ are all periodic of period $2 \pi$ in $s$. Due to this fact, the integrals over the `vertical' sides of the parallelogram cancel each other, and the integral over $\Gamma$ is equal to the integral over the interval $(-\pi + i\epsilon , + \pi+ i\epsilon)$. Then we obtain
\begin{equation}
\begin{split}
        &\int^{+ \pi+ \Delta+ i \epsilon}_{- \pi+ \Delta+ i \epsilon} ds \, \varphi(x-s+\Delta, \mathfrak{q}_{aj}) \left(T^{-1}_{ac}(s) T^{-1}_{bp}(s'-\Delta) r_{cn, pd}(s-s'+\Delta) T_{nj}(s) T_{dl}(s'-\Delta) \right)\\
        &=\int^{+ \pi+ i \epsilon}_{- \pi+ i \epsilon} ds \, \varphi(x-s+\Delta, \mathfrak{q}_{aj}) \left(T^{-1}_{ac}(s) T^{-1}_{bp}(s'-\Delta) r_{cn, pd}(s-s'+\Delta) T_{nj}(s) T_{dl}(s'-\Delta) \right) \\
        &- \text{Res}_1 \,,
        \end{split}
\end{equation}
where
\begin{multline}
    \text{Res}_1= 2 \pi i \\
    \sum_{s_i \in P^{(+)}_{\text{up}}} \text{Res} \left( \varphi(x-s+\Delta, \mathfrak{q}_{aj}) \left(T^{-1}_{ac}(s) T^{-1}_{bp}(s'-\Delta) r_{cn, pd}(s-s'+\Delta) T_{nj}(s) T_{dl}(s'-\Delta) \right)\right)\biggr|_{s=s_i} \,.
\end{multline}
Repeating a similar argument on the other terms, one gets
\begin{equation}
\label{eq:fabjl_eq5}
    \begin{split}
        &f_{abjl} (x,y)= i \frac{\delta_{aj}}{4 \pi^2}  \, \int^{+ \pi}_{- \pi} ds'  \varphi(y-s', \mathfrak{q}_{bl})  \\
        &\biggl[+ e^{-i \mathfrak{q}_{bl}} \int^{+ \pi+i \epsilon}_{- \pi+i \epsilon} ds \left(T^{-1}_{ac}(s) T^{-1}_{bp}(s'-\Delta) r_{cn, pd}(s-s'+\Delta) T_{nj}(s) T_{dl}(s'-\Delta) \right) \\
        &-\int_{-\pi-i \epsilon}^{+\pi-i \epsilon} ds  \,  \left(T^{-1}_{ac}(s) T^{-1}_{bp}(s'+\Delta) r_{cn, pd}(s-s' - \Delta) T_{nj}(s) T_{dl}(s'+\Delta) \right)\biggr]\\
        &+\frac{1}{4 \pi^2} \int^{+ \pi}_{- \pi} ds'  \varphi(y-s', \mathfrak{q}_{bl}) e^{-i \mathfrak{q}_{bl}} \left(\text{Res}_1 + e^{-i \mathfrak{q}_{aj}}\text{Res}_2 \right) \\
        &-\frac{1}{4 \pi^2} \int^{+ \pi}_{- \pi} ds'  \varphi(y-s', \mathfrak{q}_{bl})  \left(e^{-i \mathfrak{q}_{aj}}\text{Res}_3+\text{Res}_4 \right) 
    \end{split}
\end{equation}
where the remaining residues are given by
\begin{multline}
    \text{Res}_2= 2 \pi i \\
    \sum_{s_i \in P^{(+)}_{\text{down}}} \text{Res} \left( \varphi(x-s-\Delta, \mathfrak{q}_{aj}) \left(T^{-1}_{ac}(s) T^{-1}_{bp}(s'-\Delta) r_{cn, pd}(s-s'+\Delta) T_{nj}(s) T_{dl}(s'-\Delta) \right)\right)\biggr|_{s=s_i} \,.
\end{multline}
\begin{multline}
    \text{Res}_3= 2 \pi i \\
    \sum_{s_i \in P^{(-)}_{\text{down}}} \text{Res} \left( \varphi(x-s-\Delta, \mathfrak{q}_{aj}) \left(T^{-1}_{ac}(s) T^{-1}_{bp}(s'+\Delta) r_{cn, pd}(s-s'-\Delta) T_{nj}(s) T_{dl}(s'+\Delta) \right)\right)\biggr|_{s=s_i} \,.
\end{multline}
\begin{multline}
    \text{Res}_4= 2 \pi i \\
    \sum_{s_i \in P^{(-)}_{\text{up}}} \text{Res} \left( \varphi(x-s+\Delta, \mathfrak{q}_{aj}) \left(T^{-1}_{ac}(s) T^{-1}_{bp}(s'+\Delta) r_{cn, pd}(s-s'-\Delta) T_{nj}(s) T_{dl}(s'+\Delta) \right)\right)\biggr|_{s=s_i} \,.
\end{multline}

Let us now compute the residues. We start from Res$_1$. The function $\varphi(x-s+\Delta, \qq_{aj})$ has a pole at $s=x$. This pole is outside the parallelogram $P^{(+)}_{\text{up}}$ (see the green shape in figure~\ref{fig:TPoisContour}). The function $r(s-s'+\Delta)$ has a pole at $s=s'-\Delta$ which is on the lower half of the complex plane and outside $P^{(+)}_{\text{up}}$. Then,
\begin{equation}
\text{Res}_1=0 \,.
\end{equation}
In Res$_2$ there is a contributing pole of $\varphi(x-s-\Delta,q_{aj})$ at $s=x$. The pole of $r(s-s'+\Delta)$ at $s=s'-\Delta$ is below the parallelogram at a distance $i \epsilon$. Using formula~\eqref{eq:w_xtomDelta_expansion} we obtain 
\begin{equation}
    \text{Res}_2= -2 \pi i e^{i \mathfrak{q}_{aj}}
     \left(  \left(T^{-1}_{ac}(x) T^{-1}_{bp}(s'-\Delta) r_{cn, pd}(x-s'+\Delta) T_{nj}(x) T_{dl}(s'-\Delta) \right)\right) \,.
\end{equation}
In $\text{Res}_3$ there is a pole at $s=x$ due to $\varphi(x-s-\Delta,q_{aj})$. This pole is outside the parallelogram, and we obtain
\begin{equation}
    \text{Res}_3= 0 \,.
\end{equation}
Finally in $\text{Res}_4$ there is a contributing pole at $s=x$ due to $\varphi(x-s-\Delta,q_{aj}) $. Using formula~\eqref{eq:w_xtomDelta_expansion} we obtain
\begin{equation}
    \text{Res}_4= -2 \pi i 
     \left(T^{-1}_{ac}(x) T^{-1}_{bp}(s'+\Delta) r_{cn, pd}(x-s'-\Delta) T_{nj}(x) T_{dl}(s'+\Delta) \right) \,.
\end{equation}
Plugging the residues into~\eqref{eq:fabjl_eq5} we obtain
\begin{equation}
\label{eq:fabjl_eq6}
    \begin{split}
    f_{abjl} (x,y)=\text{Int}_f+\text{Res}_f
    \end{split}
\end{equation}
where we split the result into the following integral and residue terms
\begin{equation}
\label{eq:fabjl_IntPart}
    \begin{split}
        &\text{Int}_f= i \frac{\delta_{aj}}{4 \pi^2}  \, \int^{+ \pi}_{- \pi} ds'  \varphi(y-s', \mathfrak{q}_{bl})  \\
        &\biggl[+ e^{-i \mathfrak{q}_{bl}} \int^{+ \pi}_{- \pi} ds \left(T^{-1}_{ac}(s) T^{-1}_{bp}(s'-\Delta) r_{cn, pd}(s-s'+\Delta) T_{nj}(s) T_{dl}(s'-\Delta) \right) \\
        &-\int_{-\pi}^{+\pi} ds  \,  \left(T^{-1}_{ac}(s) T^{-1}_{bp}(s'+\Delta) r_{cn, pd}(s-s' - \Delta) T_{nj}(s) T_{dl}(s'+\Delta) \right)\biggr]\,,
    \end{split}
\end{equation}
\begin{equation}
\label{eq:Resf_1}
    \begin{split}
        &\text{Res}_f=\frac{1}{2 \pi i} \int^{+ \pi}_{- \pi} ds'  \varphi(y-s', \mathfrak{q}_{bl}) T^{-1}_{ac}(x) T_{nj}(x) \\
        &\biggl( e^{-i \mathfrak{q}_{bl}} T^{-1}_{bp}(s'- \Delta) r_{cn, pd} (x-s'+\Delta) T_{dl}(s'-\Delta) -   T^{-1}_{bp}(s'+ \Delta) r_{cn, pd} (x-s'-\Delta) T_{dl}(s'+\Delta) \biggr)
    \end{split}
\end{equation}
Let us develop further the term Int$_f$. We shift the $s'$ integration contour of the first term above by $\Delta-i \epsilon$, so that no poles intersect the contour. The vertical sides of the parallelogram simplify as before. Similarly we shift the $s'$ contour of the second term by $-\Delta+i \epsilon$. By doing so we obtain
\begin{equation}
    \begin{split}
        & \text{Int}_f= i \frac{\delta_{aj}}{4 \pi^2}  \, \int^{+ \pi}_{- \pi} ds \,T^{-1}_{ac}(s) \, T_{nj}(s) \,  \int^{+ \pi}_{- \pi} ds'  T^{-1}_{bp}(s') T_{dl}(s') \\
        &\biggl[+ e^{-i \mathfrak{q}_{bl}}  \varphi(y-s'-\Delta+ i \epsilon, \mathfrak{q}_{bl})    r_{cn, pd}(s-s'+i \epsilon) \\
        &- \varphi(y-s'+\Delta - i\epsilon, \mathfrak{q}_{bl})   r_{cn, pd}(s-s' - i\epsilon)   \biggr] \,.
    \end{split}
\end{equation}
We notice that the second term of the expression above has poles at $s'=y-i\epsilon$ and $s'=s-i\epsilon$. The integration over $s'$ is running over the real line, just above these poles. Now we shift the integration line as $s'\to s'-2i \epsilon$, in such a way to run below the poles and pick up the associated residues. By doing so we obtain
\begin{equation}
    \begin{split}
        & \text{Int}_f = - \frac{\delta_{aj}\delta_{bl}}{4 \pi^2}  \, \int^{+ \pi}_{- \pi} ds \int^{+ \pi}_{- \pi} ds' \,T^{-1}_{ac}(s) \, T_{nj}(s) \,   T^{-1}_{bp}(s') T_{dl}(s') r_{cn, pd}(s-s'+i\epsilon)\\
        &-\frac{\delta_{aj}}{2 \pi} \int^{+ \pi}_{- \pi} ds T^{-1}_{ac}(s) \,T_{nj}(s)  \\
        &\times  \sum_{s'_i=s-i\epsilon, y-i\epsilon} \text{Res} \left(T^{-1}_{bp}(s') T_{dl}(s') \varphi(y-s'+\Delta-i\epsilon, \mathfrak{q}_{bl}) r_{cn, pd}(s-s'-i \epsilon) \right)\Bigl|_{s'=s'_i} \,.
    \end{split}
\end{equation}
Evaluating the residues explicitly, then we obtain
\begin{equation}
    \begin{split}
        \text{Int}_f =& - \frac{\delta_{aj}\delta_{bl}}{4 \pi^2}  \, \int^{+ \pi}_{- \pi} ds \int^{+ \pi}_{- \pi} ds' \,T^{-1}_{ac}(s) \, T_{nj}(s) \,   T^{-1}_{bp}(s') T_{dl}(s') r_{cn, pd}(s-s'+i\epsilon)\\
        &+\frac{\delta_{aj}}{2 \pi} \int^{+ \pi}_{- \pi} ds T^{-1}_{ac}(s) \,T^{-1}_{bp}(y)  \, r_{cn, pd}(s-y)  T_{nj}(s) \, T_{dl}(y)\\
        &-\frac{i}{2\pi} \delta_{aj} \delta_{al} \delta_{ab} \int^{+\pi}_{-\pi} ds \varphi(y-s+\Delta, 0)
    \end{split}
\end{equation}
We should stress that despite $\varphi(y-s+\Delta,0)$ and $r_{cn, pd}(s-y)$ have a singularity when $s=y$, the combination of the second and third row is regular at $s=y$. Indeed, if we evaluate the residue at $s=y$ we obtain
\begin{equation}
    \begin{split}
        &\text{Res} \left( \delta_{aj} T^{-1}_{ac}(s) \,T^{-1}_{bp}(y)  \, r_{cn, pd}(s-y)  T_{nj}(s) \, T_{dl}(y) -i \delta_{aj} \delta_{al} \delta_{ab} \varphi(y-s+\Delta,0) \right)\Bigl|_{s=y} \\
        & -i \delta_{aj} \delta_{cd} \delta_{np} \, T^{-1}_{ac}(y) T_{nj}(y) \, T_{dl}(y)\,T^{-1}_{bp}(y)+i \delta_{aj} \delta_{al} \delta_{ab}=0 \,.
    \end{split}
\end{equation}
Since there is no pole we can shift the integration contour $s \to s + i\epsilon$ with no effect and obtain
\begin{equation}
    \begin{split}
        \text{Int}_f =& - \frac{\delta_{aj}\delta_{bl}}{4 \pi^2}  \, \int^{+ \pi}_{- \pi} ds \int^{+ \pi}_{- \pi} ds' \,T^{-1}_{ac}(s) \, T_{nj}(s) \,   T^{-1}_{bp}(s') T_{dl}(s') r_{cn, pd}(s-s'+i\epsilon)\\
        &+\frac{\delta_{aj}}{2 \pi} \int^{+ \pi}_{- \pi} ds T^{-1}_{ac}(s) \,T^{-1}_{bp}(y)  \, r_{cn, pd}(s-y+i \epsilon)  T_{nj}(s) \, T_{dl}(y)\\
        &-\frac{i}{2\pi} \delta_{aj} \delta_{al} \delta_{ab} \int^{+\pi}_{-\pi} ds \varphi(y-s+\Delta-i \epsilon, 0)\,.
    \end{split}
\end{equation}
Using now that
\begin{equation}
\int^{+ \pi}_{- \pi} ds  \varphi(y-s+\Delta - i\epsilon, 0) =\int^{+ \pi}_{- \pi} ds  \varphi(y-s, 0)=\int^{+ \pi}_{- \pi} ds  \varphi(s, 0)= i \pi\,,
\end{equation}
we obtain
\begin{equation}
\label{eq:Int_f_final}
    \begin{split}
        \text{Int}_f = &- \frac{\delta_{aj}\delta_{bl}}{4 \pi^2}  \, \int^{+ \pi}_{- \pi} ds \int^{+ \pi}_{- \pi} ds' \,T^{-1}_{ac}(s) \, T_{nj}(s) \,   T^{-1}_{bp}(s') T_{dl}(s') r_{cn, pd}(s-s'+i\epsilon)\\
        &+\frac{\delta_{aj}}{2 \pi} \int^{+ \pi}_{- \pi} ds T^{-1}_{ac}(s) \,T^{-1}_{bp}(y)  \, r_{cn, pd}(s-y+i\epsilon)  T_{nj}(s) \, T_{dl}(y) + \frac{1}{2} \delta_{aj} \delta_{al} \delta_{ab} \,.
    \end{split}
\end{equation}

Now we compute the residue term~\eqref{eq:Resf_1}. Performing two shifts $s' \to s' \pm \Delta \mp i \epsilon$ we get
\begin{equation}
\label{eq:Resf_2}
    \begin{split}
        &\text{Res}_f=\frac{1}{2 \pi i} \int^{+ \pi}_{- \pi} ds'  T^{-1}_{ac}(x) T_{nj}(x) T^{-1}_{bp}(s') T_{dl}(s') \\
        &\times \left( e^{-i \mathfrak{q}_{bl}} \varphi(y-s'-\Delta+i\epsilon, \mathfrak{q}_{bl}) r_{cn, pd} (x-s'+i \epsilon) - \varphi(y-s'+\Delta-i\epsilon, \mathfrak{q}_{bl}) r_{cn, pd} (x-s'-i \epsilon) \right)  \,.
    \end{split}
\end{equation}
Deforming the contour of the second term below and picking the residues of $\varphi$ and $r$ we end up with
\begin{equation}
\label{eq:Resf_4}
    \begin{split}
        &\text{Res}_f=i \delta_{al} \delta_{bj} \varphi(y-x+\Delta, \mathfrak{q}_{bl})-T^{-1}_{ac}(x) T^{-1}_{bp}(y) r_{cn, pd}(x-y)  T_{n j}(x) T_{dl}(y) \\
        &+\frac{1}{2 \pi} \delta_{bl} \int^{+ \pi}_{- \pi} ds'  T^{-1}_{ac}(x)  T^{-1}_{bp}(s')  r_{cn, pd} (x-s'+i \epsilon) T_{nj}(x) T_{dl}(s') \,.
    \end{split}
\end{equation}
Plugging~\eqref{eq:Int_f_final} and~\eqref{eq:Resf_4} into~\eqref{eq:fabjl_eq6} we can write
\begin{equation}
\label{eq:fabjl_final}
    \begin{split}
    f_{abjl} (x,y)&=i \delta_{al} \delta_{bj} \varphi(y-x+\Delta, \mathfrak{q}_{bl})-T^{-1}_{ac}(x) T^{-1}_{bp}(y) r_{cn, pd}(x-y)  T_{n j}(x) T_{dl}(y) \\
        &+\frac{1}{2 \pi} \delta_{bl} \int^{+ \pi}_{- \pi} ds  T^{-1}_{ac}(x)  T^{-1}_{bp}(s)  r_{cn, pd} (x-s+i \epsilon) T_{nj}(x) T_{dl}(s)\\
        &+\frac{\delta_{aj}}{2 \pi} \int^{+ \pi}_{- \pi} ds T^{-1}_{ac}(s) \,T^{-1}_{bp}(y)  \, r_{cn, pd}(s-y+i\epsilon)  T_{nj}(s) \, T_{dl}(y)\\
        &-\frac{\delta_{aj}\delta_{bl}}{4 \pi^2}  \, \int^{+ \pi}_{- \pi} ds \int^{+ \pi}_{- \pi} ds' \,T^{-1}_{ac}(s) \, T_{nj}(s) \,   T^{-1}_{bp}(s') T_{dl}(s') r_{cn, pd}(s-s'+i\epsilon) \\
        &+ \frac{1}{2} \delta_{aj} \delta_{al} \delta_{ab}\,.
    \end{split}
\end{equation}

The computation of $g_{ajl}(x)$ and $m_{ijkl}$ in~\eqref{eq:simplified_T_Pois_bracket} can be done similarly. It turns out that double poles are present in the integration contour of $m_{ijkl}$. Due to this fact, the final result contains derivatives of $T(x)$ located at the origin; these derivatives arise from the expansion of the integral around the double pole position, when taking the residues. Without these derivatives, the Jacobi identity would not be satisfied.
After summing over the $f$, $g$ and $m$ terms, all contributions containing integrals (for example the second, third and fourth line in~\eqref{eq:fabjl_final}) cancel between the different terms.
The final result is given by
\begin{equation}
\label{eq:TT_final_formula_xy}
\begin{split}
&\frac{1}{\var} \, \{T_{ij}(x), T_{kl}(y)\}=-i T_{ij}(x) T_{kl}(y) \sum^N_{a,b=1} T^{-1}_{jm}(0) T_{pa}(0) \partial_{s} \left( T^{-1}_{lp}(s) T_{mb}(s) \right)\Bigl|_{s=0}\\
&+i T_{il}(x) T_{kj}(y) \varphi(y-x+\Delta, \mathfrak{q}_{jl})- r_{in, kd}(x-y)  T_{n j}(x) T_{dl}(y) \\
&-i T_{ia}(x) T_{kl}(y) \delta_{lj} \varphi(-x+\Delta, \mathfrak{q}_{la})+i  T_{ij}(x) T_{ka}(y) \delta_{lj} \varphi(-y+\Delta, \mathfrak{q}_{ja})\\
&+ T_{kl}(y) \sum^N_{b=1}  T^{-1}_{lp}(0) r_{ic,pd}(x) T_{cj}(x) T_{db}(0)- T_{ij}(x) \sum^N_{b=1}  T^{-1}_{jp}(0) r_{kc,pd}(y) T_{cl}(y) T_{db}(0)\\
&+ T_{ij}(x) T_{kl}(y) \left( i \zeta(\mathfrak{q}_{lj}) (1-\delta_{lj}) - i \frac{\zeta(\pi)}{\pi} \mathfrak{q}_{lj}-\frac{1}{2} \sum^N_{a,b=1} T^{-1}_{jm}(0) T_{mb}(0) T^{-1}_{lc}(0)  T_{ca}(0) \sgn_{cm} \right) \,,
\end{split}
\end{equation}
and can be written in the compact notation~\eqref{eq:brackets_TT_final}.

\subsection{Evaluating the TB bracket}
\label{app:derivation_BT_bracket}

We want to compute the following bracket
\begin{equation}
    \frac{1}{\var}\{T_{ij}(x), B_{kl}(y) \}= \int^{+\pi}_{-\pi} ds \frac{\delta T_{ij}(x)}{\delta A_{mn}(s)} \, \frac{1}{\var} \{A_{mn}(s), B_{kl}(y) \}
\end{equation}
Using~\eqref{AFD} and~\eqref{eq:TovA_variation}
we obtain
\begin{equation}
\begin{split}
    &\frac{1}{\var} \{T_{ij}(x), B_{kl}(y) \}=-\frac{1}{2 \pi i} \int^{+\pi}_{-\pi} ds  \sum^N_{c=1} \Bigl[ \\
    & - e^{-i \mathfrak{q}_c} \varphi(x-s, \mathfrak{q}_{cj}) T_{ic}(x) T^{-1}_{cm}(s-\Delta) r_{ma, kb}(s-y+ i\epsilon)  A_{an}(s) T_{nj}(s+\Delta)  B_{bl}(y)\\
    & - e^{-i \mathfrak{q}_c} \varphi(x-s, \mathfrak{q}_{cj}) T_{ic}(x) T^{-1}_{cm}(s-\Delta) A_{ma}(s) B_{kb}(y) r_{an,b l}(s-y+2\Delta+i\epsilon) T_{nj}(s+\Delta)   \\
    &+e^{-i \mathfrak{q}_c} \varphi(x-s, \mathfrak{q}_{cj}) T_{ic}(x) T^{-1}_{cm}(s-\Delta) 
    B_{kb}(y) r_{ma, bl}(s-y-i\epsilon) A_{an}(s) T_{nj}(s+\Delta)\\
    &+e^{-i \mathfrak{q}_c} \varphi(x-s, \mathfrak{q}_{cj}) T_{ic}(x) T^{-1}_{cm}(s-\Delta) T_{nj}(s+\Delta) A_{ma}(s) r_{an, kb}(s-y+2\Delta+i\epsilon) B_{bl}(y) \\
    &+e^{-i \mathfrak{q}_j} \varphi(-s, \mathfrak{q}_{jc}) T_{ij}(x) T^{-1}_{jm}(s-\Delta) T_{nc}(s+\Delta) r_{ma, kb}(s-y+ i\epsilon) A_{an}(s) B_{bl}(y)\\
    &+e^{-i \mathfrak{q}_j} \varphi(-s, \mathfrak{q}_{jc}) T_{ij}(x) T^{-1}_{jm}(s-\Delta) T_{nc}(s+\Delta) A_{ma}(s) B_{kb}(y) r_{an,b l}(s-y+2\Delta+i\epsilon)\\
    &-e^{-i \mathfrak{q}_j} \varphi(-s, \mathfrak{q}_{jc}) T_{ij}(x) T^{-1}_{jm}(s-\Delta) T_{nc}(s+\Delta) B_{kb}(y) r_{ma, bl}(s-y-i\epsilon) A_{an}(s)\\
    &-e^{-i \mathfrak{q}_j} \varphi(-s, \mathfrak{q}_{jc}) T_{ij}(x) T^{-1}_{jm}(s-\Delta) T_{nc}(s+\Delta) A_{ma}(s) r_{an, kb}(s-y+2\Delta+i\epsilon) B_{bl}(y) \Bigr] \,.
     \end{split}
\end{equation}
Using that
\begin{multline}
    A(x)= T(x-\Delta) Q T^{-1}(x+\Delta) = T(x-\Delta) e^{i \mathfrak{q}} T^{-1}(x+\Delta)\\
    \implies A_{ij}(x)= \sum_n T_{in}(x-\Delta) e^{i \mathfrak{q}_n} T^{-1}_{nj}(x+\Delta) \,,
\end{multline}
and
\begin{equation}
r_{ij, kl}(x-i \epsilon)=r_{ij, kl}(x+i \epsilon) +2 \pi \delta_{il} \delta_{jk} \delta(x) \,,
\end{equation}
then the expression above can be written as follows
\begin{equation}
\label{eq:app_TB_split}
\frac{1}{\var} \{T_{ij}(x), B_{kl}(y) \}= \text{loc}+ \text{int}
\end{equation}
where we split the result into a local and integral parts
\begin{equation}
\label{eq:app_loc_BT_br}
\begin{split}
\text{loc} &=i \sum_{c} e^{-i \mathfrak{q}_{cj}} \varphi(x-y, \mathfrak{q}_{cj}) T_{ic}(x) T^{-1}_{cl}(y-\Delta) B_{kn}(y) T_{nj}(y-\Delta)\\
    &-i \sum_{c} e^{-i \mathfrak{q}_{jc}} \varphi(-y, \mathfrak{q}_{jc}) T_{ij}(x) T^{-1}_{jl}(y-\Delta) B_{kn}(y) T_{nc}(y-\Delta)\\
\end{split}
\end{equation}
and
\begin{equation}
\begin{split}
    &\text{int}=\frac{1}{2 \pi i} \int^{+\pi}_{-\pi} ds  \sum^N_{c=1} \Bigl[ \\
    & + e^{-i \mathfrak{q}_{cj}} \varphi(x-s, \mathfrak{q}_{cj}) T_{ic}(x) T^{-1}_{cm}(s-\Delta) r_{ma, kb}(s-y+ i\epsilon)  T_{aj}(s-\Delta)   B_{bl}(y)\\
    & +  \varphi(x-s, \mathfrak{q}_{cj}) T_{ic}(x) T^{-1}_{ca}(s+\Delta)  B_{kb}(y) r_{an,b l}(s-y+2\Delta+i\epsilon) T_{nj}(s+\Delta)   \\
    &-e^{-i \mathfrak{q}_{cj}} \varphi(x-s, \mathfrak{q}_{cj}) T_{ic}(x) T^{-1}_{cm}(s-\Delta) 
    B_{kb}(y) r_{ma, bl}(s-y+i\epsilon) T_{aj}(s-\Delta)\\
    &- \varphi(x-s, \mathfrak{q}_{cj}) T_{ic}(x)  T^{-1}_{ca}(s+\Delta)   r_{an, kb}(s-y+2\Delta+i\epsilon) T_{nj}(s+\Delta) B_{bl}(y) \\
    &-e^{-i \mathfrak{q}_{jc}} \varphi(-s, \mathfrak{q}_{jc}) T_{ij}(x) T^{-1}_{jm}(s-\Delta) T_{ac}(s-\Delta)  r_{ma, kb}(s-y+ i\epsilon)  B_{bl}(y)\\
    &- \varphi(-s, \mathfrak{q}_{jc}) T_{ij}(x)  T_{nc}(s+\Delta) T^{-1}_{ja}(s+\Delta)  B_{kb}(y) r_{an,b l}(s-y+2\Delta+i\epsilon)\\
    &+e^{-i \mathfrak{q}_{jc}} \varphi(-s, \mathfrak{q}_{jc}) T_{ij}(x) T^{-1}_{jm}(s-\Delta)  B_{kb}(y) r_{ma, bl}(s-y+i\epsilon) T_{ac}(s-\Delta) \\
    &+ \varphi(-s, \mathfrak{q}_{jc}) T_{ij}(x)  T_{nc}(s+\Delta)  T^{-1}_{ja}(s+\Delta)  r_{an, kb}(s-y+2\Delta+i\epsilon) B_{bl}(y) \Bigr] \,.
     \end{split}
\end{equation}
We now redefine the integration variable by suitable shifts $s \to s \pm \Delta$. Noting that for $x$ and $y$ on the real line, the integration is performed away from the singularities of $\varphi$, then we can shift the argument of $\varphi$ by $i \epsilon$ without affecting the result. In particular, we can write
\begin{equation}
\begin{split}
    \text{int}=&\scriptstyle{\frac{1}{2 \pi i}\biggl[ \int^{+\pi-\Delta+i\epsilon}_{-\pi-\Delta+i\epsilon} ds  \,  e^{-i \mathfrak{q}_{cj}} \varphi(x-s-\Delta, \mathfrak{q}_{cj}) T_{ic}(x) T^{-1}_{cm}(s) r_{ma, kb}(s-y+\Delta)  T_{aj}(s)   B_{bl}(y)}\\
    &\scriptstyle{ + \int^{+\pi+\Delta+i\epsilon}_{-\pi+\Delta+i\epsilon} ds  \, \varphi(x-s+\Delta, \mathfrak{q}_{cj}) T_{ic}(x) T^{-1}_{ca}(s)  B_{kb}(y) r_{an,b l}(s-y+\Delta) T_{nj}(s)  } \\
    &\scriptstyle{-\int^{+\pi-\Delta+i\epsilon}_{-\pi-\Delta+i\epsilon} ds  \, e^{-i \mathfrak{q}_{cj}} \varphi(x-s-\Delta, \mathfrak{q}_{cj}) T_{ic}(x) T^{-1}_{cm}(s) 
    B_{kb}(y) r_{ma, bl}(s-y+\Delta) T_{aj}(s)}\\
    &\scriptstyle{- \int^{+\pi+\Delta+i\epsilon}_{-\pi+\Delta+i\epsilon} ds  \, \varphi(x-s+\Delta, \mathfrak{q}_{cj}) T_{ic}(x)  T^{-1}_{ca}(s)   r_{an, kb}(s-y+\Delta) T_{nj}(s) B_{bl}(y)} \\
    &\scriptstyle{-\int^{+\pi-\Delta+i\epsilon}_{-\pi-\Delta+i\epsilon} ds  \, e^{-i \mathfrak{q}_{jc}} \varphi(-s-\Delta, \mathfrak{q}_{jc}) T_{ij}(x) T^{-1}_{jm}(s) T_{ac}(s)  r_{ma, kb}(s-y+\Delta)  B_{bl}(y)}\\
    &\scriptstyle{- \int^{+\pi+\Delta+i\epsilon}_{-\pi+\Delta+i\epsilon} ds  \, \varphi(-s+\Delta, \mathfrak{q}_{jc}) T_{ij}(x)  T_{nc}(s) T^{-1}_{ja}(s)  B_{kb}(y) r_{an,b l}(s-y+\Delta)}\\
    &\scriptstyle{+ \int^{+\pi-\Delta+i\epsilon}_{-\pi-\Delta+i\epsilon} ds  \, e^{-i \mathfrak{q}_{jc}} \varphi(-s-\Delta, \mathfrak{q}_{jc}) T_{ij}(x) T^{-1}_{jm}(s)  B_{kb}(y) r_{ma, bl}(s-y+\Delta) T_{ac}(s)} \\
    &\scriptstyle{+ \int^{+\pi+\Delta+i\epsilon}_{-\pi+\Delta+i\epsilon} ds  \, \varphi(-s+\Delta, \mathfrak{q}_{jc}) T_{ij}(x)  T_{nc}(s)  T^{-1}_{ja}(s)  r_{an, kb}(s-y+\Delta) B_{bl}(y) \biggr] }\,.
     \end{split}
\end{equation}
Shifting all the contours to be on the real line, we obtain
\begin{equation}
\label{eq:app_int_BT_br}
\begin{split}
    &\text{int}=\\
    &-r_{in, kb}(x-y+\Delta) T_{nj}(x) B_{bl}(y)+ 
    B_{kb}(y) r_{in, bl}(x-y+\Delta) T_{nj}(x) \\
    &+ T_{ij}(x) T^{-1}_{ja}(0)   r_{an, kb}(-y+\Delta) T_{nc}(0)  B_{bl}(y)- T_{ij}(x)   B_{kb}(y) T^{-1}_{ja}(0) r_{an, bl}(-y+\Delta) T_{nc}(0)\\
    &+\frac{1}{2 \pi i} \int^{+\pi+i\epsilon}_{-\pi+i\epsilon} ds \, \biggl[\\
    &+ \left(e^{-i \mathfrak{q}_{cj}} \varphi(x-s-\Delta, \mathfrak{q}_{cj}) - \varphi(x-s+\Delta, \mathfrak{q}_{cj}) \right) T_{ic}(x) T^{-1}_{cm}(s) r_{ma, kb}(s-y+\Delta)  T_{aj}(s)   B_{bl}(y)\\
    &- \left( e^{-i \mathfrak{q}_{cj}} \varphi(x-s-\Delta, \mathfrak{q}_{cj}) - \varphi(x-s+\Delta, \mathfrak{q}_{cj}) \right) T_{ic}(x) T^{-1}_{cm}(s) 
    B_{kb}(y) r_{ma, bl}(s-y+\Delta) T_{aj}(s)\\
    &- \left( e^{-i \mathfrak{q}_{jc}} \varphi(-s-\Delta, \mathfrak{q}_{jc}) - \varphi(-s+\Delta, \mathfrak{q}_{jc}) \right) T_{ij}(x) T^{-1}_{jm}(s) T_{ac}(s)  r_{ma, kb}(s-y+\Delta)  B_{bl}(y)\\
    &+  \left( e^{-i \mathfrak{q}_{jc}} \varphi(-s-\Delta, \mathfrak{q}_{jc}) - \varphi(-s+\Delta, \mathfrak{q}_{jc}) \right) T_{ij}(x) T^{-1}_{jm}(s)  B_{kb}(y) r_{ma, bl}(s-y+\Delta) T_{ac}(s)  \biggr]
     \end{split}
\end{equation}
Using the periodicity property of the $\varphi$ function, one can show that the remaining integrals in the expression above cancel each other. Plugging~\eqref{eq:app_loc_BT_br} and~\eqref{eq:app_int_BT_br} into~\eqref{eq:app_TB_split}, we end up with
\begin{equation}
\label{eq:app_TB_bracket}
\begin{split}
    &\frac{1}{\var} \{T_{ij}(x), B_{kl}(y) \}=\\
    &+i \sum_{c} e^{-i \mathfrak{q}_{cj}} \varphi(x-y, \mathfrak{q}_{cj}) T_{ic}(x) T^{-1}_{cl}(y-\Delta) B_{kn}(y) T_{nj}(y-\Delta)\\
    &-i \sum_{c} e^{-i \mathfrak{q}_{jc}} \varphi(-y, \mathfrak{q}_{jc}) T_{ij}(x) T^{-1}_{jl}(y-\Delta) B_{kn}(y) T_{nc}(y-\Delta)\\
    &-r_{in, kb}(x-y+\Delta) T_{nj}(x) B_{bl}(y)+ 
    B_{kb}(y) r_{in, bl}(x-y+\Delta) T_{nj}(x) \\
    &+ \sum_c T_{ij}(x) T^{-1}_{ja}(0)   r_{an, kb}(-y+\Delta) T_{nc}(0)  B_{bl}(y)- \sum_c T_{ij}(x)   B_{kb}(y) T^{-1}_{ja}(0) r_{an, bl}(-y+\Delta) T_{nc}(0)
     \end{split}
\end{equation}
In matrix notation, this bracket is
\begin{equation}
\label{eq:app_TCbracket}
\begin{split}
    &\frac{1}{\var} \{T_1(x), B_2(y) \}=\\
    &+i T_1(x)  B_{2}(y) T_{2}(y-\Delta)  \sum_{i,j} e^{-i \mathfrak{q}_{ij}} \left( \varphi(x-y, \mathfrak{q}_{ij}) E_{ij} \otimes E_{ji} -  \varphi(-y, \mathfrak{q}_{ij}) E_{ii} \otimes E_{ji} \right) T^{-1}_{2}(y-\Delta) \\
    &-r(x-y+\Delta) T_{1}(x) B_{2}(y)+ 
    B_{2}(y) r(x-y+\Delta) T_{1}(x) \\
    &+ T_{1}(x) \Bigl[ \sum_{ijkl} \left(T_1^{-1}(0)   r_{12}(-y+\Delta) T_{1}(0) \right)_{ij,kl} E_{ii} \otimes E_{kl} \Bigr] B_{2}(y)\\
    &- T_{1}(x) B_2(y) \Bigl[ \sum_{ijkl} \left(T_1^{-1}(0)   r_{12}(-y+\Delta) T_{1}(0) \right)_{ij,kl} E_{ii} \otimes E_{kl} \Bigr] \,.
     \end{split}
\end{equation}
From the expression above, one obtains
\begin{equation}
\begin{split}
&\{ B_1(x), T_2(y) \}= \{ B_{ab}(x), T_{cd}(y) \} E_{ab} \otimes E_{cd} = - \{ T_{cd}(y), B_{ab}(x) \} E_{ab} \otimes E_{cd}=\\
&- \{ T_{cd}(y), B_{ab}(x) \} P (E_{cd} \otimes E_{ab} ) P = - P \{ T_{1}(y), B_{2}(x) \}  P \,,
\end{split}
\end{equation}
from which we can write
\begin{equation}
\begin{split}
    &\frac{1}{\var} \{B_1(x), T_2(y) \}=\\
    & -  i  T_2(y)  B_{1}(x) T_{1}(x-\Delta)   \sum_{i,j} e^{-i \mathfrak{q}_{ij}} \left( \varphi(y-x, \mathfrak{q}_{ij}) E_{ji} \otimes E_{ij} -  \varphi(-x, \mathfrak{q}_{ij}) E_{ji} \otimes E_{ii} \right)  T^{-1}_{1}(x-\Delta)  \\
    & - r_{12}(x-y-\Delta) T_{2}(y) B_{1}(x)  + 
     B_{1}(x) r_{12}(x-y-\Delta) T_{2}(y) \\
    & - T_{2}(y)  \Bigl[ \sum_{ijkl} \left(T_1^{-1}(0)   r_{12}(-x+\Delta) T_{1}(0) \right)_{ij,kl} E_{kl} \otimes E_{ii} \Bigr]  B_{1}(x)\\
    & +  T_{2}(y) B_1(x) \Bigl[ \sum_{ijkl} \left(T_1^{-1}(0)   r_{12}(-x+\Delta) T_{1}(0) \right)_{ij,kl} E_{kl} \otimes E_{ii} \Bigr] \,.
     \end{split}
\end{equation}

\subsection{Brackets for mirabolic L operators}
\label{app:derivation_LL_spinless_bracket}

We want to compute the Poisson brackets for the mirabolic $L$ operators introduced in~\eqref{eq:D_mathcalL_parameterization}:
\begin{equation}
\label{eq:def_spinless_L_oper}
L(x)= T^{-1}(x-\Delta) B(x) T(x-\Delta) \,.
\end{equation}
Defining
\begin{equation}
    \begin{split}
        &G_{12}(x,y) \equiv + T^{-1}_1(x-\Delta) T^{-1}_2(y-\Delta) \frac{1}{\var} \{T_1(x-\Delta), T_2(y-\Delta) \} \,,\\
        &H_{12}(x,y) \equiv +T^{-1}_{1}(x-\Delta) T^{-1}_{2}(y-\Delta) \frac{1}{\var} \{ T_{1}(x-\Delta), B_2(y) \} T_2(y-\Delta)\\
        &H_{21}(y,x)\equiv -T^{-1}_{1}(x-\Delta) T^{-1}_{2}(y-\Delta) \frac{1}{\var} \{B_1(x),  T_{2}(y-\Delta) \} T_1(x-\Delta) 
    \end{split}
\end{equation}
we have
\begin{equation}
\begin{split}
\frac{1}{\var} \{L_1(x), L_2(y) \} = &+G_{12}(x,y)L_1(x) L_2(y)  + L_1(x) L_2(y) G_{12}(x,y)\\
&-L_1(x) G_{12}(x,y) L_2(y)- L_2(y) G_{12}(x,y) L_1(x)\\
&- H_{12}(x,y) L_1(x)+L_1(x) H_{12}(x,y)+ H_{21}(y,x) L_2(y) -L_2(y) H_{21}(y,x)\\
&+T^{-1}_1(x-\Delta) T^{-1}_2(y-\Delta) \frac{1}{\var} \{B_1(x), B_2(y) \} T_1(x-\Delta) T_2(y-\Delta) \,.
\end{split}
\end{equation}
Replacing in the expression above the brackets obtained in the previous part of this appendix, after a straightforward but long computation, we obtain
\begin{equation}
\label{eq:app_LL_bracket}
\begin{split}
\frac{1}{\var} \{L_1(x), L_2(y) \} = & k^- (x,y) L_1(x) L_2(y) +L_1(x) L_2(y) k^+(x,y) \\
&-L_1(x) s^-(x,y) L_2(y)-L_2(y) s^+(x,y) L_1(x)\,,
\end{split}
\end{equation}
where
\begin{equation}
\label{eq:app_km}
    \begin{split}
        k^-(x,y) = &\sum_{i j}  i \varphi(y-x+\Delta,\mathfrak{q}_{ji}) E_{ij} \otimes E_{ji}- \sum_{i j} i  \varphi^{\text{reg}}(\qq_{ij})  E_{ii} \otimes E_{jj}\\
    &+\sum_{ij} i  \varphi(x, \mathfrak{q}_{ij}) E_{ij} \otimes E_{jj} - \sum_{ij} i \varphi(y, \mathfrak{q}_{ij})  E_{jj} \otimes E_{ij} \\
        &- \sum_{iknm} \refl_{in,km} E_{ii} \otimes E_{kk}   \,,
    \end{split}
\end{equation}
\begin{equation}
\label{eq:app_kp}
    \begin{split}
        k^+(x,y)=&+ i \sum_{ij}   \varphi(x-y+\Delta, \mathfrak{q}_{ij}) E_{ij} \otimes E_{ji} - \sum_{i j} i \varphi^{\text{reg}}(\qq_{ij}) E_{ii} \otimes E_{jj} \\
        &- \sum_{ijkl} \refl_{ij,kl} E_{ii} \otimes E_{kk} \,,
    \end{split}
\end{equation}
\begin{equation}
\label{eq:app_sm}
    \begin{split}
        s^-(x,y)&=- \sum_{ij} i \varphi(y, \mathfrak{q}_{ij}) E_{jj} \otimes E_{ij} - \sum_{i  j} i \varphi^{\text{reg}}(\qq_{ij}) E_{ii} \otimes E_{jj}\\
        &- \sum_{ijkl} \refl_{ij,kl} E_{ii} \otimes E_{kk}  \,,
    \end{split}
\end{equation}
\begin{equation}
\label{eq:app_sp}
    \begin{split}
        s^+(x,y)&=+ \sum_{ij} i \varphi(x, \mathfrak{q}_{ij}) E_{ij} \otimes E_{jj} + \sum_{i  j} i \varphi^{\text{reg}}(\qq_{ij})   E_{jj} \otimes E_{ii}  \\
        & + \sum_{ijkl} \refl_{ij,kl} E_{kk} \otimes E_{ii}  \,.
    \end{split}
\end{equation}
This bracket matches the result previously obtained in~\cite{Arutyunov:1996uw}.

\section{Evaluation of brackets for invariant spins}
\label{app:br_inv_spins}

\subsection{Results of the computations}
In this appendix, we compute the brackets for the invariant spin variables before taking any limit or representation for the invariant oscillators. After a long computation, reported in subsections~\ref{app:aa_bracket_comput}, \ref{app:ab_bracket_comput} and~\ref{app:bb_bracket_comput} of the appendix, we obtain the following brackets for the double-loop invariant spin oscillators:
\begin{equation}
\label{eq:aa_inv_br_doubleloop}
\begin{split}
&\frac{1}{\var} \, \{\mathbf{a}_{a \alpha}(x,y) \,,\mathbf{a}_{b \beta}(z,t) \}=\scriptstyle{+t^{-1}_{a}(x) t^{-1}_{b}(z) t_b(x) t_{a}(z)}\\
&\scriptstyle{\times \Bigl( 
  i \varphi(z-x+\Delta,\mathfrak{q}_{ba})\mathbf{a}_{a \alpha}(x,y) \mathbf{a}_{b \beta}(z,t)-  i \varphi(t-x+\Delta,\mathfrak{q}_{ba})\mathbf{a}_{a \alpha}(x,y) \mathbf{a}_{a \beta}(z,t)  }\\
&\scriptstyle{- i \varphi(z-y+\Delta,\mathfrak{q}_{ba})\mathbf{a}_{b \alpha}(x,y) \mathbf{a}_{b \beta}(z,t)+  i \varphi(t-y+\Delta,\mathfrak{q}_{ba})\mathbf{a}_{b \alpha}(x,y) \mathbf{a}_{a \beta}(z,t) \Bigr)}\\
&\scriptstyle{- \mathbf{a}_{a \gamma}(x,y)\mathbf{a}_{b \delta}(z,t) \rho_{\gamma \alpha, \delta \beta}(y-t) -   \mathbf{a}_{a \gamma} (x,x)\mathbf{a}_{b \delta}(z,z) \rho_{\gamma 1, \delta 1}(x-z) \mathbf{a}_{a \alpha}(x,y) \mathbf{a}_{b \beta}(z,t)}\\
&\scriptstyle{+  \mathbf{a}_{a \gamma} (x,x)\mathbf{a}_{b \delta}(z,t) \rho_{\gamma 1, \delta \beta}(x-t) \mathbf{a}_{a \alpha}(x, y)+ \mathbf{a}_{a \gamma} (x,y)\mathbf{a}_{b \delta}(z,z)   \rho_{\gamma \alpha, \delta 1}(y-z) \mathbf{a}_{b \beta}(z, t)} \,,
\end{split}
\end{equation}
\begin{equation}
\label{eq:ab_inv_br_doubleloop}
\begin{split}
&\frac{1}{\var} \, \{\mathbf{a}_{a \alpha}(x,y), \mathbf{b}_{\beta b}(z,t)\}=\\
&\scriptstyle{+i  t^{-1}_a(x) t_a(z) \varphi(z-y+\Delta,\mathfrak{q}_{ba})   \mathbf{a}_{b \alpha}(z,y) \mathbf{b}_{\beta b}(z,t) }\\
&\scriptstyle{+\mathbf{a}_{a \gamma}(x,y) \rho_-(y-t)_{\gamma \alpha, \beta \delta} \mathbf{b}_{\delta b}(z,t)-  \mathbf{a}_{a \gamma} (x,y)\mathbf{a}_{b \delta}(x,z) \rho_{\gamma \alpha, \delta 1}(y-z) \mathbf{b}_{\beta b}(x,t)}\\
&\scriptstyle{-i  t^{-1}_a(x) t_a(z) \varphi(z-x+\Delta,\mathfrak{q}_{ba})   \mathbf{a}_{b 1}(z,x) \mathbf{b}_{\beta b}(z,t) \mathbf{a}_{a \alpha}(x,y) }\\
&\scriptstyle{-\mathbf{a}_{a \gamma}(x,x) \rho_-(x-t)_{\gamma 1, \beta \delta} \mathbf{b}_{\delta b}(z,t) \mathbf{a}_{a \alpha}(x,y)+ \mathbf{a}_{b 1}(x,z) \rho_{1 1, 1 1}(x-z) \mathbf{b}_{\beta b}(x,t) \mathbf{a}_{a \alpha}(x,y)}\\
&\scriptstyle{+\frac{2 \pi }{\var} \delta_{1 \beta} Q_b Q^{-1}_a t^{-1}_a(x) L_{ab}(t-\Delta) t_b(z) \delta(x-t)\mathbf{a}_{a \alpha}(x,y)}\\
&\scriptstyle{-\frac{2 \pi }{\var} \delta_{\alpha \beta} Q_b Q^{-1}_a t^{-1}_a(x) L_{ab}(t-\Delta) t_b(z) \delta(y-t)} \,,
\end{split}
\end{equation}
\begin{equation}
\label{eq:bb_inv_br_doubleloop}
\begin{split}
&\frac{1}{\var} \{\mathbf{b}_{\alpha a}(x,y),\mathbf{b}_{\beta b}(z,t)\} =\\
&\scriptstyle{-\rho_{\alpha \gamma, \beta \delta}(y-t)
\mathbf{b}_{\gamma a}(x,y) \mathbf{b}_{\delta b}(z,t)} \scriptstyle{+\mathbf{b}_{\beta b}(z,t)  \mathbf{a}_{b \gamma}(z,z) \rho_+(y-z)_{ \alpha \delta, \gamma 1} \mathbf{b}_{\delta a}(x,y)  }\\
&\scriptstyle{+\frac{2 \pi}{\var} \delta_{\alpha 1}  \delta(y-z)  \, \mathbf{b}_{\beta b}(x,t) Q^{-1}_b    \mathbb{L}_{ba}(x, y-\Delta) Q_a} \scriptstyle{+i \mathbf{b}_{\alpha b}(z,y) \mathbf{b}_{\beta a}(x,t)   \varphi(y-t+\Delta, \mathfrak{q}_{ba})      }\\
&\scriptstyle{+i \mathbf{b}_{\alpha a}(z,y) \varphi(z-x+\Delta,\mathfrak{q}_{ba})   \mathbf{a}_{b 1}(z,x) \mathbf{b}_{\beta b}(z,t) }\\
&\scriptstyle{+\mathbf{b}_{\alpha a}(x,y)   \mathbf{a}_{a \gamma}(x,x) \rho_-(x-t)_{\gamma 1, \beta \delta} \mathbf{b}_{\delta b}(z,t)-\frac{2 \pi }{\var} \delta_{1 \beta} \mathbf{b}_{\alpha a}(z,y) Q^{-1}_a \mathbb{L}_{ab}(z,t-\Delta) Q_b \delta(x-t)}\\
&\scriptstyle{-\mathbf{b}_{\alpha a}(x,y)  \mathbf{a}_{a \gamma} (x,x)\mathbf{a}_{b \delta}(x,z) \rho_{\gamma 1, \delta 1}(x-z) \mathbf{b}_{\beta b}(x,t) }\,.
\end{split}
\end{equation}
Subsequently we evaluate these brackets at the point $x=z=\imp^1$. Here $\imp^1$ is just the common value of $x$ and $z$. Later on we will promote this point to be dynamical but for the purpose of this appendix it is just a number in the interval $(-\pi,+\pi)$. For $x=z=\imp^1$ the brackets are expressed in terms of invariant spin oscillators alone, but derivative terms arise in the limit, returning
\begin{equation}
\label{eq:aa_bracket_final_components}
\begin{split}
&\frac{1}{\var} \, \{\mathbf{a}_{a \alpha}(x,y) \,,\mathbf{a}_{b \beta}(z,t) \}\bigl|_{x=z=\imp^1}=\\
& \scriptstyle{ + i \varphi(t-y+\Delta,\mathfrak{q}_{ba})\mathbf{a}_{b \alpha}(\imp^1,y) \mathbf{a}_{a \beta}(\imp^1,t) - i \varphi(t-\imp^1+\Delta,\mathfrak{q}_{ba})\mathbf{a}_{a \alpha}(\imp^1,y) \mathbf{a}_{a \beta}(\imp^1,t)}\\
&\scriptstyle{- i \varphi(\imp^1-y+\Delta,\mathfrak{q}_{ba})\mathbf{a}_{b \alpha}(\imp^1,y) \mathbf{a}_{b \beta}(\imp^1,t)+i \varphi^{\text{reg}}(\mathfrak{q}_{ba})\mathbf{a}_{a \alpha}(\imp^1,y) \mathbf{a}_{b \beta}(\imp^1,t)}\\
&\scriptstyle{- \mathbf{a}_{a \gamma}(\imp^1,y)\mathbf{a}_{b \delta}(\imp^1,t) \rho_{\gamma \alpha, \delta \beta}(y-t) +\mathbf{a}_{a \beta} (\imp^1,\imp^1)\mathbf{a}_{b 1}(\imp^1,t)   \rho_{\beta 1, 1 \beta}(\imp^1-t) \mathbf{a}_{a \alpha}(\imp^1, y)}\\
&\scriptstyle{+ \mathbf{a}_{a 1} (\imp^1,y)\mathbf{a}_{b \alpha}(\imp^1,\imp^1)   \rho_{1 \alpha, \alpha 1}(y-\imp^1) \mathbf{a}_{b \beta}(\imp^1, t)}\\
&\scriptstyle{-i \frac{d}{dx} ((\mathbf{a}_{a \alpha}(x,y) )\Bigr|_{x=\imp^1}   \mathbf{a}_{b \beta}(\imp^1,t)+i \frac{d}{dx} (\mathbf{a}_{b \beta}(x,t) )\Bigr|_{x=\imp^1}  \mathbf{a}_{a \alpha}(\imp^1,y)}\,,
\end{split}
\end{equation}
\begin{equation}
\label{eq:ab_bracket_with_lim2}
\begin{split}
&\frac{1}{\var} \, \{\mathbf{a}_{a \alpha}(x,y), \mathbf{b}_{\beta b}(z,t)\}\bigl|_{x=z=\imp^1}=\\
&\scriptstyle{+i \varphi(\imp^1-y+\Delta,\mathfrak{q}_{ba})   \mathbf{a}_{b \alpha}(\imp^1,y) \mathbf{b}_{\beta b}(\imp^1,t) -i \varphi^{\text{reg}}(\qq_{ba}) \mathbf{a}_{a \alpha}(\imp^1,y) \mathbf{b}_{\beta b}(\imp^1, t)}\\
&\scriptstyle{+\mathbf{a}_{a \gamma}(\imp^1,y) \rho(y-t)_{-\gamma \alpha, \beta \gamma} \mathbf{b}_{\gamma b}(\imp^1,t)-\frac{2 \pi }{\var} \delta_{\alpha \beta} Q_b Q^{-1}_a  \mathbb{L}_{ab}(t-\Delta) \delta(y-t)}\\
&\scriptstyle{-  \mathbf{a}_{a 1} (\imp^1,y)\mathbf{a}_{b \alpha}(\imp^1,\imp^1) \rho_{1 \alpha, \alpha 1}(y-\imp^1) \mathbf{b}_{\beta b}(\imp^1,t) }\\
&\scriptstyle{-\mathbf{a}_{a \gamma}(\imp^1,\imp^1) \rho(\imp^1-t)_{-\gamma 1, \beta \delta} \mathbf{b}_{\delta b}(\imp^1,t) \mathbf{a}_{a \alpha}(\imp^1,y)+\frac{2 \pi }{\var} \delta_{1 \beta} Q_b Q^{-1}_a  \mathbb{L}_{ab}(t-\Delta) \delta(\imp^1-t)\mathbf{a}_{a \alpha}(\imp^1,y)}\\
&\scriptstyle{+i \mathbf{a}_{a\alpha}(\imp^1 ,y) \frac{d}{dx} \mathbf{b}_{\beta b}(x,t)\Bigl|_{x=\imp^1} +i \mathbf{b}_{\beta b}(\imp^1 ,t) \frac{d}{dx} \mathbf{a}_{a \alpha}(x,y)\Bigl|_{x=\imp^1} } \, ,
\end{split}
\end{equation}
\begin{equation}
\label{eq:bb_bracket_final_components}
\begin{split}
&\frac{1}{\var} \{\mathbf{b}_{\alpha a}(x,y),\mathbf{b}_{\beta b}(z,t)\}\bigl|_{x=z=\imp^1} =\\
&\scriptstyle{-\rho_{\alpha \gamma, \beta \delta}(y-t)
\mathbf{b}_{\gamma a}(\imp^1,y) \mathbf{b}_{\delta b}(\imp^1,t)+i \mathbf{b}_{\alpha b}(\imp^1,y) \mathbf{b}_{\beta a}(\imp^1,t)   \varphi(y-t+\Delta, \mathfrak{q}_{ba}) } \\
&\scriptstyle{+\mathbf{b}_{\beta b}(\imp^1,t)  \mathbf{a}_{b \gamma}(\imp^1,\imp^1) \rho(y-\imp^1)_{+ \alpha \gamma, \gamma 1} \mathbf{b}_{\gamma a}(\imp^1,y)  }\\
&\scriptstyle{+\mathbf{b}_{\alpha a}(\imp^1,y)  \mathbf{a}_{a \gamma}(\imp^1,\imp^1)  \rho(\imp^1-t)_{-\gamma 1, \beta \gamma} \mathbf{b}_{\gamma b}(\imp^1,t)}\\
&\scriptstyle{+\frac{2 \pi}{\var} \delta_{\alpha 1}  \delta(y-\imp^1)  \, \mathbf{b}_{\beta b}(\imp^1, t) Q^{-1}_b    \mathbb{L}_{ba}(\imp^1, y-\Delta) Q_a-\frac{2 \pi }{\var} \delta_{1 \beta} \mathbf{b}_{\alpha a}(\imp^1,y) Q^{-1}_a \mathbb{L}_{ab}(\imp^1,t-\Delta) Q_b \delta(\imp^1-t)}\\
&\scriptstyle{+i \frac{d}{dz}\mathbf{b}_{\alpha a}(z,y) \Bigl|_{z=\imp^1} \mathbf{b}_{\beta b}(\imp^1,t)-i\mathbf{b}_{\alpha a}(\imp^1,y) \frac{d}{dz}\mathbf{b}_{\beta b}(z,t) \Bigl|_{z=\imp^1}+i \mathbf{b}_{\alpha a}(\imp^1,y) \varphi^{\text{reg}}(\mathfrak{q}_{ba})    \mathbf{b}_{\beta b}(\imp^1,t)} \,.
\end{split}
\end{equation}
In the rest of the appendix we report the explicit computation of the results listed above.

\subsection{a-a bracket}
\label{app:aa_bracket_comput}
Let us define
\begin{equation}
\mathbb{A}_{i \alpha}(x) \equiv \sum^N_{j=1}T_{ij}^{-1}(x) a_{j \alpha}(x) \,.
\end{equation}
Using the starting oscillator brackets~\eqref{eq:starting_osc_br}, we obtain
\begin{equation}
\begin{split}
\frac{1}{\var} \{ \mathbb{A}_{1}(x),\mathbb{A}_{2}(z) \} &= +\refl(x,z) \mathbb{A}_1(x)\mathbb{A}_2(z)  -\mathbb{A}_1(x)\mathbb{A}_2(z) \rho(x-z)\\
&+\frac{1}{\var} T^{-1}_1(x) T^{-1}_2(z) \{ T_1(x), T_2(z) \} \mathbb{A}_1(x)\mathbb{A}_2(z) \,.
\end{split}
\end{equation}
Substituting the T-T bracket (see equation~\eqref{eq:brackets_TT_final}), we obtain
\begin{equation}
\begin{split}
\frac{1}{\var} \{\mathbb{A}_{1}(x),\mathbb{A}_{2}(z) \} &=  +\refl(x,z) \mathbb{A}_1(x)\mathbb{A}_2(z)-\mathbb{A}_1(x)\mathbb{A}_2(z) \rho(x-z)\\
&+r_{\mathfrak{q}}(x,z) \mathbb{A}_1(x)\mathbb{A}_2(z)-\refl_{\pi}(x,z) \mathbb{A}_1(x)\mathbb{A}_2(z) \\
&+i \biggl( \sum_{ijk} \left(\X_{jk} - \X_{ik} \right) E_{ii} \otimes E_{jj} \biggr) \mathbb{A}_1(x) \mathbb{A}_2(z)\,,
\end{split}
\end{equation}
which in components is
\begin{equation}
\label{eq:AA_bracket}
\begin{split}
\frac{1}{\var} \{\mathbb{A}_{a \alpha}(x),\mathbb{A}_{b \beta}(z) \} &=  +\refl_{ac, bd}(x,z) \mathbb{A}_{c \alpha}(x)\mathbb{A}_{d \beta}(z)  - \mathbb{A}_{a \gamma}(x)\mathbb{A}_{b \delta}(z) \rho_{\gamma \alpha, \delta \beta}(x-z)\\
&+r_{\mathfrak{q}}(x,z)_{ac, bd} \mathbb{A}_{c \alpha}(x)\mathbb{A}_{d \beta}(z)-\refl_{\pi}(x,z)_{ac, bd}\mathbb{A}_{c \alpha}(x)\mathbb{A}_{d \beta}(z) \\
&+i \sum_{k} \left(\X_{bk} - \X_{ak} \right)   \mathbb{A}_{a \alpha}(x) \mathbb{A}_{b \beta}(z)\,,
\end{split}
\end{equation}

Now we introduce
\begin{equation}
    \mathbf{a}_{a \alpha}(x,y) = t^{-1}_a(x) \, \mathbb{A}_{a \alpha}(y)
\end{equation}
where
\begin{equation}
 t_a(x) = \mathbb{A}_{a 1}(x) \,.
\end{equation}

The brackets between $\mathbb{A}$ and $t$ are obtained by setting one of the two spin indices to $1$ in~\eqref{eq:AA_bracket}:
\begin{equation}
\label{eq:At_br_components}
\begin{split}
\frac{1}{\var} \{\mathbb{A}_{a \alpha}(x),t_{b}(z) \} &= + \refl_{ac, b d}(x,z) \mathbb{A}_{c \alpha}(x)t_d(z) -  \mathbb{A}_{a \gamma} (x)\mathbb{A}_{b \delta}(z) \rho_{\gamma \alpha, \delta 1}(x-z) \\
&+r_{\mathfrak{q}}(x,z)_{a c, b d} \mathbb{A}_{c \alpha}(x)t_d(z)-\refl_{\pi}(x,z)_{ac,bd} \mathbb{A}_{c \alpha}(x)t_d(z) \\
&+i \sum_{k} \left(\X_{bk} - \X_{ak} \right) \mathbb{A}_{a \alpha}(x) t_b(z)\,,
\end{split}
\end{equation}
\begin{equation}
\begin{split}
\frac{1}{\var} \{t_{a}(x), \mathbb{A}_{b \beta}(z) \} &=  + \refl(x,z)_{ac,bd} t_c(x) \mathbb{A}_{d \beta}(z) -  \mathbb{A}_{a \gamma} (x)\mathbb{A}_{b \delta}(z) \rho_{\gamma 1, \delta \beta}(x-z) \\
&+r_{\mathfrak{q}}(x,z)_{ac, bd} t_c(x) \mathbb{A}_{d \beta}(z)-\refl_{\pi}(x,z)_{ac, bd}t_c(x)\mathbb{A}_{d \beta}(z) \\
&+i  \sum_{k} \left(\X_{bk} - \X_{ak} \right)  t_a(x) \mathbb{A}_{b \beta}(z)\,,
\end{split}
\end{equation}
\begin{equation}
\label{eq:app_tt_bracket}
\begin{split}
\frac{1}{\var} \{t_{a}(x), t_{b}(z) \} &=  + \refl(x,z)_{ac,bd} t_c(x) t_{d}(z) - \mathbb{A}_{a \gamma} (x)\mathbb{A}_{b \delta}(z) \rho_{\gamma 1, \delta 1}(x-z) \\
&+r_{\mathfrak{q}}(x,z)_{ac, bd} t_c(x) t_{d}(z)-\refl_{\pi}(x,z)_{ac, bd}t_c(x)t_{d}(z) \\
&+i  \sum_{k} \left( \X_{bk} - \X_{ak} \right)  t_a(x) t_{b}(z)\,,
\end{split}
\end{equation}

The bracket for invariant spins is then given by
\begin{equation}
\begin{split}
\frac{1}{\var} \, \{\mathbf{a}_{a \alpha}(x,y) \,,\mathbf{a}_{b \beta}(z,t) \}&=   t^{-1}_{a}(x) t^{-1}_{b}(z) \frac{1}{\var} \, \{t_a(x), t_b(z) \} \mathbf{a}_{a \alpha}(x,y) \mathbf{a}_{b \beta}(z,t)\\
&-t^{-1}_{a}(x) t^{-1}_{b}(z) \frac{1}{\var} \, \{t_a(x), \mathbb{A}_{b \beta}(t) \} \mathbf{a}_{a \alpha}(x, y)\\
&-t^{-1}_{a}(x) t^{-1}_{b}(z) \frac{1}{\var} \, \{\mathbb{A}_{a \alpha}(y), t_b(z)\}  \mathbf{a}_{b \beta}(z, t)\\
&+t^{-1}_{a}(x) t^{-1}_{b}(z) \frac{1}{\var} \, \{ \mathbb{A}_{a \alpha}(y), \mathbb{A}_{b \beta}(t)\}  \,.
\end{split}
\end{equation}
After a long but straightforward computation, all derivative terms cancel out and we end up with formula~\eqref{eq:aa_inv_br_doubleloop}.

\subsection{a-b bracket}
\label{app:ab_bracket_comput}

Using that
\begin{equation}
    \mathbf{a}_{a \alpha}(x,y) = t^{-1}_a(x) \, \mathbb{A}_{a \alpha}(y) \,,
\end{equation}
we want to compute
\begin{equation}
\label{eq:ab_brackeet_starting}
\begin{split}
\frac{1}{\var} \, \{\mathbf{a}_{a \alpha}(x,y), \mathbf{b}_{\beta b}(z,t)\} &= t_a^{-1}(x) \frac{1}{\var} \, \{\mathbb{A}_{a \alpha}(y), \mathbf{b}_{\beta b}(z,t) \} - t^{-1}_a(x) \frac{1}{\var} \, \{t_a(x), \mathbf{b}_{\beta b}(z,t)\} \mathbf{a}_{a \alpha}(x,y)\\
&= t_a^{-1}(x) \frac{1}{\var} \,\{\mathbb{A}_{a \alpha}(y), \mathbf{b}_{\beta b}(z,t) \} - t^{-1}_a(x) \frac{1}{\var} \, \{\mathbb{A}_{a 1}(x), \mathbf{b}_{\beta b}(z,t)\} \mathbf{a}_{a \alpha}(x,y)\,.
\end{split}
\end{equation}
The term we must compute is then
\begin{equation}
\label{eq:building_b_ab_bracket}
\{\mathbb{A}_{a \alpha}(y), \mathbf{b}_{\beta b}(z,t) \} \,,
\end{equation}
from which the Poisson bracket can be reconstructed. 
Using that
\begin{equation}
\mathbf{b}_{\beta b}(z,t)=\left(b(t) T(t) Q^{-1} L(t-\Delta) Q t(z)\right)_{\beta b} \,,
\end{equation}
we split the computation of~\eqref{eq:building_b_ab_bracket} as follows
\begin{equation}
\label{eq:app_Ab_bracket_underbrace_I}
\begin{split}
\frac{1}{\var} \, \{\mathbb{A}_{a \alpha}(y), \mathbf{b}_{\beta b}(z,t) \} &= \underbrace{ \frac{1}{\var} \, \{\mathbb{A}_{a \alpha}(y), b_{\beta i}(t) \} T_{ij}(t) Q_b Q^{-1}_j L_{jb}(t-\Delta) t_b(z)}_{I_1}\\
&\underbrace{+ b_{\beta i}(t) \frac{1}{\var} \, \{\mathbb{A}_{a \alpha}(y),  T_{ij}(t) \}  Q_b Q^{-1}_j L_{jb}(t-\Delta) t_b(z)}_{I_2}\\
&\underbrace{+ b_{\beta i}(t) T_{ij}(t) \frac{1}{\var} \, \{\mathbb{A}_{a \alpha}(y), Q_b Q^{-1}_j  \}   L_{jb}(t-\Delta) t_b(z)}_{I_3}\\
&\underbrace{+ b_{\beta i}(t) T_{ij}(t) Q_b Q^{-1}_j \frac{1}{\var} \, \{\mathbb{A}_{a \alpha}(y), L_{jb}(t-\Delta)  \}    t_b(z)}_{I_4}\\
&\underbrace{+ b_{\beta i}(t) T_{ij}(t) Q_b Q^{-1}_j L_{jb}(t-\Delta) \frac{1}{\var} \, \{\mathbb{A}_{a \alpha}(y), t_b(z)  \} }_{I_5}\,.
\end{split}
\end{equation}
Let us compute the different terms entering the expression above one by one.

\paragraph{Evaluation of $I_1$.}
The first term is given by
\begin{equation}
I_1=T^{-1}_{ac}(y) \{a_{c \alpha}(y), b_{\beta i}(t) \} T_{ij}(t) Q_b Q^{-1}_j L_{jb}(t-\Delta) t_b(z)
\end{equation}
Using the oscillator algebra~\eqref{eq:starting_osc_br}, we obtain
\begin{equation}
\begin{split}
I_1=&\scriptstyle{-T^{-1}_{ac}(y)  b_{\beta k}(t)r_{+}(y-t)_{c l,ki}a_{l\alpha}(y) T_{ij}(t) Q_b Q^{-1}_j L_{jb}(t-\Delta) t_b(z)}\\
&\scriptstyle{+\mathbb{A}_{a \gamma}(y) \rho_-(y-t)_{\gamma \alpha, \beta \delta} \mathbf{b}_{\delta b}(z,t)-\frac{2 \pi }{\var} \delta_{\alpha \beta} Q_b Q^{-1}_a L_{ab}(t-\Delta) t_b(z) \delta(y-t)} \,.
\end{split}
\end{equation}

\paragraph{Evaluation of $I_2$.}
Using~\eqref{eq:brackets_TT_final}, we obtain
\begin{equation}
\begin{split}
I_2 =& -b_{\beta i}(t)T_{il}(t) r_\qq(y,t)_{an,lj} \mathbb{A}_{n \alpha}(y)   Q_b Q^{-1}_j L_{jb}(t-\Delta) t_b(z)\\
&+b_{\beta i}(t)T_{il}(t) \refl_\pi(y,t)_{an,lj} \mathbb{A}_{n \alpha}(y) Q_b Q^{-1}_j L_{jb}(t-\Delta) t_b(z)\\
&-i b_{\beta i}(t) \mathbb{A}_{a \alpha}(y) T_{ij}(t) \, \sum_k \left(\X_{jk} - \X_{ak} \right)  Q_b Q^{-1}_j L_{jb}(t-\Delta) t_b(z) \,.
\end{split}
\end{equation}
Plugging into the formula above the explicit expression for $r_{\mathfrak{q}}$,
we can write
\begin{equation}
\begin{split}
I_2 =& \scriptstyle{- i b_{\beta i}(t)T_{il}(t) \varphi(t-y+\Delta,\mathfrak{q}_{la})    \mathbb{A}_{l \alpha}(y)   Q_b Q^{-1}_a L_{ab}(t-\Delta) t_b(z)}\\
& \scriptstyle{+i  b_{\beta i}(t)T_{ij}(t) \varphi^{\text{reg}}(\mathfrak{q}_{aj})   \mathbb{A}_{a \alpha}(y)   Q_b Q^{-1}_j L_{jb}(t-\Delta) t_b(z)}\\
&\scriptstyle{+i b_{\beta i}(t)T_{ij}(t) \varphi(-y+\Delta, \mathfrak{q}_{ja})    \mathbb{A}_{j \alpha}(y)   Q_b Q^{-1}_j L_{jb}(t-\Delta) t_b(z)}\\
& \scriptstyle{-i b_{\beta i}(t)T_{il}(t) \varphi(-t+\Delta, \mathfrak{q}_{al})    \mathbb{A}_{a \alpha}(y)   Q_b Q^{-1}_a L_{ab}(t-\Delta) t_b(z)}\\
&\scriptstyle{+b_{\beta i}(t)T_{il}(t) \refl_\pi(y,t)_{an,lj} \mathbb{A}_{n \alpha}(y) Q_b Q^{-1}_j L_{jb}(t-\Delta) t_b(z)}\\
&\scriptstyle{-i b_{\beta i}(t) \mathbb{A}_{a \alpha}(y) T_{ij}(t) \, \sum_k \left(\X_{jk} - \X_{ak} \right)  Q_b Q^{-1}_j L_{jb}(t-\Delta) t_b(z)}\,.
\end{split}
\end{equation}

\paragraph{Evaluation of $I_3$.}
Since $Q$ and $T$ Poisson commute we get
\begin{equation}
I_3 = 0 \,.
\end{equation}

\paragraph{Evaluation of $I_4$.}

After some long manipulations, we obtain
\begin{equation}
\begin{split}
I_4&=b_{\beta i}(t) T_{ij}(t) Q_b Q^{-1}_j \frac{1}{\var} \, \{\mathbb{A}_{a \alpha}(y), L_{jb}(t-\Delta)  \}    t_b(z)=\\
& \scriptstyle{+i b_{\beta i}(t) T_{ij}(t) Q_b Q^{-1}_j  L_{jb}(t-\Delta) \varphi(-y+\Delta,\mathfrak{q}_{ba}) \mathbb{A}_{b\alpha}(y) t_b(z)}\\
& \scriptstyle{-i b_{\beta i}(t) T_{ij}(t) Q_b Q^{-1}_j 
L_{jb}(t-\Delta)  \varphi^{\text{reg}}(\mathfrak{q}_{ba})  \mathbb{A}_{a\alpha}(y) t_b(z)}\\
& \scriptstyle{+i b_{\beta i}(t) T_{ij}(t) Q_b Q^{-1}_j 
\varphi(t-y-\Delta,\mathfrak{q}_{ja})    \mathbb{A}_{j\alpha}(y) L_{ab}(t-\Delta) t_b(z)}\\
& \scriptstyle{-ib_{\beta i}(t) T_{ij}(t) Q_b Q^{-1}_j 
\varphi^{\text{reg}}(\mathfrak{q}_{aj}) \mathbb{A}_{a\alpha}(y) L_{jb}(t-\Delta) t_b(z)}\\
& \scriptstyle{-i b_{\beta i}(t) T_{ij}(t) Q_b Q^{-1}_j 
\varphi(-y+\Delta, \mathfrak{q}_{ja}) \mathbb{A}_{j\alpha}(y) L_{jb}(t-\Delta) t_b(z)}\\
& \scriptstyle{+i b_{\beta i}(t) T_{ij}(t) Q_b Q^{-1}_j 
\varphi(-t+\Delta+2 \Delta, \mathfrak{q}_{aj})  \mathbb{A}_{a\alpha}(y) L_{ab}(t-\Delta) t_b(z)}\\
&  \scriptstyle{-b_{\beta i}(t) T_{ij}(t) Q_b Q^{-1}_j L_{jb}(t-\Delta)   \sum_{l}  \refl_{as,bl}(y,0) \mathbb{A}_{s\alpha}(y)    t_b(z) }\\
& \scriptstyle{+ b_{\beta i}(t) T_{ij}(t) Q_b Q^{-1}_j L_{jb}(t-\Delta)  \sum_{nl}  \refl_{an, bl} \mathbb{A}_{a\alpha}(y)    t_b(z)}\\
& \scriptstyle{+b_{\beta i}(t) T_{ij}(t) Q_b Q^{-1}_j 
  \sum_l \refl_{as, jl}(y,0)   \mathbb{A}_{s\alpha}(y)  L_{jb}(t-\Delta) t_b(z)}\\
& \scriptstyle{-b_{\beta i}(t) T_{ij}(t) Q_b Q^{-1}_j 
\sum_{nl} \   \refl_{an,jl} \mathbb{A}_{a\alpha}(y) L_{jb}(t-\Delta) t_b(z)}\\
& \scriptstyle{-i b_{\beta i}(t) T_{ij}(t) Q_b Q^{-1}_j 
L_{jb}(t-\Delta) \sum_{k} \left( \X_{bk} - \X_{ak} \right) \mathbb{A}_{a\alpha}(y) t_b(z)}\\
&  \scriptstyle{+i  b_{\beta i}(t) T_{ij}(t) Q_b Q^{-1}_j 
\sum_{k} \left( \X_{jk} - \X_{ak} \right) \mathbb{A}_{a\alpha}(y) L_{jb}(t-\Delta) t_b(z) }\,.
\end{split}
\end{equation}

\paragraph{Evaluation of $I_5$.}

From~\eqref{eq:At_br_components} we know that
\begin{equation}
\begin{split}
\frac{1}{\var} \{\mathbb{A}_{a \alpha}(y),t_{b}(z) \} &= + \refl_{ac, b d}(y,z) \mathbb{A}_{c \alpha}(y)t_d(z) -  \mathbb{A}_{a \gamma} (y)\mathbb{A}_{b \delta}(z) \rho_{\gamma \alpha, \delta 1}(y-z) \\
&+r_{\mathfrak{q}}(y,z)_{a c, b d} \mathbb{A}_{c \alpha}(y)t_d(z)-\refl_{\pi}(y,z)_{ac,bd} \mathbb{A}_{c \alpha}(y)t_d(z) \\
&+i \sum_{k} \left(\X_{bk} - \X_{ak} \right) \mathbb{A}_{a \alpha}(y) t_b(z)\,,
\end{split}
\end{equation}
Then, using the explicit expression for $r_\qq$, we get
\begin{equation}
\begin{split}
I_5=&\scriptstyle{+b_{\beta i}(t) T_{ij}(t) Q_b Q^{-1}_j L_{jb}(t-\Delta) \frac{1}{\var} \, \{\mathbb{A}_{a \alpha}(y), t_b(z)  \}}\\
=&\scriptstyle{+b_{\beta i}(t) T_{ij}(t) Q_b Q^{-1}_j L_{jb}(t-\Delta) \refl_{ac, b d}(y,z) \mathbb{A}_{c \alpha}(y)t_d(z)}\\
&\scriptstyle{-b_{\beta i}(t) T_{ij}(t) Q_b Q^{-1}_j L_{jb}(t-\Delta) \mathbb{A}_{a \gamma} (y)\mathbb{A}_{b \delta}(z) \rho_{\gamma \alpha, \delta 1}(y-z)}\\
&\scriptstyle{-b_{\beta i}(t) T_{ij}(t) Q_b Q^{-1}_j L_{jb}(t-\Delta)\refl_{\pi}(y,z)_{ac,bd} \mathbb{A}_{c \alpha}(y)t_d(z)}\\
&\scriptstyle{+ib_{\beta i}(t) T_{ij}(t) Q_b Q^{-1}_j L_{jb}(t-\Delta) \sum_{k} \left(\X_{bk} - \X_{ak} \right) \mathbb{A}_{a \alpha}(y) t_b(z)}\\
&\scriptstyle{+i b_{\beta i}(t) T_{ij}(t) Q_b Q^{-1}_j L_{jb}(t-\Delta)  \varphi(z-y+\Delta,\mathfrak{q}_{ba})   \mathbb{A}_{b \alpha}(y)t_a(z)}\\
&\scriptstyle{-i b_{\beta i}(t) T_{ij}(t) Q_b Q^{-1}_j L_{jb}(t-\Delta) \varphi^{\text{reg}}(\mathfrak{q}_{ab})  \mathbb{A}_{a \alpha}(y)t_b(z)}\\
&\scriptstyle{-ib_{\beta i}(t) T_{ij}(t) Q_b Q^{-1}_j L_{jb}(t-\Delta) \varphi(-y+\Delta, \mathfrak{q}_{ba})    \mathbb{A}_{b \alpha}(y)t_b(z)}\\
&\scriptstyle{+i b_{\beta i}(t) T_{ij}(t) Q_b Q^{-1}_j L_{jb}(t-\Delta) \varphi(-z+\Delta, \mathfrak{q}_{ab})  \mathbb{A}_{a \alpha}(y)t_a(z) }\,.
\end{split}
\end{equation}

\paragraph{All terms together.}

Plugging the terms $I_1$, $I_2$, $I_3$, $I_4$ and $I_5$ previously obtained in~\eqref{eq:app_Ab_bracket_underbrace_I}, after a massive simplification, we get
\begin{equation}
\label{eq:res_as_sum_of_Is_final2}
\begin{split}
&\frac{1}{\var} \, \{\mathbb{A}_{a \alpha}(y), \mathbf{b}_{\beta b}(z,t) \} =\\ &\scriptstyle{+i   t_a(z) \varphi(z-y+\Delta,\mathfrak{q}_{ba})   \mathbf{a}_{b \alpha}(z,y) \mathbf{b}_{\beta b}(z,t) +i t_a(z) t_a(x) t_b^{-1}(z)   \varphi(-z+\Delta, \mathfrak{q}_{ab})  \mathbf{a}_{a \alpha}(x,y) \mathbf{b}_{\beta b}(z,t)}\\
&\scriptstyle{+\mathbb{A}_{a \gamma}(y) \rho_-(y-t)_{\gamma \alpha, \beta \delta} \mathbf{b}_{\delta b}(z,t)-\frac{2 \pi }{\var} \delta_{\alpha \beta} Q_b Q^{-1}_a L_{ab}(t-\Delta) t_b(z) \delta(y-t)}\\
&\scriptstyle{-  b_{\beta i}(t)T_{il}(t)  \sum_{s} \refl(0,t)_{as, lj} \mathbb{A}_{a \alpha}(y) Q_b Q^{-1}_j L_{jb}(t-\Delta) t_b(z)}\\
&\scriptstyle{-  \mathbb{A}_{a \gamma} (y)\mathbf{a}_{b \delta}(x,z) \rho_{\gamma \alpha, \delta 1}(y-z) \mathbf{b}_{\beta b}(x,t) + b_{\beta i}(t) T_{ij}(t) Q_b Q^{-1}_j L_{jb}(t-\Delta)\sum_{s} \refl(0,z)_{as, bd} \mathbb{A}_{a \alpha}(y)t_d(z)}\,.
\end{split}
\end{equation}
As for the $\mathbf{a}$-$\mathbf{a}$ bracket, in this case the derivatives also cancel.
Replacing this expression into~\eqref{eq:ab_brackeet_starting} and using that $\mathbf{a}_{a1}(x,x)=1$, we end up with relation~\eqref{eq:ab_inv_br_doubleloop}.

\subsection{b-b bracket}
\label{app:bb_bracket_comput}

Using
\begin{equation}
\begin{split}
&\mathbf{b}_{\alpha a}(x,y)=\left(b(y) T(y) Q^{-1} L(y-\Delta) Q t(x)\right)_{\alpha a} \,,\\
&\mathbf{b}_{\beta b}(z,t)=\left(b(t) T(t) Q^{-1} L(t-\Delta) Q t(z)\right)_{\beta b} \,,
\end{split}
\end{equation}
we compute
\begin{equation}
\label{eq:app_bb_bracket_starting}
\begin{split}
&\frac{1}{\var} \{\mathbf{b}_{\alpha a}(x,y),\mathbf{b}_{\beta b}(z,t)\} = \frac{1}{\var} \{b_{\alpha i}(y) T_{ij}(y) Q_j^{-1} L_{ja}(y-\Delta) Q_a t_a(x), \mathbf{b}_{\beta b}(z,t) \} =\\
&\underbrace{+ \frac{1}{\var} \{b_{\alpha i}(y) , \mathbf{b}_{\beta b}(z,t) \} T_{ij}(y) Q_j^{-1} L_{ja}(y-\Delta) Q_a t_a(x)}_{Y_1} \\
&\underbrace{+b_{\alpha i}(y) \frac{1}{\var} \{ T_{ij}(y) , \mathbf{b}_{\beta b}(z,t) \}  Q_j^{-1} L_{ja}(y-\Delta) Q_a t_a(x) }_{Y_2}\\
&\underbrace{+b_{\alpha i}(y) T_{ij}(y) Q_j^{-1} \frac{1}{\var} \{ L_{ja}(y-\Delta) , \mathbf{b}_{\beta b}(z,t) \}    Q_a t_a(x) }_{Y_3}\\
&\underbrace{+b_{\alpha i}(y) T_{ij}(y) Q_j^{-1} L_{ja}(y-\Delta) Q_a \frac{1}{\var} \{ t_a(x) , \mathbf{b}_{\beta b}(z,t) \} }_{Y_4}\\
&\underbrace{+b_{\alpha i}(y) T_{ij}(y)  L_{ja}(y-\Delta)  \frac{1}{\var} \{ Q_a Q_j^{-1} , \mathbf{b}_{\beta b}(z,t) \} t_a(x) }_{Y_5}\\
\end{split}
\end{equation}
We evaluate all terms composing the expression above one by one.

\paragraph{Evaluation of $Y_1$.}

We start computing
\begin{equation}
\begin{split}
Y_1&=\frac{1}{\var}\{b_{\alpha i}(y) , \mathbf{b}_{\beta b}(z,t) \} T_{ij}(y) Q_j^{-1} L_{ja}(y-\Delta) Q_a t_a(x)\\
&=\frac{1}{\var} \{b_{\alpha i}(y) , b_{\beta k}(t) T_{kl}(t) Q_l^{-1} L_{lb}(t-\Delta) Q_b t_b(z) \} T_{ij}(y) Q_j^{-1} L_{ja}(y-\Delta) Q_a t_a(x)\\
&=\frac{1}{\var}\{b_{\alpha i}(y) , b_{\beta k}(t)  \} T_{ij}(y) Q_j^{-1} L_{ja}(y-\Delta) Q_a t_a(x) T_{kl}(t) Q_l^{-1} L_{lb}(t-\Delta) Q_b t_b(z)\\
&+b_{\beta k}(t) T_{kl}(t) Q_l^{-1} L_{lb}(t-\Delta) Q_b \frac{1}{\var}\{b_{\alpha i}(y) , t_b(z)  \} T_{ij}(y) Q_j^{-1} L_{ja}(y-\Delta) Q_a t_a(x) \,.
\end{split}
\end{equation}
Replacing the brackets for the starting oscillators, we obtain
\begin{equation}
\begin{split}
Y_1&=+b_{\alpha s}(y)b_{\beta n}(t)r_{s i, n k}(y-t) T_{ij}(y) Q_j^{-1} L_{ja}(y-\Delta) Q_a t_a(x) T_{kl}(t) Q_l^{-1} L_{lb}(t-\Delta) Q_b t_b(z) \\
&-\rho_{\alpha \gamma, \beta \delta}(y-t)
\mathbf{b}_{\gamma a}(x,y) \mathbf{b}_{\delta b}(z,t) \\
&-b_{\beta k}(t) T_{kl}(t) Q_l^{-1} L_{lb}(t-\Delta) Q_b T^{-1}_{bs}(z) b_{\alpha r}(y) r_{-}(y-z)_{ri,sq } a_{q1}(z)  T_{ij}(y) Q_j^{-1} L_{ja}(y-\Delta) Q_a t_a(x)\\
&+b_{\beta k}(t) T_{kl}(t) Q_l^{-1} L_{lb}(t-\Delta) Q_b T^{-1}_{bs}(z) a_{s\gamma}(z) \rho_+(y-z)_{ \alpha \delta, \gamma 1} \mathbf{b}_{\delta a}(x,y)  \\
&+\frac{2 \pi}{\var} \delta_{\alpha 1}  \delta(y-z)  \, \mathbf{b}_{\beta b}(t) Q^{-1}_b    \mathbb{L}_{ba}(x, y-\Delta) Q_a  \,.
\end{split}
\end{equation}

\paragraph{Evaluation of $Y_2$.}

Now we compute
\begin{equation}
\begin{split}
Y_2 &= b_{\alpha i}(y) \frac{1}{\var} \{ T_{ij}(y) , \mathbf{b}_{\beta b}(z,t) \}  Q_j^{-1} L_{ja}(y-\Delta) Q_a t_a(x)\\
&=b_{\alpha i}(y) b_{\beta s}(t) \frac{1}{\var} \{ T_{ij}(y) , T_{sl}(t) \} Q^{-1}_l L_{l b}(t-\Delta) Q_b t_b(z)  Q_j^{-1} L_{ja}(y-\Delta) Q_a t_a(x) \\
&+b_{\alpha i}(y) b_{\beta s}(t) T_{sl}(t) \frac{1}{\var} \{ T_{ij}(y) ,  L_{l b}(t-\Delta) \} Q^{-1}_l  Q_b t_b(z)  Q_j^{-1} L_{ja}(y-\Delta) Q_a t_a(x)\\
&+b_{\alpha i}(y) b_{\beta s}(t) T_{sl}(t) L_{l b}(t-\Delta)\frac{1}{\var} \{ T_{ij}(y) , t_b(z)  \} Q^{-1}_l  Q_b   Q_j^{-1} L_{ja}(y-\Delta) Q_a t_a(x) \,.
\end{split}
\end{equation}
using that
\begin{equation}
L_{lb}(t-\Delta)= T_{ln}^{-1}(t-2\Delta) B_{nm}(t-\Delta) T_{mb}(t-2\Delta) \,, \qquad t_b(z)=T_{br}^{-1}(z) a_{r1}(z) \,,
\end{equation}
we can write the expression above as
\begin{equation}
\begin{split}
Y_2 =&\scriptstyle{+b_{\alpha i}(y) b_{\beta s}(t) \frac{1}{\var} \{ T_{ij}(y) , T_{sl}(t) \} Q^{-1}_l L_{l b}(t-\Delta) Q_b t_b(z)  Q_j^{-1} L_{ja}(y-\Delta) Q_a t_a(x)} \\
&\scriptstyle{-b_{\alpha i}(y) b_{\beta s}(t) T_{sl}(t) T_{lk}^{-1}(t-2\Delta)\frac{1}{\var} \{ T_{ij}(y) , T_{kr}(t-2\Delta)  \}  T_{rn}^{-1}(t-2\Delta)B_{nm}(t-\Delta) T_{mb}(t-2\Delta)Q^{-1}_l  Q_b t_b(z)  Q_j^{-1} L_{ja}(y-\Delta) Q_a t_a(x)}\\
&\scriptstyle{+b_{\alpha i}(y) b_{\beta s}(t) T_{sl}(t) T_{ln}^{-1}(t-2\Delta)\frac{1}{\var} \{ T_{ij}(y) , B_{nm}(t-\Delta)  \}   T_{mb}(t-2\Delta)Q^{-1}_l  Q_b t_b(z)  Q_j^{-1} L_{ja}(y-\Delta) Q_a t_a(x)}\\
&\scriptstyle{+b_{\alpha i}(y) b_{\beta s}(t) T_{sl}(t) T_{ln}^{-1}(t-2\Delta) B_{nm}(t-\Delta)  \frac{1}{\var} \{ T_{ij}(y) ,  T_{mb}(t-2\Delta)\}   Q^{-1}_l  Q_b t_b(z)  Q_j^{-1} L_{ja}(y-\Delta) Q_a t_a(x)}\\
&\scriptstyle{-b_{\alpha i}(y) b_{\beta s}(t) T_{sl}(t) L_{l b}(t-\Delta) T^{-1}_{bq}(z) \frac{1}{\var} \{ T_{ij}(y) , T_{qp}(z)  \} T^{-1}_{pr}(z) a_{r1}(z) Q^{-1}_l  Q_b   Q_j^{-1} L_{ja}(y-\Delta) Q_a t_a(x)} \,.
\end{split}
\end{equation}
Replacing~\eqref{eq:brackets_TT_final} into the expression above, all the derivative terms cancel, and (using the explicit expression for $r_\qq$) we obtain
\begin{equation}
\begin{split}
Y_2 =&\scriptstyle{+b_{\alpha i}(y) b_{\beta s}(t) T_{sl}(t) T_{ln}^{-1}(t-2\Delta)\frac{1}{\var} \{ T_{ij}(y) , B_{nm}(t-\Delta)  \}   T_{mb}(t-2\Delta)Q^{-1}_l  Q_b t_b(z)  Q_j^{-1} L_{ja}(y-\Delta) Q_a t_a(x)}\\
&\scriptstyle{+i b_{\alpha i}(y) b_{\beta s}(t) T_{sl}(t) L_{lj}(t-\Delta)  T_{ib}(y)  \varphi(t-y-\Delta,\mathfrak{q}_{jb})    Q^{-1}_l  Q_b t_b(z)  Q_j^{-1} L_{ja}(y-\Delta) Q_a t_a(x)}\\
&\scriptstyle{+i b_{\alpha i}(y) b_{\beta s}(t) T_{sl}(t) L_{lq}(t-\Delta)  T_{ib}(y)  \varphi(-t+\Delta+2\Delta, \mathfrak{q}_{bq})    Q^{-1}_l  Q_b t_b(z)  Q_b^{-1} L_{ba}(y-\Delta) Q_a t_a(x)}\\
&\scriptstyle{-i b_{\alpha i}(y) b_{\beta s}(t) T_{sl}(t) L_{l b}(t-\Delta) T_{ip}(y)   \varphi(z-y+\Delta,\mathfrak{q}_{bp})  T^{-1}_{pr}(z) a_{r1}(z) Q^{-1}_l  Q_b   Q_b^{-1} L_{ba}(y-\Delta) Q_a t_a(x)}\\
&\scriptstyle{-i b_{\alpha i}(y) b_{\beta s}(t) T_{sl}(t) L_{l b}(t-\Delta) T_{ip}(y)     \varphi(-z+\Delta, \mathfrak{q}_{pb})  T^{-1}_{pr}(z) a_{r1}(z) Q^{-1}_l  Q_b   Q_p^{-1} L_{pa}(y-\Delta) Q_a t_a(x)}\\
&\scriptstyle{-b_{\alpha i}(y) b_{\beta s}(t)   T_{i k}(y) T_{s n}(t) \refl_\pi(y,t)_{kj, nl}  Q^{-1}_l L_{l b}(t-\Delta) Q_b t_b(z)  Q_j^{-1} L_{ja}(y-\Delta) Q_a t_a(x)} \\
&\scriptstyle{+b_{\alpha i}(y) b_{\beta s}(t) T_{sl}(t)   T_{ip}(y)  \refl_\pi(y, t-2 \Delta)_{pj,lr}  L_{rb}(t-\Delta)Q^{-1}_l  Q_b t_b(z)  Q_j^{-1} L_{ja}(y-\Delta) Q_a t_a(x)}\\
&\scriptstyle{-b_{\alpha i}(y) b_{\beta s}(t) T_{sl}(t) L_{lq}(t-\Delta) T_{ip}(y)  \refl_\pi (y, t-2\Delta)_{pj, qb}    Q^{-1}_l  Q_b t_b(z)  Q_j^{-1} L_{ja}(y-\Delta) Q_a t_a(x)}\\
&\scriptstyle{+b_{\alpha i}(y) b_{\beta s}(t) T_{sl}(t) L_{l b}(t-\Delta)   T_{in}(y) \refl_\pi(y,z)_{nj, bp}  T^{-1}_{pr}(z) a_{r1}(z) Q^{-1}_l  Q_b   Q_j^{-1} L_{ja}(y-\Delta) Q_a t_a(x)} \,.
\end{split}
\end{equation}
Using the bracket T-B~\eqref{eq:app_TB_bracket} and the explicit form for $\refl_\pi$, after several simplifications, we obtain the following final formula 
\begin{equation}
\begin{split}
Y_2 =&\scriptstyle{-i b_{\alpha i}(y) b_{\beta s}(t) T_{sl}(t) L_{l b}(t-\Delta) T_{ip}(y)   \varphi(z-y+\Delta,\mathfrak{q}_{bp})  T^{-1}_{pr}(z) a_{r1}(z) Q^{-1}_l  Q_b   Q_b^{-1} L_{ba}(y-\Delta) Q_a t_a(x)}\\
&\scriptstyle{-i b_{\alpha i}(y) b_{\beta s}(t) T_{sl}(t) L_{l b}(t-\Delta) T_{ip}(y)     \varphi(-z+\Delta, \mathfrak{q}_{pb})  T^{-1}_{pr}(z) a_{r1}(z) Q^{-1}_l  Q_b   Q_p^{-1} L_{pa}(y-\Delta) Q_a t_a(x)}\\
&\scriptstyle{-b_{\alpha i}(y) b_{\beta s}(t)   T_{i k}(y) T_{s n}(t) \refl(y,t)_{kj, nl}  Q^{-1}_l L_{l b}(t-\Delta) Q_b t_b(z)  Q_j^{-1} L_{ja}(y-\Delta) Q_a t_a(x)} \\
&\scriptstyle{+b_{\alpha i}(y) b_{\beta s}(t)   T_{i j}(y) T_{s n}(t)  \sum_w \refl(0,t)_{jw, nl}  Q^{-1}_l L_{l b}(t-\Delta) Q_b t_b(z)  Q_j^{-1} L_{ja}(y-\Delta) Q_a t_a(x)} \\
&\scriptstyle{+b_{\alpha i}(y) b_{\beta s}(t) T_{sl}(t) L_{l b}(t-\Delta)   T_{in}(y) \refl(y,z)_{nj, bp}  T^{-1}_{pr}(z) a_{r1}(z) Q^{-1}_l  Q_b   Q_j^{-1} L_{ja}(y-\Delta) Q_a t_a(x)}\\
&\scriptstyle{-b_{\alpha i}(y) b_{\beta s}(t) T_{sl}(t) L_{l b}(t-\Delta)   T_{ij}(y) \sum_w \refl(0,z)_{jw, bp}  T^{-1}_{pr}(z) a_{r1}(z) Q^{-1}_l  Q_b   Q_j^{-1} L_{ja}(y-\Delta) Q_a t_a(x)}\,.
\end{split}
\end{equation}

\paragraph{Evaluation of $Y_3$.}

\begin{equation}
\label{eq:Y3_first_defined}
    \begin{split}
Y_3 &= b_{\alpha i}(y) T_{ij}(y) Q_j^{-1} \frac{1}{\var} \{ L_{ja}(y-\Delta) , \mathbf{b}_{\beta b}(z,t) \}    Q_a t_a(x)\\
&=b_{\alpha i}(y) T_{ij}(y) Q_j^{-1} b_{\beta k}(t)\frac{1}{\var} \{ L_{ja}(y-\Delta) ,  T_{kl}(t)  \} Q_l^{-1} L_{lb}(t-\Delta) Q_b t_b(z)    Q_a t_a(x)\\
&+b_{\alpha i}(y) T_{ij}(y) Q_j^{-1} b_{\beta k}(t) T_{kl}(t) \frac{1}{\var} \{ L_{ja}(y-\Delta) ,   L_{lb}(t-\Delta) \} Q_l^{-1}  Q_b t_b(z)    Q_a t_a(x)\\
&+b_{\alpha i}(y) T_{ij}(y) Q_j^{-1} b_{\beta k}(t) T_{kl}(t) Q_l^{-1}  L_{lb}(t-\Delta) Q_b  \frac{1}{\var} \{ L_{ja}(y-\Delta) ,  t_b(z)  \} Q_a t_a(x)     \\
&+b_{\alpha i}(y) T_{ij}(y) Q_j^{-1} b_{\beta k}(t) T_{kl}(t)   L_{lb}(t-\Delta)   \frac{1}{\var} \{ L_{ja}(y-\Delta) , Q_l^{-1}Q_b\} t_b(z) Q_a t_a(x) \,.
    \end{split}
\end{equation}
The Poisson bracket needed for the computation is
\begin{equation}
        \frac{1}{\var} \{ L_1(x), T_2(y) \}= \frac{1}{\var} \{ L_1(x), T_2(y) \}_{\text{Ell.}} + \frac{1}{\var} \{ L_1(x), T_2(y) \}_{\text{Trig.}} \,,
\end{equation}
which we split into its elliptic and trigonometric parts.
All derivative terms cancel in the sum, and we do not consider them.
It is possible to show that
\begin{equation}
    \begin{split}
        \frac{1}{\var} \{ L_{ja}(x), T_{lb}(y) \}_{\text{Ell.}}=&\scriptstyle{- i L_{j a}(x) T_{l b}(y) \varphi^{\text{reg}}(\qq_{ab})  + i  L_{j a}(x) T_{l n}(y) \delta_{ab}  \varphi(-y+\Delta, \qq_{an})  }\\
&\scriptstyle{- i T_{lm}(y) \delta_{jb}   \varphi(y-x+2\Delta,\qq_{mb}) L_{ma}(x) + i T_{lb}(y)   \varphi^{\text{reg}}(\qq_{jb})  L_{ja}(x)}\\
&\scriptstyle{+ i T_{lb}(y)  \varphi(-x+2\Delta,\qq_{bj})  L_{ba}(x) - i \delta_{bj} T_{ln}(y)     \varphi(-y+\Delta, \qq_{bn}) L_{ja}(x)} \,,
    \end{split}
\end{equation}
\begin{equation}
    \begin{split}
        \frac{1}{\var} \{ L_{ja}(x), T_{lb}(y) \}_{\text{Trig.}}=&\scriptstyle{ + L_{ja}(x) T_{lq}(y) \sum_{n} \refl(0,y)_{an,qb}  }\\
        &\scriptstyle{  - L_{ja}(x) T_{lb}(y) \sum_{nm} \refl_{an,bm} - T_{lp}(y) \sum_{n} \refl(0, y)_{jn,pb}  L_{ja}(x)  }\\
        &\scriptstyle{+ T_{lb}(y) \sum_{nm} \refl_{jn,bm}  L_{ja}(x)   }\,.
    \end{split}
\end{equation}

From these expressions, we obtain
\begin{equation}
    \begin{split}
        \frac{1}{\var} \{ L_{ja}(y-\Delta), T_{kl}(t) \}_{\text{Ell.}}=&\scriptstyle{- i L_{j a}(y-\Delta) T_{k l}(t) \varphi^{\text{reg}}(\qq_{al})  + i  L_{j a}(y-\Delta) T_{k n}(t) \delta_{al}  \varphi(-t+\Delta, \qq_{an})  }\\
&\scriptstyle{- i T_{km}(t) \delta_{jl}   \varphi(t-y+\Delta+2\Delta,\qq_{ml}) L_{ma}(y-\Delta) + i T_{kl}(t)   \varphi^{\text{reg}}(\qq_{jl})  L_{ja}(y-\Delta)}\\
&\scriptstyle{+ i T_{kl}(t)  \varphi(-y+\Delta+2\Delta,\qq_{lj})  L_{la}(y-\Delta) - i \delta_{lj} T_{kn}(t)     \varphi(-t+\Delta, \qq_{ln}) L_{ja}(y-\Delta)} \,,
    \end{split}
\end{equation}
\begin{equation}
    \begin{split}
        \frac{1}{\var} \{ L_{ja}(y-\Delta), T_{kl}(t) \}_{\text{Trig.}}=&\scriptstyle{ + L_{ja}(y-\Delta) T_{kq}(t) \sum_{n} \refl(0,t)_{an,ql}  }\\
        &\scriptstyle{  - L_{ja}(y-\Delta) T_{kl}(t) \sum_{nm} \refl_{an,lm} - T_{kp}(t) \sum_{n} \refl(0, t)_{jn,pl}  L_{ja}(y-\Delta)  }\\
        &\scriptstyle{+ T_{kl}(t) \sum_{nm} \refl_{jn,lm}  L_{ja}(y-\Delta)     }\,,
    \end{split}
\end{equation}
\begin{equation}
    \begin{split}
        \frac{1}{\var} \{ L_{ja}(y-\Delta) ,  t_b(z)  \}_{\text{Ell.}}=&+\frac{1}{\var} \{ L_{ja}(y-\Delta) ,  T^{-1}_{bp}(z)  \}_{\text{Ell.}}a_{p1}(z)\\
        =&-T^{-1}_{bk}(z) \frac{1}{\var} \{ L_{ja}(y-\Delta) ,  T_{kl}(z)  \}_{\text{Ell.}} t_l(z)\\
        =&\scriptstyle{+ i L_{j a}(y-\Delta)  \varphi^{\text{reg}}(\qq_{ab}) t_b(z)  - i  L_{j a}(y-\Delta)    \varphi(-z+\Delta, \qq_{ab}) t_a(z)  }\\
&\scriptstyle{+ i    \varphi(z-y+\Delta+2\Delta,\qq_{bj}) L_{ba}(y-\Delta) t_j(z) - i    \varphi^{\text{reg}}(\qq_{jb})  L_{ja}(y-\Delta) t_b(z)}\\
&\scriptstyle{- i   \varphi(-y+\Delta+2\Delta,\qq_{bj})  L_{ba}(y-\Delta) t_b(z) + i       \varphi(-z+\Delta, \qq_{jb}) L_{ja}(y-\Delta) t_j(z)} \,,
    \end{split}
\end{equation}
\begin{equation}
    \begin{split}
        \frac{1}{\var} \{ L_{ja}(y-\Delta) ,  t_b(z)  \}_{\text{Trig.}} =&-T^{-1}_{bk}(z) \frac{1}{\var} \{ L_{ja}(y-\Delta) ,  T_{kl}(z)  \}_{\text{Trig.}} t_l(z)\\
        &\scriptstyle{ - L_{ja}(y-\Delta)  \sum_{n} \refl(0,z)_{an,bl} t_l(z) }\\
        &\scriptstyle{  + L_{ja}(y-\Delta)  \sum_{nm} \refl_{an,bm} t_b(z) +  \sum_{n} \refl(0, z)_{jn,bl}  L_{ja}(y-\Delta) t_l(z) }\\
        &\scriptstyle{- \sum_{nm} \refl_{jn,bm}  L_{ja}(y-\Delta) t_b(z) } \,.
    \end{split}
\end{equation}
From~\eqref{eq:app_LL_bracket}, we also have
\begin{equation}
\begin{split}
\frac{1}{\var} \{L_{ja}(y-\Delta), L_{lb}(t-\Delta) \} =&+\frac{1}{\var} \{L_{ja}(y-\Delta), L_{lb}(t-\Delta) \}_{\text{Ell.}}\\
&+ \frac{1}{\var} \{L_{ja}(y-\Delta), L_{lb}(t-\Delta) \}_{\text{Trig.}}\,,
\end{split}
\end{equation}
where
\begin{equation}
\begin{split}
&\frac{1}{\var} \{L_{ja}(y-\Delta), L_{lb}(t-\Delta) \}_{\text{Ell.}} =\\ 
&\scriptstyle{   + i \varphi(t-y+\Delta,\mathfrak{q}_{lj}) L_{la}(y-\Delta) L_{jb}(t-\Delta) -   i  \varphi^{\text{reg}}(\qq_{jl}) L_{ja}(y-\Delta) L_{lb}(t-\Delta)} \\
&\scriptstyle{+ i  \varphi(y-\Delta, \mathfrak{q}_{jl}) L_{la}(y-\Delta) L_{lb}(t-\Delta)  -  i \varphi(t-\Delta, \mathfrak{q}_{lj}) L_{ja}(y-\Delta) L_{jb}(t-\Delta)}\\
&\scriptstyle{ + i L_{jb}(y-\Delta) L_{la}(t-\Delta)   \varphi(y-t+\Delta, \mathfrak{q}_{ba})  -  i L_{ja}(y-\Delta) L_{lb}(t-\Delta)  \varphi^{\text{reg}}(\qq_{ab})} \\
&\scriptstyle{
+i  \varphi(t-\Delta, \mathfrak{q}_{la}) L_{ja}(y-\Delta) L_{ab}(t-\Delta)
+  i \varphi^{\text{reg}}(\qq_{al}) L_{ja}(y-\Delta) L_{lb}(t-\Delta)
}\\
&\scriptstyle{-L_{lb}(t-\Delta)  i \varphi(y-\Delta, \mathfrak{q}_{jb}) L_{ba}(y-\Delta) - L_{lb}(t-\Delta) L_{ja}(y-\Delta)  i \varphi^{\text{reg}}(\qq_{bj}) }\,,
\end{split}
\end{equation}
and
\begin{equation}
\begin{split}
&\frac{1}{\var} \{L_{ja}(y-\Delta), L_{lb}(t-\Delta) \}_{\text{Trig.}} =\\ 
&\scriptstyle{- \sum_{pq} \refl_{jp,lq}  L_{ja}(y-\Delta) L_{lb}(t-\Delta) - L_{ja}(y-\Delta) L_{lb}(t-\Delta) \sum_{pq} \refl_{ap,bq}} \\
&\scriptstyle{+L_{ja}(y-\Delta)  \sum_{pq} \refl_{ap,lq} L_{lb}(t-\Delta)-L_{lb}(t-\Delta) \sum_{pq} \refl_{bp,jq} L_{ja}(y-\Delta)}\,.
\end{split}
\end{equation}

Then we split $Y_3$ into an elliptic, trigonometric, and extra part
\begin{equation}
Y_3 = Y^{\text{(Ell.)}}_3 + Y^{\text{(Trig.)}}_3 + Y^{\text{(Extra)}}_3 \,.
\end{equation}
We define these three contributions as follows
\begin{equation}
\label{eq:Y3_Ell_app_first_defined}
\begin{split}
Y^{\text{(Ell.)}}_3 \equiv  &b_{\alpha i}(y) T_{ij}(y) Q_j^{-1} b_{\beta k}(t)\frac{1}{\var} \{ L_{ja}(y-\Delta) ,  T_{kl}(t)  \}_{\text{Ell.}} Q_l^{-1} L_{lb}(t-\Delta) Q_b t_b(z)    Q_a t_a(x)\\
&+b_{\alpha i}(y) T_{ij}(y) Q_j^{-1} b_{\beta k}(t) T_{kl}(t) \frac{1}{\var} \{ L_{ja}(y-\Delta) ,   L_{lb}(t-\Delta) \}_{\text{Ell.}} Q_l^{-1}  Q_b t_b(z)    Q_a t_a(x)\\
&+b_{\alpha i}(y) T_{ij}(y) Q_j^{-1} b_{\beta k}(t) T_{kl}(t) Q_l^{-1}  L_{lb}(t-\Delta) Q_b  \frac{1}{\var} \{ L_{ja}(y-\Delta) ,  t_b(z)  \}_{\text{Ell.}} Q_a t_a(x)  \,,
\end{split}
\end{equation}
\begin{equation}
\begin{split}
Y^{\text{(Trig.)}}_3 \equiv  &b_{\alpha i}(y) T_{ij}(y) Q_j^{-1} b_{\beta k}(t)\frac{1}{\var} \{ L_{ja}(y-\Delta) ,  T_{kl}(t)  \}_{\text{Trig.}} Q_l^{-1} L_{lb}(t-\Delta) Q_b t_b(z)    Q_a t_a(x)\\
&+b_{\alpha i}(y) T_{ij}(y) Q_j^{-1} b_{\beta k}(t) T_{kl}(t) \frac{1}{\var} \{ L_{ja}(y-\Delta) ,   L_{lb}(t-\Delta) \}_{\text{Trig.}} Q_l^{-1}  Q_b t_b(z)    Q_a t_a(x)\\
&+b_{\alpha i}(y) T_{ij}(y) Q_j^{-1} b_{\beta k}(t) T_{kl}(t) Q_l^{-1}  L_{lb}(t-\Delta) Q_b  \frac{1}{\var} \{ L_{ja}(y-\Delta) ,  t_b(z)  \}_{\text{Trig.}} Q_a t_a(x)  \,,
\end{split}
\end{equation}
and
\begin{equation}
\begin{split}
Y^{\text{(Extra)}}_3 \equiv 
&+b_{\alpha i}(y) T_{ij}(y) Q_j^{-1} b_{\beta k}(t) T_{kl}(t)   L_{lb}(t-\Delta)   \frac{1}{\var} \{ L_{ja}(y-\Delta) , Q_l^{-1}Q_b\} t_b(z) Q_a t_a(x) \,.
\end{split}
\end{equation}

We start computing~\eqref{eq:Y3_Ell_app_first_defined}. After replacing the different pieces, after several simplifications, we get
\begin{equation}
\label{eq:Y3_elliptic_simplified}
    \begin{split}
Y^{\text{(Ell.)}}_3 =&\scriptstyle{- \mathbf{b}_{\alpha a}(x,y)   b_{\beta k}(t)            T_{k a}(t) Q^{-1}_a L_{ab}(t-\Delta) Q_b t_b(z)}\\
&\scriptstyle{+i b_{\alpha i}(y) T_{ij}(y) Q_j^{-1} b_{\beta k}(t) T_{kl}(t)  L_{jb}(y-\Delta) L_{la}(t-\Delta)   \varphi(y-t+\Delta, \mathfrak{q}_{ba}) Q_l^{-1}  Q_b t_b(z)    Q_a t_a(x)}\\
&\scriptstyle{-i b_{\alpha i}(y) T_{ij}(y) Q_j^{-1} b_{\beta k}(t) T_{kl}(t) Q_l^{-1}  L_{lb}(t-\Delta) Q_b    L_{j a}(y-\Delta)    \varphi(-z+\Delta, \qq_{ab}) t_a(z) Q_a t_a(x)}     \\
&\scriptstyle{+i b_{\alpha i}(y) T_{ij}(y) Q_j^{-1} b_{\beta k}(t) T_{kl}(t) Q_l^{-1}  L_{lb}(t-\Delta) Q_b     \varphi(z-y+\Delta+2\Delta,\qq_{bj}) L_{ba}(y-\Delta) t_j(z) Q_a t_a(x)}     \\
&\scriptstyle{+i b_{\alpha i}(y) T_{ij}(y) Q_j^{-1} b_{\beta k}(t) T_{kl}(t) Q_l^{-1}  L_{lb}(t-\Delta) Q_b        \varphi(-z+\Delta, \qq_{jb}) L_{ja}(y-\Delta) t_j(z) Q_a t_a(x)} \,.
    \end{split}
\end{equation}
Let us now compute $Y^{\text{(Trig.)}}_3$. Combining the pieces previously computed, we obtain
\begin{equation}
\label{eq:Y3_trigonometric_simplified}
\begin{split}
&Y^{\text{(Trig.)}}_3 =\\
&\scriptstyle{+ \mathbf{b}_{\alpha a}(x,y)   b_{\beta k}(t) T_{kq}(t) \sum_{n} \refl(0,t)_{an,ql}  Q_l^{-1} L_{lb}(t-\Delta) Q_b t_b(z)    }\\
&\scriptstyle{-b_{\alpha i}(y) T_{ij}(y) Q_j^{-1} L_{ja}(y-\Delta) Q_a t_a(x) b_{\beta k}(t) T_{kp}(t) \sum_{n} \refl(0, t)_{jn,pl}   Q_l^{-1} L_{lb}(t-\Delta) Q_b t_b(z)    }\\
&\scriptstyle{-\mathbf{b}_{\alpha a}(x,y)   b_{\beta k}(t) T_{kl}(t) Q_l^{-1}  L_{lb}(t-\Delta) Q_b   \sum_{n} \refl(0,z)_{an,bs} t_s(z) }\\
&\scriptstyle{+b_{\alpha i}(y) T_{ij}(y) Q_j^{-1} b_{\beta k}(t) T_{kl}(t) Q_l^{-1}  L_{lb}(t-\Delta) Q_b  \sum_{n} \refl(0, z)_{jn,bs}  L_{ja}(y-\Delta) t_s(z) Q_a t_a(x)}\,.
\end{split}
\end{equation}
Finally, using that
\begin{equation}
\label{eq:LQ_bracket}
\frac{1}{\var} \{L_{ik}(x), Q_j \} = - \delta_{kj} L_{ij}(x) Q_j 
\end{equation}
we obtain
\begin{equation}
\label{eq:Y3_extra_simplified}
    \begin{split}
        Y^{\text{(Extra)}}_3 &= \scriptstyle{-\delta_{ab} \mathbf{b}_{\alpha a}(x,y) \mathbf{b}_{\beta b}(z,t)+ \mathbf{b}_{\alpha a}(x,y)  b_{\beta k}(t) T_{ka}(t)  Q_a^{-1} L_{ab}(t-\Delta)       Q_b t_b(z) } \,.
    \end{split}
\end{equation}

Summing~\eqref{eq:Y3_elliptic_simplified}, \eqref{eq:Y3_trigonometric_simplified} and~\eqref{eq:Y3_extra_simplified} the three pieces, we obtain
\begin{equation}
\label{eq:I3_final_simplified}
    \begin{split}
Y_3 =&\scriptstyle{-\delta_{ab} \mathbf{b}_{\alpha a}(x,y) \mathbf{b}_{\beta b}(z,t)}\\
&\scriptstyle{+i b_{\alpha i}(y) T_{ij}(y) Q_j^{-1} b_{\beta k}(t) T_{kl}(t)  L_{jb}(y-\Delta) L_{la}(t-\Delta)   \varphi(y-t+\Delta, \mathfrak{q}_{ba}) Q_l^{-1}  Q_b t_b(z)    Q_a t_a(x)}\\
&\scriptstyle{-i b_{\alpha i}(y) T_{ij}(y) Q_j^{-1} b_{\beta k}(t) T_{kl}(t) Q_l^{-1}  L_{lb}(t-\Delta) Q_b    L_{j a}(y-\Delta)    \varphi(-z+\Delta, \qq_{ab}) t_a(z) Q_a t_a(x)}     \\
&\scriptstyle{+i b_{\alpha i}(y) T_{ij}(y) Q_j^{-1} b_{\beta k}(t) T_{kl}(t) Q_l^{-1}  L_{lb}(t-\Delta) Q_b     \varphi(z-y+\Delta+2\Delta,\qq_{bj}) L_{ba}(y-\Delta) t_j(z) Q_a t_a(x)}     \\
&\scriptstyle{+i b_{\alpha i}(y) T_{ij}(y) Q_j^{-1} b_{\beta k}(t) T_{kl}(t) Q_l^{-1}  L_{lb}(t-\Delta) Q_b        \varphi(-z+\Delta, \qq_{jb}) L_{ja}(y-\Delta) t_j(z) Q_a t_a(x)}\\
&\scriptstyle{+ \mathbf{b}_{\alpha a}(x,y)   b_{\beta k}(t) T_{kq}(t) \sum_{n} \refl(0,t)_{an,ql}  Q_l^{-1} L_{lb}(t-\Delta) Q_b t_b(z)    }\\
&\scriptstyle{-b_{\alpha i}(y) T_{ij}(y) Q_j^{-1} L_{ja}(y-\Delta) Q_a t_a(x) b_{\beta k}(t) T_{kp}(t) \sum_{n} \refl(0, t)_{jn,pl}   Q_l^{-1} L_{lb}(t-\Delta) Q_b t_b(z)    }\\
&\scriptstyle{-\mathbf{b}_{\alpha a}(x,y)   b_{\beta k}(t) T_{kl}(t) Q_l^{-1}  L_{lb}(t-\Delta) Q_b   \sum_{n} \refl(0,z)_{an,bs} t_s(z) }\\
&\scriptstyle{+b_{\alpha i}(y) T_{ij}(y) Q_j^{-1} b_{\beta k}(t) T_{kl}(t) Q_l^{-1}  L_{lb}(t-\Delta) Q_b  \sum_{n} \refl(0, z)_{jn,bs}  L_{ja}(y-\Delta) t_s(z) Q_a t_a(x)}\,.
    \end{split}
\end{equation}

\paragraph{Evaluation of $Y_4$.}

Here we compute
\begin{equation}
    \begin{split}
Y_4 &= b_{\alpha i}(y) T_{ij}(y) Q_j^{-1} L_{ja}(y-\Delta) Q_a \frac{1}{\var} \{ t_a(x) , \mathbf{b}_{\beta b}(z,t) \} \\
&=b_{\alpha i}(y) T_{ij}(y) Q_j^{-1} L_{ja}(y-\Delta) Q_a \frac{1}{\var} \{ \mathbb{A}_{a1}(x) , \mathbf{b}_{\beta b}(z,t) \} \,.
    \end{split}
\end{equation}
Using~\eqref{eq:res_as_sum_of_Is_final2}, we obtain
\begin{equation}
\begin{split}
&Y_4 = b_{\alpha i}(y) T_{ij}(y) Q_j^{-1} L_{ja}(y-\Delta) Q_a \frac{1}{\var} \, \{\mathbb{A}_{a 1}(x), \mathbf{b}_{\beta b}(z,t) \} =\\ 
&\scriptstyle{+i \mathbf{b}_{\alpha a}(z,y) \varphi(z-x+\Delta,\mathfrak{q}_{ba})   \mathbf{a}_{b 1}(z,x) \mathbf{b}_{\beta b}(z,t) }\\
&\scriptstyle{+i \mathbf{b}_{\alpha a}(z,y) t_a(x) t_b^{-1}(z)   \varphi(-z+\Delta, \mathfrak{q}_{ab})  \mathbf{b}_{\beta i}(z,t)}\\
&\scriptstyle{+\mathbf{b}_{\alpha a}(x,y)   \mathbf{a}_{a \gamma}(x,x) \rho_-(x-t)_{\gamma 1, \beta \delta} \mathbf{b}_{\delta b}(z,t)-\frac{2 \pi }{\var} \delta_{1 \beta} \mathbf{b}_{\alpha a}(z,y) Q^{-1}_a \mathbb{L}_{ab}(z,t-\Delta) Q_b \delta(x-t)}\\
&\scriptstyle{-b_{\alpha i}(y) T_{ij}(y) Q_j^{-1} L_{ja}(y-\Delta) Q_a   b_{\beta e}(t)T_{el}(t)  \sum_{s} \refl(0,t)_{as, lg} \mathbb{A}_{a 1}(x) Q_b Q^{-1}_g L_{gb}(t-\Delta) t_b(z)}\\
&\scriptstyle{-\mathbf{b}_{\alpha a}(x,y)  \mathbf{a}_{a \gamma} (x,x)\mathbf{a}_{b \delta}(x,z) \rho_{\gamma 1, \delta 1}(x-z) \mathbf{b}_{\beta b}(x,t) }\\
&\scriptstyle{+b_{\alpha i}(y) T_{ij}(y) Q_j^{-1} L_{ja}(y-\Delta) Q_a b_{\beta f}(t) T_{fg}(t) Q_b Q^{-1}_g L_{gb}(t-\Delta)\sum_{s} \refl(0,z)_{as, bd} \mathbb{A}_{a 1}(x)t_d(z)}\,.
\end{split}
\end{equation}

\paragraph{Evaluation of $Y_5$.}
Now we compute
\begin{equation}
    \begin{split}
        Y_5 &= b_{\alpha i}(y) T_{ij}(y)  L_{ja}(y-\Delta)  \frac{1}{\var} \{ Q_a Q_j^{-1} , \mathbf{b}_{\beta b}(z,t) \} t_a(x)\\
        &=b_{\alpha i}(y) T_{ij}(y)  L_{ja}(y-\Delta)  b_{\beta c}(t) T_{cd}(t) Q^{-1}_d \frac{1}{\var} \{ Q_a Q_j^{-1} , L_{db}(t-\Delta)  \} Q_b t_b(z) t_a(x)\,.
    \end{split}
\end{equation}
Using~\eqref{eq:LQ_bracket}, we get
\begin{equation}
    \begin{split}
        Y_5 &=\delta_{ab} \mathbf{b}_{\alpha a}(x,y)  \mathbf{b}_{\beta b}(z,t)  \\
        &-b_{\alpha i}(y) T_{ib}(y)  L_{ba}(y-\Delta)  b_{\beta c}(t) T_{cd}(t) Q^{-1}_d  Q_a   L_{db}(t-\Delta)   t_b(z) t_a(x) \,.
    \end{split}
\end{equation}

\paragraph{All terms together.}

Replacing the different contributions $Y_1$, $Y_2$, $Y_3$, $Y_4$ and $Y_5$ into~\eqref{eq:app_bb_bracket_starting} after several simplifications we obtain the following expression in~\eqref{eq:bb_inv_br_doubleloop} for the $\mathbf{b}$-$\mathbf{b}$ bracket.

\subsection{Collinear limit of the a-a bracket}
\label{app:aa_collin_lim}

Let us consider the bracket~\eqref{eq:aa_inv_br_doubleloop} in the limit $x=z=\imp^1$. Here $\imp^1$ is just the common value of $x$ and $z$. 
First we take the limit $z \to \imp^{1}$ and get
\begin{equation}
\begin{split}
&\frac{1}{\var} \, \{\mathbf{a}_{a \alpha}(x,y) \,,\mathbf{a}_{b \beta}(z,t) \}\bigl|_{z=\imp^1}=\\
&\scriptstyle{+t^{-1}_{a}(x) t^{-1}_{b}(\imp^{1}) t_b(x) t_{a}(\imp^{1})}\\
&\scriptstyle{\times \Bigl( 
i \varphi(-x+\imp^{1}+\Delta,\mathfrak{q}_{ba})\mathbf{a}_{a \alpha}(x,y) \mathbf{a}_{b \beta}(\imp^{1},t)-  i \varphi(t-x+\Delta,\mathfrak{q}_{ba})\mathbf{a}_{a \alpha}(x,y) \mathbf{a}_{a \beta}(\imp^{1},t)  }\\
&\scriptstyle{-  i \varphi(\imp^{1}-y+\Delta,\mathfrak{q}_{ba})\mathbf{a}_{b \alpha}(x,y) \mathbf{a}_{b \beta}(\imp^{1},t)+ i \varphi(t-y+\Delta,\mathfrak{q}_{ba})\mathbf{a}_{b \alpha}(x,y) \mathbf{a}_{a \beta}(\imp^{1},t) \Bigr)}\\
&\scriptstyle{- \mathbf{a}_{a \gamma}(x,y)\mathbf{a}_{b \delta}(\imp^{1},t) \rho_{\gamma \alpha, \delta \beta}(y-t) -   \mathbf{a}_{a \gamma} (x,x)\mathbf{a}_{b \delta}(\imp^{1},\imp^{1}) \rho_{\gamma 1, \delta 1}(x-\imp^{1}) \mathbf{a}_{a \alpha}(x,y) \mathbf{a}_{b \beta}(\imp^{1},t)}\\
&\scriptstyle{+ \mathbf{a}_{a \gamma} (x,x)\mathbf{a}_{b \delta}(\imp^{1},t) \rho_{\gamma 1, \delta \beta}(x-t) \mathbf{a}_{a \alpha}(x, y)+ \mathbf{a}_{a \gamma} (x,y)\mathbf{a}_{b \delta}(\imp^{1},\imp^{1})   \rho_{\gamma \alpha, \delta 1}(y-\imp^{1}) \mathbf{a}_{b \beta}(\imp^{1}, t)} \,.
\end{split}
\end{equation}
Then we send $x \to \imp^1$ and get
\begin{equation}
\label{eq:app_aa_0y0t_bracket_newg}
\begin{split}
&\frac{1}{\var} \, \{\mathbf{a}_{a \alpha}(x,y) \,,\mathbf{a}_{b \beta}(z,t) \}\bigl|_{x=z=\imp^1}=\\
&\scriptstyle{+t^{-1}_{a}(\imp^1) t^{-1}_{b}(\imp^1) t_b(\imp^1) t_{a}(\imp^1)}\\
&\scriptstyle{\times \Bigl( 
  - i \varphi(t-\imp^1+\Delta,\mathfrak{q}_{ba})\mathbf{a}_{a \alpha}(\imp^1,y) \mathbf{a}_{a \beta}(\imp^1,t)  - i \varphi(\imp^1-y+\Delta,\mathfrak{q}_{ba})\mathbf{a}_{b \alpha}(\imp^1,y) \mathbf{a}_{b \beta}(\imp^1,t)+ i \varphi(t-y+\Delta,\mathfrak{q}_{ba})\mathbf{a}_{b \alpha}(\imp^1,y) \mathbf{a}_{a \beta}(\imp^1,t) \Bigr)}\\
&\scriptstyle{- \mathbf{a}_{a \gamma}(\imp^1,y)\mathbf{a}_{b \delta}(\imp^1,t) \rho_{\gamma \alpha, \delta \beta}(y-t) +  \mathbf{a}_{a \gamma} (\imp^1,\imp^1)\mathbf{a}_{b \delta}(\imp^1,t) \rho_{\gamma 1, \delta \beta}(\imp^1-t) \mathbf{a}_{a \alpha}(\imp^1, y)+ \mathbf{a}_{a \gamma} (\imp^1,y)\mathbf{a}_{b \delta}(\imp^1,\imp^1)   \rho_{\gamma \alpha, \delta 1}(y-\imp^1) \mathbf{a}_{b \beta}(\imp^1, t)} \\
&\scriptstyle{+ \lim_{x \to \imp^1} \Bigl( t^{-1}_{a}(x) t^{-1}_{b}(\imp^1) t_b(x) t_{a}(\imp^1)}  i \varphi(-x+\imp^1+\Delta,\mathfrak{q}_{ba})\mathbf{a}_{a \alpha}(x,y) \mathbf{a}_{b \beta}(\imp^1,t) \Bigr)\\
&\scriptstyle{- \lim_{x \to \imp^1} \Bigl(  \mathbf{a}_{a \gamma} (x,x)\mathbf{a}_{b \delta}(\imp^1,\imp^1) \rho_{\gamma 1, \delta 1}(x-\imp^1) \mathbf{a}_{a \alpha}(x,y) \mathbf{a}_{b \beta}(\imp^1,t) \Bigr)} \,.
\end{split}
\end{equation}
In the last two lines, we isolated the terms containing singularities. These are the terms responsible for the derivatives.
Expanding around $x=\imp^1$ we have
\begin{equation}
\varphi(-x+\imp^1+\Delta,\mathfrak{q}_{ba})=-\frac{1}{x-\imp^1}  + \varphi^{\text{reg}}(\mathfrak{q}_{ba}) + O(x-\imp^1) \,.
\end{equation}
Both the expansion of $\hat{\rho}$ and $\check{\rho}$ return
\begin{equation}
\rho_{\gamma 1, \delta 1}(x-\imp^1)= -\frac{i}{2} \delta_{\gamma 1} \delta_{\delta 1} \cot \frac{x-\imp^1}{2}  =  -\frac{i}{x-\imp^1} \delta_{\gamma 1} \delta_{\delta 1} + O(x-\imp^1) \,,
\end{equation}
from which we obtain
\begin{equation}
\label{eq:lim_x_to_0_aa_newg}
    \begin{split}
        &\scriptstyle{+ \lim_{x \to \imp^1} \Bigl( t^{-1}_{a}(x) t^{-1}_{b}(\imp^1) t_b(x) t_{a}(\imp^1)  i \varphi(-x+\imp^1+\Delta,\mathfrak{q}_{ba})\mathbf{a}_{a \alpha}(x,y) \mathbf{a}_{b \beta}(\imp^1,t) \Bigr)}\\
&\scriptstyle{- \lim_{x \to \imp^1} \Bigl(  \mathbf{a}_{a \gamma} (x,x)\mathbf{a}_{b \delta}(\imp^1,\imp^1) \rho_{\gamma 1, \delta 1}(x-\imp^1) \mathbf{a}_{a \alpha}(x,y) \mathbf{a}_{b \beta}(\imp^1,t) \Bigr)}=\\
&\scriptstyle{-\frac{i}{x-\imp^1}\Bigl( t^{-1}_{a}(\imp^1) t^{-1}_{b}(\imp^1) t_b(\imp^1) t_{a}(\imp^1) \mathbf{a}_{a \alpha}(\imp^1,y) \mathbf{a}_{b \beta}(\imp^1,t) \Bigr)+\frac{i}{x-\imp^1} \Bigl( \sum_{\gamma \delta } \mathbf{a}_{a \gamma} (\imp^1,\imp^1)\mathbf{a}_{b \delta}(\imp^1,\imp^1) \delta_{\gamma 1} \delta_{1 \delta}  \mathbf{a}_{a \alpha}(\imp^1,y) \mathbf{a}_{b \beta}(\imp^1,t) \Bigr)}\\
&\scriptstyle{+ t^{-1}_{a}(\imp^1) t^{-1}_{b}(\imp^1) t_b(\imp^1) t_{a}(\imp^1)  i \varphi^{\text{reg}}(\mathfrak{q}_{ba})\mathbf{a}_{a \alpha}(\imp^1,y) \mathbf{a}_{b \beta}(\imp^1,t)  - \sum_{\gamma,\delta} \mathbf{a}_{a \gamma} (\imp^1,\imp^1)\mathbf{a}_{b \rho}(\imp^1,\imp^1)  \delta_{\gamma, 1} \delta_{1, \delta} \sgn_{1 \gamma}  \mathbf{a}_{a \alpha}(\imp^1,y) \mathbf{a}_{b \beta}(\imp^1,t)}\\
&\scriptstyle{-i \frac{d}{dx} \Bigl( t^{-1}_{a}(x) t^{-1}_{b}(\imp^1) t_b(x) t_{a}(\imp^1) \mathbf{a}_{a \alpha}(x,y) \mathbf{a}_{b \beta}(\imp^1,t) \Bigr)\Bigr|_{x=\imp^1}+ i \frac{d}{d x}\Bigl( \sum_{ \gamma \delta } \mathbf{a}_{a \gamma} (x,x)\mathbf{a}_{b \delta}(\imp^1,\imp^1) \delta_{\gamma 1} \delta_{1 \delta}  \mathbf{a}_{a \alpha}(x,y) \mathbf{a}_{b \beta}(\imp^1,t) \Bigr)\Bigr|_{x=\imp^1}}\,.
    \end{split}
\end{equation}
Due to the constraint
\begin{equation}
 \mathbf{a}_{a 1}(t,t)=1
\end{equation}
the singular terms cancel each other.
Performing the derivatives, we obtain
\begin{equation}
    \begin{split}
        &-i \frac{d}{dx} \Bigl( t^{-1}_{a}(x) t^{-1}_{b}(\imp^1) t_b(x) t_{a}(\imp^1) \mathbf{a}_{a \alpha}(x,y) \mathbf{a}_{b \beta}(\imp^1,t) \Bigr)\Bigr|_{x=\imp^1}=\\
        &\scriptstyle{-2 i \frac{d}{dx} (t_a^{-1}(x))\Bigr|_{x=\imp^1} t_a(\imp^1) \mathbf{a}_{a \alpha}(\imp^1,y) \mathbf{a}_{b \beta}(\imp^1,t)-i \frac{d}{dx} (t_b(x))\Bigr|_{x=\imp^1} t^{-1}_b(\imp^1) \mathbf{a}_{a \alpha}(\imp^1,y) \mathbf{a}_{b \beta}(\imp^1,t)=}\\
        &\scriptstyle{-2 i \frac{d}{dx} (t_a^{-1}(x)\mathbb{A}_{a \alpha}(y) )\Bigr|_{x=\imp^1}   \mathbf{a}_{b \beta}(\imp^1,t)-i \frac{d}{dx} (t_b(x) )\Bigr|_{x=\imp^1} t_b^{-2}(\imp^1) \mathbf{a}_{a \alpha}(\imp^1,y) \mathbb{A}_{b \beta}(t) =}\\
        &\scriptstyle{-2 i \frac{d}{dx} (t_a^{-1}(x)\mathbb{A}_{a \alpha}(y) )\Bigr|_{x=\imp^1}   \mathbf{a}_{b \beta}(\imp^1,t)+i \frac{d}{dx} (t^{-1}_b(x) )\Bigr|_{x=\imp^1}  \mathbf{a}_{a \alpha}(\imp^1,y) \mathbb{A}_{b \beta}(t) =}\\
        &\scriptstyle{-2 i \frac{d}{dx} ((\mathbf{a}_{a \alpha}(x,y) )\Bigr|_{x=\imp^1}   \mathbf{a}_{b \beta}(\imp^1,t)+i \frac{d}{dx} (\mathbf{a}_{b \beta}(x,t) )\Bigr|_{x=\imp^1}  \mathbf{a}_{a \alpha}(\imp^1,y)}\,,
    \end{split}
\end{equation}
\begin{equation}
    \begin{split}
        &i \frac{d}{d x}\Bigl( \sum_{ \gamma \delta } \mathbf{a}_{a \gamma} (x,x)\mathbf{a}_{b \delta}(\imp^1,\imp^1) \delta_{\gamma 1} \delta_{\rho 1}  \mathbf{a}_{a \alpha}(x,y) \mathbf{a}_{b \beta}(\imp^1,t) \Bigr)=\\
        &i \frac{d}{d x}\Bigl( \mathbf{a}_{a 1} (x,x) \mathbf{a}_{b 1}(\imp^1,\imp^1)   \mathbf{a}_{a \alpha}(x,y) \mathbf{a}_{b \beta}(\imp^1,t) \Bigr)=i \frac{d}{d x}\Bigl(    \mathbf{a}_{a \alpha}(x,y) \mathbf{a}_{b \beta}(\imp^1,t) \Bigr) \,.
    \end{split}
\end{equation}
Plugging the derivative contributions into~\eqref{eq:lim_x_to_0_aa_newg} we obtain
\begin{equation}
    \begin{split}
        &\scriptstyle{+ \lim_{x \to \imp^1} \Bigl( t^{-1}_{a}(x) t^{-1}_{b}(\imp^1) t_b(x) t_{a}(\imp^1)  i \varphi(-x+\imp^1+\Delta,\mathfrak{q}_{ba})\mathbf{a}_{a \alpha}(x,y) \mathbf{a}_{b \beta}(\imp^1,t) \Bigr)}\\
&\scriptstyle{- \lim_{x \to \imp^1} \Bigl(  \mathbf{a}_{a \gamma} (x,x)\mathbf{a}_{b \delta}(\imp^1,\imp^1) \rho_{\gamma 1, \delta 1}(x-\imp^1) \mathbf{a}_{a \alpha}(x,y) \mathbf{a}_{b \beta}(\imp^1,t) \Bigr)}=\\
&\scriptstyle{i \varphi^{\text{reg}}(\mathfrak{q}_{ba})\mathbf{a}_{a \alpha}(\imp^1,y) \mathbf{a}_{b \beta}(\imp^1,t)  -i \frac{d}{dx} ((\mathbf{a}_{a \alpha}(x,y) )\Bigr|_{x=\imp^1}   \mathbf{a}_{b \beta}(\imp^1,t)+i \frac{d}{dx} (\mathbf{a}_{b \beta}(x,t) )\Bigr|_{x=\imp^1}  \mathbf{a}_{a \alpha}(\imp^1,y)}\,.
    \end{split}
\end{equation}
Replacing this expression into~\eqref{eq:app_aa_0y0t_bracket_newg} we end up with equation~\eqref{eq:aa_bracket_final_components}.

\subsection{Collinear limit of the a-b bracket}
\label{app:ab_collin_lim}

First we evaluate~\eqref{eq:ab_inv_br_doubleloop} at $z=\imp^1$ and obtain
\begin{equation}
\begin{split}
&\frac{1}{\var} \, \{\mathbf{a}_{a \alpha}(x,y), \mathbf{b}_{\beta b}(z,t)\}\bigl|_{z=\imp^1}=\\
= &\scriptstyle{+i  t^{-1}_a(x) t_a(\imp^1) \varphi(\imp^1-y+\Delta,\mathfrak{q}_{ba})   \mathbf{a}_{b \alpha}(\imp^1,y) \mathbf{b}_{\beta b}(\imp^1,t) }\\
&\scriptstyle{+\mathbf{a}_{a \gamma}(x,y) \rho(y-t)_{-\gamma \alpha, \beta \delta} \mathbf{b}_{\delta b}(\imp^1,t)-\frac{2 \pi }{\var} \delta_{\alpha \beta} Q_b Q^{-1}_a t^{-1}_a(x) L_{ab}(t-\Delta) t_b(\imp^1) \delta(y-t)}\\
&\scriptstyle{-  \mathbf{a}_{a \gamma} (x,y)\mathbf{a}_{b \delta}(x,\imp^1) \rho_{\gamma \alpha, \delta 1}(y-\imp^1) \mathbf{b}_{\beta b}(x,t) }\\
&\scriptstyle{-i  t^{-1}_a(x) t_a(\imp^1) \varphi(\imp^1-x+\Delta,\mathfrak{q}_{ba})   \mathbf{a}_{b 1}(\imp^1,x) \mathbf{b}_{\beta b}(\imp^1,t) \mathbf{a}_{a \alpha}(x,y) }\\
&\scriptstyle{-\mathbf{a}_{a \gamma}(x,x) \rho(x-t)_{-\gamma 1, \beta \delta} \mathbf{b}_{\delta b}(\imp^1,t) \mathbf{a}_{a \alpha}(x,y)+\frac{2 \pi }{\var} \delta_{1 \beta} Q_b Q^{-1}_a t^{-1}_a(x) L_{ab}(t-\Delta) t_b(\imp^1) \delta(x-t)\mathbf{a}_{a \alpha}(x,y)}\\
&\scriptstyle{+ \mathbf{a}_{b 1}(x, \imp^1) \rho_{1 1, 1 1}(x-\imp^1) \mathbf{b}_{\beta b}(x,t) \mathbf{a}_{a \alpha}(x,y)}\,.
\end{split}
\end{equation}
Then we take the limit $x \to \imp^1$:
\begin{equation}
\label{eq:ab_bracket_with_lim}
\begin{split}
&\frac{1}{\var} \, \{\mathbf{a}_{a \alpha}(x,y), \mathbf{b}_{\beta b}(z,t)\}\bigl|_{x=z=\imp^1}=\\
&\scriptstyle{+i \varphi(\imp^1-y+\Delta,\mathfrak{q}_{ba})   \mathbf{a}_{b \alpha}(\imp^1,y) \mathbf{b}_{\beta b}(\imp^1,t) }\\
&\scriptstyle{+\mathbf{a}_{a \gamma}(\imp^1,y) \rho(y-t)_{-\gamma \alpha, \beta \delta} \mathbf{b}_{\delta b}(\imp^1,t)-\frac{2 \pi }{\var} \delta_{\alpha \beta} Q_b Q^{-1}_a  \mathbb{L}_{ab}(t-\Delta) \delta(y-t)}\\
&\scriptstyle{-  \mathbf{a}_{a \gamma} (\imp^1,y)\mathbf{a}_{b \delta}(\imp^1,\imp^1) \rho_{\gamma \alpha, \delta 1}(y-\imp^1) \mathbf{b}_{\beta b}(\imp^1,t) }\\
&\scriptstyle{-\mathbf{a}_{a \gamma}(\imp^1,\imp^1) \rho(\imp^1-t)_{-\gamma 1, \beta \delta} \mathbf{b}_{\delta b}(\imp^1,t) \mathbf{a}_{a \alpha}(\imp^1,y)+\frac{2 \pi }{\var} \delta_{1 \beta} Q_b Q^{-1}_a  \mathbb{L}_{ab}(t-\Delta) \delta(\imp^1-t)\mathbf{a}_{a \alpha}(\imp^1,y)}\\
&\scriptstyle{+ \lim_{x \to \imp^1} \mathbf{a}_{b 1}(x, \imp^1) \rho_{1 1, 1 1}(x-\imp^1) \mathbf{b}_{\beta b}(x,t) \mathbf{a}_{a \alpha}(x,y)}\\
&\scriptstyle{-i  \lim_{x \to \imp^1} t^{-1}_a(x) t_a(\imp^1) \varphi(\imp^1-x+\Delta,\mathfrak{q}_{ba})   \mathbf{a}_{b 1}(\imp^1,x) \mathbf{b}_{\beta b}(\imp^1,t) \mathbf{a}_{a \alpha}(x,y)} \,.
\end{split}
\end{equation}
Let us evaluate the last two terms separately. We have
\begin{equation}
\begin{split}
&\scriptstyle{\lim_{x \to \imp^1} \mathbf{a}_{b 1}(x, \imp^1) \rho_{1 1, 1 1}(x-\imp^1) \mathbf{b}_{\beta b}(x,t) \mathbf{a}_{a \alpha}(x,y)=}\\
&\scriptstyle{-\frac{i}{2}\lim_{x \to \imp^1} \mathbf{a}_{b 1}(x, \imp^1) \cot \frac{x-\imp^1}{2} \, \mathbf{b}_{\beta b}(x,t) \mathbf{a}_{a \alpha}(x,y)=}\\
& \scriptstyle{- \frac{i}{x-\imp^1} \left(\mathbf{b}_{\beta b}(\imp^1 ,t) + (x-\imp^1) \frac{d}{dx} \mathbf{b}_{\beta b}(x,t)\Bigl|_{x=\imp^1} \right)\left(\mathbf{a}_{a\alpha}(\imp^1 ,y) + (x-\imp^1) \frac{d}{dx} \mathbf{a}_{a \alpha}(x,y)\Bigl|_{x=\imp^1} \right)}\\
& \scriptstyle{\times \left(\mathbf{a}_{b 1}(x, \imp^1) + (x-\imp^1) \frac{d}{dx} \mathbf{a}_{b 1}(x, \imp^1)\Bigl|_{x=\imp^1}\right)=}\\
& \scriptstyle{- \frac{i}{x-\imp^1} \mathbf{a}_{a\alpha}(\imp^1 ,y) \mathbf{b}_{\beta b}(\imp^1 ,t) }\\
&\scriptstyle{-i \mathbf{a}_{a\alpha}(\imp^1 ,y) \frac{d}{dx} \mathbf{b}_{\beta b}(x,t)\Bigl|_{x=\imp^1} -i \mathbf{b}_{\beta b}(\imp^1 ,t) \frac{d}{dx} \mathbf{a}_{a \alpha}(x,y)\Bigl|_{x=\imp^1} -i\mathbf{b}_{\beta b}(\imp^1 ,t) \mathbf{a}_{a\alpha}(\imp^1 ,y) \frac{d}{dx} \mathbf{a}_{b 1}(x, \imp^1)\Bigl|_{x=\imp^1}} \,.
\end{split}
\end{equation}
Using that
\begin{equation}
\varphi(-x+\imp^1+\Delta,\mathfrak{q}_{ba})=-\frac{1}{x-\imp^1}  + \varphi^{\text{reg}}(\mathfrak{q}_{ba}) + \dots
\end{equation}
the other limit becomes
\begin{equation}
\begin{split}
&\scriptstyle{-i  \lim_{x \to \imp^1} t^{-1}_a(x) t_a(\imp^1) \varphi(\imp^1-x+\Delta,\mathfrak{q}_{ba})   \mathbf{a}_{b 1}(\imp^1,x) \mathbf{b}_{\beta b}(\imp^1,t) \mathbf{a}_{a \alpha}(x,y)} \\
&\scriptstyle{\left(\frac{i}{x-\imp^1} - i \varphi^{\text{reg}}(\qq_{ba}) \right)t_a(\imp^1) \mathbf{b}_{\beta b}(\imp^1,t) \biggl[t^{-1}_a(\imp^1)    \mathbf{a}_{b 1}(\imp^1,\imp^1)  \mathbf{a}_{a \alpha}(\imp^1,y) }\\
&\scriptstyle{ + (x-\imp^1) \left(\frac{d}{dx}t^{-1}_a(x)    \mathbf{a}_{b 1}(\imp^1,\imp^1)  \mathbf{a}_{a \alpha}(\imp^1,y) +  t^{-1}_a(\imp^1)    \frac{d}{dx}\mathbf{a}_{b 1}(\imp^1,x)  \mathbf{a}_{a \alpha}(\imp^1,y)+  t^{-1}_a(\imp^1) \mathbf{a}_{b 1}(\imp^1,\imp^1)  \frac{d}{dx}\mathbf{a}_{a \alpha}(x,y) \right)\Bigl|_{x=\imp^1} \biggr]=}\\ 
&\scriptstyle{ +\frac{i}{x-\imp^1} \mathbf{a}_{a \alpha}(\imp^1,y) \mathbf{b}_{\beta b}(\imp^1, t)-i \varphi^{\text{reg}}(\qq_{ba}) \mathbf{a}_{a \alpha}(\imp^1,y) \mathbf{b}_{\beta b}(\imp^1, t)+2i \mathbf{b}_{\beta b}(\imp^1,t) \frac{d}{dx}\mathbf{a}_{a \alpha}(x,y)\Bigl|_{x=\imp^1}}\\
&\scriptstyle{ +i\mathbf{b}_{\beta b}(\imp^1,t) \frac{d}{dx} \mathbf{a}_{b 1}(\imp^1,x)\Bigl|_{x=\imp^1} \mathbf{a}_{a \alpha}(\imp^1,y) } 
\end{split}
\end{equation}
Combining the derivative terms, the singularity cancels, and we obtain
\begin{equation}
\label{eq:ab_collinear_limit_firstargument}
    \begin{split}
    &\scriptstyle{+\lim_{x \to \imp^1} \mathbf{a}_{b 1}(x, \imp^1) \rho_{1 1, 1 1}(x-\imp^1) \mathbf{b}_{\beta b}(x,t) \mathbf{a}_{a \alpha}(x,y)}\\
    &\scriptstyle{-i  \lim_{x \to \imp^1} t^{-1}_a(x) t_a(\imp^1) \varphi(\imp^1-x+\Delta,\mathfrak{q}_{ba})   \mathbf{a}_{b 1}(\imp^1,x) \mathbf{b}_{\beta b}(\imp^1,t) \mathbf{a}_{a \alpha}(x,y)}\\
    &\scriptstyle{=-i \mathbf{a}_{a\alpha}(\imp^1 ,y) \frac{d}{dx} \mathbf{b}_{\beta b}(x,t)\Bigl|_{x=\imp^1} +i \mathbf{b}_{\beta b}(\imp^1 ,t) \frac{d}{dx} \mathbf{a}_{a \alpha}(x,y)\Bigl|_{x=\imp^1} -i \varphi^{\text{reg}}(\qq_{ba}) \mathbf{a}_{a \alpha}(\imp^1,y) \mathbf{b}_{\beta b}(\imp^1, t)}\\
    &\scriptstyle{ +i\mathbf{b}_{\beta b}(\imp^1,t) \mathbf{a}_{a \alpha}(\imp^1,y) \left( \frac{d}{dx} \mathbf{a}_{b 1}(\imp^1,x)\Bigl|_{x=\imp^1} -\frac{d}{dx} \mathbf{a}_{b 1}(x,\imp^1)\Bigl|_{x=\imp^1} \right)  } 
    \end{split}
\end{equation}
Let us define
\begin{equation}
\label{eq:bBdef}
\mathbb{B}_{\alpha a}(y)=b_{\alpha i}(y) T_{ij}(y) Q_j^{-1} L_{ja}(y-\Delta) Q_a \,,
\end{equation}
so that
\begin{equation}
\mathbf{b}_{\alpha a}(x,y) = \mathbb{B}_{\alpha a}(y) t_a(x) \,.
\end{equation}
Then we have
\begin{equation}
\label{eq:ab_chain_of_equalities}
    \begin{split}
&\mathbf{b}_{\beta b}(\imp^1,t)   \frac{d}{dx} \mathbf{a}_{b1}(\imp^1,x)\Bigl|_{x=\imp^1}=- \mathbf{b}_{\beta b}(\imp^1,t)   \frac{d}{dx} \mathbf{a}_{b1}(x,\imp^1)\Bigl|_{x=\imp^1}=\\
&- \mathbb{B}_{\beta b}(t)  t_b(\imp^1) \frac{d}{dx} t_b^{-1}(x)\Bigl|_{x=\imp^1}  t_{b}(\imp^1) =\\
&+ \mathbb{B}_{\beta b}(t)  \frac{d}{dx}t_b(x)\Bigl|_{x=\imp^1}  t_b^{-1}(\imp^1)  t_{b}(\imp^1)  = + \frac{d}{dx}\mathbf{b}_{\beta b}(x,t)\Bigl|_{x=\imp^1}  \,.
\end{split}
\end{equation}
Plugging this relation into~\eqref{eq:ab_collinear_limit_firstargument}, we obtain
\begin{equation}
\label{eq:ab_collinear_limit_firstargument2}
    \begin{split}
    &\scriptstyle{+\lim_{x \to \imp^1} \mathbf{a}_{b 1}(x, \imp^1) \rho_{1 1, 1 1}(x-\imp^1) \mathbf{b}_{\beta b}(x,t) \mathbf{a}_{a \alpha}(x,y)}\\
    &\scriptstyle{-i  \lim_{x \to \imp^1} t^{-1}_a(x) t_a(\imp^1) \varphi(\imp^1-x+\Delta,\mathfrak{q}_{ba})   \mathbf{a}_{b 1}(\imp^1,x) \mathbf{b}_{\beta b}(\imp^1,t) \mathbf{a}_{a \alpha}(x,y)}\\
    &\scriptstyle{=+i \mathbf{a}_{a\alpha}(\imp^1 ,y) \frac{d}{dx} \mathbf{b}_{\beta b}(x,t)\Bigl|_{x=\imp^1} +i \mathbf{b}_{\beta b}(\imp^1 ,t) \frac{d}{dx} \mathbf{a}_{a \alpha}(x,y)\Bigl|_{x=\imp^1} -i \varphi^{\text{reg}}(\qq_{ba}) \mathbf{a}_{a \alpha}(\imp^1,y) \mathbf{b}_{\beta b}(\imp^1, t)}\,.
\end{split}
\end{equation}
Plugging this expression in~\eqref{eq:ab_bracket_with_lim}, we get equation~\eqref{eq:ab_bracket_with_lim2}.

\subsection{Collinear limit of the b-b bracket}
\label{app:bb_collin_lim}

We first evaluate~\eqref{eq:bb_inv_br_doubleloop} at $x=\imp^1$:
\begin{equation}
\begin{split}
&\frac{1}{\var} \{\mathbf{b}_{\alpha a}(x,y),\mathbf{b}_{\beta b}(z,t)\}\bigl|_{x=\imp^1} =\scriptstyle{-\rho_{\alpha \gamma, \beta \delta}(y-t)
\mathbf{b}_{\gamma a}(\imp^1,y) \mathbf{b}_{\delta b}(z,t)} \\
&\scriptstyle{+\mathbf{b}_{\beta b}(z,t)  \mathbf{a}_{b \gamma}(z,z) \rho(y-z)_{+ \alpha \delta, \gamma 1} \mathbf{b}_{\delta a}(\imp^1,y)  }\\
&\scriptstyle{+\frac{2 \pi}{\var} \delta_{\alpha 1}  \delta(y-z)  \, \mathbf{b}_{\beta b}(\imp^1,t) Q^{-1}_b    \mathbb{L}_{ba}(\imp^1, y-\Delta) Q_a}\scriptstyle{+i \mathbf{b}_{\alpha b}(z,y) \mathbf{b}_{\beta a}(\imp^1,t)   \varphi(y-t+\Delta, \mathfrak{q}_{ba})      }\\
&\scriptstyle{+i \mathbf{b}_{\alpha a}(z,y) \varphi(z-\imp^1+\Delta,\mathfrak{q}_{ba})   \mathbf{a}_{b 1}(z,\imp^1) \mathbf{b}_{\beta b}(z,t) }\\
&\scriptstyle{+\mathbf{b}_{\alpha a}(\imp^1,y)  \mathbf{a}_{a \gamma}(\imp^1,\imp^1)  \rho(\imp^1-t)_{-\gamma 1, \beta \gamma} \mathbf{b}_{\gamma b}(z,t)-\frac{2 \pi }{\var} \delta_{1 \beta} \mathbf{b}_{\alpha a}(z,y) Q^{-1}_a \mathbb{L}_{ab}(z,t-\Delta) Q_b \delta(\imp^1-t)}\\
&\scriptstyle{-\mathbf{b}_{\alpha a}(\imp^1,y)  \mathbf{a}_{b 1}(\imp^1,z) \rho_{1 1, 1 1}(\imp^1-z) \mathbf{b}_{\beta b}(\imp^1,t) }\,.
\end{split}
\end{equation}
Then we take the limit $z \to \imp^1$ and get
\begin{equation}
\begin{split}
&\frac{1}{\var} \{\mathbf{b}_{\alpha a}(x,y),\mathbf{b}_{\beta b}(z,t)\}\bigl|_{x=z=\imp^1} =\scriptstyle{-\rho_{\alpha \gamma, \beta \delta}(y-t)
\mathbf{b}_{\gamma a}(\imp^1,y) \mathbf{b}_{\delta b}(\imp^1,t)} \\
&\scriptstyle{+\mathbf{b}_{\beta b}(\imp^1,t)  \mathbf{a}_{b \gamma}(\imp^1,\imp^1) \rho(y-\imp^1)_{+ \alpha \gamma, \gamma 1} \mathbf{b}_{\gamma a}(\imp^1,y)  }\\
&\scriptstyle{+\frac{2 \pi}{\var} \delta_{\alpha 1}  \delta(y-\imp^1)  \, \mathbf{b}_{\beta b}(\imp^1,t) Q^{-1}_b    \mathbb{L}_{ba}(\imp^1, y-\Delta) Q_a}\scriptstyle{+i \mathbf{b}_{\alpha b}(\imp^1,y) \mathbf{b}_{\beta a}(\imp^1,t)   \varphi(y-t+\Delta, \mathfrak{q}_{ba})      }\\
&\scriptstyle{+\mathbf{b}_{\alpha a}(\imp^1,y)  \mathbf{a}_{a \gamma}(\imp^1,\imp^1)  \rho(\imp^1-t)_{-\gamma 1, \beta \gamma} \mathbf{b}_{\gamma b}(\imp^1,t)-\frac{2 \pi }{\var} \delta_{1 \beta} \mathbf{b}_{\alpha a}(\imp^1,y) Q^{-1}_a \mathbb{L}_{ab}(\imp^1,t-\Delta) Q_b \delta(\imp^1-t)}\\
&\scriptstyle{+ \lim_{z \to \imp^1} i \mathbf{b}_{\alpha a}(z,y) \varphi(z-\imp^1+\Delta,\mathfrak{q}_{ba})   \mathbf{a}_{b 1}(z,\imp^1) \mathbf{b}_{\beta b}(z,t) }\\
&\scriptstyle{-\lim_{z \to \imp^1} \mathbf{b}_{\alpha a}(\imp^1,y)  \mathbf{a}_{b 1}(\imp^1,z) \rho_{1 1, 1 1}(\imp^1-z) \mathbf{b}_{\beta b}(\imp^1,t) }\,.
\end{split}
\end{equation}
In the last two lines, we collected terms that become singular in the limit. Taking the limit, we obtain
\begin{equation}
\begin{split}
&\scriptstyle{+ \lim_{z \to \imp^1} i \mathbf{b}_{\alpha a}(z,y) \varphi(z-\imp^1+\Delta,\mathfrak{q}_{ba})   \mathbf{a}_{b 1}(z,\imp^1) \mathbf{b}_{\beta b}(z,t) = \frac{i}{z-\imp^1} \, \mathbf{b}_{\alpha a}(\imp^1,y)     \mathbf{b}_{\beta b}(\imp^1,t) }\\
&\scriptstyle{+i \mathbf{b}_{\alpha a}(\imp^1,y) \varphi^{\text{reg}}(\mathfrak{q}_{ba})    \mathbf{b}_{\beta b}(\imp^1,t)+i\mathbf{b}_{\alpha a}(\imp^1,y) \, \mathbf{b}_{\beta b}(\imp^1,t)   \frac{d}{dz} \mathbf{a}_{b1}(z,\imp^1)\Bigl|_{z=\imp^1}}\\
&\scriptstyle{+i\mathbf{b}_{\alpha a}(\imp^1,y) \frac{d}{dz}\mathbf{b}_{\beta b}(z,t) \Bigl|_{z=\imp^1}  +i \frac{d}{dz}\mathbf{b}_{\alpha a}(z,y) \Bigl|_{z=\imp^1} \mathbf{b}_{\beta b}(\imp^1,t)}
\end{split}
\end{equation}
and
\begin{equation}
    \begin{split}
        \scriptstyle{-\lim_{z \to \imp^1} \mathbf{b}_{\alpha a}(\imp^1,y)  \mathbf{a}_{b 1}(\imp^1,z) \rho_{1 1, 1 1}(\imp^1-z) \mathbf{b}_{\beta b}(\imp^1,t) =}&\scriptstyle{-\frac{i}{z-\imp^1} \mathbf{b}_{\alpha a}(\imp^1,y)   \mathbf{b}_{\beta b}(\imp^1,t)}\\
        &\scriptstyle{-i \mathbf{b}_{\alpha a}(\imp^1,y) \mathbf{b}_{\beta b}(\imp^1,t)  \frac{d}{dz}\mathbf{a}_{b 1}(\imp^1,z)\bigl|_{z=\imp^1}} \,.
    \end{split}
\end{equation}
Using again the equality~\eqref{eq:ab_chain_of_equalities}, it holds that
\begin{equation}
\begin{split}
&\scriptstyle{+ \lim_{z \to \imp^1} i \mathbf{b}_{\alpha a}(z,y) \varphi(z-\imp^1+\Delta,\mathfrak{q}_{ba})   \mathbf{a}_{b 1}(z,\imp^1) \mathbf{b}_{\beta b}(z,t) = \frac{i}{z-\imp^1} \, \mathbf{b}_{\alpha a}(\imp^1,y)     \mathbf{b}_{\beta b}(\imp^1,t) }\\
&\scriptstyle{+i \mathbf{b}_{\alpha a}(\imp^1,y) \varphi^{\text{reg}}(\mathfrak{q}_{ba})    \mathbf{b}_{\beta b}(\imp^1,t) +i \frac{d}{dz}\mathbf{b}_{\alpha a}(z,y) \Bigl|_{z=\imp^1} \mathbf{b}_{\beta b}(\imp^1,t)}
\end{split}
\end{equation}
\begin{equation}
    \begin{split}
        \scriptstyle{-\lim_{z \to \imp^1} \mathbf{b}_{\alpha a}(\imp^1,y)  \mathbf{a}_{b 1}(\imp^1,z) \rho_{1 1, 1 1}(\imp^1-z) \mathbf{b}_{\beta b}(\imp^1,t) =}&\scriptstyle{-\frac{i}{z-\imp^1} \mathbf{b}_{\alpha a}(\imp^1,y)   \mathbf{b}_{\beta b}(\imp^1,t)}\\
        &\scriptstyle{-i \mathbf{b}_{\alpha a}(\imp^1,y) \frac{d}{dz}\mathbf{b}_{\beta b}(z,t)\Bigl|_{z=\imp^1}} \,.
    \end{split}
\end{equation}
Combining the two limits, the singularity cancels, and we obtain equation~\eqref{eq:bb_bracket_final_components}.

\section{Dynamical inhomogeneities remove derivatives}
\label{app:dyn_imp}

In this appendix, we show that it is possible to remove the derivative terms from~\eqref{eq:aa_bracket_final_components}, \eqref{eq:ab_bracket_with_lim2} and~\eqref{eq:bb_bracket_final_components}. However, to preserve the Poisson structure of the brackets, we need to introduce non-trivial Poisson brackets between the oscillators and $\imp^1$. 
We recall the previously introduced definitions
\begin{equation}
\mathbf{a}_{a \alpha}(y) \equiv \mathbf{a}_{a \alpha}(\imp^1,y) \,, \qquad \mathbf{b}_{\alpha a}(y) \equiv \mathbf{b}_{\alpha a}(\imp^1,y) \,, \qquad  \mathbb{L}_{ab}(y) \equiv \mathbb{L}_{ab}(\imp^1,y) \,. 
\end{equation}

\subsection{The gauge constraint}

Before explaining how derivative terms can be removed, we remark that the brackets for the invariant spins were derived with the gauge condition
\begin{equation}
\mathbf{a}_{a 1}(x,x)=1 \,.
\end{equation}
From this constraint, we should expect that
\begin{equation}
\frac{1}{\var} \, \{\mathbf{a}_{a 1}(y) \,, \text{any} \}\Bigl|_{y=\imp^1}= 0 \,,
\end{equation}
where by `any' we mean any element of the Poisson algebra.
Let us show how this is realized on the a-a bracket in full details.

Let us take the limit $y \to \imp^1$ of~\eqref{eq:aa_bracket_final_components} for $\alpha=1$ and set $\mathbf{a}_{a1}(\imp^1,\imp^1)=1$:
\begin{equation}
\begin{split}
&\frac{1}{\var} \, \{\mathbf{a}_{a 1}(x,y) \,,\mathbf{a}_{b \beta}(z,t) \}\bigl|_{x=z=y=\imp^1}=- \mathbf{a}_{a \beta}(\imp^1)\mathbf{a}_{b 1}(t) \rho_{\beta 1, 1 \beta}(\imp^1-t)\\
&  + i \varphi(t-\imp^1+\Delta,\mathfrak{q}_{ba}) \mathbf{a}_{a \beta}(t) - i \varphi(t-\imp^1+\Delta,\mathfrak{q}_{ba}) \mathbf{a}_{a \beta}(t)\\
&+i \varphi^{\text{reg}}(\mathfrak{q}_{ba})\mathbf{a}_{a 1}(\imp^1) \mathbf{a}_{b \beta}(t)+\mathbf{a}_{a \beta} (\imp^1)\mathbf{a}_{b 1}(t)   \rho_{\beta 1, 1 \beta}(\imp^1-t)\\
& -i \frac{d}{dx} ((\mathbf{a}_{a \alpha}(x,\imp^1) )\Bigr|_{x=\imp^1}   \mathbf{a}_{b \beta}(\imp^1,t)+i \frac{d}{dx} (\mathbf{a}_{b \beta}(x,t) )\Bigr|_{x=\imp^1} \\
&- i \lim_{y \to \imp^1}\varphi(\imp^1-y+\Delta,\mathfrak{q}_{ba})\mathbf{a}_{b 1}(y) \mathbf{a}_{b \beta}(t) - \lim_{y \to \imp^1} \mathbf{a}_{a 1} (y)   \rho_{1 1,1 1}(\imp^1-y) \mathbf{a}_{b \beta}(t)\,.
\end{split}
\end{equation}
Taking the limit,  all singularities cancel from the last line, and we obtain
\begin{equation}
\begin{split}
&\scriptstyle{- i \lim_{y \to \imp^1}\varphi(\imp^1-y+\Delta,\mathfrak{q}_{ba})\mathbf{a}_{b 1}(y) \mathbf{a}_{b \beta}(t)} \\
&\scriptstyle{- \lim_{y \to \imp^1} \mathbf{a}_{a 1} (y)   \rho_{1 1,1 1}(\imp^1-y) \mathbf{a}_{b \beta}(t)=i (\mathbf{a}'_{b1}(\imp^1)-\mathbf{a}'_{a1}(\imp^1) - \varphi^{\text{reg}}(\qq_{ba})) \mathbf{a}_{b \beta}(t)} \,.
\end{split}
\end{equation}
After simplification, we get
\begin{equation}
\begin{split}
\frac{1}{\var} \, \{\mathbf{a}_{a 1}(x,y) \,,\mathbf{a}_{b \beta}(z,t) \}\bigl|_{x=z=y=\imp^1}= &-i \frac{d}{dx} ((\mathbf{a}_{a 1}(x,\imp^1) )\Bigr|_{x=\imp^1}   \mathbf{a}_{b \beta}(\imp^1,t)+i \frac{d}{dx} (\mathbf{a}_{b \beta}(x,t) )\Bigr|_{x=\imp^1} \\
&+i (\frac{d}{dx}\mathbf{a}_{b1}(\imp^1,x)\bigl|_{x=\imp^1}-\frac{d}{dx}\mathbf{a}_{a1}(\imp^1,x))\bigl|_{x=\imp^1} \mathbf{a}_{b \beta}(t) \,.
\end{split}
\end{equation}
Using that
\begin{equation}
0 = \frac{d}{dx} \mathbf{a}_{a1}(x,x)\bigl|_{x=\imp^1}= \frac{d}{dx} \mathbf{a}_{a1}(x,\imp^1)\bigl|_{x=\imp^1} + \frac{d}{dx} \mathbf{a}_{a1}(\imp^1,x)\bigl|_{x=\imp^1}
\end{equation}
and
\begin{equation}
    \begin{split}
    \frac{d}{dx}\mathbf{a}_{b1}(\imp^1,x)\bigl|_{x=\imp^1} \mathbf{a}_{b \beta}(\imp^1, t)&=- \frac{d}{dx}\mathbf{a}_{b1}(x,\imp^1)\bigl|_{x=\imp^1} \mathbf{a}_{b \beta}(\imp^1, t)\\
    &=- \mathbf{a}_{b1}(\imp^1,\imp^1) \frac{d}{dx}\mathbf{a}_{b \beta}(x, t)\bigl|_{x=\imp^1}=- \frac{d}{dx}\mathbf{a}_{b \beta}(x, t)\bigl|_{x=\imp^1}
    \end{split}
\end{equation}
we obtain
\begin{equation}
\frac{1}{\var} \, \{\mathbf{a}_{a 1}(x,y) \,,\mathbf{a}_{b \beta}(z,t) \}\bigl|_{x=z=y=\imp^1}= 0\,,
\end{equation}
as expected.

\subsection{Removing derivative terms in a-a brackets}
\label{app:removing_D_term_in_a_a_bracket}
Now we show how to remove derivative terms in the brackets by assuming non-trivial Poisson brackets between $\imp^1$ and the invariant spins.
Suppose that $\mathbf{a}$ does not commute trivially with $\imp^1$.
Then by expanding~\eqref{eq:aa_inv_br_doubleloop} around $x=z=\imp^1$ we have
\begin{equation}
\begin{split}
&\{\mathbf{a}_{a \alpha}(x,y),\mathbf{a}_{b \beta}(z,t) \}=\{\mathbf{a}_{a \alpha}(\imp^1,y),\mathbf{a}_{b \beta}(\imp^1,t) \} \\
&+ \{  (x-\imp^1) \frac{d}{dx}\mathbf{a}_{a \alpha}(x,y)\bigl|_{x=\imp^1},\mathbf{a}_{b \beta}(\imp^1,t) \} + \{   \mathbf{a}_{a \alpha}(\imp^1,y), (z-\imp^1) \frac{d}{dz}\mathbf{a}_{b \beta}(z,t)\bigl|_{x=\imp^1} \} + \dots \,.
\end{split}
\end{equation}
If we assume $\imp^1$ does not commute with $\mathbf{a}$, then in the limit $x,z \to \imp^1$ the expression above is equal to
\begin{equation}
\begin{split}
\{\mathbf{a}_{a \alpha}(x,y),&\mathbf{a}_{b \beta}(z,t)  \}\Bigl|_{x=z=\imp^1}=\{\mathbf{a}_{a \alpha}(\imp^1,y),\mathbf{a}_{b \beta}(\imp^1,t) \} \\
&- \mathbf{a}'_{a \alpha}(\imp^1,y) \{  \imp^1 ,\mathbf{a}_{b \beta}(\imp^1,t) \} - \{   \mathbf{a}_{a \alpha}(\imp^1,y), \imp^1 \} \mathbf{a}'_{b \beta}(\imp^1,t)  + \dots \,,
\end{split}
\end{equation}
where by $\mathbf{a}'$ here we mean the derivative with respect to the first argument.
From such a relation we obtain that for $x$ and $z$ close to $\imp^1$ it must hold
\begin{equation}
\begin{split}
\{\mathbf{a}_{a \alpha}(\imp^1,y),\mathbf{a}_{b \beta}(\imp^1,t) \}&=\{\mathbf{a}_{a \alpha}(x,y),\mathbf{a}_{b \beta}(z ,t) \}\Bigl|_{x=z=\imp^1} \\
&+ \mathbf{a}'_{a \alpha}(\imp^1,y) \{  \imp^1 ,\mathbf{a}_{b \beta}(\imp^1,t) \} - \{\imp^1, \mathbf{a}_{a \alpha}(\imp^1,y) \} \mathbf{a}'_{b \beta}(\imp^1,t)  + \dots \,.
\end{split}
\end{equation}
By $\{\mathbf{a}_{a \alpha}(x,y),\mathbf{a}_{b \beta}(z ,t) \}\Bigl|_{x=z=\imp^1}$ we mean the bracket~\eqref{eq:aa_bracket_final_components} evaluated at $x=z=\imp^1$. $\{\mathbf{a}_{a \alpha}(\imp^1,y),\mathbf{a}_{b \beta}(\imp^1,t) \}$ is instead the new bracket we want to compute.
Now we replace $\{\mathbf{a}_{a \alpha}(x,y),\mathbf{a}_{b \beta}(z ,t) \}\Bigl|_{x=z=\imp^1}$ with~\eqref{eq:aa_bracket_final_components} and see that derivative terms can be cancelled in the brackets if 
\begin{equation}
\label{eq:D1_a_commut_relation_new}
\frac{1}{\var} \{\imp^1, \mathbf{a}_{a \alpha}(\imp^1,y) \} = + i \mathbf{a}_{a \alpha}(\imp^1,y) \,.
\end{equation}

We should remark anyway that for $\alpha=1$ the commutation relation above should not be imposed since it is not compatible with the gauge choice $\mathbf{a}_{a1}(x,x)=1$. Let us show explicitly why this is the case. 
Our starting bracket satisfies
\begin{equation}
\label{eq:gauge_a_constraint1}
    \{\mathbf{a}_{a 1}(x,y), \text{any} \}\bigl|_{x=y=\imp^1}=0 \,.
\end{equation}
Expanding 
\begin{equation}
\label{eq:a1D1D1_bracket_vanishing}
\begin{split}
    \{\mathbf{a}_{a 1}(\imp^1,\imp^1), \text{any} \} &- \frac{d}{dx} \mathbf{a}_{a 1}(x,\imp^1)\bigl|_{x=\imp^1} \{\imp^1, \text{any} \} \\
    &- \frac{d}{dx} \mathbf{a}_{a 1}(\imp^1,x)\bigl|_{x=\imp^1} \{\imp^1, \text{any} \}=0 
\end{split}
\end{equation}
and using that
\begin{equation}
0=\frac{d}{dx} \mathbf{a}_{a 1}(x,x)\bigl|_{x=\imp^1}=\frac{d}{dx} \mathbf{a}_{a 1}(x,\imp^1)\bigl|_{x=\imp^1}+\frac{d}{dx} \mathbf{a}_{a 1}(\imp^1,x)\bigl|_{x=\imp^1}
\end{equation}
we notice that the condition~\eqref{eq:gauge_a_constraint1} implies
\begin{equation}
\label{eq:gauge_a_constraint2}
\{\mathbf{a}_{a 1}(\imp^1,\imp^1), \text{any} \}=0 \,.
\end{equation}
Because of this we require
\begin{equation}
\label{eq:D1_a_commut_relation_new2}
\begin{split}
&\frac{1}{\var} \{\imp^1, \mathbf{a}_{a \alpha}(\imp^1,y) \} = + i \mathbf{a}_{a \alpha}(\imp^1,y) \,, \qquad \text{for} \ \alpha=2, 3, \dots, \ell\,,\\
&\frac{1}{\var} \{\imp^1, \mathbf{a}_{a 1}(\imp^1,y) \} = 0 \,.
\end{split}
\end{equation}

\subsection{Removing derivative terms in a-b and b-b brackets}

Similar considerations for the other spin brackets lead to the following commutation relation
\begin{equation}
\label{eq:D1_b_commut_relation_new}
\begin{split}
&\frac{1}{\var} \{ \imp^1, \mathbf{b}_{\alpha a}(\imp^1, y) \} = -i \mathbf{b}_{\alpha a}(\imp^1, y) \,, \qquad \text{for} \ \alpha=2, 3, \dots, \ell \,,\\
&\frac{1}{\var} \{ \imp^1, \mathbf{b}_{1 a}(\imp^1, y) \} = 0 \,. 
\end{split}
\end{equation}
In this case, it was our choice to set a trivial commutation relation for $\mathbf{b}_{1 a}(\imp^1, y)$ and $\imp^1$.
The reason is that we would like to have
\begin{equation}
\{\imp^1, f^{\alpha}_{ab}(x) \} = \{\imp^1, \mathbf{a}_{a \alpha}(x) \mathbf{b}_{\alpha b}(x)  \} =0  \,, \forall \alpha=1, \dots, \ell \,. 
\end{equation}
Once we require~\eqref{eq:D1_a_commut_relation_new2} and~\eqref{eq:D1_b_commut_relation_new} the spin brackets take the following form
\begin{equation}
\label{eq:aa_bracket_with_loop_var_app}
\begin{split}
&\frac{1}{\var} \, \{\mathbf{a}_{a \alpha}(y) \,,\mathbf{a}_{b \beta}(t) \}=- \mathbf{a}_{a \beta}(y)\mathbf{a}_{b \alpha}(t) \rho_{\beta \alpha, \alpha \beta}(y-t)\\
&  + i \varphi(t-y+\Delta,\mathfrak{q}_{ba})\mathbf{a}_{b \alpha}(y) \mathbf{a}_{a \beta}(t) - i \varphi(t-\imp^1+\Delta,\mathfrak{q}_{ba})\mathbf{a}_{a \alpha}(y) \mathbf{a}_{a \beta}(t)\\
&- i \varphi(\imp^1-y+\Delta,\mathfrak{q}_{ba})\mathbf{a}_{b \alpha}(y) \mathbf{a}_{b \beta}(t)+i \varphi^{\text{reg}}(\mathfrak{q}_{ba})\mathbf{a}_{a \alpha}(y) \mathbf{a}_{b \beta}(t)\\
&+\mathbf{a}_{a \beta} (\imp^1)\mathbf{a}_{b 1}(t)   \rho_{\beta 1, 1 \beta}(\imp^1-t) \mathbf{a}_{a \alpha}(y)- \mathbf{a}_{a 1} (y)\mathbf{a}_{b \alpha}(\imp^1)   \rho_{\alpha 1,1 \alpha}(\imp^1-y) \mathbf{a}_{b \beta}(t)\\
&-i \delta_{\beta 1} \frac{d}{dx} ((\mathbf{a}_{a \alpha}(x,y) )\Bigr|_{x=\imp^1}   \mathbf{a}_{b \beta}(\imp^1,t)+i \delta_{\alpha 1} \mathbf{a}_{a \alpha}(\imp^1,y) \frac{d}{dx} (\mathbf{a}_{b \beta}(x,t) )\Bigr|_{x=\imp^1}  \,,
\end{split}
\end{equation}
\begin{equation}
\label{eq:ab_bracket_with_loop_var_app}
\begin{split}
&\frac{1}{\var} \, \{\mathbf{a}_{a \alpha}(y), \mathbf{b}_{\beta b}(t)\}=+\mathbf{a}_{a \gamma}(y) \rho(y-t)_{-\gamma \alpha, \beta \gamma} \mathbf{b}_{\gamma b}(t)\\
&+i \varphi(\imp^1-y+\Delta,\mathfrak{q}_{ba})   \mathbf{a}_{b \alpha}(y) \mathbf{b}_{\beta b}(t) -i \varphi^{\text{reg}}(\qq_{ba}) \mathbf{a}_{a \alpha}(y) \mathbf{b}_{\beta b}(t)\\
&-  \mathbf{a}_{a 1} (y)\mathbf{a}_{b \alpha}(\imp^1) \rho_{1 \alpha, \alpha 1}(y-\imp^1) \mathbf{b}_{\beta b}(t) -\mathbf{a}_{a \gamma}(\imp^1) \rho(\imp^1-t)_{-\gamma 1, \beta \delta} \mathbf{b}_{\delta b}(t) \mathbf{a}_{a \alpha}(y)\\
&-\frac{2 \pi }{\var} \delta_{\alpha \beta} Q_b Q^{-1}_a  \mathbb{L}_{ab}(t-\Delta) \delta(y-t)+\frac{2 \pi }{\var} \delta_{1 \beta} Q_b Q^{-1}_a  \mathbb{L}_{ab}(t-\Delta) \delta(\imp^1-t)\mathbf{a}_{a \alpha}(y)\\
&+i \delta_{\beta 1}  \frac{d}{dx} \mathbf{a}_{a \alpha}(x,y)\Bigl|_{x=\imp^1} \mathbf{b}_{\beta b}(\imp^1 ,t) +i \delta_{\alpha 1} \mathbf{a}_{a\alpha}(\imp^1 ,y) \frac{d}{dx} \mathbf{b}_{\beta b}(x,t)\Bigl|_{x=\imp^1}  \,,
\end{split}
\end{equation}
\begin{equation}
\label{eq:bb_bracket_with_loop_var_app}
\begin{split}
&\frac{1}{\var} \{\mathbf{b}_{\alpha a}(y),\mathbf{b}_{\beta b}(t)\} =-\rho_{\alpha \beta, \beta \alpha}(y-t)
\mathbf{b}_{\beta a}(y) \mathbf{b}_{\alpha b}(t) \\
&+i \mathbf{b}_{\alpha b}(y) \mathbf{b}_{\beta a}(t)   \varphi(y-t+\Delta, \mathfrak{q}_{ba})   +i \mathbf{b}_{\alpha a}(y) \varphi^{\text{reg}}(\mathfrak{q}_{ba})    \mathbf{b}_{\beta b}(t) \\
&+\mathbf{b}_{\beta b}(t)  \mathbf{a}_{b \gamma}(\imp^1) \rho(y-\imp^1)_{ + \alpha \gamma, \gamma 1} \mathbf{b}_{\gamma a}(y)  -\mathbf{b}_{\alpha a}(y)  \mathbf{a}_{a \gamma}(\imp^1)  \rho(t-\imp^1)_{+ \beta \gamma, \gamma 1} \mathbf{b}_{\gamma b}(t)\\
&+\frac{2 \pi}{\var} \delta_{\alpha 1}  \delta(y-\imp^1)  \, \mathbf{b}_{\beta b}(t) Q^{-1}_b    \mathbb{L}_{ba}(y-\Delta) Q_a -\frac{2 \pi }{\var} \delta_{1 \beta} \mathbf{b}_{\alpha a}(y) Q^{-1}_a \mathbb{L}_{ab}(t-\Delta) Q_b \delta(\imp^1-t)\\
&+i \delta_{\beta 1}\frac{d}{dx}\mathbf{b}_{\alpha a}(x,y) \Bigl|_{x=\imp^1} \mathbf{b}_{\beta b}(\imp^1,t)-i \delta_{\alpha 1} \mathbf{b}_{\alpha a}(\imp^1,y) \frac{d}{dx}\mathbf{b}_{\beta b}(x,t) \Bigl|_{x=\imp^1} \,.
\end{split}
\end{equation}
Apart from the particular case $\alpha=1$ or $\beta=1$, the derivative terms disappear from the brackets above. The reason why the derivatives remain for the cases $\alpha, \beta=1$ comes from the fact that $\imp^1$ trivially commutes with $\mathbf{a}_{i1}$ and $\mathbf{b}_{1i}$. In the next appendix, we will show that these residual derivatives play an important role in providing a finite-dimensional realization of the spin brackets.

\section{Finite dimensional representations with inhomogeneities}
\label{app:fin_dim_with_imp}

In this appendix, we search for finite-dimensional representations of the brackets~\eqref{eq:aa_bracket_with_loop_var_app}, \eqref{eq:ab_bracket_with_loop_var_app} and~\eqref{eq:bb_bracket_with_loop_var_app} in the presence of inhomogeneities $\{\imp^\alpha\}^\ell_{\alpha=1}$. We evaluate the spin oscillators $\mathbf{a}$ at the inhomogeneities\footnote{We recall that the first loop variable has already been evaluated at $\imp^1$.}
\begin{equation}
\label{eq:app_a_representation}
\mathbf{a}_{a \alpha} \equiv \mathbf{a}_{a \alpha}(\imp^\alpha)
\end{equation}
and represent the oscillators $\mathbf{b}$ as
\begin{equation}
\label{eq:app_b_representation}
\mathbf{b}_{\alpha a}(x) \equiv \mathbf{b}_{\alpha a} \delta_\epsilon(x-\imp^\alpha) \,.
\end{equation}
Here $\delta_\epsilon$ is a regularized periodic Dirac delta function:
\begin{equation}
\label{eq:reg_Dirac_Delta}
\delta_\epsilon(x)= \frac{1}{2 \pi} \frac{\sinh \epsilon}{\cosh \epsilon - \cos{x}} \,.
\end{equation}
Throughout this appendix, we assume that the impurity positions $\imp^\alpha$ are real and pairwise distinct modulo $2\pi$, and that the loop-dependent $\mathbf{a}$ oscillators are smooth near these positions. The regulator $\epsilon>0$ is kept finite while evaluating the brackets, and the limit $\epsilon\to0^+$ is taken only after combining all contributions and canceling their divergences. In particular, regularized delta functions multiplying terms of order $1/\epsilon$ must be retained until this cancellation has occurred, since replacing them prematurely by their limits may change the finite part. Boundary values denoted by $\pm i0$ are always taken at fixed $\epsilon$, before the limit $\epsilon\to0^+$. For coincident evaluation points in the $\mathbf{a}$-$\mathbf{a}$ bracket, we similarly apply the same point-splitting prescription to all terms and remove the splitting only after combining them. All distributional asymptotic expansions below are understood in $\mathcal D'$; a remainder denoted by $o_{\mathcal D'}(1)$ tends to zero when paired with any smooth test function, without implying a specific power-law rate in $\epsilon$.

For the rest of this section we choose the $\rho$ matrix to be $\check{\rho}$, as defined in Appendix~\ref{app:rho_matrices}. The only non-zero element of this matrix are
\begin{equation}
\label{eq:app_check_rho_matrices}
\begin{split}
&\check{\rho}_{\alpha \beta, \beta \alpha}(x)=  \pi \delta(x)  \sgn_{\beta \alpha}  - \frac{i}{2} \, \delta_{\alpha \beta} \, \pv \, \cot \frac{x}{2}\,,\\
&\check{\rho}_{\pm \alpha \beta, \beta \alpha}(x)=  \pi \delta(x)  (\sgn_{\beta \alpha} \pm 1) - \frac{i}{2} \, \delta_{\alpha \beta} \, \pv \, \cot \frac{x}{2}\,.
\end{split}
\end{equation}
It turns out that the matrix $\check{\rho}$ is compatible with the representations in~\eqref{eq:app_a_representation} and~\eqref{eq:app_b_representation}, while the matrix $\hat{\rho}$ is not.

At coincident real arguments, the singular kernels
$\varphi(y-t+\Delta,s)$ and $\check{\rho}(y-t)$ are interpreted in the
principal-value sense. This prescription is distinct from the boundary
value $\mathbb L(t-\Delta+i0)$ used in section~\ref{app:bb_bracket_finite}: the latter is taken from the interior
of the strip of analyticity, as explained below. These prescriptions
are part of the definition of the localized realization, and the
cancellation of singular terms is established with these choices held
fixed.

\subsection{Finite dimensional realization of the a-a bracket}
\label{app:aa_com_finite_relations}

We want to evaluate the bracket~\eqref{eq:aa_bracket_with_loop_var_app} at $y=\imp^\alpha$ and $t=\imp^\beta$.
Formula~\eqref{eq:D1_a_commut_relation_new2} was given for a generic loop parameter $y$:
\begin{equation}
\frac{1}{\var} \{\imp^1, \mathbf{a}_{a \alpha}(y) \} = + i \mathbf{a}_{a \alpha}(y) (1-\delta_{\alpha 1})\,.
\end{equation}
Now we want to take $y=\imp^\alpha$. Let us expand the RHS and LHS of the expression above around $y=\imp^\alpha$. Then we obtain
\begin{equation}
\label{eq:expansion_D1_aalphaDalpha}
\frac{1}{\var} \{\imp^1, \mathbf{a}_{a \alpha}(\imp^\alpha) \}  - \frac{1}{\var} \{\imp^1, \imp^\alpha \} \mathbf{a}'_{a \alpha}(\imp^\alpha)=i (1-\delta_{\alpha 1}) \mathbf{a}_{a \alpha} (\imp^\alpha)\,.
\end{equation}
The requirement of not getting derivatives in the Poisson brackets of the localised oscillators, imposes
\begin{equation}
\label{eq:comD1aalpha}
\frac{1}{\var} \{\imp^1, \mathbf{a}_{a \alpha} \} 
=i (1-\delta_{\alpha 1})\mathbf{a}_{a \alpha} \,, \qquad \{ \imp^1, \imp^\alpha\} = 0 \,.
\end{equation}
From condition~\eqref{eq:gauge_a_constraint1} and the discussion in section~\ref{app:removing_D_term_in_a_a_bracket} we know that
\begin{equation}
\frac{1}{\var} \, \{\mathbf{a}_{a 1}(\imp^1, \imp^1) \,,\mathbf{a}_{b \beta}(\imp^1,t) \} = \frac{1}{\var} \, \{\mathbf{a}_{a \alpha}(\imp^1, y) \,,\mathbf{a}_{b 1}(\imp^1,\imp^1) \}=0 \,.
\end{equation}
The only nontrivial Poisson relations between $\mathbf{a}$ oscillators are those involving $\alpha \ne 1$ and $\beta \ne 1$. 

We start writing~\eqref{eq:aa_bracket_with_loop_var_app} for $\alpha, \beta \ne 1$:
\begin{equation}
\label{eq:aa_bracket_with_loop_var_app2}
\begin{split}
&\frac{1}{\var} \, \{\mathbf{a}_{a \alpha}(y) \,,\mathbf{a}_{b \beta}(t) \}=- \mathbf{a}_{a \beta}(y)\mathbf{a}_{b \alpha}(t) \check{\rho}_{\beta \alpha, \alpha \beta}(y-t)\\
&  + i \varphi(t-y+\Delta,\mathfrak{q}_{ba})\mathbf{a}_{b \alpha}(y) \mathbf{a}_{a \beta}(t) - i \varphi(t-\imp^1+\Delta,\mathfrak{q}_{ba})\mathbf{a}_{a \alpha}(y) \mathbf{a}_{a \beta}(t)\\
&- i \varphi(\imp^1-y+\Delta,\mathfrak{q}_{ba})\mathbf{a}_{b \alpha}(y) \mathbf{a}_{b \beta}(t)+i \varphi^{\text{reg}}(\mathfrak{q}_{ba})\mathbf{a}_{a \alpha}(y) \mathbf{a}_{b \beta}(t)\\
&+\mathbf{a}_{a \beta} (\imp^1)\mathbf{a}_{b 1}(t)   \check{\rho}_{\beta 1, 1 \beta}(\imp^1-t) \mathbf{a}_{a \alpha}(y)- \mathbf{a}_{a 1} (y)\mathbf{a}_{b \alpha}(\imp^1)   \check{\rho}_{\alpha 1,1 \alpha}(\imp^1-y) \mathbf{a}_{b \beta}(t) \,.
\end{split}
\end{equation}
Using the matrices in~\eqref{eq:app_check_rho_matrices}, for $y=\imp^\alpha$ and $t=\imp^\beta$, with $\alpha, \beta \ne 1$, we have
\begin{equation}
\label{eq:app_expandion_rhoaa_zero}
\mathbf{a}_{a \beta} (\imp^1)\mathbf{a}_{b 1}(t)   \check{\rho}_{\beta 1, 1 \beta}(\imp^1-t) \mathbf{a}_{a \alpha}(y)- \mathbf{a}_{a 1} (y)\mathbf{a}_{b \alpha}(\imp^1)   \check{\rho}_{\alpha 1,1 \alpha}(\imp^1-y) \mathbf{a}_{b \beta}(t) =0 \,.
\end{equation}
and
\begin{equation}
- \mathbf{a}_{a \beta}(y)\mathbf{a}_{b \alpha}(t) \check{\rho}_{\beta \alpha, \alpha \beta}(y-t)= +\frac{i}{2} \delta_{\alpha \beta} \mathbf{a}_{a \beta}(y)\mathbf{a}_{b \alpha}(t) \pv \cot \frac{y-t}{2} \,.
\end{equation}
The latter expression is nonzero only for $\alpha=\beta$. Therefore we should only care about the situation $y=t=\imp^{\alpha}$. Since this point is singular, to evaluate the expression we  first set $t=\imp^\alpha$. Then we define $y=\imp^\alpha + \epsilon$, where $\epsilon$ is a finite small regulator parameter.
Then we obtain
\begin{equation}
\label{eq:app_expandion_rhoaa_eps}
\begin{split}
- \mathbf{a}_{a \beta}(y)\mathbf{a}_{b \alpha}(t) \check{\rho}_{\beta \alpha, \alpha \beta}(y-t)&= +\frac{i}{2} \delta_{\alpha \beta} \mathbf{a}_{a \alpha}(\imp^\alpha+\epsilon)\mathbf{a}_{b \alpha} \cot \frac{\epsilon}{2}=+ \frac{i}{\epsilon} \delta_{\alpha \beta} \mathbf{a}_{a \alpha}(\imp^\alpha + \epsilon) \mathbf{a}_{b \alpha} + O(\epsilon)\\
&=+ \frac{i}{\epsilon} \delta_{\alpha \beta} \mathbf{a}_{a \alpha} \mathbf{a}_{b \alpha} + i \delta_{\alpha \beta} \mathbf{a}'_{a \alpha} \mathbf{a}_{b \alpha} + O(\epsilon) \,.
\end{split}
\end{equation}
The same limit (i.e. setting first $t=\imp^{\alpha}$ and then $y=\imp^{\alpha}+\epsilon$) must be applied on all the other singular terms in the bracket. By doing so we get
\begin{equation}
\begin{split}
+ i \varphi(t-y+\Delta,\mathfrak{q}_{ba})\mathbf{a}_{b \alpha}(y) \mathbf{a}_{a \beta}(t)=&+ i (1-\delta_{\alpha \beta}) \varphi(\imp^{\beta \alpha}+\Delta,\mathfrak{q}_{ba})\mathbf{a}_{b \alpha} \mathbf{a}_{a \beta}\\
&+\delta_{\alpha \beta} i \varphi(-\epsilon+\Delta,\mathfrak{q}_{ba})\mathbf{a}_{b \alpha}(\imp^\alpha+\epsilon) \mathbf{a}_{a \alpha} \,.
\end{split}
\end{equation}
After expanding in $\epsilon$ we obtain
\begin{equation}
\label{eq:app_expandion_phiaa_eps}
\begin{split}
+ i \varphi(t-y+\Delta,\mathfrak{q}_{ba})\mathbf{a}_{b \alpha}(y) \mathbf{a}_{a \beta}(t)=&+ i (1-\delta_{\alpha \beta}) \varphi(\imp^{\beta \alpha}+\Delta,\mathfrak{q}_{ba})\mathbf{a}_{b \alpha} \mathbf{a}_{a \beta}\\
&-\frac{i}{\epsilon} \delta_{\alpha \beta} \mathbf{a}_{b \alpha} \mathbf{a}_{a \alpha} -i \delta_{\alpha \beta} \mathbf{a}'_{b \alpha} \mathbf{a}_{a \alpha}+\delta_{\alpha \beta} i \varphi^{\text{reg}}(\mathfrak{q}_{ba})\mathbf{a}_{b \alpha} \mathbf{a}_{a \alpha} + O(\epsilon)\,.
\end{split}
\end{equation}
Plugging~\eqref{eq:app_expandion_rhoaa_zero}, \eqref{eq:app_expandion_rhoaa_eps} and~\eqref{eq:app_expandion_phiaa_eps} in~\eqref{eq:aa_bracket_with_loop_var_app2}, all singularities cancel, and we obtain
\begin{equation}
\label{eq:aa_bracket_with_loop_var_app3}
\begin{split}
&\frac{1}{\var} \, \{\mathbf{a}_{a \alpha}(y) \,,\mathbf{a}_{b \beta}(t) \}\Bigl|_{y=\imp^\alpha, t=\imp^\beta}=+ i \delta_{\alpha \beta} \mathbf{a}'_{a \alpha} \mathbf{a}_{b \alpha}-i \delta_{\alpha \beta} \mathbf{a}'_{b \alpha} \mathbf{a}_{a \alpha}\\
&  + i (1-\delta_{\alpha \beta})\varphi(\imp^{\beta \alpha}+\Delta,\mathfrak{q}_{ba})\mathbf{a}_{b \alpha} \mathbf{a}_{a \beta}+i (1+\delta_{\alpha \beta}) \varphi^{\text{reg}}(\mathfrak{q}_{ba})\mathbf{a}_{a \alpha} \mathbf{a}_{b \beta}  \\
&- i \varphi(\imp^{\beta1}+\Delta,\mathfrak{q}_{ba})\mathbf{a}_{a \alpha} \mathbf{a}_{a \beta}- i \varphi(\imp^{1\alpha}+\Delta,\mathfrak{q}_{ba})\mathbf{a}_{b \alpha} \mathbf{a}_{b \beta} \,.
\end{split}
\end{equation}
Leaving open the possibility for $\imp^\alpha$ and $\imp^\beta$ to be dynamical variables, the bracket above is given by
\begin{equation}
\begin{split}
\frac{1}{\var} \, \{\mathbf{a}_{a \alpha}(y) \,,\mathbf{a}_{b \beta}(t) \}\Bigl|_{y=\imp^\alpha, t=\imp^\beta}&=\frac{1}{\var} \, \{\mathbf{a}_{a \alpha} \,,\mathbf{a}_{b \beta} \}\\
&-\frac{1}{\var} \, \{\imp^\alpha \,,\mathbf{a}_{b \beta} \} \mathbf{a}'_{a \alpha}-\frac{1}{\var} \, \{\mathbf{a}_{a \alpha} \,,\imp^\beta \}\mathbf{a}'_{b \beta}+\frac{1}{\var} \, \{\imp^\alpha \,,\imp^\beta \} \mathbf{a}'_{a \alpha} \mathbf{a}'_{b \beta} \,.
\end{split}
\end{equation}
Combining this expansion with the result in~\eqref{eq:aa_bracket_with_loop_var_app3}, we obtain
\begin{equation}
\label{eq:aa_bracket_with_loop_var_app4}
\begin{split}
&\frac{1}{\var} \, \{\mathbf{a}_{a \alpha} \,,\mathbf{a}_{b \beta} \}=-\frac{1}{\var} \, \{\imp^\alpha \,,\imp^\beta \} \mathbf{a}'_{a \alpha} \mathbf{a}'_{b \beta}\\
&+ \left( i \delta_{\alpha \beta}  \mathbf{a}_{b \alpha}+\frac{1}{\var} \, \{\imp^\alpha \,,\mathbf{a}_{b \beta} \} \right) \mathbf{a}'_{a \alpha}- \left(i \delta_{\alpha \beta}  \mathbf{a}_{a \alpha}-\frac{1}{\var} \, \{\mathbf{a}_{a \alpha} \,,\imp^\beta \} \right) \mathbf{a}'_{b \beta}\\
&  + i (1-\delta_{\alpha \beta})\varphi(\imp^{\beta \alpha}+\Delta,\mathfrak{q}_{ba})\mathbf{a}_{b \alpha} \mathbf{a}_{a \beta}+i (1+\delta_{\alpha \beta}) \varphi^{\text{reg}}(\mathfrak{q}_{ba})\mathbf{a}_{a \alpha} \mathbf{a}_{b \beta}  \\
&- i \varphi(\imp^{\beta1}+\Delta,\mathfrak{q}_{ba})\mathbf{a}_{a \alpha} \mathbf{a}_{a \beta}- i \varphi(\imp^{1\alpha}+\Delta,\mathfrak{q}_{ba})\mathbf{a}_{b \alpha} \mathbf{a}_{b \beta} \,.
\end{split}
\end{equation}
To have a finite number of degrees of freedom we require the the coefficients in front of derivatives vanish; then we require for $\alpha, \beta \ne 1$:
\begin{equation}
\label{eq:aalpha_Dbeta_com}
\frac{1}{\var} \{\imp^\alpha, \mathbf{a}_{b \beta} \} =- i \delta_{\alpha \beta} \mathbf{a}_{b \beta} \,, \qquad \{\imp^\alpha, \imp^\beta\}=0 \,.
\end{equation}
In this manner, we obtain
\begin{equation}
\label{eq:aa_bracket_with_loop_var_app5}
\begin{split}
\frac{1}{\var} \, \{\mathbf{a}_{a \alpha} \,,\mathbf{a}_{b \beta} \}=&+ i (1-\delta_{\alpha \beta})\varphi(\imp^{\beta \alpha}+\Delta,\mathfrak{q}_{ba})\mathbf{a}_{b \alpha} \mathbf{a}_{a \beta}+i (1+\delta_{\alpha \beta}) \varphi^{\text{reg}}(\mathfrak{q}_{ba})\mathbf{a}_{a \alpha} \mathbf{a}_{b \beta}  \\
&- i \varphi(\imp^{\beta1}+\Delta,\mathfrak{q}_{ba})\mathbf{a}_{a \alpha} \mathbf{a}_{a \beta}- i \varphi(\imp^{1\alpha}+\Delta,\mathfrak{q}_{ba})\mathbf{a}_{b \alpha} \mathbf{a}_{b \beta} \,.
\end{split}
\end{equation}
For $\alpha \ne \beta$ this expression corresponds to eq. (B1) in~\cite{Arutyunov:2026yjy}. For $\alpha=\beta$ we obtain the regular part of that equation. 

Expanding~\eqref{eq:D1_a_commut_relation_new2} around $y=\imp^\alpha$, we also have
\begin{equation}
\label{eq:app_D1a_Dalpha_Dbeta}
\begin{split}
&\frac{1}{\var} \{\imp^1, \mathbf{a}_{a \alpha} \} = + i \mathbf{a}_{a \alpha} \,, \qquad \text{for} \ \alpha=2, 3, \dots, \ell\,,\\
&\frac{1}{\var} \{\imp^1, \mathbf{a}_{a 1} \} = 0 \,.
\end{split}
\end{equation}

\subsection{Finite dimensional representation of the b-b bracket}
\label{app:bb_bracket_finite}

We search now for finite-dimensional representations of the $\mathbf{b}$ oscillators of the form~\eqref{eq:app_b_representation}.
Indicating by `any' a generic element of the Poisson algebra, then by the Leibniz rule, it must hold
\begin{equation}
\{\text{any}, \mathbf{b}_{\beta b} (t) \} = \{\text{any}, \mathbf{b}_{\beta b} \} \delta_\epsilon(t-\imp^\beta)- \{\text{any}, \imp^\beta \}\mathbf{b}_{\beta b} \delta_\epsilon'(t-\imp^\beta) \,,
\end{equation}
and we obtain
\begin{equation}
\label{eq:any_b_Delta_repr}
\{\text{any}, \mathbf{b}_{\beta b} \} \delta_\epsilon(t-\imp^\beta)=\{\text{any}, \mathbf{b}_{\beta b} \delta_\epsilon(t-\imp^\beta) \} + \{\text{any}, \imp^\beta \}\mathbf{b}_{\beta b} \delta_\epsilon'(t-\imp^\beta) \,.
\end{equation}
If we want that~\eqref{eq:app_b_representation}
is a representation of the original Poisson algebra, then the RHS of~\eqref{eq:ab_bracket_with_loop_var_app}, after the substitution
\begin{equation}
\mathbf{b}_{\beta b}(t) \to \mathbf{b}_{\beta b} \delta_\epsilon(t-\imp^\beta) \,,
\end{equation}
must coincide with
\begin{equation}
\{\mathbf{a}_{a \alpha}(y), \mathbf{b}_{\beta b} \delta_\epsilon(t-\imp^\beta) \} \,.
\end{equation}

To evaluate the $\mathbf{b}$-$\mathbf{b}$ bracket, we start from the expression in~\eqref{eq:bb_bracket_with_loop_var_app}:
\begin{equation}
\label{eq:bb_bracket_with_loop_var_app2}
\begin{split}
&\frac{1}{\var} \{\mathbf{b}_{\alpha a}(y),\mathbf{b}_{\beta b}(t)\} =A^{\text{bb}}_{\alpha 1}+B^{\text{bb}}_{\beta 1}-\check{\rho}_{\alpha \beta, \beta \alpha}(y-t)
\mathbf{b}_{\beta a}(y) \mathbf{b}_{\alpha b}(t) \\
&+i \mathbf{b}_{\alpha b}(y) \mathbf{b}_{\beta a}(t)   \varphi(y-t+\Delta, \mathfrak{q}_{ba})   +i \mathbf{b}_{\alpha a}(y) \varphi^{\text{reg}}(\mathfrak{q}_{ba})    \mathbf{b}_{\beta b}(t) \,.
\end{split}
\end{equation}
where we defined
\begin{equation}
\label{eq:app_Aalpha1_first_defined}
\begin{split}
A^{\text{bb}}_{\alpha 1} =&+\mathbf{b}_{\beta b}(t)  \mathbf{a}_{b \gamma}(\imp^1) \check{\rho}(y-\imp^1)_{+ \alpha \gamma, \gamma 1} \mathbf{b}_{\gamma a}(y)  \\
&+\frac{2 \pi}{\var} \delta_{\alpha 1}  \delta(y-\imp^1)  \, \mathbf{b}_{\beta b}(t) Q^{-1}_b    \mathbb{L}_{ba}(y-\Delta+i0) Q_a -i \delta_{\alpha 1} \mathbf{b}_{\alpha a}(\imp^1,y) \frac{d}{dx}\mathbf{b}_{\beta b}(x,t) \Bigl|_{x=\imp^1} \,,
\end{split}
\end{equation}
\begin{equation}
\label{eq:app_Bbeta1_first_defined}
\begin{split}
B^{\text{bb}}_{\beta 1} =&-\mathbf{b}_{\alpha a}(y)  \mathbf{a}_{a \gamma}(\imp^1)  \check{\rho}(t-\imp^1)_{+\beta \gamma, \gamma 1} \mathbf{b}_{\gamma b}(t)  \\
& -\frac{2 \pi }{\var} \delta_{1 \beta} \mathbf{b}_{\alpha a}(y) Q^{-1}_a \mathbb{L}_{ab}(t-\Delta+i0) Q_b \delta(\imp^1-t)  +i \delta_{\beta 1}\frac{d}{dx}\mathbf{b}_{\alpha a}(x,y) \Bigl|_{x=\imp^1} \mathbf{b}_{\beta b}(\imp^1,t) \,.
\end{split}
\end{equation}
The $+i0$ prescription follows from the boundary value used to define the shifted Lax matrix. With $\operatorname{Im}\Delta>0$, we approach the lower boundary of the strip of analyticity from its interior, defining
\begin{equation}
\mathbb{L}(t-\Delta+i0) \equiv \lim_{\sigma\to0^+}\mathbb{L}(t-\Delta+i\sigma)
\end{equation}
in the sense of distributions. This is the same prescription adopted in the Green's function equation~\eqref{eq:eq_for_Green_fun}, and fixes the sign of the infinitesimal shift. The limit $\sigma\to0^+$ is taken at fixed $\epsilon>0$, before removing the regulator of the source. In particular, $\sigma$ and $\epsilon$ are independent regulators and should not be identified.

\paragraph{Evaluating $A^{\text{bb}}_{\alpha 1}$ and $B^{\text{bb}}_{\beta 1}$.}

With the definition in~\eqref{eq:app_check_rho_matrices}, we have
\begin{equation}
\begin{split}
&+\mathbf{b}_{\beta b}(t)  \mathbf{a}_{b \gamma}(\imp^1) \check{\rho}(y-\imp^1)_{+ \alpha \gamma, \gamma 1} \mathbf{b}_{\gamma a}(y)=\\
&+\delta_{\alpha 1} \mathbf{b}_{\beta b}(t)  \check{\rho}(y-\imp^1)_{+ 1 1, 1 1} \mathbf{b}_{1 a}(y)+\delta_{\alpha 1} \sum_{\gamma>1} \mathbf{b}_{\beta b}(t)  \mathbf{a}_{b \gamma}(\imp^1) \check{\rho}(y-\imp^1)_{+ 1 \gamma, \gamma 1} \mathbf{b}_{\gamma a}(y)
\end{split}
\end{equation}
The second term on the RHS of the equation above vanishes when we choose 
\begin{equation}
\mathbf{b}_{\alpha a} (y)=\mathbf{b}_{\alpha a} \delta_\epsilon(y-\imp^\alpha) \,.
\end{equation}
Indeed, for $\gamma \ne 1$, we have
\begin{equation}
\check{\rho}(y-\imp^1)_{ + 1 \gamma, \gamma 1} \mathbf{b}_{\gamma a}(y) =  2 \pi \delta(y-\imp^1) \mathbf{b}_{\gamma a} \delta_\epsilon(y-\imp^\gamma) \xrightarrow[\epsilon \to 0]{} 0  \,,
\end{equation}
and
\begin{equation}
\begin{split}
&\mathbf{b}_{\beta b}(t)  \mathbf{a}_{b \gamma}(\imp^1) \check{\rho}(y-\imp^1)_{+ \alpha \gamma, \gamma 1} \mathbf{b}_{\gamma a}(y)=\\
& \left(\pi  \delta(y-\imp^1)-\frac{i}{2} \ \pv \cot \frac{y-\imp^1}{2} \,    \right) \delta_{\alpha 1} \mathbf{b}_{\beta b}   \mathbf{b}_{1 a} \delta_\epsilon(y-\imp^1) \delta_\epsilon(t-\imp^\beta)=\\
& -\frac{i}{2} \  \cot \frac{y-\imp^1 - i 0}{2} \,     \delta_{\alpha 1} \mathbf{b}_{\beta b}   \mathbf{b}_{1 a} \delta_\epsilon(y-\imp^1) \delta_\epsilon(t-\imp^\beta) \,.
\end{split}
\end{equation}
In the limit $\epsilon \to 0$, this contribution is singular. However, we can compute it assuming $\epsilon$ to be a small finite parameter. Because $\epsilon$ is small, the nonvanishing support of the expression above is located around $y=\imp^1$; performing the expansion of the $\cot$, we can then write
\begin{equation}
\label{eq:app_checkrho_alpha1_dist}
\begin{split}
&\mathbf{b}_{\beta b}(t)  \mathbf{a}_{b \gamma}(\imp^1) \check{\rho}(y-\imp^1)_{+ \alpha \gamma, \gamma 1} \mathbf{b}_{\gamma a}(y)=-\frac{i}{y-\imp^1-i 0}   \, \delta_{\alpha 1} \mathbf{b}_{\beta b}   \mathbf{b}_{1 a} \delta_\epsilon(y-\imp^1) \delta_\epsilon(t-\imp^\beta)\,.
\end{split}
\end{equation}
To evaluate the expression above we need to check how this distribution acts on a test function.
Deforming the integration contour by surrounding the pole at $y=\imp^1-i \epsilon$ from below, we can write
\begin{equation}
\begin{split}
    \int^{+\pi}_{-\pi} dy\delta_\epsilon (y-\imp^1) \frac{1}{y-\imp^1 - i 0} f(y)&=
    \int_{D^-_1} dy \delta_\epsilon (y-\imp^1) \frac{1}{y-\imp^1 - i 0} f(y)\\
    &-\frac{1}{i \epsilon} f(\imp^1 - i \epsilon) \,,
\end{split}
\end{equation}
where $D^-_\alpha$ is the contour connecting $-\pi$ and $+\pi$ surrounding the pole $\imp^1-i \epsilon$ of $\delta_\epsilon$ from below. The RHS expression in the first line cancels in the limit $\epsilon \to 0$ since all singularities lie on the same contour size and do not trap the contour in the limit.
For small values of $\epsilon$, we obtain
\begin{equation}
    \int^{+\pi}_{-\pi} dy\delta_\epsilon (y-\imp^1) \frac{1}{y-\imp^1 - i 0} f(y)=-\frac{1}{i \epsilon} f(\imp^1)+  f'(\imp^1) \,.
\end{equation}
Then we identify
\begin{equation}
\label{eq:app_Dirac_1ovxpieps}
\delta_\epsilon (y-\imp^1) \frac{1}{y-\imp^1 - i 0} = \frac{i}{ \epsilon} \delta(y-\imp^1)  -  \delta'(y-\imp^1) \,.
\end{equation}
Substituting this expression in~\eqref{eq:app_checkrho_alpha1_dist}, we obtain
\begin{equation}
\label{eq:app_checkrho_alpha1_dist2}
\begin{split}
&\mathbf{b}_{\beta b}(t)  \mathbf{a}_{b \gamma}(\imp^1) \check{\rho}(y-\imp^1)_{+ \alpha \gamma, \gamma 1} \mathbf{b}_{\gamma a}(y)=\\
&\frac{1}{\epsilon} \delta(y-\imp^\alpha)    \, \delta_{\alpha 1} \mathbf{b}_{\beta b}   \mathbf{b}_{1 a} \delta_\epsilon(t-\imp^\beta)+i \delta'(y-\imp^\alpha)    \, \delta_{\alpha 1} \mathbf{b}_{\beta b}   \mathbf{b}_{1 a} \delta_\epsilon(t-\imp^\beta)\,.
\end{split}
\end{equation}

The other term to be computed is the one proportional to
\begin{equation}
\begin{split}
&\frac{2 \pi}{\var} Q^{-1}_b \mathbb{L}_{ba}(y-\Delta+ i0) Q_a=\\
&\frac{2 \pi}{\var} \tilde{\var} Q^{-1}_b \sum^\ell_{\gamma=1} \int^{+\pi}_{-\pi}\frac{dz}{2 \pi i} \varphi(y-z-\Delta+ i0, \qq_{ba} - \eta) \mathbf{a}_{b \gamma}(z) \mathbf{b}_{\gamma a}(z) Q_a=\\
&\frac{1}{i}  \sum^\ell_{\gamma=1} \int^{+\pi}_{-\pi} dz \varphi(y-z+\Delta+ i0, \qq_{ba} - \eta) \mathbf{a}_{b \gamma}(z) \mathbf{b}_{\gamma a} \delta_\epsilon(z-\imp^\gamma) \,.
\end{split}
\end{equation}
In the last equality, we used the quasi-periodicity of $\varphi$ together with the definition $\tilde{\var}/\var=e^{i \eta}$ and the representation of the b oscillators.
This term appears multiplied by an additional Dirac delta function:
\begin{equation}
\label{eq:app_Leps_sing_value}
\begin{split}
&\frac{2 \pi}{\var} Q^{-1}_b \mathbb{L}_{ba}(y-\Delta+ i0) Q_a \delta(y-\imp^\alpha)=\\
&+\frac{1}{i}  \sum^\ell_{\gamma=1} \int^{+\pi}_{-\pi} dz \varphi(y-z+\Delta+ i0, \qq_{ba} - \eta) \mathbf{a}_{b \gamma}(z) \mathbf{b}_{\gamma a} \delta_\epsilon(z-\imp^\gamma)\delta(y-\imp^\alpha)=\\
&+\frac{1}{i}  \sum_{\gamma \ne \alpha}  \varphi(\imp^{\alpha\gamma}+\Delta+ i0, \qq_{ba} - \eta) \mathbf{a}_{b \gamma} \mathbf{b}_{\gamma a} \delta(y-\imp^\alpha)\\
&+\frac{1}{i}  \int^{+\pi}_{-\pi} dz \varphi(y-z+\Delta+ i0, \qq_{ba} - \eta) \mathbf{a}_{b \alpha}(z) \mathbf{b}_{\alpha a} \delta_\epsilon(z-\imp^\alpha) \delta(y-\imp^\alpha) \,.
\end{split}
\end{equation}
For small values of $\epsilon$ the non-vanishing support of the integrand is around $z=y=\imp^\alpha$. We can then expand
\begin{equation}
\varphi(y-z+\Delta+ i0, \qq_{ba} - \eta)= \frac{1}{y-z + i0} + \varphi^{\text{reg}}(\qq_{ba} - \eta) \,.
\end{equation}
To evaluate the integral in the last line of~\eqref{eq:app_Leps_sing_value}, we use
the identity~\eqref{eq:app_Dirac_1ovxpieps}, leading to
\begin{equation}
\label{eq:app_lim_deltadeltaeps_sing2}
 \frac{\delta(y-\imp^\alpha) \delta_\epsilon(z-\imp^\alpha) }{y-z + i0} = - \frac{i}{ \epsilon} \delta(y-\imp^\alpha) \delta(z-\imp^\alpha)  + \delta(y-\imp^\alpha) \delta'(z-\imp^\alpha)  +o_{\mathcal D'}(1)\,.
\end{equation}
Replacing this identity in the last line of~\eqref{eq:app_Leps_sing_value}, we obtain
\begin{equation}
\begin{split}
&\frac{2 \pi}{\var} Q^{-1}_b \mathbb{L}_{ba}(y-\Delta+ i0) Q_a \delta(y-\imp^\alpha)=\\
&+\frac{1}{i}  \sum_{\gamma \ne \alpha}  \varphi(\imp^{\alpha\gamma}+\Delta+ i0, \qq_{ba} - \eta) \mathbf{a}_{b \gamma} \mathbf{b}_{\gamma a} \delta(y-\imp^\alpha)\\
&+\frac{1}{i}  \varphi^{\text{reg}}(\qq_{ba} - \eta) \mathbf{a}_{b \alpha} \mathbf{b}_{\alpha a}  \delta(y-\imp^\alpha)- \frac{1}{\epsilon}  \int^{+\pi}_{-\pi} dz  \mathbf{a}_{b \alpha}(z) \mathbf{b}_{\alpha a} \delta(z-\imp^\alpha) \delta(y-\imp^\alpha)\\
&+ \frac{1}{i}  \int^{+\pi}_{-\pi} dz  \mathbf{a}_{b \alpha}(z) \mathbf{b}_{\alpha a}  \delta(y-\imp^\alpha) \delta'(z-\imp^\alpha) 
\end{split}
\end{equation}
Defining
\begin{equation}
\label{eq:app_Lreg_ba_alpha}
\mathbb{L}^{\text{reg},\alpha}_{ba} \equiv \frac{1}{i}  \sum_{\gamma \ne \alpha}  \varphi(\imp^{\alpha\gamma}+\Delta, \qq_{ba} - \eta) \mathbf{a}_{b \gamma} \mathbf{b}_{\gamma a}+\frac{1}{i}  \varphi^{\text{reg}}(\qq_{ba} - \eta) \mathbf{a}_{b \alpha} \mathbf{b}_{\alpha a} \,,
\end{equation}
after integration, we get
\begin{equation}
\label{eq:app_L_sing_formula}
\begin{split}
\frac{2 \pi}{\var} Q^{-1}_b \mathbb{L}_{ba}(y-\Delta+ i0) Q_a \delta(y-\imp^\alpha)&=\mathbb{L}^{\text{reg},\alpha}_{ba} \delta(y-\imp^\alpha)- \frac{1}{\epsilon}    \mathbf{a}_{b \alpha} \mathbf{b}_{\alpha a}  \delta(y-\imp^\alpha)\\
&+i \mathbf{a}'_{b \alpha} \mathbf{b}_{\alpha a} \delta(y-\imp^\alpha)\,.
\end{split}
\end{equation}

If we replace~\eqref{eq:app_checkrho_alpha1_dist2} and~\eqref{eq:app_L_sing_formula} in~\eqref{eq:app_Aalpha1_first_defined}, all singularities cancel and we get
\begin{equation}
\begin{split}
&A^{\text{bb}}_{\alpha 1} =i \delta'(y-\imp^\alpha)    \, \delta_{\alpha 1} \mathbf{b}_{\beta b}   \mathbf{b}_{1 a} \delta(t-\imp^\beta)+\delta_{\alpha 1} \mathbb{L}^{\text{reg},\alpha}_{ba} \mathbf{b}_{\beta b} \delta(y-\imp^\alpha)\delta(t-\imp^\beta)\\
& +i  \delta_{\alpha 1}  \mathbf{a}'_{b \alpha} \mathbf{b}_{\alpha a} \mathbf{b}_{\beta b} \delta(y-\imp^\alpha) \delta(t-\imp^\beta)-i \delta_{\alpha 1} \mathbf{b}_{\alpha a}(\imp^1,y) \frac{d}{dx}\mathbf{b}_{\beta b}(x,t) \Bigl|_{x=\imp^1} \,.
\end{split}
\end{equation}
Using that
\begin{equation}
\mathbf{a}'_{b1}= \frac{d}{dx}\mathbf{a}_{b1}(\imp^1, x)\bigr|_{x=\imp^1}=\frac{t'_b(\imp^1)}{t_b(\imp^1)} \,, \qquad \frac{d}{dx}\mathbf{b}_{\beta b}(x,t) \Bigl|_{x=\imp^1}= \frac{t'_b(\imp^1)}{t_b(\imp^1)} \mathbf{b}_{\beta b}(t) \,,
\end{equation}
the last line of the expression above also vanishes and we end up with
\begin{equation}
\begin{split}
&A^{\text{bb}}_{\alpha 1} =i  \, \delta_{\alpha 1} \mathbf{b}_{\beta b}   \mathbf{b}_{1 a} \delta'(y-\imp^\alpha)    \delta(t-\imp^\beta)+\delta_{\alpha 1} \mathbb{L}^{\text{reg},\alpha}_{ba} \mathbf{b}_{\beta b} \delta(y-\imp^\alpha)\delta(t-\imp^\beta) \,.
\end{split}
\end{equation}
A similar computation leads to
\begin{equation}
\begin{split}
&B^{\text{bb}}_{\beta 1} =-i  \, \delta_{\beta 1} \mathbf{b}_{\alpha a}   \mathbf{b}_{1 b} \delta'(t-\imp^\beta)    \delta(y-\imp^\alpha)-\delta_{\beta 1} \mathbb{L}^{\text{reg},\beta}_{ab} \mathbf{b}_{\alpha a} \delta(t-\imp^\beta)\delta(y-\imp^\alpha) \,.
\end{split}
\end{equation}
Plugging these expressions in~\eqref{eq:bb_bracket_with_loop_var_app2}, we obtain
\begin{equation}
\label{eq:bb_bracket_with_loop_var_app3}
\begin{split}
&\frac{1}{\var} \{\mathbf{b}_{\alpha a}(y),\mathbf{b}_{\beta b}(t)\} =+i \mathbf{b}_{\alpha a}(y) \varphi^{\text{reg}}(\mathfrak{q}_{ba})    \mathbf{b}_{\beta b}(t) \\
&+i  \, \delta_{\alpha 1} \mathbf{b}_{\beta b}   \mathbf{b}_{1 a} \delta'(y-\imp^\alpha)    \delta(t-\imp^\beta)+\delta_{\alpha 1} \mathbb{L}^{\text{reg},\alpha}_{ba} \mathbf{b}_{\beta b} \delta(y-\imp^\alpha)\delta(t-\imp^\beta)\\
&-i  \, \delta_{\beta 1} \mathbf{b}_{\alpha a}   \mathbf{b}_{1 b} \delta'(t-\imp^\beta)    \delta(y-\imp^\alpha)-\delta_{\beta 1} \mathbb{L}^{\text{reg},\beta}_{ab} \mathbf{b}_{\alpha a} \delta(t-\imp^\beta)\delta(y-\imp^\alpha)\\
&-\check{\rho}_{\alpha \beta, \beta \alpha}(y-t)
\mathbf{b}_{\beta a}(y) \mathbf{b}_{\alpha b}(t) +i \mathbf{b}_{\alpha b}(y) \mathbf{b}_{\beta a}(t)   \varphi(y-t+\Delta, \mathfrak{q}_{ba})   \,.
\end{split}
\end{equation}

\paragraph{Evaluating the remaining terms.}

We start computing
\begin{equation}
\label{eq:app_combination_rho_varphi_bb}
\begin{split}
&-\check{\rho}_{\alpha \beta, \beta \alpha}(y-t)
\mathbf{b}_{\beta a}(y) \mathbf{b}_{\alpha b}(t) +i \mathbf{b}_{\alpha b}(y) \mathbf{b}_{\beta a}(t)   \varphi(y-t+\Delta, \mathfrak{q}_{ba})=\\   &+(1-\delta_{\alpha \beta}) i \mathbf{b}_{\alpha b} \mathbf{b}_{\beta a}   \varphi(\imp^{\alpha \beta}+\Delta, \mathfrak{q}_{ba}) \delta(y-\imp^\alpha) \delta(t-\imp^\beta) \\
&+\delta_{\alpha \beta} \Bigl(\frac{i}{2} \pv \, \cot \frac{y-t}{2} 
+i \pv \, \varphi(y-t+\Delta, \mathfrak{q}_{ba})    \Bigr) \mathbf{b}_{\alpha a} \mathbf{b}_{\alpha b}  \delta_\epsilon(y-\imp^\alpha)\delta_\epsilon(t-\imp^\alpha) \,.
\end{split}
\end{equation}
Notice that in the last line of~\eqref{eq:app_combination_rho_varphi_bb} we cannot take the limit $\epsilon \to 0$ immediately since on the non-vanishing support of the Dirac delta functions the terms in parentheses are singular.

For collinear values, we have
\begin{equation}
\begin{split}
&i \pv \, \varphi(y-t+\Delta, \qq_{ba}) = \pv \frac{i}{y-t} + i \varphi^{\text{reg}}(\qq_{ba})+O(y-t)\,,\\
&\frac{i}{2}  \, \pv \cot \frac{y-t}{2} = \pv \frac{i}{y-t} +O(y-t)\,.
\end{split}
\end{equation}
Using the following distribution identity
\begin{equation}
\lim_{\epsilon \to 0} \pv \frac{1}{y-t}  \delta_\epsilon(y-\imp^\alpha) \delta_\epsilon(t-\imp^\alpha) = -\frac{1}{2} \left(\delta'(y-\imp^\alpha) \delta(t-\imp^\alpha) - \delta(y-\imp^\alpha) \delta'(t-\imp^\alpha) \right)\,,
\end{equation}
we obtain
\begin{equation}
\begin{split}
&-\check{\rho}_{\alpha \beta, \beta \alpha}(y-t)
\mathbf{b}_{\beta a}(y) \mathbf{b}_{\alpha b}(t) +i \mathbf{b}_{\alpha b}(y) \mathbf{b}_{\beta a}(t)   \varphi(y-t+\Delta, \mathfrak{q}_{ba})=\\ 
&+(1-\delta_{\alpha \beta}) i \mathbf{b}_{\alpha b} \mathbf{b}_{\beta a}   \varphi(\imp^{\alpha \beta}+\Delta, \mathfrak{q}_{ba}) \delta(y-\imp^\alpha) \delta(t-\imp^\beta) \\
&+i \delta_{\alpha \beta}  
  \varphi^{\text{reg}}( \mathfrak{q}_{ba})   \mathbf{b}_{\alpha a} \mathbf{b}_{\alpha b}  \delta(y-\imp^\alpha)\delta(t-\imp^\alpha)\\
&-i\delta_{\alpha \beta} \mathbf{b}_{\alpha a} \mathbf{b}_{\alpha b} \left(\delta'(y-\imp^\alpha) \delta(t-\imp^\alpha) - \delta(y-\imp^\alpha) \delta'(t-\imp^\alpha) \right) \,.
\end{split}
\end{equation}
Replacing this result in~\eqref{eq:bb_bracket_with_loop_var_app3}, we have
\begin{equation}
\label{eq:bb_bracket_with_loop_var_app4}
\begin{split}
&\frac{1}{\var} \{\mathbf{b}_{\alpha a}(y),\mathbf{b}_{\beta b}(t)\} =+i (1+ \delta_{\alpha \beta})  
  \varphi^{\text{reg}}( \mathfrak{q}_{ba})   \mathbf{b}_{\alpha a} \mathbf{b}_{\beta b}  \delta(y-\imp^\alpha)\delta(t-\imp^\beta) \\
&+i  \, \delta_{\alpha 1} \mathbf{b}_{\beta b}   \mathbf{b}_{1 a} \delta'(y-\imp^\alpha)    \delta(t-\imp^\beta)+\delta_{\alpha 1} \mathbb{L}^{\text{reg},\alpha}_{ba} \mathbf{b}_{\beta b} \delta(y-\imp^\alpha)\delta(t-\imp^\beta)\\
&-i  \, \delta_{\beta 1} \mathbf{b}_{\alpha a}   \mathbf{b}_{1 b} \delta'(t-\imp^\beta)    \delta(y-\imp^\alpha)-\delta_{\beta 1} \mathbb{L}^{\text{reg},\beta}_{ab} \mathbf{b}_{\alpha a} \delta(t-\imp^\beta)\delta(y-\imp^\alpha)\\
&+(1-\delta_{\alpha \beta}) i \mathbf{b}_{\alpha b} \mathbf{b}_{\beta a}   \varphi(\imp^{\alpha \beta}+\Delta, \mathfrak{q}_{ba}) \delta(y-\imp^\alpha) \delta(t-\imp^\beta) \\
&-i\delta_{\alpha \beta} \mathbf{b}_{\alpha a} \mathbf{b}_{\alpha b} \left(\delta'(y-\imp^\alpha) \delta(t-\imp^\alpha) - \delta(y-\imp^\alpha) \delta'(t-\imp^\alpha) \right)
\end{split}
\end{equation}

If we use formula~\eqref{eq:any_b_Delta_repr}, we obtain
\begin{equation}
\label{eq:bb_bracket_with_loop_var_app5}
\begin{split}
&\frac{1}{\var} \{\mathbf{b}_{\alpha a},\mathbf{b}_{\beta b}\} \delta(y-\imp^\alpha)\delta(t-\imp^\beta) =+i (1+ \delta_{\alpha \beta})  
  \varphi^{\text{reg}}( \mathfrak{q}_{ba})   \mathbf{b}_{\alpha a} \mathbf{b}_{\beta b}  \delta(y-\imp^\alpha)\delta(t-\imp^\beta)\\
&+i  \, \delta_{\alpha 1} \mathbf{b}_{\beta b}   \mathbf{b}_{1 a} \delta'(y-\imp^\alpha)    \delta(t-\imp^\beta)+\delta_{\alpha 1} \mathbb{L}^{\text{reg},\alpha}_{ba} \mathbf{b}_{\beta b} \delta(y-\imp^\alpha)\delta(t-\imp^\beta)\\
&-i  \, \delta_{\beta 1} \mathbf{b}_{\alpha a}   \mathbf{b}_{1 b} \delta'(t-\imp^\beta)    \delta(y-\imp^\alpha)-\delta_{\beta 1} \mathbb{L}^{\text{reg},\beta}_{ab} \mathbf{b}_{\alpha a} \delta(t-\imp^\beta)\delta(y-\imp^\alpha)\\
&+(1-\delta_{\alpha \beta}) i \mathbf{b}_{\alpha b} \mathbf{b}_{\beta a}   \varphi(\imp^{\alpha \beta}+\Delta, \mathfrak{q}_{ba}) \delta(y-\imp^\alpha) \delta(t-\imp^\beta) \\
&-i\delta_{\alpha \beta} \mathbf{b}_{\alpha a} \mathbf{b}_{\alpha b} \left(\delta'(y-\imp^\alpha) \delta(t-\imp^\alpha) - \delta(y-\imp^\alpha) \delta'(t-\imp^\alpha) \right)\\
&+ \frac{1}{\var} \{\mathbf{b}_{\alpha a}, \imp^\beta \}\mathbf{b}_{\beta b} \delta(y-\imp^\alpha)\delta'(t-\imp^\beta)+ \frac{1}{\var} \{\imp^\alpha , \mathbf{b}_{\beta b} \} \mathbf{b}_{\alpha a}  \delta'(y-\imp^\alpha)\delta(t-\imp^\beta) \,.
\end{split}
\end{equation}
Terms proportional to $\{\imp^\alpha, \imp^\beta \} \, \delta_{\epsilon}'(y-\imp^\alpha) \delta_{\epsilon}'(t-\imp^\beta)$ vanish because of the second equation in~\eqref{eq:aalpha_Dbeta_com}.
From this expression, we see that to cancel all terms proportional to $\delta' \delta$ we need to require
\begin{equation}
\label{eq:app_D1b_Dalpha_Dbeta}
\begin{split}
\frac{1}{\var} \{\imp^\alpha , \mathbf{b}_{\beta b} \} &= +i \delta_{\alpha \beta} \mathbf{b}_{\beta b} \qquad (\alpha, \beta \ne 1) \,,\\
\frac{1}{\var} \{\imp^1 , \mathbf{b}_{\beta b} \} &= -i  \mathbf{b}_{\beta b} \hspace{14mm} ( \beta \ne 1)\,,\\
\frac{1}{\var} \{\imp^\alpha , \mathbf{b}_{1 b} \} &= 0\,.
\end{split}
\end{equation}

Then we get
\begin{equation}
\label{eq:bb_bracket_with_loop_var_app6}
\begin{split}
&\frac{1}{\var} \{\mathbf{b}_{\alpha a},\mathbf{b}_{\beta b}\} \delta(y-\imp^\alpha)\delta(t-\imp^\beta) =+i (1+ \delta_{\alpha \beta})  
  \varphi^{\text{reg}}( \mathfrak{q}_{ba})   \mathbf{b}_{\alpha a} \mathbf{b}_{\beta b}  \delta(y-\imp^\alpha)\delta(t-\imp^\beta) \\
&+\delta_{\alpha 1} \mathbb{L}^{\text{reg},\alpha}_{ba} \mathbf{b}_{\beta b} \delta(y-\imp^\alpha)\delta(t-\imp^\beta)-\delta_{\beta 1} \mathbb{L}^{\text{reg},\beta}_{ab} \mathbf{b}_{\alpha a} \delta(t-\imp^\beta)\delta(y-\imp^\alpha)\\
&+(1-\delta_{\alpha \beta}) i \mathbf{b}_{\alpha b} \mathbf{b}_{\beta a}   \varphi(\imp^{\alpha \beta}+\Delta, \mathfrak{q}_{ba}) \delta(y-\imp^\alpha) \delta(t-\imp^\beta) \,.
\end{split}
\end{equation}
We can now simplify the Dirac delta functions and get the following final dimensional realisation of the bracket
\begin{equation}
\label{eq:bb_bracket_with_loop_var_app7}
\begin{split}
\frac{1}{\var} \{\mathbf{b}_{\alpha a},\mathbf{b}_{\beta b}\}  =&+i (1+\delta_{\alpha \beta})  
  \varphi^{\text{reg}}( \mathfrak{q}_{ba})   \mathbf{b}_{\alpha a} \mathbf{b}_{\beta b}  +(1-\delta_{\alpha \beta}) i \mathbf{b}_{\alpha b} \mathbf{b}_{\beta a}   \varphi(\imp^{\alpha \beta}+\Delta, \mathfrak{q}_{ba})\\
&+\delta_{\alpha 1} \mathbb{L}^{\text{reg},\alpha}_{ba} \mathbf{b}_{\beta b} -\delta_{\beta 1} \mathbb{L}^{\text{reg},\beta}_{ab} \mathbf{b}_{\alpha a}   \,.
\end{split}
\end{equation}
This relation corresponds to eq. (B3) in~\cite{Arutyunov:2026yjy}.

\subsection{Finite dimensional representation of the a-b bracket}

We start with the expression in~\eqref{eq:ab_bracket_with_loop_var_app} and assume $\alpha \ne 1$; with $\alpha=1$ indeed the bracket vanishes. Defining
\begin{equation}
\begin{split}
B_{\beta1}^{\text{ab}} \equiv &+\mathbf{a}_{a \gamma}(\imp^1) \check{\rho}(t-\imp^1)_{+\beta \gamma,\gamma 1} \mathbf{b}_{\gamma b}(t) \mathbf{a}_{a \alpha}(y) +\frac{2 \pi }{\var} \delta_{1 \beta} Q_b Q^{-1}_a  \mathbb{L}^{\epsilon}_{ab}(t-\Delta) \delta(\imp^1-t)\mathbf{a}_{a \alpha}(y) \\
&+ i \delta_{\beta 1}  \frac{d}{dx} \mathbf{a}_{a \alpha}(x,y)\Bigl|_{x=\imp^1} \mathbf{b}_{\beta b}(\imp^1 ,t) \,,
\end{split}
\end{equation}
we have
\begin{equation}
\label{eq:ab_bracket_with_loop_var_app2}
\begin{split}
&\frac{1}{\var} \, \{\mathbf{a}_{a \alpha}(y), \mathbf{b}_{\beta b}(t)\}=B_{\beta1}^{\text{ab}} -\mathbf{a}_{a \gamma}(y) \check{\rho}(t-y)_{+\beta \gamma,\gamma \alpha} \mathbf{b}_{\gamma b}(t)\\
&+i \varphi(\imp^1-y+\Delta,\mathfrak{q}_{ba})   \mathbf{a}_{b \alpha}(y) \mathbf{b}_{\beta b}(t) -i \varphi^{\text{reg}}(\qq_{ba}) \mathbf{a}_{a \alpha}(y) \mathbf{b}_{\beta b}(t)\\
&-  \mathbf{a}_{a 1} (y)\mathbf{a}_{b \alpha}(\imp^1) \check{\rho}_{1 \alpha, \alpha 1}(y-\imp^1) \mathbf{b}_{\beta b}(t) -\frac{2 \pi }{\var} \delta_{\alpha \beta} Q_b Q^{-1}_a  \mathbb{L}_{ab}(t-\Delta) \delta(y-t) \,.
\end{split}
\end{equation}
After replacing $y=\imp^\alpha \ne \imp^1$ and using
\begin{equation}
\mathbf{b}_{\beta b}(t) = \mathbf{b}_{\beta b} \delta_\epsilon(t-\imp^\beta)
\end{equation}
we obtain
\begin{equation}
    \mathbf{a}_{a 1} (y)\mathbf{a}_{b \alpha}(\imp^1) \check{\rho}_{1 \alpha, \alpha 1}(y-\imp^1) \mathbf{b}_{\beta b}(t)=0 \,,
\end{equation}
\begin{equation}
\begin{split}
    -\mathbf{a}_{a \gamma}(y) \check{\rho}(t-y)_{+\beta \gamma,\gamma \alpha} \mathbf{b}_{\gamma b}(t) &= \delta_{\alpha \beta} \left(-\pi \delta(t-\imp^\alpha)+ \frac{i}{2}    \text{Pv }\cot \left(\frac{t-\imp^\alpha}{2} \right) \,   \right) \mathbf{a}_{a \alpha}\mathbf{b}_{\beta b} \delta_\epsilon(t-\imp^\beta)\\
    &= \delta_{\alpha \beta} \frac{i}{t-\imp^\alpha-i0} \mathbf{a}_{a \alpha}\mathbf{b}_{\beta b} \delta_\epsilon(t-\imp^\beta) \,.
\end{split}
\end{equation}
From these relations, it follows that
\begin{equation}
\label{eq:ab_bracket_with_loop_var_app3}
\begin{split}
&\frac{1}{\var} \, \{\mathbf{a}_{a \alpha}, \mathbf{b}_{\beta b}\} \delta(t-\imp^\beta)=B_{\beta1}^{\text{ab}} + \delta_{\alpha \beta} \frac{i}{t-\imp^\alpha-i0} \mathbf{a}_{a \alpha}\mathbf{b}_{\beta b} \delta_\epsilon(t-\imp^\beta)\\
&+i \varphi(\imp^{1\alpha}+\Delta,\mathfrak{q}_{ba})   \mathbf{a}_{b \alpha} \mathbf{b}_{\beta b} \delta(t-\imp^\beta) -i \varphi^{\text{reg}}(\qq_{ba}) \mathbf{a}_{a \alpha} \mathbf{b}_{\beta b} \delta(t-\imp^\beta)\\
&-\frac{2 \pi }{\var} \delta_{\alpha \beta} Q_b Q^{-1}_a  \mathbb{L}_{ab}(t-\Delta) \delta(\imp^\alpha-t)\\
&+ \frac{1}{\var} \, \{\mathbf{a}_{a \alpha}, \imp^\beta \}\mathbf{b}_{\beta b} \delta'(t-\imp^\beta)+ \frac{1}{\var} \, \{\imp^\alpha, \mathbf{b}_{\beta b}  \} \mathbf{a}'_{a \alpha}  \delta(t-\imp^\beta)
\end{split}
\end{equation}
The terms in the last line come from expanding $\mathbf{a}$ around the dynamical value $\imp^\alpha$ (as it was done in Appendix~\ref{app:aa_com_finite_relations}) and by using~\eqref{eq:any_b_Delta_repr}.

\paragraph{Evaluating $B^{\text{ab}}_{\beta 1}$.}
Let us evaluate
\begin{equation}
\label{eq:app_Bbeta1_ab_bracket}
\begin{split}
&B_{\beta1}^{\text{ab}} = - \frac{i}{2} \delta_{\beta 1} \cot \left( \frac{t-\imp^1-i0}{2}\right) \mathbf{b}_{1b} \mathbf{a}_{a \alpha} \delta_\epsilon (t-\imp^1) \\
&+\frac{2 \pi }{\var} \delta_{\beta 1}  Q^{-1}_a  \mathbb{L}^{\epsilon}_{ab}(t-\Delta+ i0)Q_b \delta(\imp^1-t)\mathbf{a}_{a \alpha} + i \delta_{\beta 1}  \frac{d}{dx} \mathbf{a}_{a \alpha}(x,\imp^\alpha)\Bigl|_{x=\imp^1} \mathbf{b}_{\beta b} \delta(t-\imp^\beta) \,.
\end{split}
\end{equation}
Using the results of the previous section, we get
\begin{equation}
\begin{split}
&- \frac{i}{2} \delta_{\beta 1} \cot \left( \frac{t-\imp^1-i0}{2}\right) \mathbf{b}_{1b} \mathbf{a}_{a \alpha} \delta_\epsilon (t-\imp^1) = - \delta_{\beta 1} \frac{i}{t - \imp^1 - i0}  \mathbf{b}_{1b} \mathbf{a}_{a \alpha} \delta_\epsilon (t-\imp^1)\\
&= +\delta_{\beta 1} \frac{1}{\epsilon} \delta(t-\imp^1)\mathbf{b}_{1b} \mathbf{a}_{a \alpha} + i \delta_{\beta 1} \delta'(t-\imp^1)\mathbf{b}_{1b} \mathbf{a}_{a \alpha}
\end{split}
\end{equation}
and
\begin{equation}
\begin{split}
&\frac{2 \pi}{\var} \delta_{\beta 1} Q^{-1}_a \mathbb{L}^\epsilon_{ab}(t-\Delta+ i0) Q_b \delta(t-\imp^1)\mathbf{a}_{a \alpha} =\delta_{\beta 1} \mathbb{L}^{\text{reg},1}_{ab} \delta(t-\imp^1)\mathbf{a}_{a \alpha} \\
&- \delta_{\beta 1} \frac{1}{\epsilon}    \mathbf{b}_{1 b}  \delta(t-\imp^1)\mathbf{a}_{a \alpha} +i \delta_{\beta 1} \mathbf{a}'_{a 1} \mathbf{b}_{1 b} \mathbf{a}_{a \alpha}  \delta(t-\imp^1)\,.
\end{split}
\end{equation}
Plugging these expressions into~\eqref{eq:app_Bbeta1_ab_bracket}, all singularities cancel and we obtain
\begin{equation}
\label{eq:app_Bbeta1_ab_bracket2}
\begin{split}
&B_{\beta1}^{\text{ab}} = + i \delta_{\beta 1} \delta'(t-\imp^1)\mathbf{b}_{1b} \mathbf{a}_{a \alpha} +i \delta_{\beta 1} \mathbf{a}'_{a 1} \mathbf{b}_{1 b} \mathbf{a}_{a \alpha}  \delta(t-\imp^1)+\delta_{\beta 1} \mathbb{L}^{\text{reg},1}_{ab} \delta(t-\imp^1)\mathbf{a}_{a \alpha}\\
&+ i \delta_{\beta 1}  \frac{d}{dx} \mathbf{a}_{a \alpha}(x,\imp^\alpha)\Bigl|_{x=\imp^1} \mathbf{b}_{\beta b} \delta(t-\imp^\beta) \,.
\end{split}
\end{equation}
We also notice that
\begin{equation}
\begin{split}
\frac{d}{dx} \mathbf{a}_{a \alpha}(x,\imp^\alpha) &= \frac{d}{dx} t^{-1}_a(x)\bigr|_{x=\imp^1} \mathbb{A}_{a \alpha}(\imp^\alpha) =  \frac{d}{dx} t^{-1}_a(x)\bigr|_{x=\imp^1} t_a(\imp^1) \mathbf{a}_{a \alpha}\\
&=-\frac{d}{dx} t_a(x)\bigr|_{x=\imp^1} t^{-1}_a(\imp^1) \mathbf{a}_{a \alpha}=- \mathbf{a}'_{a1} \mathbf{a}_{a \alpha} \,.
\end{split}
\end{equation}
Using this fact, we obtain
\begin{equation}
\label{eq:app_Bbeta1_ab_bracket3}
\begin{split}
&B_{\beta1}^{\text{ab}} = + i \delta_{\beta 1} \delta'(t-\imp^1)\mathbf{b}_{1b} \mathbf{a}_{a \alpha} +\delta_{\beta 1} \mathbb{L}^{\text{reg},1}_{ab} \delta(t-\imp^1)\mathbf{a}_{a \alpha}\,.
\end{split}
\end{equation}
Plugging this expression in~\eqref{eq:ab_bracket_with_loop_var_app3} we get
\begin{equation}
\label{eq:ab_bracket_with_loop_var_app4}
\begin{split}
&\frac{1}{\var} \, \{\mathbf{a}_{a \alpha}, \mathbf{b}_{\beta b}\} \delta(t-\imp^\beta)= + i \delta_{\beta 1} \delta'(t-\imp^1)\mathbf{b}_{1b} \mathbf{a}_{a \alpha} +\delta_{\beta 1} \mathbb{L}^{\text{reg},1}_{ab} \delta(t-\imp^1)\mathbf{a}_{a \alpha} \\
&+ \delta_{\alpha \beta} \frac{i}{t-\imp^\alpha-i0} \mathbf{a}_{a \alpha}\mathbf{b}_{\beta b} \delta_\epsilon(t-\imp^\beta)\\
&+i \varphi(\imp^{1\alpha}+\Delta,\mathfrak{q}_{ba})   \mathbf{a}_{b \alpha} \mathbf{b}_{\beta b} \delta(t-\imp^\beta) -i \varphi^{\text{reg}}(\qq_{ba}) \mathbf{a}_{a \alpha} \mathbf{b}_{\beta b} \delta(t-\imp^\beta)\\
&-\frac{2 \pi }{\var} \delta_{\alpha \beta} Q_b Q^{-1}_a  \mathbb{L}^\epsilon_{ab}(t-\Delta) \delta(\imp^\alpha-t)\\
&+ \frac{1}{\var} \, \{\mathbf{a}_{a \alpha}, \imp^\beta \}\mathbf{b}_{\beta b} \delta'(t-\imp^\beta)+ \frac{1}{\var} \, \{\imp^\alpha, \mathbf{b}_{\beta b}  \} \mathbf{a}'_{a \alpha}  \delta(t-\imp^\beta)
\end{split}
\end{equation}

\paragraph{Evaluating the remaining terms.}

Let us evaluate the remaining singular terms. Using again the results of the previous section we have
\begin{equation}
\begin{split}
&+ \delta_{\alpha \beta} \frac{i}{t-\imp^\alpha-i0} \mathbf{a}_{a \alpha}\mathbf{b}_{\beta b} \delta_\epsilon(t-\imp^\beta)=\delta_{\alpha \beta} \mathbf{a}_{a \alpha}\mathbf{b}_{\beta b} \left(-\frac{1}{\epsilon} \delta(t-\imp^\beta) - i \delta'(t-\imp^\beta) \right)
\end{split}
\end{equation}
\begin{equation}
\begin{split}
-\frac{2 \pi }{\var} \delta_{\alpha \beta}  Q^{-1}_a  \mathbb{L}^\epsilon_{ab}(t-\Delta+i0)Q_b  \delta(t-\imp^\alpha)=&-\delta_{\alpha \beta} \mathbb{L}^{\text{reg},\alpha}_{ab} \delta(t-\imp^\alpha)+\delta_{\alpha \beta} \frac{1}{\epsilon}    \mathbf{a}_{a \alpha} \mathbf{b}_{\alpha b}  \delta(t-\imp^\alpha)\\
&-i \delta_{\alpha \beta} \mathbf{a}'_{a \alpha} \mathbf{b}_{\alpha b} \delta(t-\imp^\alpha) \,.
\end{split}
\end{equation}
Plugging these formulas in~\eqref{eq:ab_bracket_with_loop_var_app4} we obtain
\begin{equation}
\label{eq:ab_bracket_with_loop_var_app5}
\begin{split}
&\frac{1}{\var} \, \{\mathbf{a}_{a \alpha}, \mathbf{b}_{\beta b}\} \delta(t-\imp^\beta)= + i \delta_{\beta 1} \delta'(t-\imp^1)\mathbf{b}_{1b} \mathbf{a}_{a \alpha} +\delta_{\beta 1} \mathbb{L}^{\text{reg},1}_{ab} \delta(t-\imp^1)\mathbf{a}_{a \alpha} \\
&-i \delta_{\alpha \beta} \mathbf{a}_{a \alpha}\mathbf{b}_{\beta b} \delta'(t-\imp^\beta)\\
&+i \varphi(\imp^{1\alpha}+\Delta,\mathfrak{q}_{ba})   \mathbf{a}_{b \alpha} \mathbf{b}_{\beta b} \delta(t-\imp^\beta) -i \varphi^{\text{reg}}(\qq_{ba}) \mathbf{a}_{a \alpha} \mathbf{b}_{\beta b} \delta(t-\imp^\beta)\\
&-\delta_{\alpha \beta} \mathbb{L}^{\text{reg},\alpha}_{ab} \delta(t-\imp^\alpha)-i \delta_{\alpha \beta} \mathbf{a}'_{a \alpha} \mathbf{b}_{\alpha b} \delta(t-\imp^\alpha) \\
&+ \frac{1}{\var} \, \{\mathbf{a}_{a \alpha}, \imp^\beta \}\mathbf{b}_{\beta b} \delta'(t-\imp^\beta)+\frac{1}{\var} \, \{\imp^\alpha, \mathbf{b}_{\beta b}  \} \mathbf{a}'_{a \alpha}  \delta(t-\imp^\beta)\\
\end{split}
\end{equation}
Using the relations in~\eqref{eq:aalpha_Dbeta_com}, \eqref{eq:app_D1a_Dalpha_Dbeta},and~\eqref{eq:app_D1b_Dalpha_Dbeta} all terms containing $\delta'$ and $\mathbf{a}'$ cancel:
\begin{equation}
\label{eq:ab_bracket_with_loop_var_app6}
\begin{split}
&\frac{1}{\var} \, \{\mathbf{a}_{a \alpha}, \mathbf{b}_{\beta b}\} \delta(t-\imp^\beta)=\\
&+i \varphi(\imp^{1\alpha}+\Delta,\mathfrak{q}_{ba})   \mathbf{a}_{b \alpha} \mathbf{b}_{\beta b} \delta(t-\imp^\beta) -i \varphi^{\text{reg}}(\qq_{ba}) \mathbf{a}_{a \alpha} \mathbf{b}_{\beta b} \delta(t-\imp^\beta)\\
&-\delta_{\alpha \beta} \mathbb{L}^{\text{reg},\alpha}_{ab} \delta(t-\imp^\alpha)+\delta_{\beta 1} \mathbb{L}^{\text{reg},1}_{ab} \delta(t-\imp^1)\mathbf{a}_{a \alpha} \,.
\end{split}
\end{equation}
We can now simplify the Dirac delta functions everywhere and write
\begin{equation}
\label{eq:ab_bracket_with_loop_var_app7}
\begin{split}
\frac{1}{\var} \, \{\mathbf{a}_{a \alpha}, \mathbf{b}_{\beta b}\}=&+i \varphi(\imp^{1\alpha}+\Delta,\mathfrak{q}_{ba})   \mathbf{a}_{b \alpha} \mathbf{b}_{\beta b}  -i \varphi^{\text{reg}}(\qq_{ba}) \mathbf{a}_{a \alpha} \mathbf{b}_{\beta b} \\
&-\delta_{\alpha \beta} \mathbb{L}^{\text{reg},\alpha}_{ab} +\delta_{\beta 1} \mathbb{L}^{\text{reg},1}_{ab} \mathbf{a}_{a \alpha} \,.
\end{split}
\end{equation}
This formula corresponds to equation (B2) in~\cite{Arutyunov:2026yjy}.

\subsection{Comparison with the Coulomb branch result}
\label{app:AH_comparison}

We now compare the finite-dimensional brackets obtained above with the
Poisson brackets of Appendix~B of Ref.~\cite{Arutyunov:2026yjy}.
To make the identification precise, we distinguish the normalization
of the elliptic functions and of the Poisson bracket. In this subsection
we relabel the spin indices of Ref.~\cite{Arutyunov:2026yjy} by $\alpha=1,\ldots,\ell$, so
that their distinguished index $0$ is denoted by $1$.

The elliptic functions $E_1$ and $\Phi$ of Ref.~\cite{Arutyunov:2026yjy} are defined on the
curve with periods $1$ and $\tau$, whereas the periods used in this
paper are $2\pi$ and $2\Delta$. We therefore introduce
\begin{equation}
\label{eq:F4-AH-variables}
 x_a=\frac{q_a}{2\pi},
 \qquad
 d^\alpha=\frac{\imp^\alpha}{2\pi},
 \qquad
 \eta_{\mathrm{AH}}=\frac{\eta}{2\pi},
 \qquad
 \tau=\frac{\Delta}{\pi},
\end{equation}
and identify the spin variables as
\begin{equation}
\label{eq:F4-AH-spins}
 a_a^\alpha=\mathbf{a}_{a\alpha},
 \qquad
 c_a^\alpha=\mathbf{b}_{\alpha a}.
\end{equation}
We write $x_{ab}=x_a-x_b$ and $d^{\alpha\beta}=d^\alpha-d^\beta$.
Throughout this subsection, $E_1$ and $\Phi$ have precisely the
normalization of Ref.~\cite{Arutyunov:2026yjy}. Their relation to our functions is
\begin{equation}
\label{eq:F4-AH-functions}
 \Phi\!\left(\frac{u}{2\pi},\frac{v}{2\pi}\right)
 =2\pi\,\varphi(u+\Delta,v),
 \qquad
 E_1\!\left(\frac{v}{2\pi}\right)
 =2\pi\,\varphi^{\mathrm{reg}}(v).
\end{equation}
These relations follow from the definition of $\varphi$ in Appendix~\ref{app:elliptic_Green}
and the theta-function conventions in Section~S\,2 of Ref.~\cite{Arutyunov:2026yjy}.
In particular, both $\Phi(z,u)$ and $\varphi(z+\Delta,u)$ have residue
one at $z=0$ in their respective coordinates. We also adopt the
regular-part prescriptions used in that reference,
\begin{equation}
\label{eq:F4-AH-zero-arguments}
 E_1(0):=0,
 \qquad
 \Phi(0,u)=\Phi(u,0):=E_1(u).
\end{equation}

The corresponding normalization of the Poisson bracket is
\begin{equation}
\label{eq:F4-AH-Poisson-normalization}
 \{F,G\}_{\mathrm{AH}}
 :=\frac{2\pi i}{\var}\{F,G\},
\end{equation}
where the bracket on the right-hand side is the one used in the
preceding subsections. This rescaling absorbs the factors of $i$
in our spin brackets, together with the change of elliptic periods.

In these conventions, the Lax matrix in equation~(22) of Ref.~\cite{Arutyunov:2026yjy} reads
\begin{equation}
\label{eq:F4-AH-Lax}
 L^{\mathrm{AH}}_{ab}(z)
 =\sum_{\gamma=1}^{\ell}
 \Phi(z+d^\gamma,x_{ba}+\eta_{\mathrm{AH}})
 a_a^\gamma c_b^\gamma,
\end{equation}
where $z$ is now a dimensionless spectral parameter. Its regular
part at $z=-d^\alpha$ is defined by
\begin{equation}
\label{eq:F4-AH-Lax-expansion}
 L^{\mathrm{AH}}_{ab}(-d^\alpha+w)
 =\frac{a_a^\alpha c_b^\alpha}{w}
   +\mathcal{L}_{ab}^{\alpha}+O(w),
 \qquad w\to0.
\end{equation}
For distinct inhomogeneities, this gives
\begin{equation}
\label{eq:F4-AH-Lax-regular}
 \mathcal{L}_{ab}^{\alpha}
 =\sum_{\gamma\ne\alpha}
 \Phi(d^{\gamma\alpha},x_{ba}+\eta_{\mathrm{AH}})
 a_a^\gamma c_b^\gamma
 +E_1(x_{ba}+\eta_{\mathrm{AH}})a_a^\alpha c_b^\alpha.
\end{equation}
Using
\begin{equation}
\label{eq:F4-AH-reflection}
 \varphi(\Delta-u,-v)=-\varphi(\Delta+u,v),
 \qquad
 \varphi^{\mathrm{reg}}(-v)=-\varphi^{\mathrm{reg}}(v),
\end{equation}
and comparing with our definition~\eqref{eq:app_Lreg_ba_alpha}, we obtain
\begin{equation}
\label{eq:F4-AH-Lax-identification}
 \mathcal{L}_{ab}^{\alpha}
 =-2\pi i\,\mathbb{L}_{ab}^{\mathrm{reg},\alpha}.
\end{equation}
The sign in this relation comes from reversing both
$\imp^{\alpha\gamma}$ and $q_{ab}-\eta$; no transposition of the
matrix indices is involved.

With these identifications, equation~\eqref{eq:aa_bracket_with_loop_var_app5} becomes
\begin{equation}
\label{eq:F4-AH-aa}
\begin{aligned}
 \{a_a^\alpha,a_b^\beta\}_{\mathrm{AH}}
 ={}& E_1(x_{ab})a_a^\alpha a_b^\beta
      -\Phi(d^{\beta\alpha},x_{ba})a_b^\alpha a_a^\beta\\
    &+\Phi(d^{1\alpha},x_{ba})a_b^\alpha a_b^\beta
      -\Phi(d^{1\beta},x_{ab})a_a^\alpha a_a^\beta.
\end{aligned}
\end{equation}
Similarly, equations~\eqref{eq:ab_bracket_with_loop_var_app7} and~\eqref{eq:bb_bracket_with_loop_var_app7} give, respectively,
\begin{equation}
\label{eq:F4-AH-ab}
\begin{aligned}
 \{a_a^\alpha,c_b^\beta\}_{\mathrm{AH}}
 ={}&-E_1(x_{ab})a_a^\alpha c_b^\beta
      -\Phi(d^{1\alpha},x_{ba})a_b^\alpha c_b^\beta\\
    &+\delta_{\alpha\beta}\mathcal{L}_{ab}^{\alpha}
      -\delta_{\beta1}\mathcal{L}_{ab}^{1}a_a^\alpha,
\end{aligned}
\end{equation}
and
\begin{equation}
\label{eq:F4-AH-bb}
\begin{aligned}
 \{c_a^\alpha,c_b^\beta\}_{\mathrm{AH}}
 ={}& E_1(x_{ab})c_a^\alpha c_b^\beta
      +\Phi(d^{\beta\alpha},x_{ab})c_b^\alpha c_a^\beta\\
    &-\delta_{\alpha1}\mathcal{L}_{ba}^{1}c_b^\beta
      +\delta_{\beta1}\mathcal{L}_{ab}^{1}c_a^\alpha.
\end{aligned}
\end{equation}
In obtaining these expressions we used
$E_1(-u)=-E_1(u)$ and $\Phi(-u,-v)=-\Phi(u,v)$.
Equations~\eqref{eq:F4-AH-aa}--\eqref{eq:F4-AH-bb} are precisely
equations~(B1)--(B3) of Ref.~\cite{Arutyunov:2026yjy}, with
$L_{ab}^{\mathrm{AH}}(-d^\alpha)$ understood as the regular value
$\mathcal{L}_{ab}^{\alpha}$.

The agreement includes $\alpha=\beta$. Indeed, the prescription
$\Phi(0,u)=E_1(u)$ combines the two contributions proportional to
$E_1$ in the $a$--$a$ and $c$--$c$ brackets and reproduces the factors
$1+\delta_{\alpha\beta}$ in our formulas. It also includes coincident
particle indices through $E_1(0)=0$ and $\Phi(u,0)=E_1(u)$.
With the gauge condition $a_a^1=1$, equations
\eqref{eq:F4-AH-aa} and \eqref{eq:F4-AH-ab} consistently give zero
whenever an entry of this first spin column appears in the bracket.
These regular-part prescriptions concern coincident arguments within
the displayed formulas; the impurity positions remain distinct.

The same normalization also identifies the brackets involving the
dynamical inhomogeneities. Our relations~\eqref{eq:aalpha_Dbeta_com}, \eqref{eq:app_D1a_Dalpha_Dbeta}, and~\eqref{eq:app_D1b_Dalpha_Dbeta} imply
\begin{equation}
\label{eq:F4-AH-impurity-brackets}
\begin{aligned}
 \{d^\alpha,a_a^\beta\}_{\mathrm{AH}}
   &=(\delta_{\alpha\beta}-\delta_{\alpha1})a_a^\beta,\\
 \{d^\alpha,c_a^\beta\}_{\mathrm{AH}}
   &=-(\delta_{\alpha\beta}-\delta_{\alpha1})c_a^\beta,
 \qquad
 \{d^\alpha,d^\beta\}_{\mathrm{AH}}=0.
\end{aligned}
\end{equation}
For $\alpha\ne1$, these are equations~(B6)--(B7) of Ref.~\cite{Arutyunov:2026yjy}.
For the full comparison with the particle-position brackets, we take
$\{\imp^\alpha,q_a\}=0$ and $\{q_a,q_b\}=0$.
Using $Q_a=e^{i \q_a}$, equation~\eqref{eq:LQ_bracket}, and the definitions of the
invariant spins gives
\begin{equation}
\label{eq:F4-original-position-brackets}
 \frac{1}{\var}\{\q_a,\mathbf{a}_{b\alpha}\}=0,
 \qquad
 \frac{1}{\var}\{\q_a,\mathbf{b}_{\alpha b}\}
   =-i\delta_{ab}\mathbf{b}_{\alpha b}.
\end{equation}
Equivalently,
\begin{equation}
\label{eq:F4-AH-position-brackets}
\begin{gathered}
 \{x_a,a_b^\alpha\}_{\mathrm{AH}}=0,
 \qquad
 \{x_a,c_b^\alpha\}_{\mathrm{AH}}=\delta_{ab}c_b^\alpha,\\
 \{x_a,x_b\}_{\mathrm{AH}}=0,
 \qquad
 \{d^\alpha,x_a\}_{\mathrm{AH}}=0,
\end{gathered}
\end{equation}
which reproduces equations~(B4), (B5), (B8), and~(B9) in their paper.

Finally, Ref.~\cite{Arutyunov:2026yjy} imposes the constraint
\begin{equation}
\label{eq:F4-AH-impurity-constraint}
 d^1=-\sum_{\alpha=2}^{\ell}d^\alpha.
\end{equation}
Our brackets make $\mathcal{C}:=\sum_{\alpha=1}^{\ell}d^\alpha$
central, so this constraint selects the level $\mathcal{C}=0$.
The brackets with $d^1$ in
\eqref{eq:F4-AH-impurity-brackets} then follow from those with
$d^\alpha$, $\alpha\ne1$, exactly as in Ref.~\cite{Arutyunov:2026yjy}.
More generally, the central shift
$\widehat d^\alpha=d^\alpha-\mathcal{C}/\ell$
sets $\sum_\alpha\widehat d^\alpha=0$ without changing any impurity
difference or spin bracket. In the Lax matrix, this shift is accompanied
by $\widehat z=z+\mathcal{C}/\ell$, so that $z+d^\alpha$ is unchanged.
Thus the finite-dimensional algebra obtained here reproduces the algebra obtained from the Coulomb branch on the level $\mathcal{C}=0$.

\section{The homogeneous limit}
\label{app:hom_limit}

In this appendix we search for finite-dimensional representations of the brackets~\eqref{eq:aa_bracket_with_loop_var_app}, \eqref{eq:ab_bracket_with_loop_var_app} and~\eqref{eq:bb_bracket_with_loop_var_app} in the homogeneous limit 
\begin{equation}
\label{eq:app_a_representation_hom}
\mathbf{a}_{a \alpha} \equiv \mathbf{a}_{a \alpha}(\imp^1) \,, \qquad \mathbf{b}_{\alpha a}(x) \equiv \mathbf{b}_{\alpha a} \delta_\epsilon(x-\imp^1) \,.
\end{equation}
The matrix allowing for the cancellation of singularities in this case is $\hat{\rho}(x)$, as defined in Appendix~\ref{app:rho_matrices} (in contrast to the inhomogeneous case, where we had to select $\check{\rho}$). The nonzero entries of this matrix are
\begin{equation}
\label{eq:app_check_rho_matrices2}
\begin{split}
&\hat{\rho}_{\alpha \beta, \beta \alpha}(x)=  \frac{1}{2} \sgn_{\beta \alpha}  - \frac{i}{2}  \, \pv \, \cot \frac{x}{2}\,.
\end{split}
\end{equation}
Even if singularities cancel, we could not find a way to remove residual derivatives arising in the limit. We check this fact explicitly in the computation of the $\mathbf{a}$-$\mathbf{a}$ bracket.

\subsection{Homogeneous limit of the a-a bracket}
Let us consider~\eqref{eq:aa_bracket_with_loop_var_app} for $\alpha, \beta \ne 1$ (we recall that for $\alpha, \beta=1$ the $\mathbf{a}$-$\mathbf{a}$ bracket vanishes):
\begin{equation}
\begin{split}
&\frac{1}{\var} \, \{\mathbf{a}_{a \alpha}(y) \,,\mathbf{a}_{b \beta}(t) \}=- \mathbf{a}_{a \beta}(y)\mathbf{a}_{b \alpha}(t) \rho_{\beta \alpha, \alpha \beta}(y-t)\\
&  + i \varphi(t-y+\Delta,\mathfrak{q}_{ba})\mathbf{a}_{b \alpha}(y) \mathbf{a}_{a \beta}(t) - i \varphi(t-\imp^1+\Delta,\mathfrak{q}_{ba})\mathbf{a}_{a \alpha}(y) \mathbf{a}_{a \beta}(t)\\
&- i \varphi(\imp^1-y+\Delta,\mathfrak{q}_{ba})\mathbf{a}_{b \alpha}(y) \mathbf{a}_{b \beta}(t)+i \varphi^{\text{reg}}(\mathfrak{q}_{ba})\mathbf{a}_{a \alpha}(y) \mathbf{a}_{b \beta}(t)\\
&+\mathbf{a}_{a \beta} (\imp^1)\mathbf{a}_{b 1}(t)   \rho_{\beta 1, 1 \beta}(\imp^1-t) \mathbf{a}_{a \alpha}(y)- \mathbf{a}_{a 1} (y)\mathbf{a}_{b \alpha}(\imp^1)   \rho_{\alpha 1,1 \alpha}(\imp^1-y) \mathbf{a}_{b \beta}(t) \,.
\end{split}
\end{equation}
First we take the limit $y \to \imp^1$ and get
\begin{equation}
\begin{split}
&\frac{1}{\var} \, \{\mathbf{a}_{a \alpha}(y) \,,\mathbf{a}_{b \beta}(t) \}\bigl|_{y=\imp^1}=- \mathbf{a}_{a \beta} \mathbf{a}_{b \alpha}(t) \rho_{\beta \alpha, \alpha \beta}(\imp^1-t)\\
&  + i \varphi(t-\imp^1+\Delta,\mathfrak{q}_{ba})\mathbf{a}_{b \alpha} \mathbf{a}_{a \beta}(t) - i \varphi(t-\imp^1+\Delta,\mathfrak{q}_{ba})\mathbf{a}_{a \alpha} \mathbf{a}_{a \beta}(t)\\
&+i \varphi^{\text{reg}}(\mathfrak{q}_{ba})\mathbf{a}_{a \alpha} \mathbf{a}_{b \beta}(t)+\mathbf{a}_{a \beta} \mathbf{a}_{b 1}(t)   \rho_{\beta 1, 1 \beta}(\imp^1-t) \mathbf{a}_{a \alpha}\\
&- \lim_{y \to \imp^1} \left(\mathbf{a}_{a 1} (y)\mathbf{a}_{b \alpha}   \rho_{\alpha 1,1 \alpha}(\imp^1-y) + i \varphi(\imp^1-y+\Delta,\mathfrak{q}_{ba})\mathbf{a}_{b \alpha}(y) \right)\mathbf{a}_{b \beta}(t)
\,,
\end{split}
\end{equation}
In the last line we collected terms that develop singularities in the limit. After performing the limit we obtain
\begin{equation}
\begin{split}
&\scriptstyle{\lim_{y \to \imp^1} \left(\mathbf{a}_{a 1} (y)\mathbf{a}_{b \alpha}   \rho_{\alpha 1,1 \alpha}(\imp^1-y) + i \varphi(\imp^1-y+\Delta,\mathfrak{q}_{ba})\mathbf{a}_{b \alpha}(y) \right)=}\\
&\scriptstyle{\left(\frac{i}{y - \imp^1} + \frac{1}{2} \sgn_{1 \alpha} \right) ( \mathbf{a}_{b \alpha} + (y-\imp^1) \mathbf{a}'_{a 1} \mathbf{a}_{b \alpha})+ i \varphi^{\text{reg}}(\mathfrak{q}_{ba})\mathbf{a}_{b \alpha} + \frac{i}{\imp^1-y}  \mathbf{a}_{b \alpha} -i  \mathbf{a}'_{b \alpha}=}\\
&\scriptstyle{\frac{1}{2} \sgn_{1 \alpha}  \mathbf{a}_{b \alpha}+ i \varphi^{\text{reg}}(\mathfrak{q}_{ba})\mathbf{a}_{b \alpha} + i \mathbf{a}'_{a 1} \mathbf{a}_{b \alpha} - i \mathbf{a}'_{b \alpha} } \,,
\end{split}
\end{equation}\
where we defined
\begin{equation}
\mathbf{a}'_{b \alpha} \equiv \frac{d}{dx} \mathbf{a}_{b \alpha}(\imp^1,x)\bigl|_{x=\imp^1} \,.
\end{equation}
Substituting into the expression above, we obtain
\begin{equation}
\begin{split}
&\frac{1}{\var} \, \{\mathbf{a}_{a \alpha}(y) \,,\mathbf{a}_{b \beta}(t) \}\bigl|_{y=\imp^1}=- \mathbf{a}_{a \beta} \mathbf{a}_{b \alpha}(t) \rho_{\beta \alpha, \alpha \beta}(\imp^1-t)\\
&  + i \varphi(t-\imp^1+\Delta,\mathfrak{q}_{ba})\mathbf{a}_{b \alpha} \mathbf{a}_{a \beta}(t) - i \varphi(t-\imp^1+\Delta,\mathfrak{q}_{ba})\mathbf{a}_{a \alpha} \mathbf{a}_{a \beta}(t)\\
&+i \varphi^{\text{reg}}(\mathfrak{q}_{ba})\mathbf{a}_{a \alpha} \mathbf{a}_{b \beta}(t)+\mathbf{a}_{a \beta} \mathbf{a}_{b 1}(t)   \rho_{\beta 1, 1 \beta}(\imp^1-t) \mathbf{a}_{a \alpha}\\
&-\frac{1}{2} \sgn_{1 \alpha}  \mathbf{a}_{b \alpha}\mathbf{a}_{b \beta}(t)- i \varphi^{\text{reg}}(\mathfrak{q}_{ba})\mathbf{a}_{b \alpha}\mathbf{a}_{b \beta}(t) - i \mathbf{a}'_{a 1} \mathbf{a}_{b \alpha} \mathbf{a}_{b \beta}(t) + i \mathbf{a}'_{b \alpha}  \mathbf{a}_{b \beta}(t)
\,,
\end{split}
\end{equation}
Now we take the limit $t \to \imp^1$ and get
\begin{equation}
\label{eq:app_a_a_full_coll}
\begin{split}
&\frac{1}{\var} \, \{\mathbf{a}_{a \alpha}(y) \,,\mathbf{a}_{b \beta}(t) \}\bigl|_{y=t=\imp^1}=\\
&+i \varphi^{\text{reg}}(\mathfrak{q}_{ba}) \left(\mathbf{a}_{a \alpha} \mathbf{a}_{b \beta}- \mathbf{a}_{b \alpha}\mathbf{a}_{b \beta} -  \mathbf{a}_{a \beta} \mathbf{a}_{a \alpha}+\mathbf{a}_{b \alpha} \mathbf{a}_{a \beta} \right)\\
&- \frac{1}{2} \sgn_{\alpha \beta} \mathbf{a}_{a \beta} \mathbf{a}_{b \alpha}+\frac{1}{2} \mathbf{a}_{a \beta} \mathbf{a}_{a \alpha} \sgn_{1 \beta}-\frac{1}{2} \sgn_{1 \alpha}  \mathbf{a}_{b \alpha}\mathbf{a}_{b \beta}\\
&- i \mathbf{a}'_{a 1} \mathbf{a}_{b \alpha} \mathbf{a}_{b \beta} + i \mathbf{a}'_{b \alpha}  \mathbf{a}_{b \beta}+i \mathbf{a}_{a \beta} \mathbf{a}_{a \alpha} \mathbf{a}'_{b 1} - i \mathbf{a}'_{a \beta} \mathbf{a}_{a \alpha}+i \mathbf{a}_{b \alpha} \mathbf{a}'_{a \beta}  - i \mathbf{a}_{a \beta} \mathbf{a}'_{b \alpha} \,.
\end{split}
\end{equation}
Using that $\imp^1$ is dynamical
\begin{equation}
\begin{split}
\frac{1}{\var} \, \{\mathbf{a}_{a \alpha}(y) \,,\mathbf{a}_{b \beta}(t) \}\bigl|_{y=t=\imp^1}&=\frac{1}{\var} \, \{\mathbf{a}_{a \alpha} \,,\mathbf{a}_{b \beta} \} - \frac{1}{\var} \, \{\imp^1 \,,\mathbf{a}_{b \beta} \} \mathbf{a}'_{a \alpha} + \frac{1}{\var} \, \{\imp^1 \,,\mathbf{a}_{a \alpha} \} \mathbf{a}'_{b \beta}\\
&=\frac{1}{\var} \, \{\mathbf{a}_{a \alpha} \,,\mathbf{a}_{b \beta} \} - i \mathbf{a}_{b \beta}  \mathbf{a}'_{a \alpha} + i \, \mathbf{a}_{a \alpha} \mathbf{a}'_{b \beta} \,,
\end{split}
\end{equation}
we end up with
\begin{equation}
\begin{split}
&\frac{1}{\var} \, \{\mathbf{a}_{a \alpha} \,,\mathbf{a}_{b \beta} \}=i \varphi^{\text{reg}}(\mathfrak{q}_{ba}) \left(\mathbf{a}_{a \alpha} -\mathbf{a}_{b \alpha} \right) \left( \mathbf{a}_{b \beta}- \mathbf{a}_{a \beta} \right) - \frac{1}{2} \sgn_{\alpha \beta} \mathbf{a}_{a \beta} \mathbf{a}_{b \alpha}-\frac{1}{2} \mathbf{a}_{a \beta} \mathbf{a}_{a \alpha} +\frac{1}{2}  \mathbf{a}_{b \alpha}\mathbf{a}_{b \beta}\\
&+i \mathbf{a}_{a \beta} \mathbf{a}_{a \alpha} \mathbf{a}'_{b 1} - i  \mathbf{a}_{b \alpha} \mathbf{a}_{b \beta} \mathbf{a}'_{a 1} +i \mathbf{a}_{b \alpha} \mathbf{a}'_{a \beta}  - i \mathbf{a}_{a \beta} \mathbf{a}'_{b \alpha}+ i \mathbf{a}_{b \beta} ( \mathbf{a}'_{a \alpha} + \mathbf{a}'_{b \alpha} )- i \mathbf{a}_{a \alpha}  (\mathbf{a}'_{b \beta}+\mathbf{a}'_{a \beta}) \,.
\end{split}
\end{equation}
In the final expression, there are residual derivatives that do not cancel. We could not find a $\rho$ matrix for the starting oscillators able to remove such terms.

\section{Further structures related to the 
Poisson bracket (\ref{eq:brackets_TT_final})}
\def\TB{\varPhi}
\def\PB{\varPsi}
\def\OB{\varOmega}
\def\OS{\varSigma}
\def\td{\mbox{\textdelta}}

In this appendix, we investigate additional structures associated with
the Poisson bracket (\ref{eq:brackets_TT_final}). Although these results are not needed for
the main developments of the paper, they may be useful for future
investigations.

\subsection{Constant elliptic \texorpdfstring{$r$}{r_mat}-matrix}
In the limit $x,y\to 0$, the singularities in the bracket (\ref{eq:brackets_TT_final})
cancel, leaving finite contributions involving derivatives of $T$.
The resulting bracket is
\bea
\la{TTat2}
\begin{aligned}
\frac{1}{\var}\{\T_1,\T_2\}&=\T_{1}\T_{2} r_{\mathfrak{q}}-\T_1\T_2\, \refl_{\pi}\\
  &\hspace{-1.5cm}+i\,  \T_1\T_2\bigg[ \sum_{ij} \big(F_{ij}\otimes \X F_{ji}- \X F_{ij}\otimes F_{ji}\big)
 - \sum_{ijk}\X_{ij}\big(F_{ij}\otimes F_{kj}- F_{kj}\otimes F_{ij} \big)     \bigg]\, ,
\end{aligned}
\eea
where we set $\T:=T(0)$.
The matrix $\refl_{\pi}$ has components
\bea
(\refl_{\pi })_{ij,kl}=\refl_{ij,kl}-\delta_{ij}\sum_{n=1}^N \refl_{in,kl}-\delta_{kl}\sum_{n=1}^N \refl_{ij,kn}+\delta_{ij}\delta_{kl} \sum_{n,m=1}^N\refl_{in,km}\, ,
\eea
where the constant matrix $\refl$ is defined by
\bea
\refl=\T_1^{-1}\T_2^{-1}r \T_1 \T_2\, . 
\eea
Like $r$, the matrix $\refl$ is skew-symmetric and satisfies the modified Yang-Baxter equation 
\bea
[\refl_{12},\refl_{13}]+[\refl_{12},\refl_{23}]+[\refl_{13},\refl_{23}]=\frac{1}{4}[C_{12},C_{23}]\, .
\eea
Equivalently, $\refl_{\pi }$ can be written in the Frobenius basis as
\bea
\la{reflpi}
\refl_{\pi }=\sum_{ijkl}(\refl_{\pi })_{ij,kl} E_{ij}\otimes E_{kl}=\sum_{ijkl}\refl_{ij,kl} F_{ij}\otimes F_{kl}\, .
\eea


Finally, the matrix $r_{\q}$ defined in  (\ref{eq:rq_def}) takes the explicit form
\begin{equation}
\label{cer}
r_{\q}=
-i\sum_{i  j} \varphi^{\text{reg}}(\mathfrak{q}_{ij}) F_{ij} \otimes F_{ji}
=- i \sum_{i\neq j} \bigg[\zeta(\q_{ij})-\frac{\zeta(\pi)}{\pi}\q_{ij} \bigg] F_{ij}\otimes F_{ji}\, .
\end{equation}
We call $r_\q$ the {\it constant elliptic $r$-matrix}, where
``constant'' refers to its independence of the spectral parameters;
it still depends on the dynamical variables $\q_i$.
Since $F_{ij}e=0$, the bracket (\ref{TTat2}) is compatible with
the Frobenius condition $\T e=e$.

We now derive an equation satisfied by $r_{\q}$.
To this end, we return to the two-parameter elliptic
$r$-matrix (\ref{eq:rqxy_def}):
\bea
\begin{aligned}
r_{\q}(x,y)&=\sum_{ij}i \varphi(y-x+\Delta,\q_{ji}) E_{ij}\otimes E_{ji}\\
&-\sum_{ij}i \varphi(-x+\Delta,\q_{ji}) E_{ij}\otimes E_{jj}+\sum_{ij} i \varphi(-y+\Delta,\q_{ji}) E_{jj}\otimes E_{ij}\\
&-\sum_{i\neq j} i\Big(\zeta(\q_{ij})-\frac{\zeta(\pi)}{\pi}\q_{ij} \Big)E_{ii}\otimes E_{jj}.
\end{aligned}                
\eea
This $r$-matrix solves the shifted classical Yang-Baxter equation 
\bea
\begin{aligned}
&\hspace{-0.3cm}[r_{\q 12}(x_1,x_2),r_{\q 13}(x_1,x_3)]+[r_{\q 12}(x_1,x_2),r_{\q 23}(x_2,x_3)]+[r_{\q 13}(x_1,x_3),r_{\q 23}(x_2,x_3)]\\
&-i (\pa_1+\pa_2)r_{\q 12}(x_1,x_2)+i (\pa_1+\pa_3)r_{\q 13}(x_1,x_3)-i (\pa_2+\pa_3)r_{\q 23}(x_2,x_3)=0\, .
\end{aligned}
\eea
 In the limit $y\to 0$, it expands as 
 \bea
r_{\q}(x,y)
 =-\frac{i}{y}\OB+r_{\q}(x)+y s_{\q}(x)\ldots\, , ~~~~\OB:=\sum_{ij}E_{jj}\otimes E_{ij}\, ,
 \eea
where 
\bea
r_{\q}(x)=-\sum_{ij}i \varphi(-x+\Delta,\q_{ji}) E_{ij}\otimes F_{ji}
-\sum_{i\neq j} i \Big(\zeta(\q_{ij})-\frac{\zeta(\pi)}{\pi}\q_{ij} \Big)E_{ii}\otimes F_{ji}
\eea
and
\bea
\begin{aligned}
s_{\q12}(x)&=\sum_{i\neq j}i \varphi(-x+\Delta,\q_{ji})\Big(\zeta(\q_{ji}-x)+\zeta(x)-\frac{\zeta(\pi)}{\pi}\q_{ji}\Big) E_{ij}\otimes E_{ji}\\
&+\frac{i}{2}\sum_{i\neq j}\bigg[\wp(\q_{ij})-\Big(\zeta(\q_{ij})-\frac{\zeta(\pi)}{\pi}\q_{ij}\Big)^2\bigg] E_{jj}\otimes E_{ij}
-i \wp(x)\sum_{i}E_{ii}\otimes E_{ii}\, .
\end{aligned}
\eea
Using skew-symmetry, we find an analogous expression for $x\to 0$
\bea
r_{\q}(x,y)=\frac{i}{x}\OB_{21}-r_{\q 21}(y)-x s_{\q 21}(y)\ldots
\eea
Expanding the shifted classical Yang--Baxter equation as
$x_2\to 0$, with $x_1$ and $x_3$ fixed, we find that the
double and simple poles in $x_2$ cancel by the identities

\bea
[\OB_{12},\OB_{32}]+\OB_{12}-\OB_{32}=0\, 
\eea
and 
\bea
[\OB_{12}+\OB_{32},r_{\q 13}(x,y)]-[\OB_{12},r_{\q 32}(y)]-[r_{\q 12}(x),\OB_{32}]=0\, ,
\eea
 respectively. The finite part then yields, upon setting $x:=x_1$ and $y:=x_3$,
 \bea
 \label{fpart_YB}
 \begin{aligned}
 &\hspace{-0.5cm}[r_{\q 12}(x)+r_{\q 32}(y),r_{\q13}(x,y)]-[r_{\q 12}(x),r_{\q 32}(y)]+i [\OB_{12},s_{\q 32}(y)]+i [s_{\q 12}(x),\OB_{32}]\\
 &-i \pa_{x}r_{\q 12}(x)+i \pa_{y} r_{\q 32}(y)+i (\pa_{x}+\pa_{y})r_{\q 13}(x,y) 
 -i s_{\q 12}(x)+i s_{q 32}(y)=0\, .
 \end{aligned}
 \eea
\smallskip
 
 We further note that
 in the limit $x\to 0$ the function $s_{\q12}(x)$ expands as
\bea
\begin{aligned}
s_{\q12}(x)&=-\frac{i}{x^2}C_{12}
+\frac{i}{2}\sum_{i\neq j}\bigg[\wp(\q_{ij})-\Big(\zeta(\q_{ij})-\frac{\zeta(\pi)}{\pi}\q_{ij}\Big)^2\bigg] F_{ji}\otimes E_{ij}+\ldots ,
\end{aligned}
\eea
while in the same limit 
the matrix $r_{\q}(x)$ behaves as 
\bea
\label{H17}
\begin{aligned}
r_{\q}(x)&=\frac{i}{x}\OS+r_{\q}+\frac{x}{2} \td r_{\q}+\ldots\, , ~~~~\OS=\sum_{ij} E_{ij}\otimes F_{ji}\\
\end{aligned}
\eea
where 
\bea
\label{H18}
\begin{aligned}
\td r_{\q}&=-i\sum_{i\neq j}\bigg[\wp(\q_{ij})-\Big(\zeta(\q_{ij})-\frac{\zeta(\pi)}{\pi}\q_{ij}\Big)^2 \bigg] E_{ij}\otimes F_{ji}\, .
\end{aligned}
\eea
Thus, 
\bea
\label{H19}
\begin{aligned}
s_{\q12}(x)&=-\frac{i}{x^2}C_{12}-\frac{1}{2}\td r_{q 21}+\ldots
\end{aligned}
\eea

We now take $y\to 0$ in (\ref{fpart_YB}), keeping $x$ fixed. The double pole cancels by the identity
\bea
[\OS_{32},\OB_{13}]+[\OB_{12},C_{23}]+\OS_{32}-\OB_{13}+C_{23}=0\, ,
\eea
while the single pole vanishes due to 
\bea
[\OS_{32},r_{\q 12}(x)+r_{\q 13}(x)]+[\OB_{13},r_{\q 12}(x)]=[r_{\q 32},\OB_{13}]\, ,
\eea 
which is equivalent to 
\bea
[\OS_{23},r_{\q 12}(x)+r_{\q 13}(x)]+[\OB_{12},r_{\q 13}(x)]=[r_{\q 23},\OB_{12}]\, .
\eea 
In both identities, the $x$-dependent terms on the left-hand side cancel. The remaining finite part yields
\bea
\label{pfinite_x}
\begin{aligned}
&[r_{\q 12}(x)+r_{\q 32},r_{\q 13}(x)]-[r_{\q 12}(x),r_{\q 32}]\\
&~~~ +i [\OS_{32},s_{\q 13}(x)]-i [\OB_{32},s_{\q 12}(x)]-\frac{i}{2}[\td r_{\q 32},\OB_{13}]-\frac{i}{2}[\OB_{12},\td r_{\q 23}]
\\
&~~~ -i \pa_x r_{\q 12}(x)+i \pa_x r_{\q 13}(x)+\frac{i}{2}\td r_{\q 32}-\frac{i}{2}\td r_{\q 23}-i s_{\q 12}(x)+i s_{\q 13}(x)=0\, .
\end{aligned}
\eea
Next, we take $x\to 0$ in (\ref{pfinite_x}),
using the expansions (\ref{H17}) and (\ref{H19}). The double pole cancels by the identity
\bea
-[\OS_{12},\OS_{13}]+[\OS_{32},C_{13}]-[\OB_{32},C_{12}]-\OS_{12}+\OS_{13}-C_{12}+C_{13}=0\, ,
\eea
while the single pole cancels due to the identity
\bea
[\OS_{12},r_{\q 13}+r_{\q 23}]+[r_{\q 12}-r_{\q 23},\OS_{13}]=0\, .
\eea
The remaining finite part gives the following equation for the constant elliptic $r$-matrix
\bea
\la{eq_for_rq}
\begin{aligned}
[r_{\q 12},r_{\q 13}]&+[r_{\q 12},r_{\q 23}]+[r_{\q 13},r_{\q 23}]=\\
&-\frac{i}{2}[\OS_{12}, \td r_{\q 13}]+\frac{i}{2}[\OS_{13}, \td r_{\q 12}]+\frac{i}{2}[\OS_{32}, \td r_{\q 31}]\\
&-\frac{i}{2}[\OB_{32}, \td r_{\q 21}]-\frac{i}{2}[\OB_{13}, \td r_{\q 32}]+\frac{i}{2}[\OB_{12}, \td r_{\q 23}]\\
&+\frac{i }{2}(\td r_{\q 12}-\td r_{\q 21})-\frac{i }{2}(\td r_{\q 13}-\td r_{\q 31})+\frac{i }{2}(\td r_{\q 23}-\td r_{\q 32})\, .
\end{aligned}
\eea
We introduce the following function
 \bea
 f_{ij}=\wp(\q_{ij})-\Big(\zeta(\q_{ij})-\frac{\zeta(\pi)}{\pi}\q_{ij}\Big)^2\, 
 \eea
and the tensor 
\bea
\TB_{123}^{A}=\sum_{\substack{i,j,k\\ \text{pairwise distinct}}} A_{ijk}(F_{ji}\otimes F_{kj}\otimes F_{ik}-F_{ij}\otimes F_{jk}\otimes F_{ki})
\eea
with coefficients 
\bea
A_{ijk}=\frac{1}{2}(f_{ij}+f_{jk}+f_{ki})\, .
\eea
Substituting (\ref{H18}) into (\ref{eq_for_rq}),
expanding the commutators, and cyclicly relabelling the summation indices, we obtain
\bea
[r_{\q 12},r_{\q 13}]+[r_{\q 12},r_{\q 23}]+[r_{\q 13},r_{\q 23}]=\TB_{123}^{A}\, .
\eea
This is the Yang-Baxter equation for $r_{\q}$.

We now take the trigonometric degeneration
$\operatorname{Im}\Delta\to+\infty$, keeping the real period
$2\pi$ and the coordinates $\q_i$ fixed. In this limit,
\[
\zeta(z)-\frac{\zeta(\pi)}{\pi}z
\rightarrow \frac12\cot\frac z2,
\qquad
f_{ij}\rightarrow\frac16,
\]
and hence $A_{ijk}\rightarrow\frac14$.
The matrix $r_{\q}$ turns into
\bea
r_{\q}\to r_{\q}^{\rm trig}=-\frac{i}{2}  \sum_{i\neq j}\cot\frac{ \q_{ij}}{2}F_{ij}\otimes F_{ji}\, .
\eea
In the trigonometric limit the Yang-Baxter equation boils down to 
\bea
\begin{aligned}
&[r^{\rm trig}_{\q 12},r^{\rm trig}_{\q 13}]+[r^{\rm trig}_{\q 12},r^{\rm trig}_{\q 23}]+[r^{\rm trig}_{\q 13},r^{\rm trig}_{\q 23}]\\
&\hspace{2cm}=\frac{1}{4}
\sum_{\substack{i,j,k\\ \text{pairwise distinct}}} (F_{ji}\otimes F_{kj}\otimes F_{ik}-F_{ij}\otimes F_{jk}\otimes F_{ki})\, .
\end{aligned}
\eea

\subsection{Reduction to the hyperbolic case}
\def\r{{\mathtt r}}
\def\YB{\mathsf{YB}}
The finite-dimensional Heisenberg double associated with the hyperbolic model \cite{Arutyunov:2019alw} is obtained by restricting the relations of the affine double 
to constant loops, taking the limit $\Im\Delta\to \infty$, and performing the analytic continuation $\q_i\to -\i q_i$.
Under this restriction, $T(x)$ is independent of $x$ and can be identified with $\T=T(0)$, while $\X=\pa T(0)$ vanishes. 
The Poisson bracket (\ref{TTat2}) therefore reduces to (for simplicity, we set $\var=1$)
\bea
\la{br_trig}
\{\T_1,\T_2\}=\T_1\T_2 (\r_{q}-\refl_{\pi})\, ,
\eea
where $\refl_{\pi}$ is still given by (\ref{reflpi}), while 
$\r_{q}$ is obtained from the constant elliptic $r$-matrix by the degeneration and analytic continuation 
\bea
\la{trig_deg}
\r_{q}=-i \lim_{\Im\Delta\to \infty }\sum_{i\neq j} \bigg[\zeta(\q_{ij})-\frac{\zeta(\pi)}{\pi}\q_{ij} \bigg] _{\q_i\to -\i q_i}\hspace{-10mm}F_{ij}\otimes F_{ji}=
\frac{1}{2}  \sum_{i\neq j}\coth\frac{q_{ij}}{2} F_{ij}\otimes F_{ji}
\, .
\eea

We now verify the Jacobi identity for (\ref{br_trig}) explicitly, 
thereby identifying the mechanism that ensures its validity. Since $\refl_{\pi}$ depends on the entries of $\T$, the Jacobi identity takes the form
\bea
\la{YBconst}
\YB(\r_{q}-\refl_{\pi})-\T_1^{-1}\{\T_1,\refl_{\pi 23}\} + \T_2^{-1}\{\T_2,\refl_{\pi 13}\} -\T_3^{-1}\{\T_3,\refl_{\pi 12}\}=0.
\eea
Here, for any $r$-matrix, we define the Yang-Baxter 3-tensor by
\bea
\YB(r)=[r_{12},r_{13}]+[r_{12},r_{23}]+[r_{13},r_{23}]\, .
\eea
Evaluating the Poisson brackets involving $\refl_{\pi}$, we obtain 
\bea
\la{YBexpanded}
\begin{aligned}
\YB(\r_{q})+\YB(\refl_{\pi})+[\refl_{\pi 23},\r_{q 12}+\r_{q 13}]-[\refl_{\pi 12},\r_{q 13}+\r_{q 23}]+[\refl_{\pi 13}, \r_{q 12}-\r_{q 23}]&\\
-\pi_{23}([\refl_{23},\r_{q 12}+\r_{q 13}]-[\refl_{23},\refl_{\pi 12}+\refl_{\pi 13}])&\\
-\pi_{13}([\refl_{13},\r_{q 12}-\r_{q 23}]-[\refl_{13},\refl_{\pi 12}-\refl_{\pi 23}])&\\
+\pi_{12}([\refl_{12},\r_{q13}+\r_{q 23}]-[\refl_{12},\refl_{\pi 13}+\refl_{\pi 23}])&=0\, .\\
\end{aligned}
\eea 
The first line is the expansion of $\YB(\r_{q}-\refl_{\pi})$. Using the definition of the projector on $\mathcal{F}\otimes {\mathcal F}$, where $\mathcal F$ denotes the Frobenius Lie algebra spanned by $F_{ij}$,
 we find 
\bea
\la{batch1}
\begin{aligned}
\pi_{23}\big([\refl_{23},\r_{q 12}]\big)&=[\refl_{\pi 23},\r_{q 12}]
-(\refl_{ka,mn}-\refl_{la,mn})\r_{q ij,kl}F_{ij}\otimes F_{kl}\otimes F_{mn}\, , \\
\pi_{23}\big([\refl_{23},\r_{q 13}]\big)&=[\refl_{\pi 23},\r_{q 13}]
-(\refl_{kl,ma}-\refl_{kl,na})\r_{q ij,mn}F_{ij}\otimes F_{kl}\otimes F_{mn}\, , \\
\pi_{13}\big([\refl_{13},\r_{q 12}]\big)&=[\refl_{\pi 13},\r_{q 12}]
-(\refl_{ia,mn}-\refl_{ja,mn})\r_{q ij,kl}F_{ij}\otimes F_{kl}\otimes F_{mn}\, , \\
\pi_{13}\big([\refl_{13},\r_{q 23}]\big)&=[\refl_{\pi 13},\r_{q 23}]
-(\refl_{ij,ma}-\refl_{ij,na})\r_{q kl,mn}F_{ij}\otimes F_{kl}\otimes F_{mn}\, , \\
\pi_{12}\big([\refl_{12},\r_{q 13}]\big)&=[\refl_{\pi 12},\r_{q 13}]
-(\refl_{ia,kl}-\refl_{ja,kl})\r_{q ij,mn}F_{ij}\otimes F_{kl}\otimes F_{mn}\, , \\
\pi_{12}\big([\refl_{12},\r_{q 23}]\big)&=[\refl_{\pi 12},\r_{q 23}]
-(\refl_{ij,ka}-\refl_{ij,la})\r_{q kl,mn}F_{ij}\otimes F_{kl}\otimes F_{mn}\, .
\end{aligned}
\eea
Here the summation over repeated indices and index $a$ is assumed. 
Substituting these expressions into (\ref{YBexpanded}) and 
using formula (\ref{trig_deg}), we observe that all terms containing the matrix $\r_{q}$ cancel, and we obtain
\bea
\begin{aligned}
\YB(\r_q)&+\YB(\refl_{\pi})\\
&\hspace{-0.5cm}+\pi_{23}([\refl_{23},\refl_{\pi 12}+\refl_{\pi 13}])
+\pi_{13}([\refl_{13},\refl_{\pi 12}-\refl_{\pi 23}])
-\pi_{12}([\refl_{12},\refl_{\pi 13}+\refl_{\pi 23}])=0\, .
\end{aligned}
\eea 
The following expressions, analogous to eqs.(\ref{batch1}), hold 
\bea
\la{batch2}
\begin{aligned}
\pi_{23}\big([\refl_{23},\refl_{\pi 12}]\big)&=[\refl_{\pi 23},\refl_{\pi 12}]
-(\refl_{ka,mn}-\refl_{la,mn})\refl_{ij,kl}F_{ij}\otimes F_{kl}\otimes F_{mn}\, , \\
\pi_{23}\big([\refl_{23},\refl_{\pi 13}]\big)&=[\refl_{\pi 23},\refl_{\pi 13}]
-(\refl_{kl,ma}-\refl_{kl,na})\refl_{ij,mn}F_{ij}\otimes F_{kl}\otimes F_{mn}\, , \\
\pi_{13}\big([\refl_{13},\refl_{\pi 12}]\big)&=[\refl_{\pi 13},\refl_{\pi 12}]
-(\refl_{ia,mn}-\refl_{ja,mn})\refl_{ ij,kl}F_{ij}\otimes F_{kl}\otimes F_{mn}\, , \\
\pi_{13}\big([\refl_{13},\refl_{\pi 23}]\big)&=[\refl_{\pi 13},\refl_{\pi 23}]
-(\refl_{ij,ma}-\refl_{ij,na})\refl_{kl,mn}F_{ij}\otimes F_{kl}\otimes F_{mn}\, , \\
\pi_{12}\big([\refl_{12},\refl_{\pi 13}]\big)&=[\refl_{\pi 12},\refl_{\pi 13}]
-(\refl_{ia,kl}-\refl_{ja,kl})\refl_{ij,mn}F_{ij}\otimes F_{kl}\otimes F_{mn}\, , \\
\pi_{12}\big([\refl_{12},\refl_{\pi 23}]\big)&=[\refl_{\pi 12},\refl_{\pi 23}]
-(\refl_{ij,ka}-\refl_{ij,la})\refl_{kl,mn}F_{ij}\otimes F_{kl}\otimes F_{mn}\, .
\end{aligned}
\eea
Using these formulae and the skew symmetry of $\refl$, we obtain
\bea
\begin{aligned}
\YB(\r_{q})&+\YB(\refl_{\pi})-2 \YB(\refl_{\pi})\\
&+(\refl_{ia,kl}-\refl_{ja,kl})\refl_{ij,mn}\big(F_{mn}\otimes F_{ij}\otimes F_{kl}-F_{mn}\otimes F_{kl}\otimes F_{ij}\\
&-F_{ij}\otimes F_{mn}\otimes F_{kl}+F_{kl}\otimes F_{mn}\otimes F_{ij}+F_{ij}\otimes F_{kl}\otimes F_{mn}
-F_{kl}\otimes F_{ij}\otimes F_{mn})=0\, ,
\end{aligned}
\eea 
so that 
\bea
\la{YBexpanded_2}
\begin{aligned}
\YB(\r_{q})& - \YB(\refl_{\pi})+(\refl_{ia,kl}-\refl_{ja,kl})\refl_{ij,mn}\ F_{ij}\wedge  F_{kl}\wedge F_{mn}=0\, .
\end{aligned}
\eea 

Now we evaluate $\YB(\refl_{\pi})$.
Using the Lie algebra relations of $\mathcal{F}$, we find 
\bea
\begin{aligned}
[\refl_{\pi 12},\refl_{\pi 13}]&=\refl_{ijkl}\refl_{mnsp}(\delta_{in}F_{mj}-\delta_{jm}F_{in})\otimes F_{kl}\otimes F_{sp}\\
&+(\refl_{in kl}-\refl_{jn kl})\refl_{ijsp} \big(F_{ij}\otimes F_{kl}\otimes F_{sp}-F_{ij}\otimes F_{sp}\otimes F_{kl} \big)\, ,\\
[\refl_{\pi 12},\refl_{\pi 23}]&=\refl_{ijkl}\refl_{mnsp}F_{ij}\otimes (\delta_{kn}F_{ml}-\delta_{ml}F_{kn})\otimes F_{sp}\\
&+(\refl_{ij kn}-\refl_{ij ln})\refl_{kl sp} F_{ij}\otimes F_{kl}\otimes F_{sp}
-(\refl_{kn ij}-\refl_{ln ij})\refl_{sp kl} F_{sp}\otimes F_{kl}\otimes F_{ij}\, .
\end{aligned}
\eea
Using the skew-symmetry of the $\refl$-matrix, we can rewrite the last formula as 
\bea
\begin{aligned}
[\refl_{\pi 12},\refl_{\pi 23}]&=\refl_{ijkl}\refl_{mnsp}F_{ij}\otimes (\delta_{kn}F_{ml}-\delta_{ml}F_{kn})\otimes F_{sp}\\
&+(\refl_{ij kn}-\refl_{ij ln})\refl_{kl sp}\big( F_{ij}\otimes F_{kl}\otimes F_{sp} -F_{sp}\otimes F_{kl}\otimes F_{ij}\big)\, .
\end{aligned}
\eea
Finally, we have 
\bea
\begin{aligned}
[\refl_{\pi 13},\refl_{\pi 23}]&=\refl_{ijkl}\refl_{mnsp}F_{ij}\otimes F_{mn}\otimes (\delta_{kp}F_{sl}-\delta_{ls}F_{kp})\\
&+(\refl_{ij kp}-\refl_{ij lp})\refl_{mn kl}\big( F_{ij}\otimes F_{mn}\otimes F_{kl} -F_{mn}\otimes F_{ij}\otimes F_{kl}\big)\, .
\end{aligned}
\eea
Summing up, we obtain 
\bea
\begin{aligned}
\YB(\refl_{\pi})&=-F_{ij}\otimes F_{kl}\otimes F_{mn}\\
&\times (\refl_{ia, kl}\refl_{aj, mn}-\refl_{ia, mn}\refl_{aj, kl}+\refl_{ij, ka}\refl_{al, mn}-\refl_{ka, mn}\refl_{ij, al}
+\refl_{ij, ma}\refl_{kl, an}-\refl_{kl, ma}\refl_{ij, an}\big)\\
&+(\refl_{ia, kl}-\refl_{ja, kl})\refl_{ij,mn} \big(F_{ij}\otimes F_{kl}\otimes F_{mn}-F_{ij}\otimes F_{mn}\otimes F_{kl} \big)\\
&+(\refl_{ij, ka}-\refl_{ij, la})\refl_{kl ,mn}\big( F_{ij}\otimes F_{kl}\otimes F_{mn} -F_{mn}\otimes F_{kl}\otimes F_{ij} \big)\\
&+(\refl_{ij, ka}-\refl_{ij, la})\refl_{mn, kl}\big( F_{ij}\otimes F_{mn}\otimes F_{kl} -F_{mn}\otimes F_{ij}\otimes F_{kl}\big)\, .
\end{aligned}
\eea
The Yang-Baxter equation for $\refl$ implies that 
\bea
\begin{aligned}
&\refl_{ia, kl}\refl_{aj, mn}-\refl_{ia, mn}\refl_{aj, kl}+\refl_{ij, ka}\refl_{al, mn}-\refl_{ka, mn}\refl_{ij, al}
+\refl_{ij, ma}\refl_{kl, an}-\refl_{kl, ma}\refl_{ij, an}\\
&=\frac{1}{4}[C_{12},C_{23}]_{ij,kl,mn}=\frac{1}{4}\sum_{abc}(E_{ab}\otimes E_{bc}\otimes E_{ca}-
E_{ba}\otimes E_{cb}\otimes E_{ac})_{ij,kl,mn}\\
&=\frac{1}{4}(\delta_{in}\delta_{jk}\delta_{ml}-\delta_{il}\delta_{mj}\delta_{kn})\, .
\end{aligned}
\eea
Thus, 
\bea
\la{YB!1}
\begin{aligned}
&\YB(\refl_{\pi})=\frac{1}{4}\TB_{123}+(\refl_{ia, kl}-\refl_{ja, kl})\refl_{ij,mn} F_{ij}\wedge F_{kl}\wedge F_{mn}\, .
\end{aligned}
\eea
where 
\bea
\TB_{123}=\sum_{\substack{i,j,k\\ \text{pairwise distinct}}}(F_{ji}\otimes F_{kj}\otimes F_{ik}-F_{ij}\otimes F_{jk}\otimes F_{ki})\, 
\eea
is a constant, totally antisymmetric tensor in $\mathcal{F}^{\otimes 3}$. Finally, the hyperbolic $r$-matrix (\ref{trig_deg}) satisfies 
\bea
\la{YB!2}
\YB(\r_{q})=\frac{1}{4} \TB_{123}\, .
\eea
Substituting (\ref{YB!1}) and (\ref{YB!2}) into (\ref{YBexpanded_2}), we verify the fulfilment of the Jacobi identity.

\bibliographystyle{JHEP}
\bibliography{refs}

\end{document}